\documentclass[]{aa}           

\usepackage{graphicx,natbib,amssymb,amsmath,txfonts,curves,graphics,epsfig}
\bibpunct{(}{)}{;}{a}{}{,}

\newcommand{\hlf}{\ensuremath{\frac{1}{2}}}
\newcommand{\thrhlf}{\ensuremath{\frac{3}{2}}}
\newcommand{\nux}{{\nu_x}}
\newcommand{\nue}{{\nu_\mathrm{e}}}
\newcommand{\nuae}{{\bar\nu_\mathrm{e}}}
\newcommand{\ye}{{Y_\mathrm{e}}}
\newcommand{\equ}{\equiv}
\newcommand{\dlin}[1]{\mathrm{d}#1\,}

\newcommand{\Real}{{\rm I\mathchoice{\kern-0.70mm}{\kern-0.70mm}{\kern-0.65mm}{\kern-0.50mm}R}}
\newcommand{\Natural}{{\rm I\mathchoice{\kern-0.55mm}{\kern-0.55mm}{\kern-0.50mm}{\kern-0.40mm}N}}
\newcommand{\Complex}{\rm C\kern-.42em\vrule width.03em height.58em
  depth-.02em\kern.4em}

\newcommand{\gcm}{~\ensuremath{\mathrm{g}\,\mathrm{cm}^{-3}}}

\newcommand{\mev}{\ensuremath{\mathrm{MeV}}}
\newcommand{\msol}{\ensuremath{M_{\sun}}}
\newcommand{\simgt}{\,\rlap{\lower 3.5 pt \hbox{$\mathchar \sim$}} \raise
1pt \hbox {$>$}\,}
\newcommand{\simlt}{\,\rlap{\lower 3.5 pt \hbox{$\mathchar \sim$}} \raise
1pt \hbox {$<$}\,}

\newcommand{\pd}[2]{\frac{\partial #1}{\partial #2}}
\def\be{\begin{equation}}
\def\ee{\end{equation}}
\def\ba{\begin{eqnarray}}
\def\ea{\end{eqnarray}}
\def\nn{\nonumber}
\def\l{\left}
\def\r{\right}
\def\ot{\frac{1}{2}}
\def\oc{\frac{1}{c}}
\def\ofp{\frac{1}{4\pi}}

\def\br{\beta_{r}}
\def\bt{\beta_\vartheta}
\def\bp{\beta_\varphi}
\def\Hr{H_{r}}
\def\Ht{H_\vartheta}
\def\Hp{H_\varphi}
\def\Prr{P_{rr}}
\def\Prt{P_{{r}\vartheta}}
\def\Prp{P_{{r}\varphi}}
\def\Ptr{P_{\vartheta{r}}}
\def\Ptt{P_{\vartheta\vartheta}}
\def\Ptp{P_{\vartheta\varphi}}
\def\Ppr{P_{\varphi{r}}}
\def\Ppt{P_{\varphi\vartheta}}
\def\Ppp{P_{\varphi\varphi}}
\def\Nrrr{N_{rrr}}
\def\Nrrt{N_{{rr}\vartheta}}
\def\Nrrp{N_{{rr}\varphi}}
\def\Nrtr{N_{{r}\vartheta{r}}}
\def\Nrtt{N_{{r}\vartheta\vartheta}}
\def\Nrtp{N_{{r}\vartheta\varphi}}
\def\Nrpr{N_{{r}\varphi{r}}}
\def\Nrpt{N_{{r}\varphi\vartheta}}
\def\Nrpp{N_{{r}\varphi\varphi}}
\def\Ntrr{N_{\vartheta{rr}}}
\def\Ntrt{N_{\vartheta{r}\vartheta}}
\def\Ntrp{N_{\vartheta{r}\varphi}}
\def\Nttr{N_{\vartheta\vartheta{r}}}
\def\Nttt{N_{\vartheta\vartheta\vartheta}}
\def\Nttp{N_{\vartheta\vartheta\varphi}}
\def\Ntpr{N_{\vartheta\varphi{r}}}
\def\Ntpt{N_{\vartheta\varphi\vartheta}}
\def\Ntpp{N_{\vartheta\varphi\varphi}}
\def\Nprr{N_{\varphi{rr}}}
\def\Nprt{N_{\varphi{r}\vartheta}}
\def\Nprp{N_{\varphi{r}\varphi}}
\def\Nptr{N_{\varphi\vartheta{r}}}
\def\Nptt{N_{\varphi\vartheta\vartheta}}
\def\Nptp{N_{\varphi\vartheta\varphi}}
\def\Nppr{N_{\varphi\varphi{r}}}
\def\Nppt{N_{\varphi\varphi\vartheta}}
\def\Nppp{N_{\varphi\varphi\varphi}}
\def\ene{\epsilon}
\def\Cj{C^{(0)}}
\def\C1{C^{(1)}}
\def\Cr{C^{(1)}_{r}}
\def\Ct{C^{(1)}_\vartheta}
\def\Cp{C^{(1)}_\varphi}

\def\ir{\frac{1}{r}}
\def\irst{\frac{1}{r\sin\vartheta}}
\def\rst{r\sin\vartheta}
\def\st{\sin\vartheta}
\def\ct{\cos\vartheta}

\begin{document}

\title{Two-dimensional hydrodynamic core-collapse supernova
   simulations with spectral neutrino transport}
 \subtitle{I.~Numerical method and results for a $15~\msol$ star}

   \author{R. Buras \inst{1,2} \and
           M. Rampp \inst{1}\thanks{\emph{Present address:} Rechenzentrum der Max-Planck-Gesellschaft am Max-Planck-Institut f\"ur Plasmaphysik, Boltzmannstr.~2, D-85748 Garching, Germany} \and
           H.-Th. Janka \inst{1} \and
           K. Kifonidis \inst{1}}

   \offprints{H.-Th. Janka, \email{thj@mpa-garching.mpg.de}}

   \institute{Max-Planck-Institut f\"ur Astrophysik, 
              Karl-Schwarzschild-Str.\ 1, D-85741 Garching, Germany
                 \and
              Max-Planck-Institut f\"ur  Physik,
                 F\"ohringer Ring 6, D-80805 M\"unchen, Germany
                }

   \date{Received 7 July 2005 / Accepted 18 October 2005 }

\abstract{
Supernova models with a full spectral treatment of the neutrino
transport are presented, employing the \textsc{Prometheus/Vertex}
neutrino-hydrodynamics code with a variable Eddington factor closure
of the ${\cal O}(v/c)$ moments equations of neutrino number, energy,
and momentum.  Our ``ray-by-ray plus'' approximation developed for
two- (or three-) dimensional problems assumes that the local neutrino
distribution function is azimuthally symmetric around the radial
direction, which implies that the nonradial flux components
disappear. Other terms containing the angular velocity components are
retained in the moments equations and establish a coupling of the
transport at different latitudes by lateral derivatives. Also lateral
components of the neutrino pressure gradients are included in the
hydrodynamics equations. This approximative approach for neutrino
transport in multi-dimensional environments is motivated and
critically assessed with respect to its capabilities, limitations, and
inaccuracies in the context of supernova simulations.
In this first paper of a series, one- (1D) and two-dimensional (2D)
core-collapse calculations for a (nonrotating) $15\,\msol$ star are
discussed, uncertainties in the treatment of the equation of
state -- numerical and physical -- are tested, Newtonian results are
compared with simulations using a general relativistic potential,
bremsstrahlung and interactions of neutrinos of different flavors are
investigated, and the standard approximation in neutrino-nucleon
interactions with zero energy transfer is replaced by rates that
include corrections due to nucleon recoil, thermal motions, weak
magnetism, and nucleon correlations.
Models with the full implementation of the ``ray-by-ray plus''
spectral transport were found not to explode, neither in spherical
symmetry nor in 2D when the computational grid is constrained to a
lateral wedge ($< \pm 45\degr$) around the equator. The success
of previous two-dimensional simulations with grey, flux-limited
neutrino diffusion can therefore not be confirmed. An explosion is
obtained in 2D for the considered $15\,\msol$ progenitor, when the
radial velocity terms in the neutrino momentum equation are
omitted. This manipulation increases the neutrino energy density in
the convective gain layer by about 20--30\% and thus the integral
neutrino energy deposition in this region by about a factor of two
compared to the non-exploding 2D model with the full transport. The
spectral treatment of the transport and detailed description of
charged-current processes leads to proton-rich neutrino-heated ejecta,
removing the problem that previous explosion models with approximate
neutrino treatment overproduced $N=50$ closed neutron shell nuclei by
large factors.

}

\maketitle
%

\section{Introduction}
\label{sec:intro}

Convective processes and hydrodynamic instabilities play an important
role in different regions of collapsing and exploding stars.
(1) Inside the nascent neutron star, i.e.\ below and around the
neutrinosphere, convection enhances the transport and release of
neutrinos and thus can raise the neutrino luminosities with
potentially helpful consequences for the delayed shock revival and the
neutrino-driven explosion mechanism
\cite[e.g.,][]{bur87,burlat88,mirpon00,ponred99,wilmay88,wilmay93,
keijan96,kei97}.
(2) In the neutrino-heating region between gain radius and stalled
supernova shock, convective overturn was found to enhance the energy
deposition by neutrinos and the energy transport to the
shock. Therefore it allows for neutrino-driven explosions even if
models in spherical symmetry fail
(\citealp{herben94,burhay95,janmue96}).
(3) Multi-dimensional processes in the outer layers of the exploding
star destroy the onion-shell structure of the progenitor and are
responsible for the anisotropies and macroscopic mixing of chemical
elements which was observed in Supernova~1987A and other supernovae,
and which possibly is the origin of many morphological features of
supernova remnants (e.g., \citealp{kifple03} and references therein).

The existence of convectively unstable stratifications in the
supernova core had long been suspected on grounds of spherically
symmetric models \cite[e.g.,][]{eps79,betbro87,bur87,bet90} before
confirmation became possible by multi-dimensional modeling with
sufficient resolution.

The first such multi-dimensional simulations
(\citealp{burfry92,janmue93,mue93}) still ignored neutrino effects,
were able to follow the evolution for a very limited period of time
only, and started from post-bounce models that were constructed
artificially or were adopted from simulations based on different
assumptions about the input physics.  Therefore they were only
suitable to demonstrate the possibility in principle, but otherwise
showed the transient decay of an initially unstable postshock
stratification and thus could not be considered as meaningful models
of supernovae \cite[cf.][]{brumez94}. However, simulations including
the effects of neutrino heating and cooling \cite[][]{herben92} and
true core collapse environments \cite[][]{janmue94,janmue95,janmue96},
neutrino transport by grey, flux-limited diffusion
\cite[][]{herben94,burhay95}, multi-group 1D transport
\cite[][]{mezcal98:ndconv}, rotation \cite[][]{fryheg00}, and
three-dimensional hydrodynamics \cite[][]{frywar02,frywar04} confirmed
robustly the development of convective overturn in the neutrino-heated
layer just behind the stalled shock, although it is still a matter of
debate whether the associated convective effects are sufficient for
driving delayed explosions \cite[][]{mezcal98:ndconv}.

The situation is more controversial for convective activity below the
neutrinosphere. A detailed investigation of neutrino transport
properties at the conditions inside nascent neutron stars
\cite[][]{brural04} revealed no evidence for neutron finger
instability as assumed by the Livermore group
\cite[][]{wilmay88,wilmay93} to trigger neutrino-driven explosions by
convectively enhanced neutrino fluxes.  In the simulations of
\cite{janmue96} the use of an inner boundary condition with imposed
neutrino luminosities certainly played a role, causing neutrino
absorption and heating and thus convective energy transport in the
neutron star layers around the inner boundary (neutrino diffusion was
ignored!). This convection below the neutrinosphere extended outward
to a region where cooling by neutrino losses was fast enough to take
up the energy delivered by the convective transport.  However,
full-scale hydrodynamic models of nascent neutron stars in two
dimensions including a simple description of grey, radial neutrino
diffusion and its effects on lepton number and energy transport showed
the rapid development of convection inside the proto-neutron star
\cite[][]{burfry93} and its persistence in a growing volume until at
least one second in nonrotating as well as rotating cases
\cite[][whose simulations covered the evolution until at most 1.2
seconds after bounce]{keijan96,kei97,jankei98,jan04:sydney}. A careful
analysis \cite[see][]{kei97} revealed that the exchange of energy and
in particular of lepton number between moving fluid elements and their
surroundings was important so that the convective mode could not be
described by ideal Ledoux convection. Instead, it had the character of
a doubly diffusive instability, which can occur at Ledoux stable
conditions as discussed by \cite{brudin96,brumez95:phyrep} and more
recently in much detail by \cite{brural04}.

This importance of entropy and lepton exchange of buoyant fluid
elements with their surroundings via neutrino transport was also
pointed out by \cite{mezcal98:pnconv}. The latter authors, moreover,
argued on the basis of their numerical and analytic models and
timescale analyses that fast equilibration by neutrino transport tends
to damp or even suppress entropy- and lepton-driven proto-neutron star
convection around and somewhat below the neutrinosphere, making it
inefficient in the transport of lepton number and energy. The authors,
however, admit that their investigations lack multi-dimensional
``realistic'' multi-group neutrino transport, coupled
self-consistently to multi-dimensional hydrodynamics, which have to be
applied in time-dependent evolution simulations of supernova cores
before final conclusions can be drawn about the development of
convection in optically thick regions. \cite{brural04} in particular
stress the importance of a careful inclusion of neutrino transport
effects for a reliable description of doubly diffusive instabilities.

While the basic results of the models of \cite{keijan96} might
therefore still withstand critical assessment by future, better
simulations, quantitative aspects (e.g., the speed at which the
convective layer grows inward towards the stellar center and enhances
the deleptonization and energy loss) were undoubtedly affected by the
simplicity of the treatment of the neutrino physics. \cite{keijan96}
used the assumption that neutrinos are in equilibrium with the stellar
fluid (neutrino energy and pressure were therefore added to the
corresponding gas quantities), that neutrino transport can be
approximated by first-order grey, (flux-limited) radial equilibrium
diffusion \cite[cf.][\S 97]{mihmih84}, and that the diffusive fluxes
of lepton number and energy can be combined for electron neutrinos and
antineutrinos into a gradient form with a single, average diffusion
coefficient \cite[cf.\ Appendix A of][]{keijan95}. The need for
removing these deficiencies has motivated new efforts for improved
modeling, which were recently reported by \cite{swemyr05}.

In addition to Ledoux convection and doubly diffusive instabilities
another form of multi-dimensional hydrodynamic instability seems to
play a role in the supernova core: While one-dimensional,
time-dependent hydrodynamic simulations (without neutrino transport)
found that the standing accretion shock is stable against radial
perturbations, two- and three-dimensional (adiabatic) simulations
demonstrated that it is unstable to non-radial modes, which lead to
rapid growth of turbulence behind the shock and thus shock expansion
and deformation \cite[][]{blomez03,mezblo03}.  Such a standing
accretion shock instability, SASI, was actually predicted for
accreting black holes by analytic considerations by
\cite{fog01,fog02}, who described its origin from an amplifying
feedback cycle of entropy and vorticity perturbations, which are
advected inward and trigger acoustic waves that propagate outward and
distort the shock, thus closing the cycle. The existence of this
phenomenon at non-adiabatic conditions in the postshock flow was
confirmed by supernova simulations including a simplified treatment of
neutrino transport \cite[][]{secple04,sec05}. Its action can clearly
be seen in particular if neutrino heating is too weak to drive strong
postshock convection. In agreement with the adiabatic models of
\cite{fog01,fog02} and \cite{blomez03}, \cite{secple04} find highest
growth rates for low ($l = 1,\,2$) modes, which produce large, global
asymmetries of the supernova explosion and might explain the observed
pulsar kick velocities and grossly asymmetric distribution of elements
seen in supernova remnants like Cassiopeia~A \cite[see][]{jansec05}.

Advancing core collapse supernova modeling to the next level of
complexity and reliability beyond the stage of merely demonstrating
possibilities in principle, requires improvements in the handling of
neutrino transport coupled to multi-dimensional hydrodynamic
simulations. Such improvements are necessary to clarify which role
hydrodynamic instabilities play in the supernova core and how these
are linked to the mechanism of the explosion. Crucial questions in
this context are: Does convection take place below the neutrinosphere
and if so, which kind of instability is it? Does proto-neutron star
convection raise the neutrinospheric luminosities and thus the
neutrino heating behind the shock on timescales relevant for the
explosion? Does neutrino-driven convection in the heating layer become
sufficiently strong to trigger the explosion? Can the positive results
of previous simulations with simplified neutrino transport by grey,
flux-limited diffusion
\cite[][]{herben94,fry99,fryheg00,frywar02,frywar04} be confirmed?  Is
the accretion shock instability found by \cite{blomez03} important and
how does it interact with convective instabilities and rotation?

With the aim to address these questions in a next generation of core
collapse studies, we have developed a new method for treating time-
and frequency-dependent neutrino transport in multi-dimensional (in
this work two-dimensional) environments. Our transport code is a
generalization of the 1D version \textsc{Vertex} of \cite{ramjan02},
adopting the approximations to general relativity described in the
latter paper and tested for spherically symmetric problems against
fully relativistic calculations in \cite{lieram05} and
\cite{mardim05}.  The code employs a detailed spectral description of
neutrino-matter interactions with fully implicit energy-bin coupling
and with a set of processes which was updated from \cite{ramjan02} by
including interactions between neutrinos of different flavors and
improving neutrino-nucleon interactions such that the effects of
nucleon thermal motions, recoil, weak magnetism, and correlations in
charged and neutral current processes are taken into account.  The
high dimensionality of the spatial and momentum dependence of the
phase space distribution function is reduced by assuming the neutrino
intensity to be axially symmetric around the radial
direction. Consistent with this, we ignore lateral flux components and
effects of neutrino viscosity and come up with a set of moments
equations for neutrino number density, energy density and radial flux,
which include all velocity-dependent terms of ${\cal O}(v/c)$ for
radial {\em and lateral} velocity components. These equations are
solved in each angular bin of the spatial grid, taking into account
the lateral (and in 3D also azimuthal) coupling by corresponding
derivative terms in an operator splitting step (``ray-by-ray plus''
approach). Closure of the set of moments equations is achieved by
using variable Eddington factors, which are computed from a model
Boltzmann equation as described in \cite{ramjan02}.  The efficiency of
the code performance is increased by determining these normalized
moments of the neutrino intensity, which usually exhibit relatively
weak variability with lateral direction, only once for all latitudes
on grounds of an angularly averaged stellar background. Because of the
fact that we refer to the solution of the model Boltzmann equation for
closing the set of moments equations, we named our code
\textsc{MuDBaTH} ({\bf Mu}lti-{\bf D}imensional {\bf B}oltzm{\bf a}nn
{\bf T}ransport and {\bf H}ydrodynamics).

Our approach to deal with neutrino transport in two- (and in an
analogous way in three-) dimensional environments is intended to be a
valid approximation for situations where only local macroscopic
inhomogeneities are present, e.g. due to convection.  It is, however,
likely to become poor when a large global deformation (e.g. by rapid
rotation) develops and the neutrino flow is therefore not well
described by an at least on average dominant radial component. Our
approach is complementary concerning the employed approximations to
that of \cite{livbur04}, who introduced a scheme for solving the
frequency-dependent Boltzmann transport equation in axially symmetric
problems, but sacrificed energy-bin coupling and the majority of the
terms with dependence on the motion of the stellar plasma.  It also
differs from the two-dimensional multi-group flux-limited diffusion
approximation employed by \cite{walbur05}, who abandon energy-bin
coupling and velocity-dependent terms, too.

In this Paper~I and a subsequent one (\citealp{burram05:II}; Paper~II)
we present the first two-dimensional hydrodynamic simulations with a
full spectral implementation of neutrino transport, applied to the
core collapse and post-bounce evolution of progenitor stars with
different masses (in the range from 11.2$\,\msol$ to 25$\,\msol$),
compared to corresponding simulations in spherical symmetry.  In
Paper~I we shall describe in detail our method for neutrino transport
in two spatial dimensions, discuss numerical aspects and tests of the
code and input physics, and try to assess the limitations of our
method. We will concentrate on the presentation of numerical results
for the evolution of a 15$\,\msol$ progenitor star \cite[Model
s15s7b2 of][]{woowea95} with varied input physics. For the 2D (axially
symmetric) simulations we confine ourselves to using a roughly 90
degree angular wedge around the equatorial plane and will focus our
discussion on neutrino-driven convection in the hot bubble region and
its effect on the supernova dynamics.  In Paper~II we shall present
and compare 1D and 2D simulations for different progenitors, using
different angular wedges and resolution (including a model with a full
180$\degr$ polar grid) and considering also core
rotation. Paper~II will contain a detailed analysis of the importance
of convection below the neutrinosphere relative to convection in the
neutrino-heating layer, and of the role of low-mode instabilities,
pre-collapse perturbations in the stellar core, and rotation in the
investigated models. Since our spectral treatment of neutrino
transport in the 2D simulations, which cover evolution periods of up
to 300$\,$ms after core bounce, is very CPU-time consuming -- one model
requires $(2-5)\times 10^{17}$ floating point operations, depending on
the angular resolution and wedge size --, computer resources available
to us prevented us from performing simulations with a systematic
variation of important but not finally determined degrees of freedom,
e.g., of the magnitude and distribution of angular momentum in the
progenitor core. We therefore are forced to constrain our discussion
on some selective cases.

The present paper is organized as follows. Section~\ref{sec:general}
(supplemented by Appendices~\ref{sec:neuopa}--\ref{app:noff}) contains
a description of the employed equations, numerical algorithm, and
tests of the numerical scheme and input physics, in particular of the
equation of state. Section~\ref{sec:results} presents results for 1D
(Sect.~\ref{sec:ssm}) and 2D (Sect.~\ref{sec:tdm}) core collapse and
post-bounce simulations of a 15$\,\msol$ progenitor with Newtonian
gravity and with our approximative treatment of the effects of
relativistic gravity, varied neutrino-matter interactions, and
neutrino transport with and without taking into account the
velocity-dependent terms in the neutrino momentum equation. Thus
retreating from our most complete implementation of neutrino transport
by neglecting these velocity-dependent terms, we get cases with
interesting behaviour, which once more demonstrate the sensitivity of
the supernova evolution to supposedly secondary changes of neutrino
transport: In the 1D simulation (Sect.~\ref{sec:osci}) we observe a
large-amplitude oscillation of the neutrino-decoupling layer between
neutrinosphere and accretion shock, which grows in amplitude until it
leads to an explosion; this reminds us of the $\kappa$ mechanism of
vibrational instability of stars. In the corresponding 2D simulation
(Sect.~\ref{sec:expl}) strong post-shock convection develops and
triggers a neutrino-powered supernova explosion, whose properties will
be compared to observational constraints. In contrast, both the 1D and
2D models with the full transport implementation (Sects.~\ref{sec:std}
and \ref{sec:acme}, respectively) do not develop explosions.  In
Sect.~\ref{sec:summary} we give a summary of our main results and draw
conclusions.

\section{Governing equations and numerical implementation}
\label{sec:general}

In order to be able to exploit symmetries of the problem we work in a
system of spherical coordinates with radius $r$, latitudinal angle
$\vartheta$, and longitudinal angle $\varphi$. For all simulations
performed so far we have assumed that azimuthal symmetry holds with
respect to the polar axis, but a generalization of the presented
method to three-dimensional situations is straightforward.

Like the one-dimensional version documented by \cite{ramjan02} the
algorithm relies on an operator splitting approach which means that
the coupled system of evolution equations is processed in two
independent steps, a hydrodynamic step and a
neutrino-transport/interaction step. In each timestep these two steps
are solved subsequently.

\subsection{Hydrodynamics}
\label{sec:hydro}

For an ideal (i.e. nonviscous) fluid characterized by the mass density
$\rho$, the radial, lateral, and azimuthal components of the velocity
vector $(v_r, v_\vartheta, v_\varphi)$, specific energy $\varepsilon =
e + \frac{1}{2} (v_r^2+v_\vartheta^2+v_\varphi^2)$ with $e$ being the
specific internal energy, and gas pressure $p$, the Eulerian,
nonrelativistic equations of hydrodynamics in spherical coordinates
and azimuthal symmetry read:
\begin{eqnarray}
\frac{\partial}{\partial t} \rho &+&
\frac{1}{r^2}\frac{\partial}{\partial r}\left(r^2\rho\,v_r\right) +
\frac{1}{r\st}\frac{\partial}{\partial \vartheta}\Big(\rho \st
\,v_\vartheta\Big) = 0
\,,  \label{eq:hydro.rho} \\
\frac{\partial}{\partial t} \left(\rho v_r\right) &+&
\frac{1}{r^2}\frac{\partial}{\partial r}\left(r^2\rho v_r\,v_r\right) +
\frac{1}{r\st}\frac{\partial}{\partial \vartheta}\Big(\rho \st \,
v_\vartheta \,v_r \Big) \nn\\ &-&
\rho\frac{v_\vartheta^2 + v_\varphi^2}{r}+
\frac{\partial p}{\partial r}
=
-\rho\frac{\partial\Phi}{\partial r}
+{Q_\mathrm{M}}_r
\,, \label{eq:hydro.v_r} \\
\frac{\partial}{\partial t} \left(\rho v_\vartheta\right) &+&
\frac{1}{r^2}\frac{\partial}{\partial r}\left(r^2\rho v_r\,v_\vartheta\right) +
\frac{1}{r\st}\frac{\partial}{\partial \vartheta}\Big(\rho \st \, 
v_\vartheta \,v_\vartheta\Big) \nn\\ &+&
\rho\frac{v_r v_\vartheta-v_\varphi^2\cot{\vartheta}}{r}+
\frac{1}{r}\frac{\partial p}{\partial \vartheta}
=
-\frac{\rho}{r}\frac{\partial\Phi}{\partial \vartheta}
+{Q_\mathrm{M}}_\vartheta
\,,\label{eq:hydro.v_t} \\
\frac{\partial}{\partial t} \left(\rho v_\varphi\right) &+&
\frac{1}{r^2}\frac{\partial}{\partial r}\left(r^2\rho v_r\,v_\varphi\right) +
\frac{1}{r\st}\frac{\partial}{\partial \vartheta}\Big(\rho \st \, 
v_\vartheta \,v_\varphi \Big) \nn\\ &+&
\rho\frac{v_r v_\varphi + v_\vartheta v_\varphi\cot{\vartheta}}{r} =
0
\,,\label{eq:hydro.v_p} \\
\frac{\partial}{\partial t} \left(\rho \varepsilon\right) &+&
\frac{1}{r^2}\frac{\partial}{\partial r}\left(r^2\left(\rho \varepsilon + p\right)\,v_r\right) \nn\\ &+&
\frac{1}{r\st}\frac{\partial}{\partial \vartheta}\Big(\left(\rho \varepsilon + p\right) \st \, v_\vartheta\Big)
= \nn\\
&&-\rho\left(v_r\frac{\partial\Phi}{\partial r}
      + \frac{v_\vartheta}{r} \frac{\partial\Phi}{\partial \vartheta}
    \right)
+Q_\mathrm{E} + v_r{Q_\mathrm{M}}_r + v_\vartheta{Q_\mathrm{M}}_\vartheta
\,,\label{eq:hydro.e}
\end{eqnarray}
where $\Phi$ denotes the gravitational potential of the fluid, and
$\vec{Q}_\mathrm{M}=({Q_\mathrm{M}}_r, {Q_\mathrm{M}}_\vartheta)$ and
$Q_\mathrm{E}$ are the neutrino source terms for momentum transfer and
energy exchange,
respectively. Eqs.~(\ref{eq:hydro.rho}--\ref{eq:hydro.e}) are closed
by the equation of state (EoS) which yields the pressure $p$ for given
$\rho$, $e$, and composition. In the case of nuclear statistical
equilibrium (NSE) a third independent variable, the electron fraction
$\ye$, is sufficient to characterize the composition. For this
variable the evolution is computed according to a conservation
equation
\begin{equation}\label{eq:hydro.ye}
\frac{\partial}{\partial t} \left(\rho \ye\right) +
\frac{1}{r^2}\frac{\partial}{\partial r}\left(r^2\rho \ye\,v_r\right) +
\frac{1}{r\st}\frac{\partial}{\partial \vartheta}\left(\st \,
\rho \ye\,v_\vartheta\right) =
Q_\mathrm{N}
\,,
\end{equation}
where the source term $Q_\mathrm{N}/(\rho/m_\mathrm{by})$ is the rate of
change of the net electron fraction (i.e.~the number fraction of
electrons minus that of positrons) due to emission and absorption of
electron-flavor neutrinos, and $m_\mathrm{by}$ is the baryon mass. In
case the medium is not in NSE, an equation like
Eq.~(\ref{eq:hydro.ye}) has to be solved also for the abundance of
each nucleus $k$, $Y_k \equiv n_k/n_\mathrm{by}$, using
\begin{equation}\label{eq:hydro.x_k}
\frac{\partial}{\partial t} \left(\rho Y_k\right) +
\frac{1}{r^2}\frac{\partial}{\partial r}\left(r^2\rho Y_k\,v_r\right) +
\frac{1}{r\st}\frac{\partial}{\partial \vartheta}\left(\st \,
\rho Y_k\,v_\vartheta\right) = R_{k}
\, ,
\end{equation}
where $n_k$ and $n_\mathrm{by}$ are the number density of nucleus $k$ and
the baryon number density, respectively, and $R_{k} \equiv \rho\,
\delta Y_k/\delta t$, where $\delta Y_k/\delta t$ is a source term that
describes the rate of composition changes by nuclear reactions for
species $k$.

For the numerical integration of
Eqs.~(\ref{eq:hydro.rho}--\ref{eq:hydro.x_k}) we employ the Newtonian
finite-volume code PROMETHEUS (\citealp{frymue89,fryols00}), which was
supplemented by additional problem specific features \cite[]{kei97}
and the improvements described in \cite{kifple03}.
PRO\-ME\-THE\-US is a direct Eulerian implementation of the Piecewise
Pa\-ra\-bol\-ic Method (PPM) of \cite{colwoo84}.  As a time-explicit,
third-order in space, second order in time Godunov scheme with a
Riemann solver it is particularly well suited for following
discontinuities in the fluid flow like shocks, contact
discontinuities, or boundaries between layers of different chemical
composition. A notable advantage in the present context is its
capability of solving multi-dimensional problems with high
computational efficiency and numerical accuracy. Our code makes use of
the ``Consistent Multifluid Advection'' (CMA) method \cite[]{plemue99}
for ensuring an accurate advection of different chemical components of
the fluid, and switches from the original PPM method to the HLLE
solver of \cite{ein88} in the vicinity of strong shocks to avoid
spurious oscillations (the so-called ``odd-even decoupling'', or
``carbunkel'', phenomenon) when such shocks are aligned with one of
the coordinate directions in multidimensional simulations
\cite[]{qui94,lio00,kifple03,sutbis03}.

Although our hydrodynamic scheme is Newtonian, we have included
effects of general relativistic (GR) gravity approximately in the
following way: The gravitational potential used in our simulations can
be symbolically written as
$\Phi(r,\vartheta)=\Phi_{2\mathrm{D}}^\text{Newt}(r,\vartheta)+
\left(\Phi_{1\mathrm{D}}^\text{GR}(r)
-\Phi_{1\mathrm{D}}^\text{Newt}(r)\right)$.  We compute the Newtonian
gravitational potential $\Phi_{2\mathrm{D}}^\text{Newt}$ for the
two-dimensional axisymmetric mass distribution by expanding the
integral solution of the Poisson equation into a Legendre series,
truncated at $l=10$ \cite[cf.][]{mueste95}.  General relativistic
effects are approximately taken into account by the spherically
symmetric correction term
$\Phi_{1\mathrm{D}}^\text{GR}-\Phi_{1\mathrm{D}}^\text{Newt}$, where
$\Phi_{1\mathrm{D}}^\text{GR}$ denotes an effective general
relativistic gravitational potential as employed for spherically
symmetric simulations \cite[see][Eq.~53]{ramjan02} and
$\Phi_{1\mathrm{D}}^\text{Newt}$ is its Newtonian counterpart.
The general relativistic potential $\Phi_{1\mathrm{D}}^\text{GR}$ is
deduced from a comparison of the Newtonian and relativistic equations
of motion in spherical symmetry and includes terms due to the pressure
and energy of the stellar medium and neutrinos \cite[see][]{ramjan02}.
Both, $\Phi_{1\mathrm{D}}^\text{GR}(r)$ and
$\Phi_{1\mathrm{D}}^\text{Newt}(r)$ are computed using angular
averages of the evolved variables.
In two-dimensional simulations which cover only a limited range of
latitudes $0 < \vartheta_\text{min} \le \vartheta \le
\vartheta_\text{max} < \pi$ around the equatorial plane we set
$\Phi(r,\vartheta)=\Phi_{1\mathrm{D}}^\text{GR}(r)$.

The source terms ${Q_\mathrm{M}}_r$, ${Q_\mathrm{M}}_\vartheta$,
$Q_\mathrm{E}$, and $Q_\mathrm{N}$ on the right-hand sides of
Eqs.~(\ref{eq:hydro.v_r},\ref{eq:hydro.v_t},\ref{eq:hydro.e},\ref{eq:hydro.ye})
are determined by the solution of the neutrino transport
equations. The source terms $R_{k}$ depend on changes of the
composition according to nuclear burning. Unless stated otherwise, the
EoS we apply is the same as described in detail in \cite[][Appendix
B]{ramjan02}.

Note that PROMETHEUS only solves the left-hand sides of the
hydrodynamic Eqs.~(\ref{eq:hydro.rho}--\ref{eq:hydro.x_k}) and that
the EoS is not evaluated during this procedure. The computation of the
terms on the right-hand sides, i.e~the gravitational, neutrino, and
burning effects, as well as the evaluation of the EoS and, if
necessary, the determination of the NSE composition, are done in
operator split steps.

\subsection{Neutrino transport}
\label{sec:2dneutra}

Solving the full two-dimensional neutrino transport equation would be
the most precise way of simulating supernovae in two dimensions. On
the level of the radiation moments equations this would, compared to
the 1D case, lead to several new degrees of freedom and one additional
moments equation, which would require five more closure relations to
be introduced. Therefore we limit ourselves to a less ambitious
extension of the one-dimensional moments equations to two dimensions
in which new degrees of freedom (including, in case of the considered
azimuthal symmetry, the lateral flux and off-diagonal pressure tensor
terms which account for neutrino viscosity) are set to zero, but the
additional moments equation (in our case for the lateral flux) is
taken into account (in the sense that we retain the terms
corresponding to the lateral neutrino pressure gradients at optically
thick conditions). In the next section, we will justify why we believe
this approximation is sufficient for simulating two-dimensional
neutrino transport in the considered models of core-collapse
supernovae, but also explain why further simplifications are not
allowed.

The great advantage of our ``ray-by-ray'' 2D neutrino transport is
that the neutrino moments equations at different latitudes (except for
some terms which can be accounted for explicitly in an operator split)
decouple from each other. Therefore, for each ``radial ray'', i.e.~for
all zones of same polar angle, the moments equations can be solved
independently. Except for some additional terms this problem is
identical to solving $N_\vartheta$ times the moments equations for a
spherically symmetric star with $N_\vartheta$ being the number of grid
zones in polar direction.

We further make the following standard assumptions: First, we ignore
neutrino oscillations in the SN core, where neutrino effects are
treated by solving the transport problem. This is justified if one
ignores the results of LSND; then the parameters for atmospheric and
solar neutrino oscillations predict the resonant MSW effect to be at
densities far below $10^5\gcm$ so that this effect has no influence on
the region of interest (for the possible importance of neutrino flavor
conversions at high densities, see \citealp{fulqia05}). Furthermore,
non-resonant oscillations are strongly suppressed in the proto neutron
star (PNS) due to first and second order refractive effects, see
\cite{hanjan00}. Second, the medium, even in the PNS, initially does
not contain any muons and temperatures and densities are always too
low to produce tauons, which implies small or vanishing chemical
potentials for the $\mu$ and $\tau$ type neutrinos. Further, the
opacities are nearly equal for $\nu_\mu$, $\bar\nu_\mu$, $\nu_\tau$,
and $\bar\nu_\tau$. Therefore, we treat these four neutrino types
identically and set $\mu_{\nu_\mathrm{\mu}} \equiv
\mu_{\nu_\mathrm{\tau}} \equiv 0$. We will notate them collectively as
``$\nu_\mathrm{x}$''.

\subsubsection{Moments equations}
\label{sec:2dmomequ}

In a two-dimensional transport scheme assuming azimuthal symmetry, the
specific intensity ${\cal
I}(t,r,\vartheta,\varphi,\ene,\boldsymbol{n})$ does not depend on the
azimuth $\varphi$. We describe the direction of propagation
$\boldsymbol{n}$ by the angle cosine $\mu \equ
\boldsymbol{n}\cdot\boldsymbol{r}/|\boldsymbol{r}|$, measured with
respect to the radius vector $\vec{r}$, and the angle $\omega$. Then
azimuthal symmetry implies ${\cal I}(\dots,\mu,\omega)={\cal
I}(\dots,\mu,-\omega)$, see also Appendix \ref{app:momeq3d}. Making the
additional assumption that ${\cal I}$ is independent of $\omega$, each
of the angular moments of the specific intensity can be expressed by
one scalar, namely
\be
\{J,H,K,L,\dots\}(t,r,\vartheta,\ene)\equ
\frac{1}{2}\int\limits_{-1}^{+1}\!\dlin{\mu}\mu^{\{0,1,2,3,\dots\}}
\mathcal{I}(t,r,\vartheta,\ene,\mu)
\,,
\ee
where we have used
Eqs.~(\ref{appeq:tensorrelations}--\ref{appeq:1drels}) -- which follow
from our assumptions -- to reduce the number of independent variables
in the angular moments of the neutrino intensity as defined in
Eqs.~(\ref{appeq:moments}). As usual, $\ene$ denotes the energy of the
neutrinos. As a consequence of the afore mentioned assumptions the set
of moments equations for describing the evolution of neutrino energy
and flux in the comoving frame, given by
Eqs.~(\ref{J_spher}--\ref{Hp_spher}) in the Newtonian,
$\mathcal{O}(v/c)$ approximation, simplifies to
Eqs.~(\ref{eq:momeqe1}) and (\ref{eq:momeqe2}) for the remaining
independent variables $J$ and $H\equiv\Hr$. In our approximation
Eqs.~(\ref{Ht_spher}) and (\ref{Hp_spher}) can be ignored because the
variables the evolution of which they describe, i.e.~$H_\vartheta$ and
$H_\varphi$, are strictly set to zero. With ${\cal J}=J/\ene$, ${\cal
H}=H/\ene$, ${\cal K}=K/\ene$, and ${\cal L}=L/\ene$, the moments
equations describing the evolution of neutrino number are given by
Eqs.~(\ref{eq:momeqn1}) and (\ref{eq:momeqn2}).

The velocity-dependent terms, in the order of their
appearance and grouping in Eqs.~(\ref{eq:momeqe1}--\ref{eq:momeqn2}), 
correspond 
to the physical effects of radiation advection with the moving stellar
fluid (expressed by terms containing $\beta_r\partial/\partial r$ 
and $(\beta_\vartheta/r)\partial/\partial\vartheta$ for the 
velocity components $v_r$ and $v_\vartheta$, respectively),
radiation compression, Doppler shifting (within the energy
derivatives), and finally the fluid acceleration. 
Note that in the current version of our computer code
the $\br\,\partial/\partial t$-derivatives of the moments, which
appear at the end of the respective first lines of
Eqs.~(\ref{eq:momeqe1}--\ref{eq:momeqn2}), cf.~\cite{ramjan02},
are ignored.

The system of moments equations (\ref{eq:momeqe1}--\ref{eq:momeqn2})
is very similar to the Newtonian, $\mathcal{O}(v/c)$ moments equations
in spherical symmetry \cite[see][Eqs.~7,8,30,31]{ramjan02}. We have
set the additional terms arising from our approximative generalization
to two dimensions in boldface. Adding general relativistic (GR)
effects in the spirit of \cite{ramjan02} to our approximative 2D
transport proves not to alter the boldface terms, so that the
corresponding equations are Eqs.~(54--57) in the latter paper, which
handle the GR effects for 1D transport, plus those terms in
Eqs.~(\ref{eq:momeqe1}--\ref{eq:momeqn2}) that are typeset in
boldface. The equations are closed by substituting $K=f_K\cdot J$ and
$L=f_L\cdot J$, where $f_K$ and $f_L$ are the variable Eddington
factors.

\medskip
In order to discretize Eqs.~(\ref{eq:momeqe1}--\ref{eq:momeqn2}), the
computational domain
$[0,r_\text{max}]\times[\vartheta_\text{min},\vartheta_\text{max}] $
is covered by $N_r$ radial and $N_\vartheta$ angular zones, where
$\vartheta=0,\pi$ correspond to the polar axis and $\vartheta=\pi/2$
to the equatorial plane of the spherical grid.
All neutrino variables are defined on the angular centres of the zones
with the coordinate $\vartheta_{k+\hlf}\equ {{1}\over{2}}\,
(\vartheta_{k}+\vartheta_{k+1})$ being defined as the arithmetic mean
of the corresponding interface values. The equations are solved in two
operator-split steps corresponding to a lateral and a radial sweep.

In a first step, we treat the boldface terms in the respectively first
lines of Eqs.~(\ref{eq:momeqe1}--\ref{eq:momeqn2}), which describe the
lateral advection of the neutrinos with the stellar fluid, and thus
couple the angular moments of the neutrino distribution of
neighbouring angular zones. For this purpose we solve the equation
\be\label{eq:latadvec}
\frac{1}{c}\frac{\partial \Xi}{\partial t}
 + \frac{1}{r \st}\frac{\partial(\st \,\bt \,\Xi)}{\partial\vartheta}
=0
\,,
\ee
where $\Xi$ represents one of the moments $J$, $H$, $\mathcal{J}$, or
$\mathcal{H}$.
After integration over the volume of a zone $(i+\hlf,k+\hlf)$, where i
and k are the indices of the radial and lateral zones, the
finite-differenced version of Eq.~(\ref{eq:latadvec}) reads:
\ba
\frac{\Xi_{i+\hlf,k+\hlf}^{n+1}-\Xi_{i+\hlf,k+\hlf}^{n}}
     {ct^{n+1}-ct^{n}} &&\nn\\
     +\frac{1}{\Delta V_{i+\hlf,k+\hlf}} &&
     \left(\Delta A_{i+\hlf,k+1}\,{\bt}_{i+\hlf,k+1}
           \Xi^n_{i+\hlf,\kappa_{i+1/2}(k+1)} \r.\nn\\&&-\l.~
           \Delta A_{i+\hlf,k  }\,{\bt}_{i+\hlf,k}
           \Xi^n_{i+\hlf,\kappa_{i+1/2}(k)} \right) =0 
\,,
\label{eq:latadvec_fd}
\ea
with the volume element $\Delta
V_{i+\hlf,k+\hlf}={{2\pi}\over{3}}\,(r_{i+1}^3-r_{i}^3)\,(\ct_{k}-\ct_{k+1})$
and the surface element $\Delta
A_{i+\hlf,k}=\pi\,(r_{i+1}^2-r_{i}^2)\st_{k}$. Note that additional
indices of $\Xi$ which label energy bins and the different types of
neutrinos are suppressed for clarity. To guarantee monotonicity,
upwind values of the moments $\Xi^n_{i+\hlf,\kappa_{i+1/2}(k)}$, with
\begin{equation}
\kappa_{i+1/2}(k)\equ
\begin{cases}
k-\hlf &\text{for\quad} {\bt}_{i+\hlf,k} > 0 \,, \\
k+\hlf &\text{else}\,, \\
\end{cases} 
\end{equation}
are used for computing the lateral fluxes across the interfaces of the
angular zones. The time-step limit enforced by the
Courant-Friedrichs-Lewy (CFL) condition is
\be \Delta t_\mathrm{CFL} = \min_{i,k} \frac{\Delta
               x_{\vartheta;i+\hlf,k+\hlf}}{\l|v_{\vartheta;i+\hlf,k+\hlf}\r|},
\ee
where $\Delta x_\vartheta$ is the zone width in lateral direction. It
turns out that this condition is not restrictive in our
simulations. In practice, the numerical time-step is always limited by
other constraints.

Now, in the second step, the radial sweep for solving
Eqs.~(\ref{eq:momeqe1}--\ref{eq:momeqn2}) is performed. Considering a
radial ray with given $\vartheta_{k+\hlf}$, the radial discretization
of the equations (and of their general relativistic counterparts)
proceeds exactly as detailed in \cite{ramjan02}. The terms in boldface
not yet taken into account in the lateral sweep do not couple the the
neutrino distribution function of neighbouring angular zones and thus
can be included into the discretization scheme of the radial sweep in
a straightforward way.

\subsubsection{Eddington factors}
\label{sec:2deddfac}

In analogy to the treatment of the moments equations in the previous
subsection the calculation of the variable Eddington factors could in
principle also be done on (almost) decoupled radial rays. However,
such a procedure would require a sizeable amount of computer time. On
the other hand, the Eddington factors are normalized moments of the
neutrino phase space distribution function and thus, in the absence of
persistent global deformation of the star, should not show significant
variation with the angular coordinate
\cite[cf.~][]{ram00,ramjan02}. Therefore we have decided to determine
the variable Eddington factors only once for an ``angularly averaged''
radial ray instead of computing them for each radial ray separately.
The corresponding reduction of the computational load can be up to a
factor of 10 \cite[][]{ramjan02}.

In analogy to the spherically symmetric case the code solves a
time-step of the one-dimensional neutrino transport on a spherically
symmetric image of the stellar background. The latter is defined as
the angular averages of the structure variables
$\xi\in\{\rho,T,Y_\mathrm{e},\beta_r,\dots\}$ according to
$\xi(t,r)\equ {{1}\over{b-a}} \int_{a}^{b} \dlin{\cos\vartheta}
\xi(t,r,\vartheta)~w(t,r,\vartheta)$, where
$a=\cos(\vartheta_\text{max})$ and $b=\cos(\vartheta_\text{min})$ and
$w$ is a weighting function which can be either $\rho$ or 1 depending
on which choice is more appropriate. The computation of the 1D
transport step proceeds exactly as described for the spherically
symmetric version of the code \cite[see][]{ramjan02}, i.e.~the coupled
set of the 1D moments equations and a 1D model Boltzmann equation are
iterated to convergence to obtain solutions for the variable Eddington
factors $f_\mathrm{K}$ and $f_\mathrm{L}$. These variable Eddington
factors are used for all latitudes $\vartheta$ of the multidimensional
transport grid when solving the two-dimensional moments equations. We
will discuss and try to estimate the possible errors associated with
our approximate 2D transport treatment in Sect.~\ref{sec:codetest}.

\subsubsection{Coupling to the hydrodynamics}
\label{sec:2dcoupling}

\medskip
The system of neutrino transport equations
(\ref{eq:momeqe1}--\ref{eq:momeqn2}) is coupled with the equations of
hydrodynamics (\ref{eq:hydro.rho}--\ref{eq:hydro.ye}) by virtue of the
source terms
\begin{eqnarray}
Q_\mathrm{N}&=&
    \sum_\nu-4\pi\,m_\mathrm{by} ~\mathrm{sgn}(\nu)
    \int_0^\infty\dlin{\ene}\mathfrak{C}^{(0)}_\nu(\ene)\,,
    \label{eq:sourceterm_N}\\
Q_\mathrm{E}&=&
    \sum_\nu-4\pi
    \int_0^\infty\dlin{\ene}C^{(0)}_\nu(\ene)\,,
    \label{eq:sourceterm_E}\\
{Q_\mathrm{M}}_r&=&
    \sum_\nu-\frac{4\pi}{c}
    \int_0^\infty\dlin{\ene}C^{(1)}_{r,\nu}(\ene)\,,
    \label{eq:sourceterm_Mr}\\
{Q_\mathrm{M}}_\vartheta&=&
    \sum_\nu-\frac{4\pi}{c}
    \int_0^\infty\dlin{\ene}C^{(1)}_{\vartheta,\nu}(\ene)\,,
    \label{eq:sourceterm_Mt}
\end{eqnarray}
where $m_\mathrm{by}$ denotes the baryon mass,
$\mathfrak{C}^{(0)}_\nu(\ene)\equ\ene^{-1}C^{(0)}_\nu(\ene)$,
and $C^{(0)}_\nu(t,r,\vartheta,\ene)\equ (4\pi)^{-1} \int\dlin{\Omega}
C_\nu(t,r,\vartheta,\ene,\vec{n})$ and
$\boldsymbol{C}^{(1)}_\nu(t,r,\vartheta,\ene) =
\left(C^{(1)}_{r,\nu},C^{(1)}_{\vartheta,\nu}\right)= (4\pi)^{-1}
\int\dlin{\Omega} \boldsymbol{n} ~C_\nu(t,r,\vartheta,\ene,\vec{n})$
are angular moments of the collision integral of the Boltzmann
equation, $C_\nu$. Note that in
Eqs.~(\ref{eq:sourceterm_N}--\ref{eq:sourceterm_Mt}) the moments of
the collision integral are summed over all neutrino types
$\nu\in\{\nu_\mathrm{e}, \bar \nu_\mathrm{e}, \nu_\mu, \bar \nu_\mu,
\nu_\tau, \bar \nu_\tau\}$, and $\mathrm{sgn}(\nu)={}+1$ for neutrinos
and ${}-1$ for antineutrinos. Remember that we treat $\nu_\mu$, $\bar
\nu_\mu$, $\nu_\tau$, and $\bar \nu_\tau$ identically because their
matter interactions are nearly equal. They do not transport electron
lepton number and therefore do not contribute to $Q_\mathrm{N}$. In
the following we suppress the index $\nu$.

Our simplification of the neutrino transport equations enforces a
\emph{radial} flux vector, i.e.~the angular flux component
$H_\vartheta\equ 0$.  However, as we shall demonstrate in
Sect.~\ref{sec:latadvec}, the corresponding lateral component of the
momentum transfer from neutrinos to the stellar medium, described by
the source term ${Q_\mathrm{M}}_\vartheta$, can not be neglected in
the Euler equation of the stellar fluid, Eq.~(\ref{eq:hydro.v_t}),
when the neutrinos are tightly coupled to the medium. This implies
that we should solve the moments equation for the lateral transport of
neutrino momentum (\ref{Ht_spher}) which, using the assumptions
Eqs.~(\ref{appeq:tensorrelations}--\ref{appeq:1drels}), simplifies to
\ba
\Ct(\ene) &=& \l( \frac{\partial \beta_\vartheta}{\partial r}
                 - \frac{\beta_\vartheta}{r} \r) H
  + \frac{1}{2r} \frac{\partial (J-K)}{\partial \vartheta}
       + \frac{1}{2c} \frac{\partial \beta_\vartheta}{\partial t}
               ( 3J-K ) \nn\\&&{}
       + \frac{\beta_\vartheta}{2c} \frac{\partial
                        (J-K)}{\partial t} 
       - \frac{\partial}{\partial \ene} \l\{ \ene \l[
              \frac{1}{2c}\frac{\partial \beta_\vartheta}{\partial t}
               (J-K) \r.\r. \nn\\&&{}\l.\l.
             +\frac{1}{2} \l(
                 \frac{\partial \beta_\vartheta}{\partial r}
                +\frac{1}{r} \frac{\partial \beta_r}{\partial
                                                        \vartheta}
                - \frac{\beta_\vartheta}{r} \r) (H-L) \r] \r\}\,.
\label{eq:Ct}\ea
On the other hand, at the conditions present in the optically thick
PNS an isotropic neutrino distribution (i.e.~$J\equiv 3K$) and
diffusion are good approximations. Concerning the velocity dependent
first term in Eq.~(\ref{eq:Ct}), which is of order
$\frac{v}{c}\frac{\lambda}{r}$ where $\lambda$ is the neutrino mean
free path, see the discussion about first order and 2nd order
diffusion in \citet[\S97, p.~458ff]{mihmih84}. Omitting this term
means that we ignore all effects of neutrino viscosity. Thus the above
equation simplifies considerably to (assuming also stationary
conditions, i.e.~$\partial/\partial t\equ 0$)
\be\label{eq:collint_theta}
\int_0^\infty \dlin{\ene} \Ct(\ene)
    =  \int_0^\infty \dlin{\ene} \frac{1}{2r}\pd{(J-K)}{\vartheta}
    =  \int_0^\infty \dlin{\ene} \frac{1}{3r}\pd{J}{\vartheta}
\,,
\ee
where the terms with the energy derivative in Eq.~(\ref{eq:Ct}) vanish 
when integrating over the neutrino energy. Given $J(r,\vartheta)$ as the
solution of the moments equations (\ref{eq:momeqe1},
\ref{eq:momeqe2}), Eq.~(\ref{eq:collint_theta}) together with the
definition Eq.~(\ref{eq:sourceterm_Mt}) allows us to compute an
approximation for the momentum exchange rate,
${Q_\mathrm{M}}_\vartheta$, between neutrinos and the stellar fluid.

Finally, Eq.~(\ref{Hp_spher}) would result in a momentum transfer from
the neutrinos to the medium in azimuthal direction, $C_\varphi^{(1)}$, in
the presence of rotation, $\beta_\varphi \neq 0$. Using our usual
assumptions Eqs.~(\ref{appeq:tensorrelations}--\ref{appeq:1drels}) we
obtain for Eq.~(\ref{Hp_spher})
\ba
\Cp(\ene) &=& \l( \frac{\partial \beta_\varphi}{\partial r}
                 - \frac{\beta_\varphi}{r} \r) H
       + \frac{1}{2c} \frac{\partial \beta_\varphi}{\partial t}
               ( 3J-K )
       + \frac{\beta_\varphi}{2c} \frac{\partial
                        (J-K)}{\partial t} \nn\\&&{}
       - \frac{\partial}{\partial \ene} \l\{ \ene \l[
              \frac{1}{2c}\frac{\partial \beta_\varphi}{\partial t}
               (J-K) \r.\r. \nn\\&&{}\l.\l.
             +\frac{1}{2} \l(
                 \frac{\partial \beta_\varphi}{\partial r}
                - \frac{\beta_\varphi}{r} \r) (H-L) \r] \r\}
\ea
However, again assuming diffusion, stationary conditions, and ignoring
neutrino viscosity, all terms of this expression vanish.

\medskip
The numerical discretization of Eq.~(\ref{eq:collint_theta}) reads
\be
{\Ct}_{i+\hlf,k+\hlf}=
\frac{J_{i+\hlf,k+\thrhlf}-J_{i+\hlf,k-\hlf}}
  {3\,r_{i+\hlf}(\vartheta_{k+\thrhlf}-\vartheta_{k-\hlf})}.
\ee
Since Eq.~(\ref{eq:collint_theta}) is valid only in the limit of an
optically thick medium we set ${\Ct}_{i+\hlf,k+\hlf}=0$ if the density
$\rho_{i+\hlf,k+\hlf}$ in a zone drops below $10^{12}\gcm$. The
chosen cut-off value is obviously specific to the core-collapse
supernova problem where outside of this density the neutrino pressure
gradients, in particular in the lateral direction, turn out to be
negligibly small.

\subsection{Discussion and tests of the numerical scheme}
\label{sec:codetest}

In the last subsection we described our implementation of an
approximative neutrino transport scheme for two-dimensional
configurations using spherical coordinates and azimuthal symmetry.

Besides adopting the approximations of general relativistic effects
from the spherically symmetric VERTEX code of \cite{ramjan02} we made
two major approximations of the transport equations. First, the
dependence of the specific intensity on the direction of propagation
$\vec{n}$ is replaced by a dependence on only one angle cosine
$\mu$. Secondly and closely related to the first approximation, we use
scalar variable Eddington factors.  These are obtained from the
solution of the one-dimensional transport equations on a spherically
symmetric image of the stellar background. Note, however, that our
treatment described here has been extended considerably by additional
terms (cf. Sects.~\ref{sec:2dmomequ}, \ref{sec:2dcoupling}) compared
to the simpler ray-by-ray transport scheme suggested in
\citet[][Sect.~3.8]{ramjan02}.

In the following we point out limitations of our approach and try to
critically assess their influence on the results obtained with our
method.

\subsubsection{Treatment of general relativity}
\label{sec:genrel}

The radial neutrino transport contains gravitational redshift and time
dilation, but ignores the distinction between coordinate radius and
proper radius. This simplification is necessary for consistently
coupling the transport code to our basically Newtonian
hydrodynamics. Of course, one would ultimately have to work in a
genuinely multi-dimensional GR framework which, among other
complications \cite[see e.g.~][]{carmez03}, entails abandoning the use
of the Lindquist metric.

Tests showed that in spherically symmetric simulations our
approximations seem to work satisfactorily well
\cite[]{lieram05,mardim05}, at least as long as the infall velocities
do not reach more than 10--20\% of the speed of light in decisive
phases of the evolution. Unless very extreme conditions are considered
(e.g.~very rapid rotation) gravity in supernovae is dominated by
radial gradients. We therefore expect that effects from a fully
multi-dimensional GR treatment are small and disregarding them is
acceptable in view of the other approximations made in the
multi-dimensional treatment of the transport. Some tests of
applicability of our 2D approximation of relativistic gravity can be
found in \cite{mardim05}.

\begin{figure}[!]
  \begin{tabular}{c}
    \put(0.9,0.3){{\Large\bf a}}
    \resizebox{0.9\hsize}{!}{\includegraphics{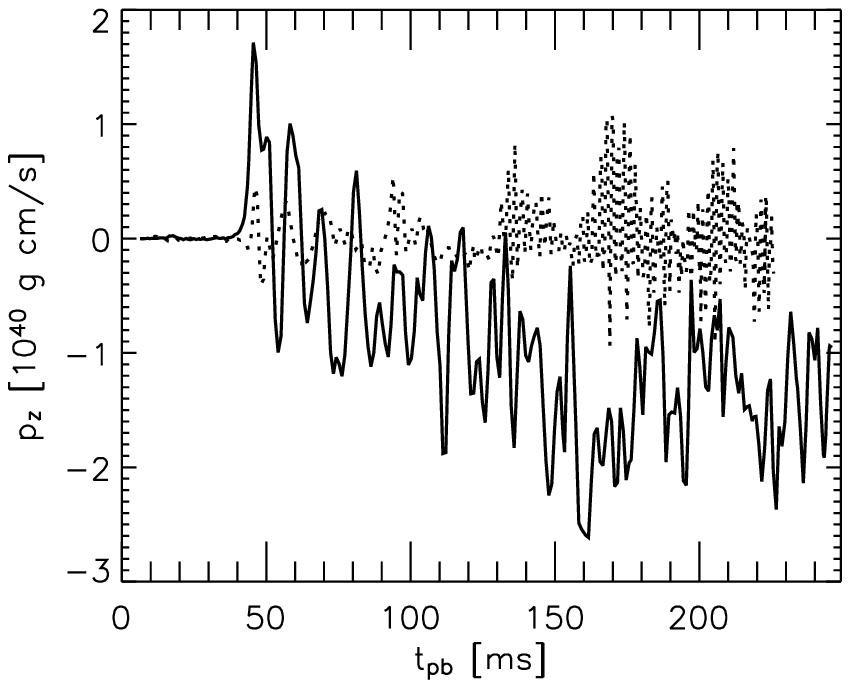}} \\ 
    \put(0.9,0.3){{\Large\bf b}}
    \resizebox{0.9\hsize}{!}{\includegraphics{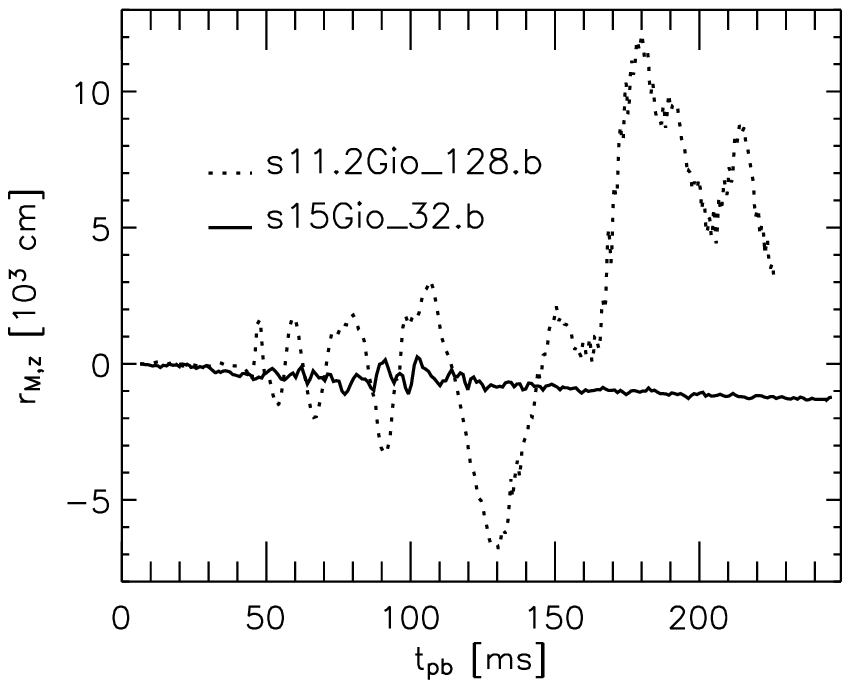}} 
  \end{tabular}

  \caption[]{
  {\bf a} Component of the linear momentum in polar axis direction of
  all matter on the computational grid. The solid line shows its time
  evolution for Model s15Gio\_32.b, in which a lateral wedge around
  the equator and with periodic boundary conditions was used (this
  model is described in detail in Sect.~\ref{sec:acme}). The dotted
  line belongs to Model s11.2Gio\_128.b, which is a model with a full
  180$\degr$ grid and thus with reflecting boundary conditions at the
  poles (in lateral direction); this model will be described in
  \cite{burram05:II}.
  {\bf b} Distance between the center of the grid and the center of
  mass for the same two models.
  }\label{fig:pz_viol}
\end{figure}

It should be noted that the effective relativistic gravitational
potential which we use in multi-dimensional simulations (see
Sect.~\ref{sec:hydro}) cannot guarantee strict conservation of
momentum (for a detailed explanation see \citealp{mardim05}). However,
in practical applications the size of the violation turns out to be
very small, see Fig.~\ref{fig:pz_viol}. In a 2D model with lateral
periodic boundary conditions, Model s15Gio\_32.b, the total momentum
of the matter on the computational grid (which should remain zero)
reaches moderate values (corresponding to velocities of at most
100km/s for a mass of $1\msol$). However, due to the periodic boundary
conditions the center of mass does not move significantly away from
the center of the grid because the matter which flows out of the grid
through one lateral boundary reappears at the opposite grid
boundary. In a model with lateral reflecting boundary conditions,
Model s11.2Gio\_128.b (in this model the lateral grid reaches from
pole to pole; this model will be described in detail in
\citealp{burram05:II}), no such shift of mass occurs, so that the
center of mass moves according to the total momentum. However, as can
be seen in Fig.~\ref{fig:pz_viol}, for this model the total momentum
remains much smaller than in Model s15Gio\_32.b (corresponding to
velocities of at most 50km/s for $1\msol$), and the time average of
the velocity is close to zero. Consequently, the distance between the
center of the grid and the center of mass remains far below 1km. This
behaviour is explained by the fact that the direction of the
artificial ``force'' creating the violation of momentum will usually
be from the center of mass towards the center of the grid. In summary,
the nonconservation of total momentum due to the employed treatment of
the effective relativistic gravitational potential remains a very
small perturbation which has no influence on our simulations.

\subsubsection{Lateral coupling and neutrino pressure gradients}
\label{sec:latadvec}

Including the terms printed in boldface in
Eqs.~(\ref{eq:momeqe1}--\ref{eq:momeqn2}) and taking into account the
acceleration of the stellar fluid in the lateral direction by neutrino
pressure gradients, we extend our non-equilibrium transport scheme
beyond a simple ray-by-ray description. This extension is consistent
with our fundamental assumption of azimuthal symmetry of the
intensity ${\cal I}$ (which implies $H_\vartheta \equiv 0$) and makes
sure that our transport description produces the correct behaviour in
important limiting cases.

First, in optically thick regions where neutrinos are tightly coupled
to the stellar fluid, neutrinos must be allowed to be carried along
with (laterally) moving fluid elements. This assures the conservation
of the total lepton number ($Y_\mathrm{lep} = \ye + Y_\nu$) in these
fluid elements in the absence of neutrino transport relative to the
medium ($DY_\mathrm{lep}/Dt = 0$ for the Lagrangian time
derivative). Below we will demonstrate that if this conservation is
violated by omitting the terms describing lateral advection,
radiation compression, and Doppler effects in the
neutrino moments equations, fluctuations in the lepton number are
artificially induced, which grow and trigger macroscopic fluid
motions, e.g., by buoyancy-driven rise of high-$\ye$
domains. Secondly, when neutrinos yield a significant contribution to
the pressure (as is the case in the dense interior of the hot, nascent
neutron star) the inclusion of lateral neutrino pressure gradients is
again important to prevent artificial acceleration of the fluid by
gradients of the gas pressure and to provide a restoring force which
damps the motion of the fluid when it transports neutrinos from one
place to another.

In contrast, the omission of angular flux components ($H_\vartheta =
0$) means the disregard of ``active propagation'' of neutrinos
relative to the stellar fluid. Although it is not clear that this is a
good assumption (for example, the effects of neutrino diffusion might
be underestimated, see below), it is unlikely to lead to fundamental
inconsistencies, because it is the correct physical limit for
situations where the opacity is very high. In the same spirit also
off-diagional terms of the neutrino pressure tensor ($P_{ij}$ with
$i\neq j$) can be dropped, implying that effects of neutrino viscosity
are ignored.

We point out here that another inconsistency is imported into our
treatment of 2D transport. The omission of angular flux components
causes the problem that the correct limit at large radii and small
optical depth may not be accurately reproduced because the evolution
of the radiation moments $J$ and $H$ even at large radius depends on
lateral gradient terms that include $\beta_\vartheta$. Fortunately,
the specifics of the supernova problem help us justifying our
approach: At large radii and low optical depth the lateral component
$v_\vartheta$ of the fluid velocity is usually small (in particular
relative to the radial component of the velocity, which determines
relevant timescales) so that all terms scaling with
$\beta_\vartheta=v_\vartheta /c$ become of minor importance. Secondly,
multi-dimensional effects of neutrino transport usually lead to a
neutrino distribution outside of the neutrinosphere, which is more
isotropic than the stellar structure (\citealp{livbur04}), an effect
which our transport treatment tends to underestimate anyway. Thirdly,
at large radii the non-cancelling $\beta_\vartheta$-terms are
suppressed by a factor $1/r$, and the small remaining lateral
redistribution of the neutrino flux is irrelevant because the coupling
between neutrinos and stellar medium (e.g.~neutrino heating and
cooling) vanishes.

\paragraph{Criterion for instability}

In the absense of neutrino diffusion the hydrodynamic stability in the
neutrino trapping regime of the PNS is tested by the Ledoux criterion,
\be
C_\mathrm{L} \equiv  \l(\frac{\partial\rho}
		          {\partial s}\r)_{Y,P} \frac{ds}{dr}
	  +	  \l(\frac{\partial\rho}
	                  {\partial Y}\r)_{s,P} \frac{dY}{dr}\,\,,
\label{eq:ledoux}\ee
with $s=s+s_\nu$ being the entropy \emph{including} the neutrino
entropy $s_\nu$, and $Y=Y_\mathrm{lep}=\ye + Y_\nu$ being the total
lepton number. Stability means $C_\mathrm{L}<0$. For 2D models
Eq.~(\ref{eq:ledoux}) is evaluated on a spherically symmetric image of
the star as defined in Sect.~\ref{sec:2deddfac}. The corresponding
Brunt-V\"ais\"al\"a frequency
\be
\omega_\mathrm{BV} \equiv \mathrm{sign}(C_\mathrm{L}) \sqrt{- \frac{g}{\rho}
 \l|C_\mathrm{L}\r|}\,\,,
\label{eq:omega_BV}\ee
where $g=-\mathrm{d}\Phi/\mathrm{d}r$ is the gravitational
acceleration, is closely related to $C_\mathrm{L}$ and denotes the growth
rate of fluctuations, if positive (unstable modes), and the negative
of the oscillation frequency, if negative (stable modes).

Recently, \cite{brural04} presented a more elaborate discussion of
hydrodynamic stability in the PNS including the effects of neutrino
diffusion (an extension of a previous analysis by
\citealp{brudin96}). They argue that local perturbations in the lepton
number will be reflected in the neutrino phase space and thus cause a
net neutrino diffusion which tries to wash out the perturbation, an
effect which can be accounted for by a ``response function''. Since
neutrinos also carry entropy the neutrino diffusion that smoothes the
lepton number perturbation will create an entropy perturbation. This
effect is characterized by a ``cross response function''. Of course,
entropy perturbations will analogously induce an equilibrating net
neutrino diffusion which at the same time carries lepton number,
corresponding to another ``response function'' and a ``cross response
function''. \cite{brural04} found in a numerical analysis that
perturbation-induced neutrino diffusion transports lepton number more
efficiently than entropy, and that the transport of lepton number
reacts faster to entropy perturbations than to lepton number
perturbations. For such a situation convective instability should set
in at most stellar conditions, even when the fluid is Ledoux
stable. In particular, \cite{brural04} describe two kinds of
instabilities in the presence of neutrino diffusion, ``lepto-entropy
finger'' (LEF) convection and ``lepto-entropy semi-convection''
(LESC).

Unfortunately, \cite{brural04} did not provide detailed information
about the values of the (cross) response functions in the different
regions of the PNS, and the calculation of these response functions
appears to be quite involved. Although one might use their approximate
values for the response functions as an estimate, \cite{brural04} show
that these depend on which interaction rates are taken into
account. They further mention that the existence of LESC is sensitive
to the exact values of the cross response functions. Due to these
uncertainties we refrain from applying their stability analysis here.
Instead, in order to account for efficient transport of lepton number
happening in convective 2D models between buoyant fluid elements and
their surroundings, we extend Eq.~(\ref{eq:ledoux}) to a
``Quasi-Ledoux'' criterion, following \cite{wilmay93} and
\cite{kei97}:
\be
C_\mathrm{QL} \equiv  \l(\frac{\partial\rho}
		          {\partial s}\r)_{\l<Y\r>,\l<P\r>}
                      \frac{d\l<s\r>}{dr}
	  +        \l(\frac{\partial\rho}
	                  {\partial Y}\r)_{\l<s\r>,\l<P\r>} \l(
                      \frac{d\l<Y\r>}{dr} -
                      \beta \frac{dY}{dr} \r )\,.
\label{eq:quasi_ledoux}\ee
Here, the brackets $\l<\r>$ denote angular averaging to emphasize that
the last term, $\mathrm{d}Y/\mathrm{d}r$, is evaluated
\emph{locally}. The additional parameter $\beta$ has been introduced
as a measure of the efficiency of the exchange of lepton number
between a perturbed fluid element and its surroundings. From
simulations it was found that values of $\beta\simeq 0.5$--$1.5$ are
typical, and that the criterion is not very sensitive to the exact
value of $\beta$ \cite[][]{kei97}. Therefore, we will use $\beta=1$.

Different from convection inside the PNS (i.e.~below the
neutrinosphere) exchange of entropy and lepton number by neutrino
transport between moving fluid elements and their surroundings does
not play an important role in the gain layer, where neutrinos and
stellar fluid are much less tightly coupled. Therefore we apply
Eq.~(\ref{eq:ledoux}) using the medium entropy $s$ and $Y=\ye$ when
evaluating the stability criterion, or the Brunt-V\"ais\"al\"a
frequency, at densities lower than $10^{12}\gcm$, considering the
latter as the density below which neutrinos begin decoupling. 

\paragraph{Stability analysis}

In order to assess the importance of the lateral gradient terms which
we take into account in our neutrino-hydrodynamics treatment, we
performed three test calculations which were started by imposing
random seed perturbations in the density with a maximum amplitude of
$\pm 2.5\%$ on an early post-bounce model which was taken from a
spherically symmetric simulation.  The perturbed model was then
evolved for a few milliseconds (corresponding to a multiple of the
relevant dynamical timescale) in order to test different variants of
the 2D transport code described above.

\begin{figure}[!]
  \begin{tabular}{c}
    \put(0.9,0.3){{\Large\bf a}}
    \resizebox{0.9\hsize}{!}{\includegraphics{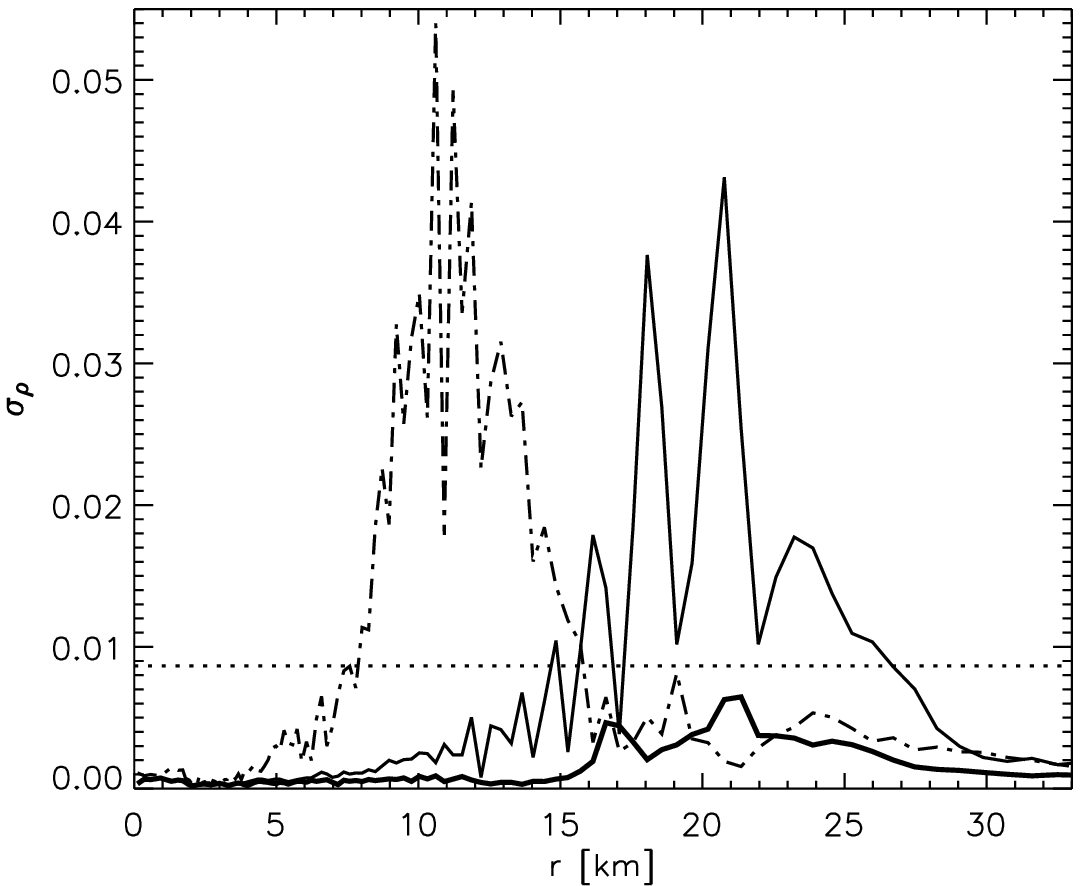}} \\ 
    \put(0.9,0.3){{\Large\bf b}}
    \resizebox{0.9\hsize}{!}{\includegraphics{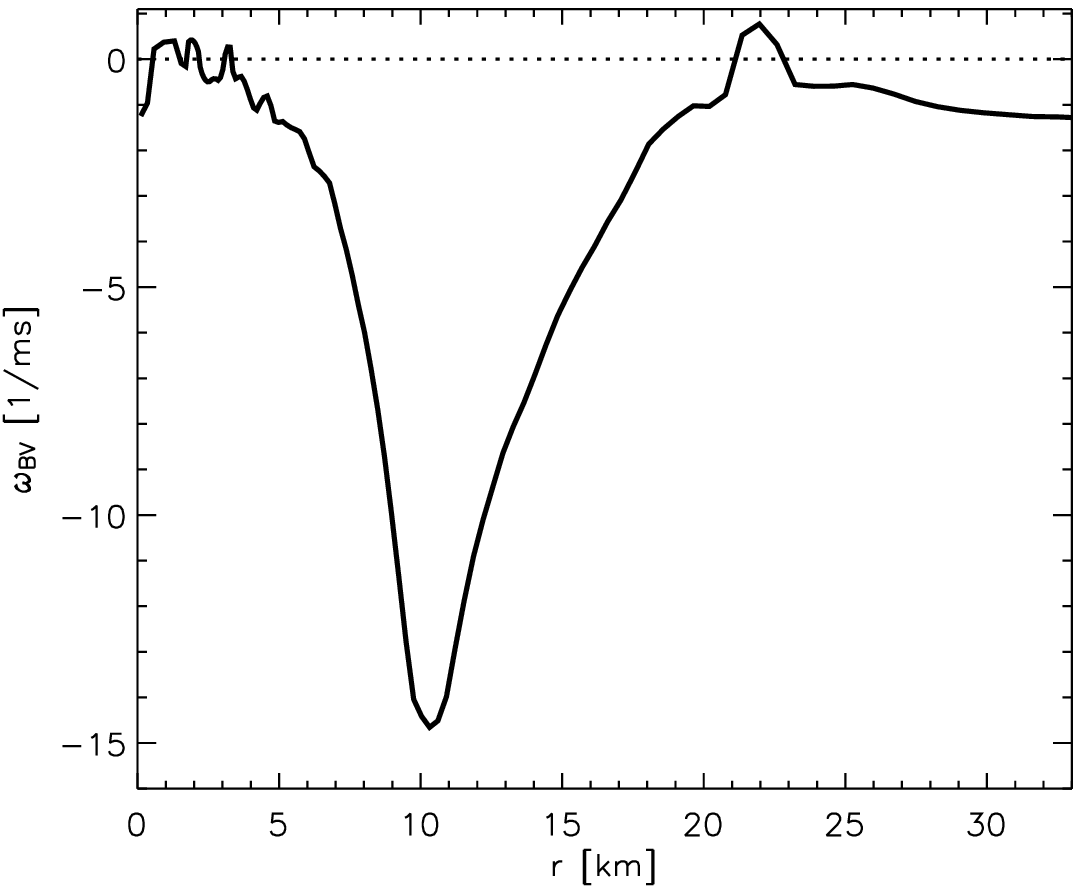}} 
  \end{tabular}

  \caption[]{
  {\bf a} Standard deviation of the density $\sigma_\rho$ (see
  Eq.~\ref{eq:sigma}) indicating convective activity inside the
  neutron star. The 1D Model s15Gio\_1d.b was mapped to 2D (16 zones
  with a resolution of $2.7\degr$) $27\mathrm{ms}$ after core bounce
  and the density distribution $\rho$ was perturbed. The plot shows
  the situation after 3.6ms of dynamical evolution computed with
  different implementations of the 2D transport equations. The dotted
  line is the initial value of $\sigma_\rho$, the thick solid line
  shows the standard deviation $\sigma_\rho$ when the lateral terms
  are included in our scheme as described in Sects.~\ref{sec:2dmomequ}
  and \ref{sec:2dcoupling}.  For comparison, the thin solid line shows
  $\sigma_\rho$ when running with pure ray-by-ray transport (i.e., 
  without all boldface terms in Eqs.~(\ref{eq:momeqe1}--\ref{eq:momeqn2})
  and without the lateral component of neutrino pressure gradients), and
  the dash-dotted line corresponds to a simulation where the lateral terms
  of Sect.~\ref{sec:2dmomequ} and Appendix~\ref{app:momeq3d} 
  were taken into account but not the
  neutrino momentum transfer to the fluid discussed in
  Sect.~\ref{sec:2dcoupling}. {\bf b} Brunt-V\"ais\"al\"a frequency
  for the same model at the beginning of the test calculations,
  derived from the Quasi-Ledoux criterion,
  Eq.~(\ref{eq:quasi_ledoux}), using $s=s+s_\nu$ and
  $Y=Y_\mathrm{lep}$. Negative $\omega_\mathrm{BV}$ indicates
  convectively stable regions.
  }\label{fig:ledoux}
\end{figure}

Figure \ref{fig:ledoux}a shows the standard deviation of density
fluctuations in lateral direction,
\be
\sigma_{\rho}(r) \equ \sqrt{ \frac{1}{N} \sum_k
  \Delta\mathrm{c}\vartheta_k \cdot
  \l[\frac{\rho_k(r)-\l<\rho(r)\r>_\vartheta}{\l<\rho(r)\r>_\vartheta}\r]^2}
\,,\label{eq:sigma}\ee
where $\Delta\mathrm{c}\vartheta_k \equ
\cos(\vartheta_{k-\ot})-\cos(\vartheta_{k+\ot})$, $N \equ \sum_{k'}
\Delta\mathrm{c}\vartheta_{k'}$, and $\l<\rho(r)\r>_\vartheta \equ
{{1}\over{N}} \sum_{k'} \rho_{k'}(r) \Delta\mathrm{c}\vartheta_{k'}$, as
a function of radius after 3.6 ms of 2D evolution. The quantity
$\sigma_{\rho}$ serves as a convenient measure to specify the
magnitude of the density fluctuations.  Employing the full
implementation of our neutrino treatment, which takes into account
both lateral gradients in the neutrino moments equations as well as
lateral gradients of the neutrino pressure in the fluid equations, the
initial perturbations do not grow to large instabilities anywhere
inside the PNS, which is in accordance with the prediction by our
Quasi-Ledoux criterion. The residual fluctuations which can be
discerned at radii between 15km and 27km seem to be of physical origin
because the Quasi-Ledoux criterion predicts instability in part of
this region, see Fig.~\ref{fig:ledoux}b.

When switching off the effects of the lateral component of the
neutrino pressure gradient we notice a strong amplification of the
initial perturbations in a region where the damping should be
strongest according to the Quasi-Ledoux criterion. A naive ray-by-ray
scheme, which in addition to lateral neutrino pressure differences
also disregards the lateral gradients in the neutrino moments
equations (including those terms which account for the lateral
advection of neutrinos), produces spurious convective activity in a
broad region between 15km and 25km. The latter phenomenon is not
Ledoux convection because it transports lepton number and entropy in
the wrong direction (i.e.~against the gradients). While lateral fluid
motions (adequately modelled by the two-dimensional hydrodynamics
scheme) tend to damp fluctuations, the resulting advection of electron
number produces artificial flucuations of lepton number and therefore
obviously instigates buoyancy motions, if the advection of electrons
is not accompanied by the corresponding advection of neutrinos
(i.e.~if the total lepton number is not preserved in the moving fluid
elements even when diffusion of neutrinos is unimportant). This is the
case when the terms accounting for lateral advection in the neutrino
moments equations are neglected. Neutrino pressure differences, on the
other hand, produce restoring forces which help damping lateral
motions after local fluctuations have been reduced by fluid advection.

We have not attempted to perform a very detailed analysis of all
effects of lateral advection and pressure gradients of neutrinos to
elaborate our understanding beyond the more qualitative insights
described above. However, we interprete our tests as a demonstration
that approximations of multi-dimensional transport schemes must be
tested carefully for the possibility of producing spurious convective
activity (or suppression of convection) in the newborn neutron star,
where neutrinos contribute significantly to the total pressure and
total lepton number density.

\subsubsection{Lateral propagation of neutrinos}

A number of our two-dimensional supernova simulations, in particular
those which produced lively hot-bubble convection, showed transient
neutrino-bursts when narrow downflows of accreted stellar gas entered
the cooling region and penetrated down to the vicinity of the
neutrinosphere. In our simulations such bursts occurred roughly every
20~ms, typically persisting for a few ms. The neutrinos of such
bursts, which are the result of locally enhanced neutrino emission,
would naturally propagate in all directions in a fully
multi-dimensional treatment and would therefore illuminate the
surroundings in all directions. Our code ignores the lateral
propagation of these neutrinos, thus the burst is radiated away in the
radially outward direction and its lateral width is essentially
constrained to the layers above the hot, radiating area. Here we shall
argue that although locally and transiently the neutrino heating rates
can be incorrect by up to a factor of two (in very rare, extreme
cases) due to the disregard of lateral neutrino propagation, the
transfer of energy between neutrinos and stellar gas is not
significantly changed on larger spatial and temporal scales, and
therefore the global dynamics of our supernova simulations is not
likely to be affected significantly.

Basically, our simplified neutrino transport overestimates the heating
in the radial direction at polar latitudes where ``hot spots'' occur,
while adjacent ``rays'' experience less heating than in a 2D neutrino
transport which includes lateral neutrino fluxes. Truly
two-dimensional transport tends to redistribute neutrinos in lateral
direction, in particular in the semi-transparent and transparent
regimes where the mean free path becomes large. This can lead to a
more uniform spatial distribution of neutrinos exterior to the
neutrinosphere than in case of our ray-by-ray treatment, in particular
in the presence of local hot spots \cite[see also the discussion
in][]{livbur04}.

To test the implications of such a lateral neutrino redistribution for
the neutrino heating behind the shock, we performed a post-processing
analysis in which we averaged the frequency-dependent neutrino
densities over all polar angles at a given radius and time and used
the result to recalculate the local net heating rates. Then we
compared the recalculated heating rate with the actual heating rate of
the simulation, both integrated over the respective gain layer, for
different times. Hereby the gain layer was defined for each grid value
of $\vartheta$ separately as the region between the shock and the gain
radius, i.e.~the innermost radial point where net heating occurs. For
this purpose, the position of the gain radius was also redetermined
for the recalculated heating rates.

\begin{figure*}[!]
\sidecaption
\centering
\includegraphics[width=10cm]{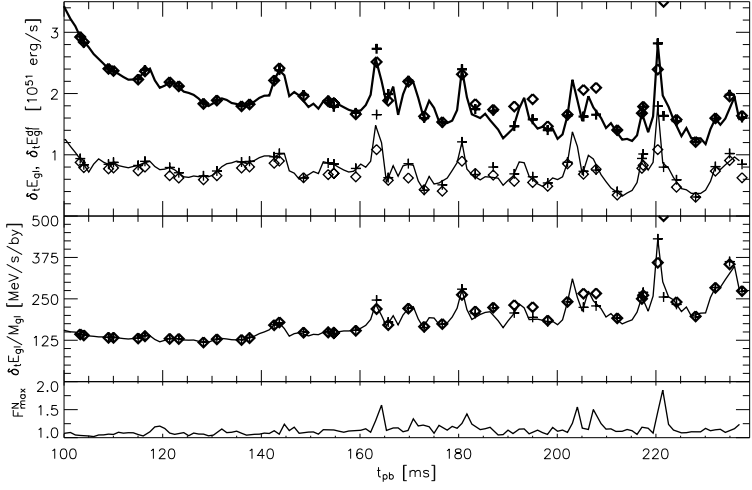} 
  \caption[]{
  The upper panel compares total (volume integrated) neutrino heating
  rates in the gain layer $\delta_t E_\mathrm{gl}$.  The bold upper solid
  line shows the evolution of $\delta_t E_\mathrm{gl}$ for the
  $11.2\msol$ supernova model of \cite{burram03}. The post-processed
  heating rate obtained by averaging the neutrino densities over all
  latitudes is represented by diamond-like symbols.  For a consistency
  check crosses are drawn for the post-processed neutrino heating rate
  without using the lateral averages of the specific neutrino
  density. Ideally the latter post-processed results should be equal
  to the values returned from the model run. Remaining differences
  result from the fact that the post-processing may yield a gain
  radius shifted by one radial zone. For drawing the second curve
  (thin solid line) and the corresponding symbols the analysis was
  restricted to downflows in the gain layer, i.e.~for regions with
  velocity $v<1.5\l<v\r>_{\vartheta}<0$.
  The middle panel shows the average net heating rate per baryon (the
  symbols have the same meaning as in the upper panel), i.e.~the total
  neutrino heating rate in the gain layer divided by the total mass
  $M_\mathrm{gl}$ contained in this region.  The lower panel shows the
  maximum neutrino flux as measured slightly above the gain radius
  normalized to the flux average over all latitudes at the same
  radius, $F_\mathrm{max}^\mathrm{N} = \max_{\vartheta} \l(F\r) /
  \l<F\r>_\vartheta$. The values indicate the relative strength of
  localized luminosity outbursts.
  }\label{fig:average}
\end{figure*}

We have carried out this analysis for an $11.2\msol$ progenitor
\cite[s11.2,][]{wooheg02} computed with 32 angular zones and $\sim
90\degr$ lateral wedge during its post-bounce evolution
\cite[][]{burram03}, a model which showed powerful hot bubble
convection. The evaluation was started at
$t_\mathrm{pb}=100\mathrm{ms}$; before that time hot bubble convection
was weak and no strong local bursts of accretion luminosity
happened. We found that the angular averaging of the neutrino
densities hardly changes the total net heating rate in the gain layer,
$\delta_t E_\mathrm{gl}$, see Fig.~\ref{fig:average}, upper panel. At
very few instants of evaluation we discovered a significant increase
of $\delta_t E_\mathrm{gl}$ by at most 30\%. In the temporal average
this difference is reduced to only a few percent.  For the average net
heating per baryon in the gain layer (Fig.~\ref{fig:average}, middle
panel), which, concerning global supernova dynamics, we consider to be
the more decisive quantity \cite[][]{jan01}, the values for the two
heating rates are almost indistinguishable.

Interestingly, the differences are somewhat larger if we separately
analyze downflows of cold material and hot bubbles in the gain layer;
a zone of the numerical grid is attributed to a downflow if the
negative velocity is more than 1.5 times the angle-averaged (negative)
velocity $\l<v\r>_{\vartheta}$ at a given radius $r$, otherwise we
define the zone to belong to a high-entropy bubble. For the downflows
(see also Fig.~\ref{fig:average}, upper panel, lower curve), the
time-averaged total net heating decreases by 14\% in case of the
angle-averaged neutrino density. On the other hand it increases by
12\% for the high-entropy bubbles.

The results from this analysis should be read carefully: It does not
take into account the dynamical consequences of the altered heating
rates. Also, the angle-averaging of the neutrino density overestimates
the spreading of the burst neutrinos in all directions, especially
close to the gain radius, where we obtain by far most of the
heating. The nevertheless modest sensitivity of the integral
quantities to the described averaging of the neutrino distribution,
however, gives us some confidence that our approximation of the
moments equations is a reasonable step towards fully consistent
multidimensional simulations.

Even more difficult to assess quantitatively is the effect of
replacing the Eddington tensors (c.f. Appendix \ref{app:momeq3d}) by
scalar variable Eddington factors which are calculated from a solution
of the one-dimensional transport equations using a spherically
symmetric image of the star. We have to rely on the fact that these
variables are normalized moments of the neutrino phase space
distribution and thus, in the absence of persistent global deformation
of the star, should not show significant variation with the angular
coordinate \cite[cf.][]{ram00,ramjan02}. Anyway, as we have seen,
ignoring the lateral propagation of neutrinos does not seem to largely
affect the neutrino-matter coupling in the semi-transparent
region. Further, \cite{thoqua05} have recently shown that neutrino
viscocity, which is connected with the non-diagonal elements of the
Eddington tensors, is of minor importance. Therefore we hope that the
transport is also sufficiently insensitive to our approximation
concerning the variable Eddington factors. Final answers can only be
expected from simulations employing a fully multidimensional treatment
of the neutrino transport.

\subsection{Equation of state}
\label{sec:eos}

All calculations presented in this paper were run with the equation of
state (EoS) described in detail in \citet[][Appendix
B]{ramjan02}. The EoS of \cite{latswe91} (LS EoS) is employed to treat
matter at densities above $6\times 10^7\gcm$. Recently it became
clear that this EoS underestimates the fraction of alpha particles
(due to an error in the definition of the alpha particle rest mass;
J.~Lattimer, personal communication) and concerns were expressed that
deficiencies of this EoS in the density regime between $10^8\gcm$
and $10^{11}\gcm$ could have consequences for the possibility of
an explosion, which we fail to get with the LS EoS (C.~Fryer, personal
communication).

In order to test this we have implemented a new EoS in our code. This
EoS is similar to the low-density EoS \cite[][]{jan99} described by
\citet[][Appendix B]{ramjan02}, but the very approximative handling
of nuclear dissociation and recombination in nuclear statistical
equilibrium (NSE) in that EoS is now replaced by a composition table
in the $(\rho,T,\ye)$-space. This table was computed assuming matter
to be in NSE and solving the Saha equations for equilibrium of a
mixture of free neutrons and protons, $^4$He, and $^{54}$Mn as a
representative heavy nucleus \cite[][]{jan91}. We use the table for
$T>5\times10^9$K, where NSE is a reasonably good assumption. Given the
nuclear composition, variables such as pressure and entropy can be
calculated from the EoS of \cite{jan99}.

\begin{figure}[!]

  \resizebox{0.9\hsize}{!}{\includegraphics{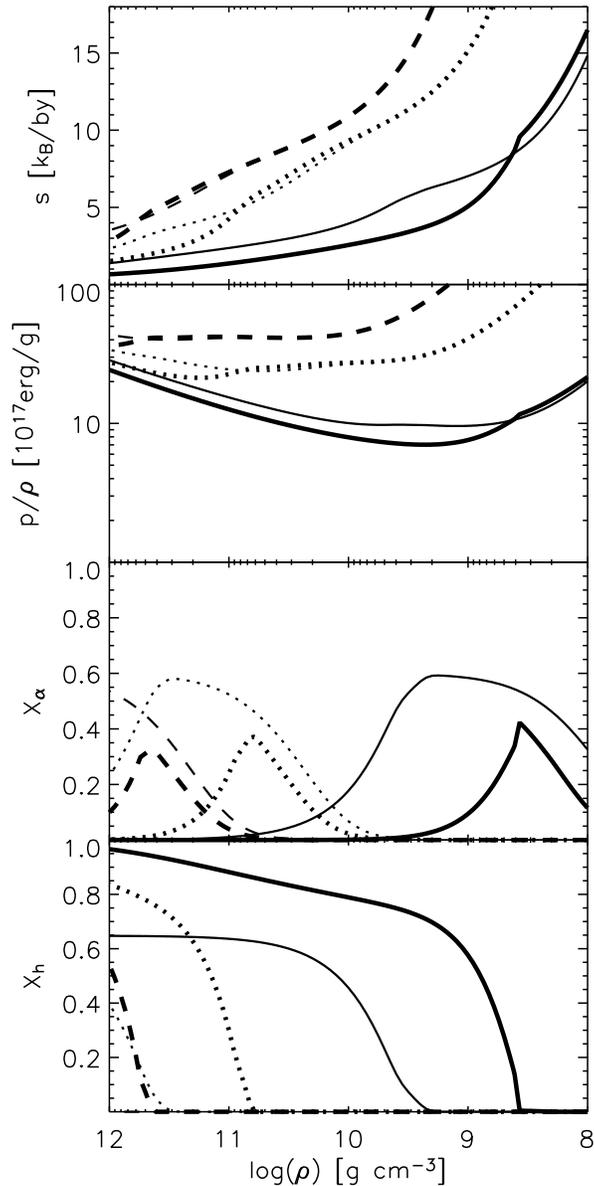}} 

  \caption[]{
  Comparison of the LS EoS (thick lines) with the four-species NSE
  table introduced here (thin lines), for $\ye=0.3$ and $T=1$ (solid),
  2 (dotted), and 3 (dashed) MeV. $X_\alpha$ and $X_\mathrm{h}$ are the
  mass fractions of $\alpha$-particles and of the representative heavy
  nucleus, respectively.
  }\label{fig:kokls_stat}
\end{figure}

A comparison of this simple 4-species NSE table with a sophisticated
NSE solver that takes into account 32 different species of heavy
nuclei showed excellent agreement in entropy, pressure, and helium
mass fraction for typical conditions met in the post-shock layer. Note
that the 4-species table overestimates the mass fraction of free
nucleons in parameter regions where heavy nuclei dominate the
composition and $\ye$ is far from $25/54\simeq 0.463$ (e.g.~for
$T=1$MeV, $\ye=0.3$ this is the case at $\rho>3\times10^{10}\gcm$,
see Fig.~\ref{fig:kokls_stat}). This problem originates from the
restriction to one fixed, representative heavy nucleus and the
necessity to fulfill charge neutrality. However, NSE in the post-shock
layer of a supernova is characterized by conditions where essentially
all heavy nuclei are dissociated into free neutrons, protons, and
$\alpha$-particles.

Comparing our composition table with data from the LS EoS
(Fig.~\ref{fig:kokls_stat}) we indeed observe a significantly larger
helium mass fraction for certain combinations of density and
temperature. Moreover, the region where helium contributes
significantly to the composition is larger. This also affects the
pressure and entropy, especially at low temperatures and entropies
(Fig.~\ref{fig:kokls_stat}).

To test whether the sizeable differences between the EoS using our NSE
table at densities below $10^{11}\gcm$ and the LS EoS have any
influence on supernova simulations we ran the spherically symmetric
$11.2\msol$ model of \cite{burram03} with both equations of
state. This comparison directly tests the influence of the different
EoS implementations on the dynamical evolution of the supernova.

\begin{figure}[!]
  \resizebox{0.9\hsize}{!}{\includegraphics{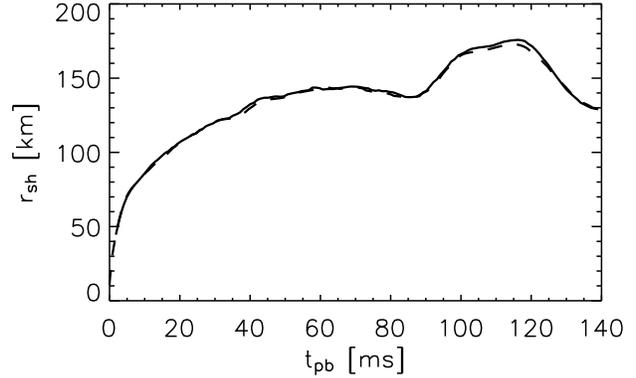}} 
  \caption[]{
  Shock positions for the two dynamical simulations of the post-bounce
  evolution of an $11.2\msol$ star with the LS EoS (dashed) and the
  four-species NSE table (solid) used in the density regime below
  $10^{11}\gcm$.
  }\label{fig:eos_sh}
\end{figure}

\begin{figure}[!]

  \resizebox{0.9\hsize}{!}{\includegraphics{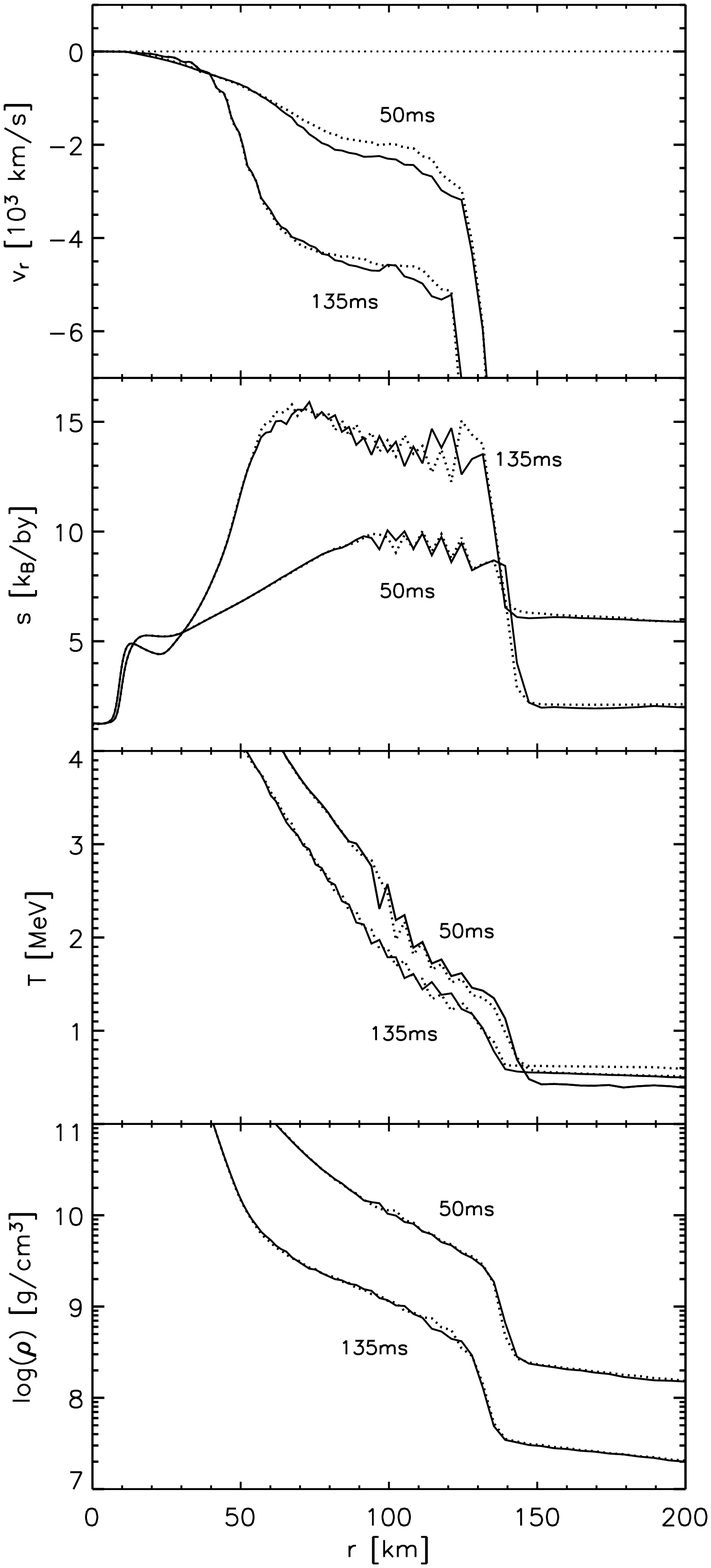}} 

  \caption[]{
  Comparison of the profiles for the two dynamical simulations of the
  post-bounce evolution of an $11.2\msol$ star with the LS EoS
  (dotted) and the four-species NSE table (solid) used in the density
  regime below $10^{11}\gcm$ at two post-bounce times.
  }\label{fig:eos_cp}
\end{figure}

\begin{figure}[!]
  \resizebox{0.9\hsize}{!}{\includegraphics{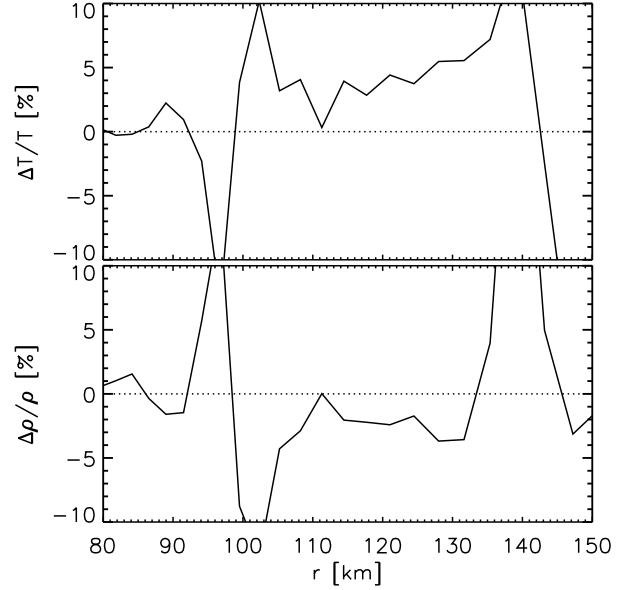}} 
  \caption[]{
  Relative difference of temperature and density between the
  simulations described in Fig.~\ref{fig:eos_cp} at $t=50$ms
  post-bounce. $\Delta T>0$ means higher $T$ in the simulation with
  the four-species NSE table (solid line in Fig.~\ref{fig:eos_cp})
  used in the density regime below $10^{11}\gcm$, dito for $\Delta
  \rho$.
  }\label{fig:eos_delt}
\end{figure}

\begin{figure}[!]
  \resizebox{0.9\hsize}{!}{\includegraphics{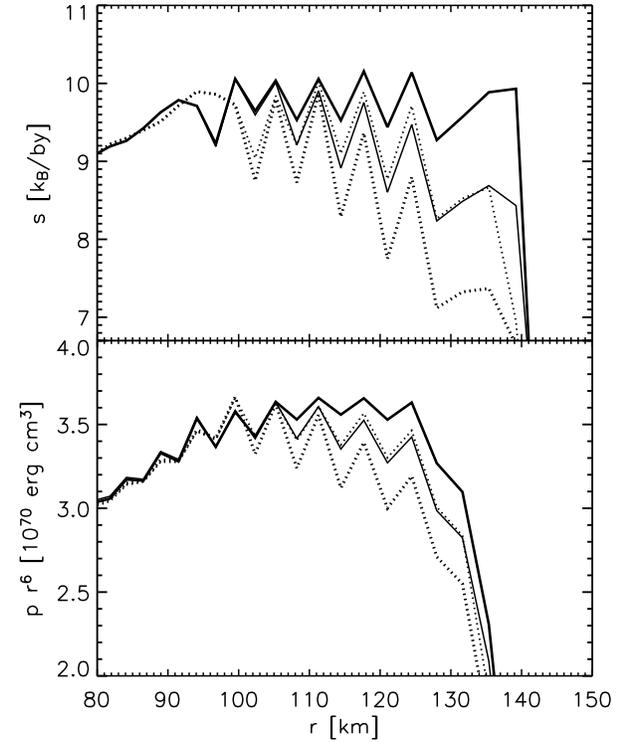}} 
  \caption[]{
  Enlarged comparison of the entropy and the pressure (renormalized
  with $r^6$) between the two simulations shown in
  Fig.~\ref{fig:eos_cp} at $t=50$ms. The thin lines correspond to the
  actual values, the thick lines correspond to a post-processing
  analysis of the profiles using the respectively other variant of the
  EoS.
  }\label{fig:eos_cc}
\end{figure}

Surprisingly, both runs showed virtually no difference in the
post-bounce history of the supernova (Fig.~\ref{fig:eos_sh}) and only
small differences in the post-shock structure (Figs.~\ref{fig:eos_cp}
and \ref{fig:eos_delt}), even though a post-processing analysis where
values of $T$, $\rho$, $\ye$ were crosswise fed into the two variants
of the EoS revealed differences in entropy and pressure of up to 15\%
and 10\%, respectively, below the shock (Fig.~\ref{fig:eos_cc}). This
phenomenon has a simple explanation: On the one hand the relatively
high entropies behind the shock in our models imply that the influence
of $\alpha$-particles is in general moderate
(Fig.~\ref{fig:kokls_stat}). On the other hand the accretion layer
behind the supernova shock can be considered as an approximately
hydrostatic situation where the velocities of the shocked medium and
of the shock itself are much smaller than the sound speed
$c_\mathrm{s}$. In this case the pressure profile behind the shock is
determined by constraining boundary conditions, i.e.~the mass and
radius of the proto neutron star (and therefore its gravitational
field) on the one hand, and the mass accretion rate at the shock on
the other \cite[][]{jan01}. Keeping in mind that the Rankine-Hugoniot
conditions connect the flow in the infall region with the conditions
in the post-shock layer, the pressure profile in the accretion layer
is fully determined.

Thus, for a given pressure profile other variables, such as $\rho$ and
$T$, may vary for different EoSs (see Fig.~\ref{fig:eos_delt}) without
changing the dynamical evolution of the supernova. Only if these
variables have an immediate influence on the evolution, e.g.~by
temperature-dependent neutrino emission, do we have to worry about the
differences and consequences of the composition. Neutrino heating,
however, is dynamically not very important in the gain layer of our
non-exploding models, while in the cooling region the composition
differences between the EoSs are negligibly small.

We conclude that, although the LS EoS has problems with the
$\alpha$-mass fraction and other dependent variables in the
low-density regime at supernova conditions, we do not find a major
effect on the dynamics of our simulations.

o%
\begin{figure}[!]
  \begin{tabular}{c}
    \put(0.9,0.3){{\Large\bf a}}
    \resizebox{0.9\hsize}{!}{\includegraphics{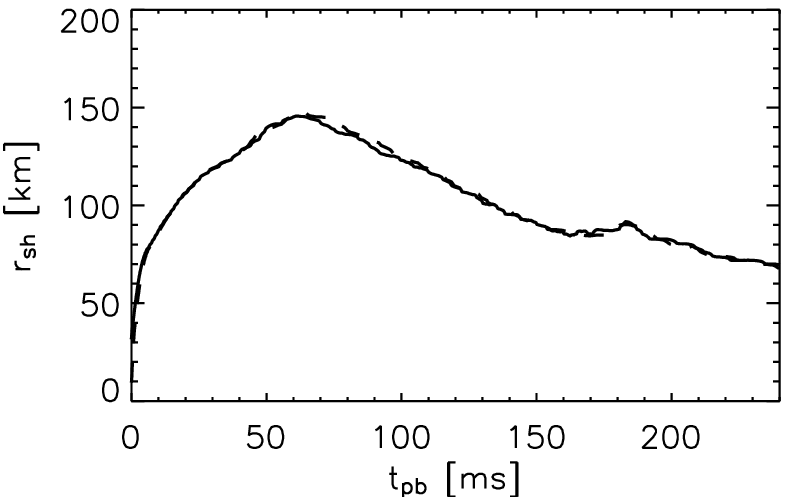}} \\ 
    \put(0.9,0.3){{\Large\bf b}}
    \resizebox{0.9\hsize}{!}{\includegraphics{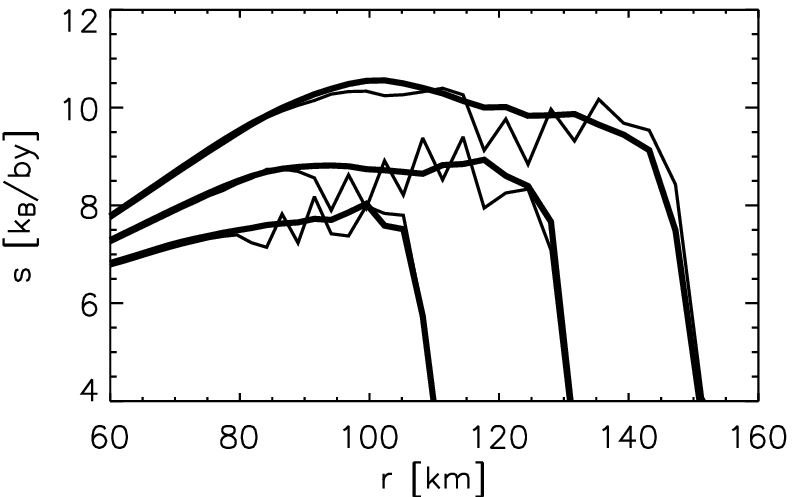}} 
  \end{tabular}
  \caption[]{
  {\bf a} Shock trajectories of Model s15Gio\_1d.b, described in
  Sect.~\ref{sec:std} (solid), and a calculation with identical
  micro-physics but with our new definition of the energy $e$
  (dashed). Note that the visible very small differences actually   
  originate from slight differences of the gravitational potentials  
  used in both simulations. {\bf b} Entropy profiles for Model
  s15Gio\_1d.b (thin) and the simulation with our new definition of
  the energy $e$ (thick) at the times 20ms, 40ms, and 60ms after
  bounce (from bottom to top).
  }\label{fig:ent_wig}
\end{figure}
\begin{figure}[!]

  \resizebox{0.8\hsize}{!}{\includegraphics{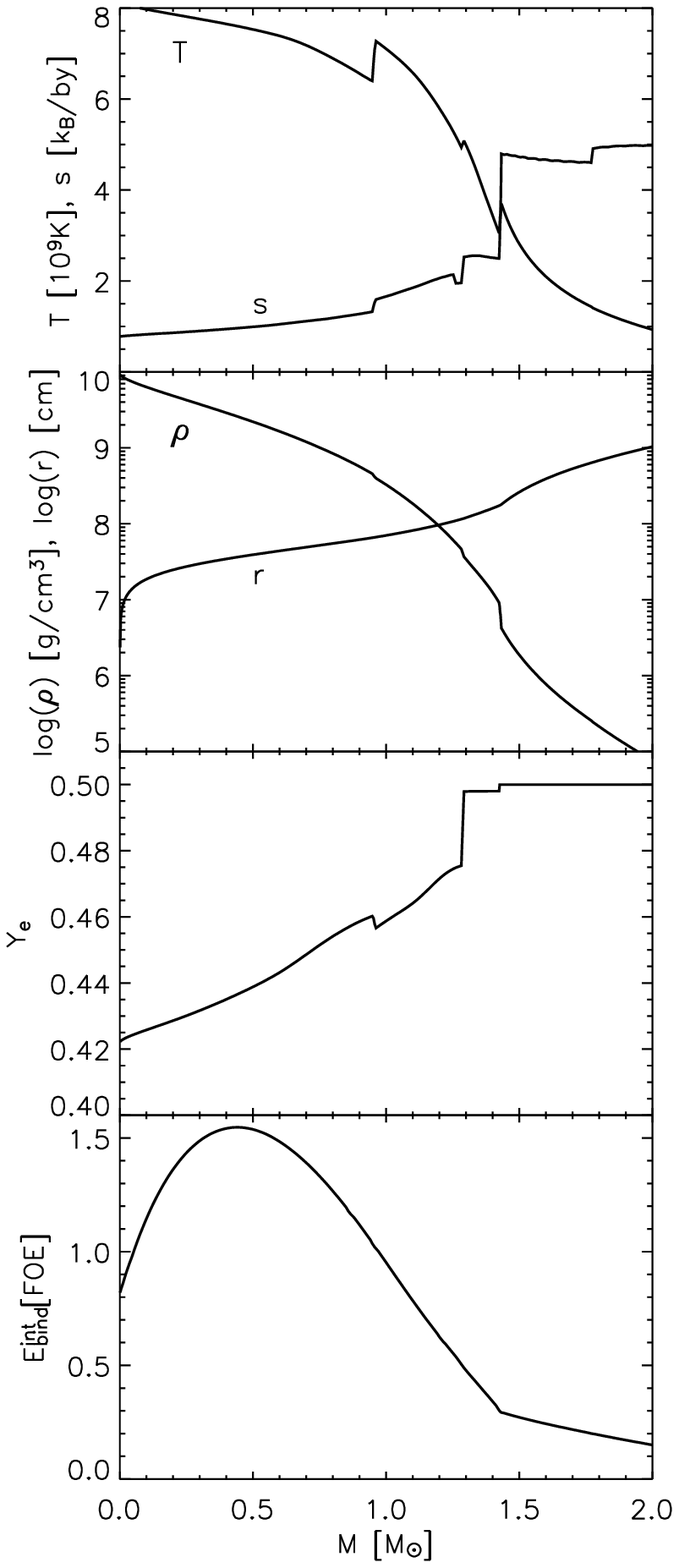}} 

  \caption[]{
  Progenitor structure. Temperature $T$, entropy per baryon $s$,
  electron fraction $\ye$, density $\rho$, radius $r$, and shell
  binding energy $E_\mathrm{bind}^\mathrm{shell}$, defined in
  Eq.~(\ref{eq:eexpl_expl_def}), as functions of the enclosed mass. At   
  $1.42~\msol$ the interface between the silicon shell and the
  oxygen-rich silicon shell is located. $T$, $\ye$, and $\rho$ are
  taken from model s15s7b2 of \cite{woowea95}, while the $s$ and
  $E_\mathrm{bind}^\mathrm{shell}$ are computed by applying our EoS.
  }\label{fig:prog_prof}
\end{figure}

\subsection{Entropy wiggles}
\label{sec:swiggles}

All the models presented here feature a numerical problem, the origin
of which we found only recently. In all profiles of 1D calculations,
and also partly in 2D calculations (when convective effects did not
wash out the effect), artificial wiggles appear behind the
shock. These wiggles are most pronounced in the entropy and pressure
profiles with amplitudes of up to 10\%, see
e.g.~Fig.~\ref{fig:eos_cc}.

The problem is connected with the use of the specific energy 
$\varepsilon = e + e_\mathrm{kin}$ in the hydrodynamics equations
(\ref{eq:hydro.rho}--\ref{eq:hydro.x_k}), which are solved in the
PROMETHEUS code with a Riemann solver technique, see Sect.~\ref{sec:hydro},
and the corresponding calculation of the required index 
$\widetilde{\Gamma}_e \equiv {{p}\over{e\rho}} +1$ with an
arbitrarily normalized energy $e$
instead of the internal energy $e_\mathrm{int}$. The Riemann
solver expects $\widetilde{\Gamma}_e$ to be calculated 
with the internal energy $e_\mathrm{int}$, because for thermodynamical
consistency between $\widetilde{\Gamma}_e$ and the adiabatic index
$\Gamma_{\mathrm{ad}} = ({\mathrm{d}}\ln p/{\mathrm{d}}\ln \rho)_s$,
the energy $e$ and the pressure $p$ should go to zero simultaneously.
On the other hand, the physics should only depend on energy differences,
and the energy normalization should not play a role, i.e., the EoS
as well as the hydrodynamics should remain unchanged, independent of 
the chosen zero point of the energy. Therefore it is allowed to add an
arbitrary normalization to $e$, still retaining the validity of
Eqs.~(\ref{eq:hydro.rho}--\ref{eq:hydro.x_k}). In our code, we
use the definition
\be
e = e_\mathrm{int} + e_\mathrm{rm} + e_0 \ ,
\label{eq:def_e_code}\ee
where $e_\mathrm{rm}$ is the specific baryon and electron rest mass
energy and $e_0$ is a constant offset chosen to be in accordance with
the energy definition in our high-density EoS provided by
\cite{latswe91}. 

$\widetilde{\Gamma}_e$ as defined above is therefore different from the
thermodynamical index $\Gamma_e = {{p}\over{e_\mathrm{int}\rho}} +1$, 
which is expected to be fed into the Riemann
solver. In particular, in regions with relatively low internal energy
and large value of ($e_\mathrm{rm} + e_0$) (e.g., just behind the shock 
where the matter consists mostly of free nucleons),
$\widetilde{\Gamma}_e$ with our chosen normalization of $e$ can have values
which are significantly below the physically meaningful lower bound of
4/3. Empirically, we found that the numerical noise appears whenever
$\widetilde{\Gamma}_e < 4/3$, which turned out to be the case in
the post-shock region. However, we did not make efforts to 
locate the source of these numerical fluctuations more precisely 
than linking it to the use of a Riemann solver technique in cases 
where $\widetilde{\Gamma}_e$ adopts unphysical values.

We have solved this problem by renormalizing the energy 
used in the integration scheme of the hydrodynamics equations relative
to $e$ by subtracting from $e$ the additional terms $e_\mathrm{rm} + e_0$
of Eq.~(\ref{eq:def_e_code}). The numerical procedure is
described in detail in Appendix \ref{app:noff}. In a test calculation
with this improved handling, the wiggles turned out to disappear, see
Fig.~\ref{fig:ent_wig}b.

Figure \ref{fig:ent_wig} also shows that the dynamics and global
evolution of the models are not affected by the wiggles. This confirms
that the hydrodynamics solutions given by PROMETHEUS are not globally
flawed even if $e$ from Eq.~(\ref{eq:def_e_code}) and $e_\mathrm{int}$
deviate significantly from each other. Therefore the simulations
presented in this paper are sound despite the wiggles in the postshock
profiles (see also \citealp{lieram05} for a confirmation). Of course,
future calculations will be performed using the improved treatment of
Appendix \ref{app:noff}.

%
\begin{table*}[htb]
\centering
\begin{minipage}[t]{12cm}
  \caption{Input physics for the sample of computed models presented
  here.}
  \begin{tabular}{lllllll}
  \hline\hline
  \it Model & \it Dim. & \it Gravity & $\nu$ \it Reactions & \it
      Transport & \it Wedge$^\mathrm{a}$ & \it $\vartheta$-zones\cr
  \hline
  s15Nso\_1d.b        & 1D & Newtonian  & standard       & Case B  &
       &    \cr
  s15Gso\_1d.b        & 1D & approx. GR & standard       & Case B  &
       &    \cr
  s15Gso\_1d.b$^\ast$ & 1D & approx. GR & standard$^\mathrm{b}$   & Case B  &
           &    \cr
  s15Gio\_1d.a        & 1D & approx. GR & improved$^\mathrm{c}$   & Case A  &
                &    \cr
  s15Gio\_32.a        & 2D & approx. GR & improved       & Case A  & $\pm 43.2\d
egr$& 32 \cr
  s15Gio\_1d.b        & 1D & approx. GR & improved       & Case B  & 
       &    \cr
  s15Gio\_32.b        & 2D & approx. GR & improved       & Case B  & $\pm 43.2\d
egr$& 32 \cr
  \hline
  \end{tabular}
  \begin{list}{}{}
    \item[$^\mathrm{a}$] Angular wedge of the spherical coordinate
      grid around the equatorial plane.
    \item[$^\mathrm{b}$] Calculation without neutrino-pair creation
      by nucleon-nucleon bremsstrahlung.  
    \item[$^\mathrm{c}$] Calculation without the
      neutrino-antineutrino processes of \cite{burjan03:nunu}.
  \end{list}
  \label{table:1}
\end{minipage}
\end{table*}
%

\section{Results}
\label{sec:results}

Here we present results from one-dimensional and from first
two-dimensional simulations with our ``coupled ray-by-ray''
approximation of spectral neutrino transport and with an
implementation of neutrino opacities which is improved compared to
\cite{bru85}. First, we will discuss the differences resulting from
our enlarged set of opacities in the context of spherically symmetric
(1D) models. We will also discuss the influence of general
relativistic corrections on the core collapse and post-bounce
evolution and will investigate the importance of velocity-dependent
terms in the neutrino momentum equation, especially for a
quantitatively correct description of neutrino heating and cooling
between proto neutron star (PNS) and supernova shock. In this context
we shall present an interesting 1D model which shows large-amplitude
oscillations of the shock position and neutrino luminosities. Then we
will elaborate on our first two-dimensional (2D)
neutrino-hydrodynamical simulation with spectral neutrino
transport. Finally, we present a 2D model that develops an explosion
when we retreat from the most complete version of our neutrino
transport treatment by neglecting the velocity-dependent terms in the
neutrino momentum equation.

\begin{figure*}[!]
\sidecaption
  \includegraphics[width=11cm]{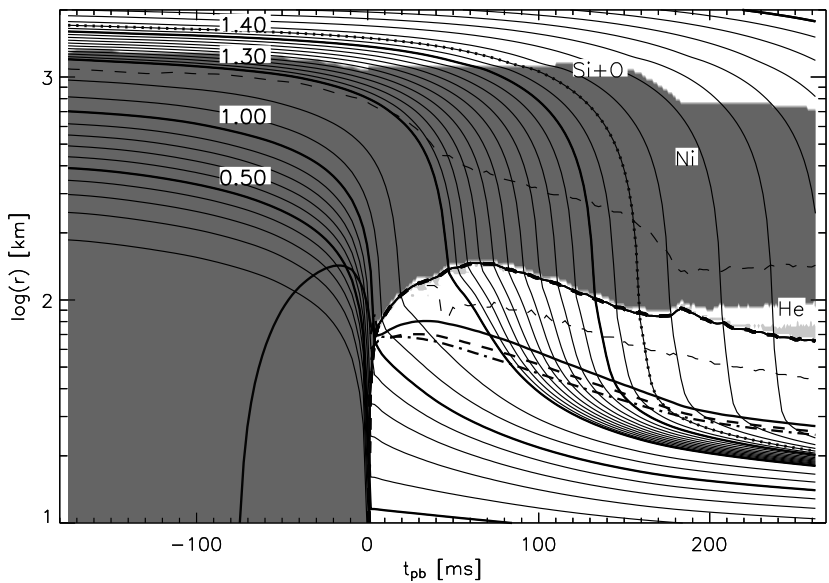} 
  \caption[]{
  Trajectories of mass shells for Model s15Gio\_1d.b. The plot also
  shows the neutrinospheres for $\nu_\mathrm{e}$ (thick solid line),
  $\bar \nu_\mathrm{e}$ (thick dashed), and $\nu_\mathrm{x}$ (thick
  dash-dotted), the mass shell at which the silicon shell becomes
  oxygen-rich (knotted solid line, at 1.42$\msol$), and the shock
  (thick solid line with superimposed dashes). Further we have marked
  the regions with a mass fraction of more than 60\% in iron-group
  elements (dark shaded). Also shown are regions with a mass fraction
  between 30\% and 60\% in $^4\mathrm{He}$ (light shaded). Note that the
  white region at $t_\mathrm{pb}>200$ms and $r\sim 90$km corresponds to
  $X_\mathrm{Fe}<0.6$ and $X_\mathrm{He}<0.3$. Finally, the lower thin
  dashed line marks the gain radius, while the upper one marks the
  interface between our high-density and low-density EoSs (i.e.~the
  upper thin dashed line corresponds to a density of
  $\rho=6\times10^7\gcm$).
  }\label{fig:mas_std}
\end{figure*}

\paragraph{Progenitor and model notation:}

The names of our models are chosen with the aim to provide information
about the employed input physics. All models in this paper were
started from the progenitor ``s15s7b2'', a star with a main-sequence
mass of $15~\msol$ kindly provided to us by S.Woosley
\cite[][]{woowea95}, see Fig.~\ref{fig:prog_prof}.
Thus our model names start with ``s15''. The simulations were
performed either with Newtonian gravity (``N'') or with our
approximative implementation of general relativistic gravity
(``G''). While most simulations included the most advanced description
of neutrino interactions (``io''), we performed a number of
calculations with the standard opacities of \cite{bru85} plus
nucleon-nucleon bremsstrahlung (``so''). One-dimensional models are
labelled with ``\_1d'', the names of the two-dimensional models give
information about the number of lateral zones $N_\vartheta$ (in this
paper only ``\_32''). Combining this with the chosen size of the
angular wedge allows one to infer the equidistant angular zoning of
the model. Finally, it is very important whether a model employed our
full implementation of the neutrino transport, Case B (``.b''), or
whether the radial velocity-dependent terms (except for the advection
terms) in the neutrino momentum equation were omitted, Case A
(``.a''). The models presented in this paper are listed in Table
\ref{table:1}.
The collapse phase was calculated with a Lagrangian grid, the
post-bounce phase with a Eulerian grid, and the simulations were run
with 400 radial hydro zones and 230 radial neutrino transport zones,
both with geometric spacing. The number of tangent rays employed was
250. Inside of 400km, the hydro and transport zones were chosen to be
identical, outside of 400km only a few transport zones were added to
correctly simulate the neutrino propagation out to the outer boundary
of the grid at 10$^4$km. All models were run with 17 geometrically
spaced neutrino energy bins between 2MeV and 380MeV. When necessary,
the radial grids were refined to account for the steepening density
gradient at the surface of the PNS.

\paragraph{Definitions:}

We define the shock radius $r_\mathrm{shock}$ as the radial position
where the radial velocity in the shock ``discontinuity'' is half of
the minimum preshock value of the velocity. The gain radius
$r_\mathrm{gain}$ is the zone interface below the innermost radial
zone (at given value of $v$) outside of which no zones with net
cooling exist. Thus the gain layer is between $r_\mathrm{gain}$ and
$r_\mathrm{shock}$ while the cooling region is below $r_\mathrm{gain}$
and includes the PNS. The total net energy transfer rates between
neutrinos and medium, $\delta_t E_\mathrm{cool}$ and $\delta_t
E_\mathrm{gl}$ in the cooling region and gain layer are defined as
integrals over the respective regions. An interesting quantity is the
``shell binding energy'' $E_\mathrm{bind}^\mathrm{shell}$, i.e.~the
energy needed to lift all material above a considered radius out of
the gravitational potential of the mass enclosed by this radius and
thus move the material to infinity. In the approximation with
Newtonian 1D gravity (which is a very good assumption for the
progenitor) its definition is
\ba
E_\mathrm{bind}^\mathrm{shell}(r) &\equ&
  \frac{4\pi}{\cos{\vartheta_\mathrm{max}}-\cos{\vartheta_\mathrm{min}}} \nn\\
  &&\int_r^\infty \int_{\cos{\vartheta_\mathrm{min}}}^{\cos{\vartheta_\mathrm{max}}}
  \varepsilon_\mathrm{bind}^\mathrm{shell}(r',\vartheta) \rho(r',\vartheta)
 r'^2 \dlin{\cos{\vartheta}} \dlin{r'},
\label{eq:eexpl_expl_def}
\ea
where $\varepsilon_\mathrm{bind}^\mathrm{shell}$ is our so-called ``local
specific binding energy'',
\be
\varepsilon_\mathrm{bind}^\mathrm{shell}(r,\vartheta) \equ e_\mathrm{int}(r,\vartheta) +
   \frac{1}{2}(v_r^2+v_\vartheta^2+v_\varphi^2)(r,\vartheta)
    + \Phi_\mathrm{1D}^\mathrm{enclosed}(r)\,.
\label{eq:eexpl_bind_def}
\ee
The gravitational potential $\Phi_\mathrm{1D}^\mathrm{enclosed}(r)$ is
calculated taking into account only the mass inside of the radius $r$
and assuming it to be distributed spherically symmetrical. These
energies, as well as the neutrino luminosities, are generally given in
the unit of FOE, the abbreviation for ``ten to the fifty-one
erg''. Finally, the spectrally averaged transport optical depth is
evaluated according to the formula
\ba
\tau_\nu(r) &\equ& \int_r^\infty \l<\kappa_\nu\r>(r') \dlin{r'}; \nn\\
\l<\kappa_\nu\r>(r') &\equ& \int_0^\infty 
\lambda_\nu^{-1}(r',\ene) H_\nu(r',\ene)
\dlin{\ene} \Bigg/ 
\int_0^\infty H_\nu(r',\ene) \dlin{\ene}\,,
\label{eq:tau}\ea
where $\lambda_\nu^{-1}(r',\ene)= \sum_i \lambda_{\nu,i}^{-1}$ is the
inverse mean free path for neutrinos of energy $\ene$ and type $\nu$
including all scattering and absorption processes. The different
contributions to the sum are taken from the RHS of the neutrino
momentum equation as implemented in our transport code. We define an
``energy-averaged'' neutrinosphere (radius) by the condition
$\tau_\nu(r_\nu)=1$. Where needed, the energy-dependent neutrinosphere
is defined by $\tau_\nu(r_\nu(\ene),\ene) = \int_r^\infty
\lambda_\nu^{-1}(r',\ene) \dlin{r'} \equiv 1$. Many properties of 2D
models are evaluated from 1D radial profiles, obtained by laterally
averaging the 2D data, see Sect.~\ref{sec:2deddfac} for the definition
of this procedure. This also includes the definition of ``mass
shells'' in 2D, i.e.~a mass shell corresponds to the radius of the
sphere which encloses this mass.  For 2D simulations, the criterion of
convective instability (see Sect.~\ref{sec:latadvec}) is also
evaluated with laterally averaged variables.

\begin{figure}[!]
  \resizebox{0.9\hsize}{!}{\includegraphics{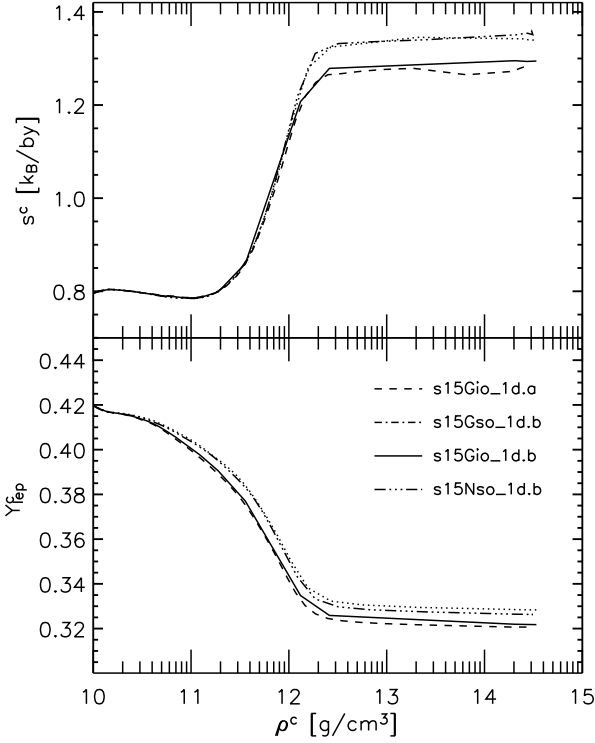}} 
  \caption[]{
  Evolution of the central values of the entropy (top) and lepton
  number (bottom) as functions of the central density during the
  collapse of Models s15Nso\_1d.b (dash-triple-dotted), s15Gso\_1d.b
  (dotted), s15Gio\_1d.b (solid), and s15Gio\_1d.a (dashed).
  }\label{fig:corecoll}
\end{figure}

\begin{figure}[!]
  \resizebox{\hsize}{!}{\includegraphics{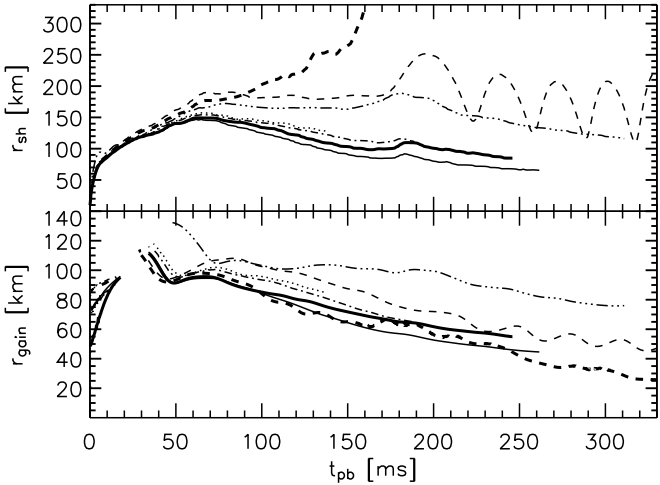}} 
  \caption[]{
  The top plot shows the shock positions versus time for all presented
  models. Thin lines are for 1D models, thick lines for 2D, see
  Fig.~\ref{fig:dtE_all} for the labelling. The plot below shows the
  gain radius which separates the neutrino cooling from the neutrino
  heating layer farther out. Note that the curves in the lower plot
  were smoothed over intervals of 5ms. The evolution of Models
  s15Gio\_1d.a and s15Gio\_32.a was followed to later times. Their
  complete evolution can be seen in
  Figs.~\ref{fig:osci_mecha},~\ref{fig:mas_expl}, and
  \ref{fig:lum_expl}.
  }\label{fig:spos_all}
\end{figure}

\begin{figure}[!]
  \resizebox{\hsize}{!}{\includegraphics{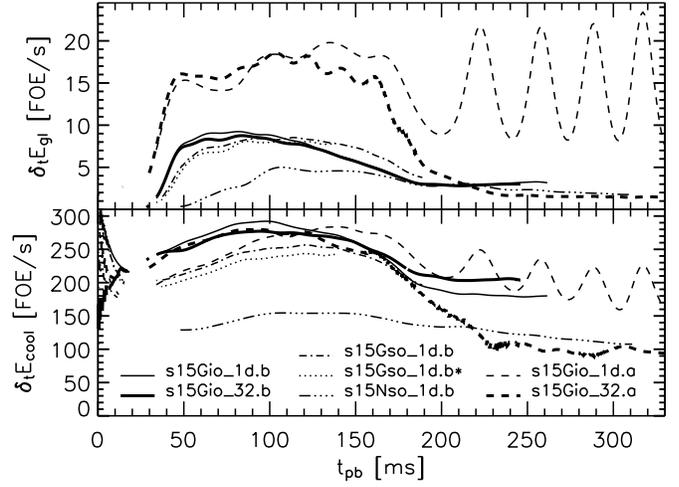}} 
  \caption[]{
  The top plot shows the total net heating rate in the gain layer,
  i.e.~between the gain radius and the shock. The lower plot shows the
  total cooling rate below the gain radius.  Note that the curves were
  smoothed over intervals of 5ms.
  }\label{fig:dtE_all}
\end{figure}

\begin{figure}[!]
  \resizebox{\hsize}{!}{\includegraphics{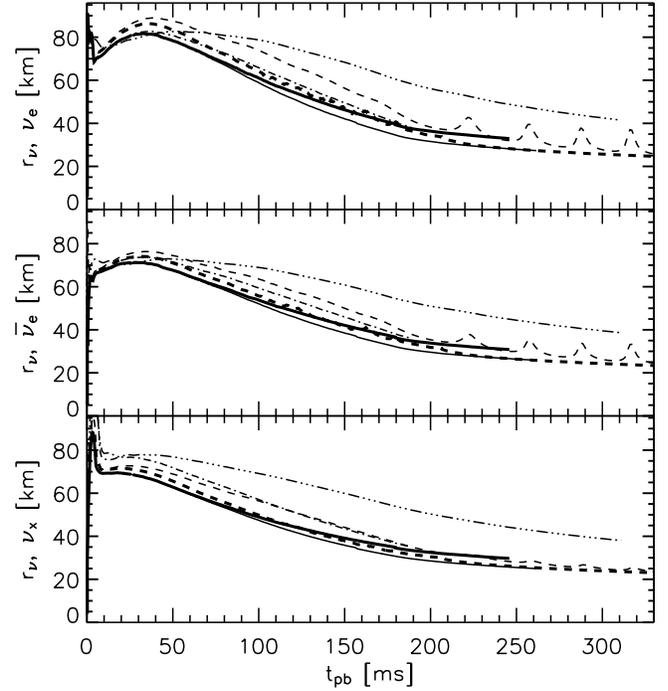}} 
  \caption[]{
  The three plots show the neutrinosphere radii of all presented
  models for $\nue$ (top), $\nuae$ (middle), and $\nu_{\mu,\tau}$ and
  $\bar\nu_{\mu,\tau}$ (bottom), see Fig.~\ref{fig:dtE_all} for the
  labelling. Note that Model s15Gso\_1d.b$^\ast$ is missing in these
  plots (because of data loss).
  }\label{fig:rnu_all}
\end{figure}

\begin{figure}[!]
  \resizebox{\hsize}{!}{\includegraphics{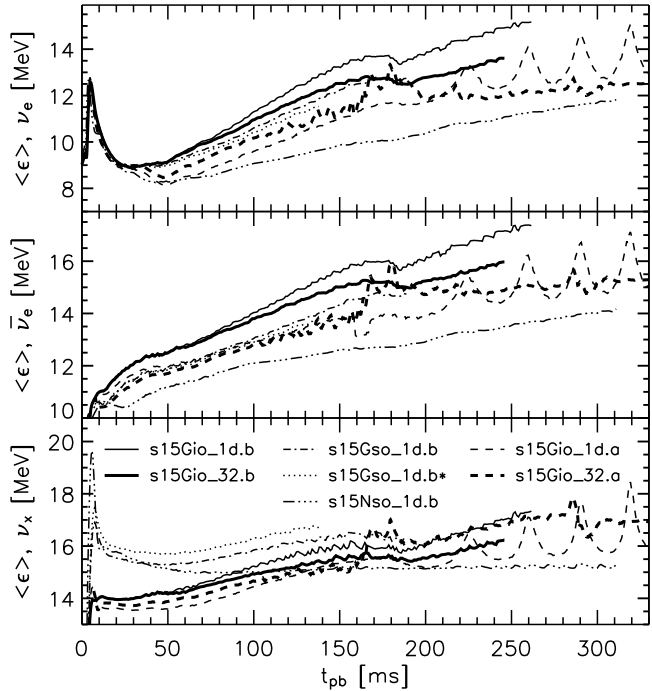}} 
  \caption[]{
  The three plots show the average energies of the emitted $\nue$
  (top), $\nuae$ (middle), and $\nu_{\mu,\tau}$ or
  $\bar\nu_{\mu,\tau}$ (bottom) as measured by an observer at rest at
  400km.
  }\label{fig:eav_all}
\end{figure}

\begin{figure}[!]
  \resizebox{0.8\hsize}{!}{\includegraphics{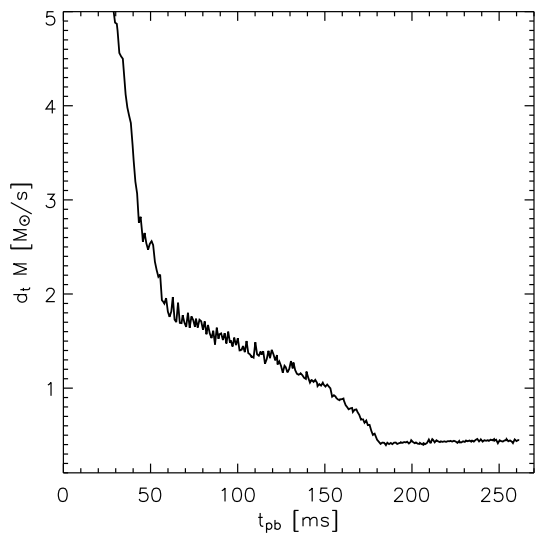}} 
  \caption[]{
  Mass accretion rate through the shock for Model s15Gio\_1d.b. For
  the other models its evolution is nearly the same, unless an
  explosion occurs.
  }\label{fig:mdot}
\end{figure}

\begin{figure}[!]
  \resizebox{\hsize}{!}{\includegraphics{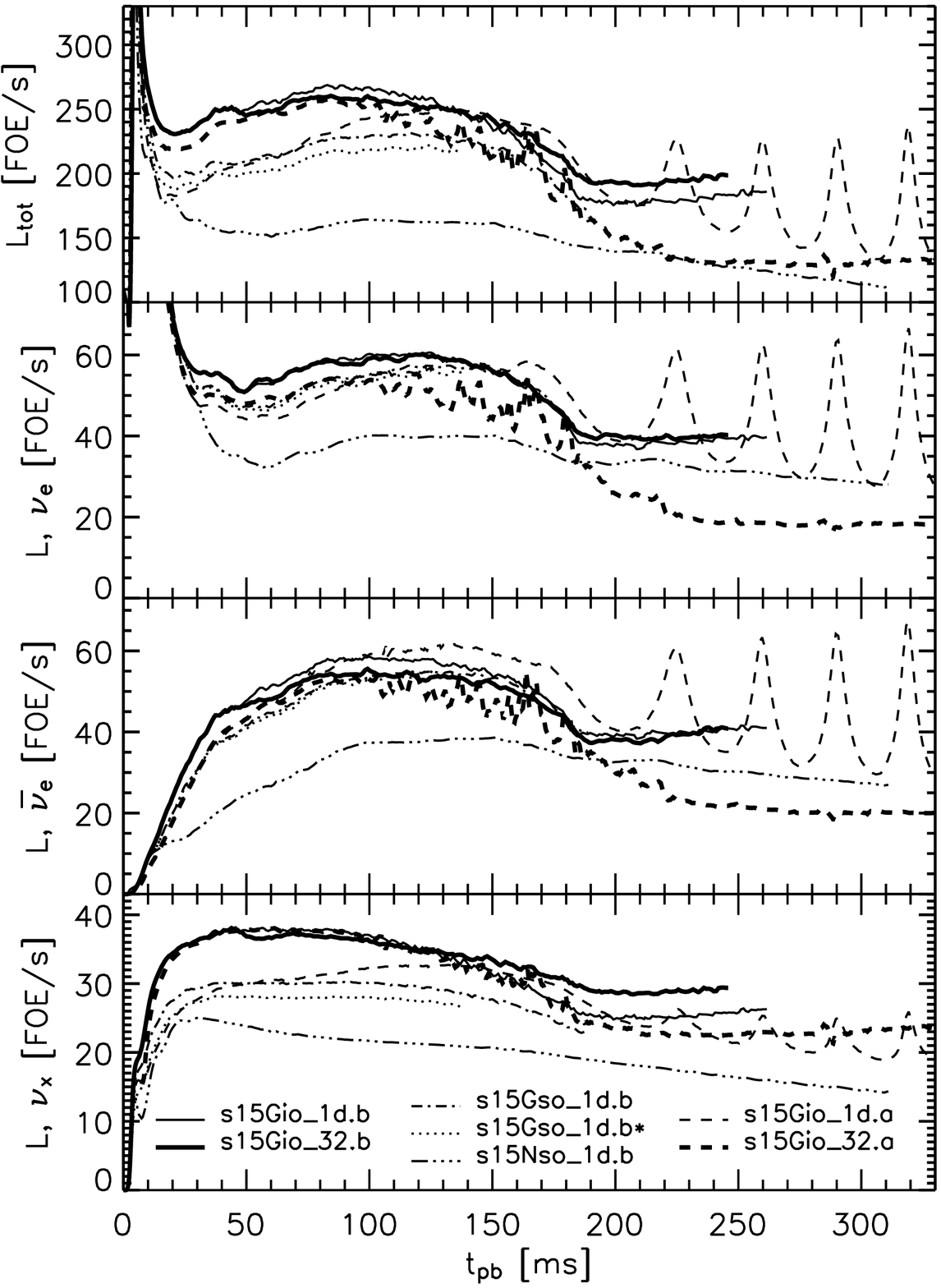}} 
  \caption[]{
  Comparison of the neutrino luminosities for all models for an
  observer at rest at 400km. The plot on top shows the total
  luminosities, the three lower ones those for the different $\nu$
  types. The plot at the bottom gives the luminosities of $\nu_\mu$,
  $\bar\nu_\mu$, $\nu_\tau$, or $\bar\nu_\tau$, which are all assumed
  to be equal.
  }\label{fig:lum_all}
\end{figure}

\subsection{Spherically symmetric models}
\label{sec:ssm}

\subsubsection{The reference model}
\label{sec:std}

This one-dimensional model, s15Gio\_1d.b, was run with our most
elaborate ``state-of-the-art'' description of neutrino opacities
described in Appendix \ref{sec:neuopa} and references therein. This
treatment of neutrino-matter interactions is improved compared to the
rates of \cite{bru85}, which we consider as the ``standard case'', by
including effects of nucleon recoil and thermal motions,
weak magnetism corrections (\citealp{hor02}), and nucleon-nucleon
correlations (\citealp{bursaw98,bursaw99}) in neutrino-nucleon
interactions. Moreover, the reduction of the effective nucleon mass at
nuclear densities (\citealp{redpra99}) and the quenching of the
axial-vector coupling in nuclear matter (\citealp{carpra02}) are taken
into account. In addition, we include nucleon-nucleon bremsstrahlung
and neutrino-neutrino interactions like scattering as well as pair
creation processes between neutrinos of different flavors. The full
list of included rates and corresponding references can be found in
Table \ref{tab:reactions}.

We applied the approximations of general relativity described in
Sects.~\ref{sec:hydro} and \ref{sec:genrel} (see also
\citealp{ramjan02} and \citealp{mardim05}). Finally, the simulation
was done including the velocity-dependent terms in the neutrino
momentum equation, i.e.~the terms proportional to $\br$ in
Eqs.~(\ref{eq:momeqe2},\ref{eq:momeqn2}). Despite of being formally
small ($\beta_r \leq 0.1$) these terms turn out to be important in
supernova simulations as we will elaborate in Sect.~\ref{sec:veloc}.

The evolution of this model can be seen in Fig.~\ref{fig:mas_std}. The
progenitor needs 174ms to reach core bounce. The typical time
evolution of the central values of $Y_\mathrm{lep}$ and $s$ during
collapse can be seen as functions of the central density in
Fig.~\ref{fig:corecoll}. The shock is created at an enclosed mass of
$M_\mathrm{sc}=0.49\msol$ and a radius of $r_\mathrm{sc}=10.6$km. Its
prompt expansion stalls after about 1ms at $r=32$km, turning the shock
into an accretion front. Initially the high mass accretion rate
through the shock leads to matter being accumulated between the PNS
and the shock; the energy via neutrino cooling is not quick enough to
allow this material to settle quickly so that the shock continues to
move outward. After 6ms, when the shock reaches 80km, this expansion
slows down because the mass accretion rate drops and neutrino cooling
has become more efficient, leading to an ``equilibrium'' between mass
accretion through the shock and mass settling onto the PNS. The shock
now slowly expands to 140km at 60ms, see
Fig.~\ref{fig:spos_all}. Then, however, the shock slowly retreats
again as a consequence of the high ram pressure. The neutrino heating
in a relatively narrow gain layer (see also $\delta_t E_\mathrm{gl}$
in Fig.~\ref{fig:dtE_all}) is too weak to support shock expansion to
larger radii. At 170ms, when the shock has reached a radius of less
than 80km, we see a transient reexpansion of the shock, which is
stopped again after a few km. This feature results from a sudden drop
of the density in the progenitor (see Fig.~\ref{fig:prog_prof}) and
thus of the mass accretion rate (Fig.~\ref{fig:mdot}) and of the ram
pressure, and is connected with the ``interface'' between the silicon
shell and the oxygen-rich silicon shell (i.e.~the Si-SiO interface at
$1.42\msol$). Such a phase of transient shock expansion can be seen in
all our models, with strongly differing consequences (even explosions,
see Sect.~\ref{sec:expl}). In Model s15Gio\_1d.b the shock finds a new
equilibrium radius at only 90km to then follow the subsequent
contraction of the nascent neutron star (visible from the decreasing
neutrinosphere radii, see Fig.~\ref{fig:rnu_all}).

\begin{figure}[!]
  \resizebox{\hsize}{!}{\includegraphics{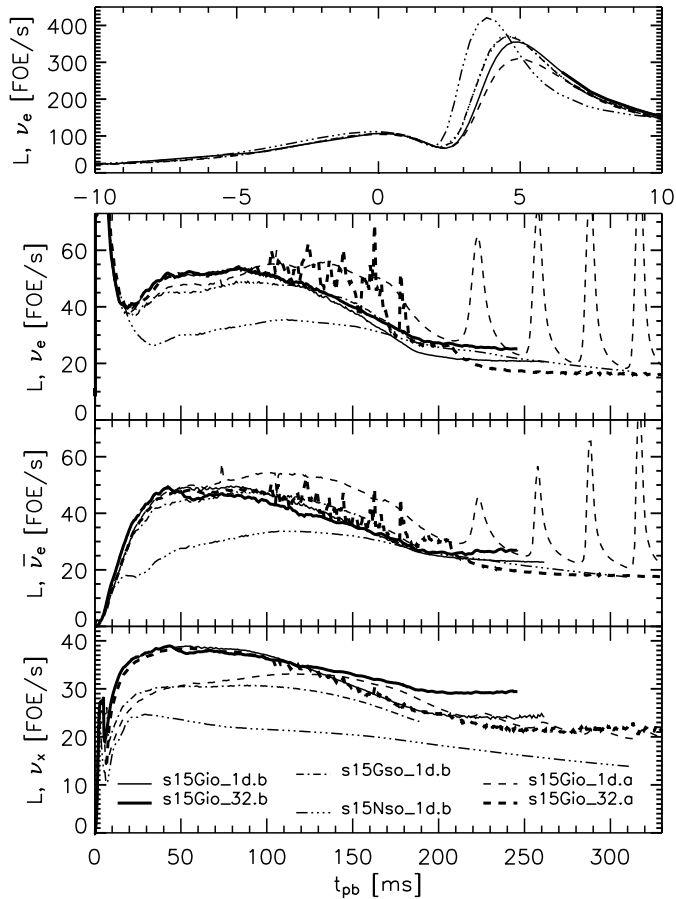}} 
  \caption[]{
  The uppermost plot shows the electron neutrino burst, again for an
  observer at rest at 400km. Note the different scale on the abscissa
  of this plot. The lower three plots display the neutrino
  luminosities of $\nue$, $\nuae$, and heavy-lepton neutrinos
  individually, evaluated at their respective neutrinospheres (for a
  comoving observer). Note that the differences compared to the
  luminosities shown in Fig.~\ref{fig:lum_all} come mostly from
  neutrino emission and absorption in the cooling and heating layers
  outside of the neutrinosphere (and not from observer frame
  motion). Also note that Model s15Gso\_1d.b$^\ast$ is missing in the
  lower three plots (due to missing data).
  }\label{fig:lum2_all}
\end{figure}

The neutrino emission reflects the different phases of the evolution:
After the collapse phase, during which only $\nu_\mathrm{e}$ are
emitted in significant numbers, the prompt $\nu_\mathrm{e}$ burst is
created few ms after bounce (see Fig.~\ref{fig:lum2_all}; note that
the signal is delayed by approximately 1ms due to the
time-of-flight). The burst lasts approximately 25ms, has a fwhm of
6ms, and reaches a peak luminosity of about 350FOE/s (FOE = ten to the
fifty-one ergs), see Fig.~\ref{fig:lum2_all}. Simultaneously, the
newly created PNS starts emitting $\bar\nu_\mathrm{e}$ and
$\nu_\mathrm{x}$. Their rise in luminosity takes 35ms and 15ms,
respectively. In the following accretion phase, the emitted neutrinos
come from two regions, from a layer below the neutrinosphere
``inside'' the cooling PNS and to a smaller part from the layer of
newly accreted material between the PNS surface and the gain
layer. The luminosity produced by the latter source depends
sensitively on the mass accretion rate as can be seen in the drop of
the luminosity when the Si-SiO interface enters the cooling
region. Interestingly, the mean energies of $\bar\nu_\mathrm{e}$ and
$\nu_\mathrm{x}$ become very similar after about 100ms
post-bounce. This fact has been discussed in detail in \cite{keiraf03}
and will be addressed in more detail below (Sect.~\ref{sec:vars}).

\begin{figure*}[!]
  \centering
  \includegraphics[width=17cm]{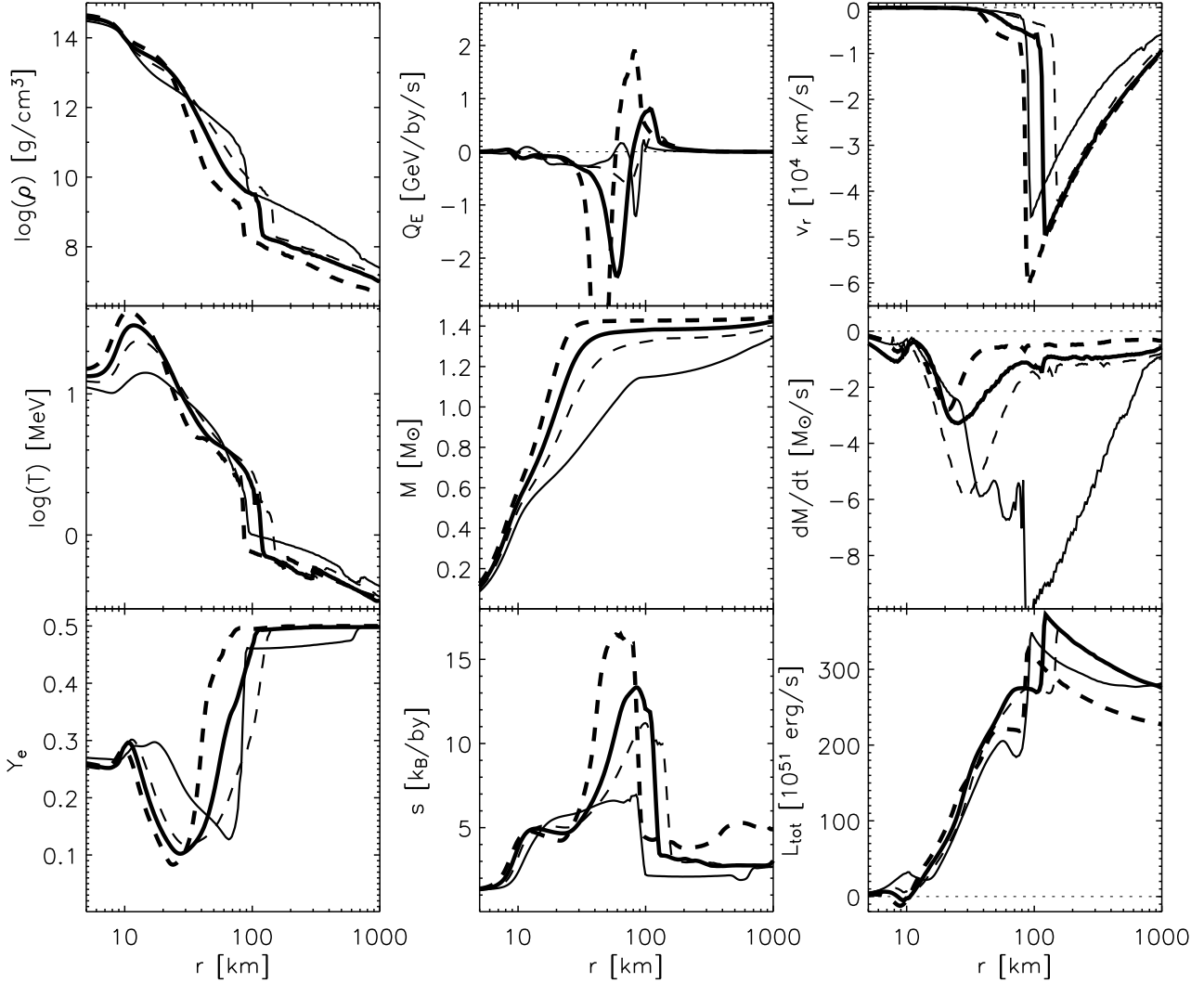} 
  \caption[]{
  Stellar profiles of different quantities at post-bounce times of
  11.6ms (solid), 72.5ms (dashed), 113.7ms (thick solid), and 170.7ms
  (thick dashed) versus radius for our reference Model
  s15Gio\_1d.b. For the latest time the cooling rate reaches a maximum
  value of $Q_\mathrm{E} = -10\mathrm{GeV/s}$ per baryon.
  }\label{fig:1d_snaps_std}
\end{figure*}

Looking more closely at our simulation of Model s15Gio\_1d.b, we find
that its outcome is not very surprising. After shock stagnation and
during the whole subsequent evolution we see that the shock is unable
to stop the rapid infall of the accreted material: Behind the shock,
the matter still has negative velocities of several 1000km/s, thus
falling quickly through the narrow gain layer. As an example, at the
time of maximal shock expansion at 72.5ms, the infall velocity behind
the shock is still $4\times10^{3}\mathrm{km/s}$ (see
Fig.~\ref{fig:1d_snaps_std}), and the short distance between shock and
gain radius of at most 50km corresponds to an advection timescale
through the gain layer of less than 15ms. With a moderate neutrino
heating rate of $300\mathrm{MeV/s}$ per baryon only about
$4.5\mathrm{MeV}$ per baryon can be deposited in these 15ms, an amount
which is clearly insufficient to promote shock expansion: At the high
entropies ($>10\mathrm{k}_\mathrm{B}/\mathrm{by}$) behind the shock
nuclei are nearly completely dissociated to nucleons. Only a small
fraction of $\alpha$-particles (less than 20\%) survive the shock
passage, and are quickly dissociated during infall. Thus, in the shock
the infalling matter loses 8--9MeV per baryon due to nuclear
photo-dissociation.

\subsubsection{Variations of the input physics}
\label{sec:vars}

\paragraph{``Standard'' opacities vs.~improved opacities}

We have performed most of our calculations either using a set of
neutrino interactions which we call ``standard opacities'' (Case so)
or our fully updated (``state-of-the-art'') description of neutrino
interactions which we call ``improved opacities'' (Case
io)\footnote{The new treatment of electron captures on nuclei during
core collapse with rates from \cite{lanmar03}, however, is not yet
included in any of our models presented here.}. The treatment of
neutrino interactions in Case io is described in Appendix
\ref{sec:neuopa}. Case so uses the neutrino interactions from
\cite{bru85} and \cite{mezbru93:code,mezbru93:nes}, implemented as
described by \cite{ramjan02}, but supplemented by neutrino pair
creation and annihilation due to nucleon-nucleon bremsstrahlung. In
neither of our simulations we take into account neutrino pair creation
by the plasmon process because this process is negligible for $\nue$
and $\nuae$, and for $\nux$ it is either dominated by nucleon-nucleon
bremsstrahlung (at high densities) or by the flavor coupling
$\nu\bar\nu$ pair process
($\nux\bar\nu_\mathrm{x}\leftrightarrow\nue\nuae$; at low
densities). For a recent calculation of the plasmon process at
SN-relevant conditions see the paper by \cite{ratdut03}.

\begin{figure*}[!]
  \centering

  \includegraphics[width=13cm]{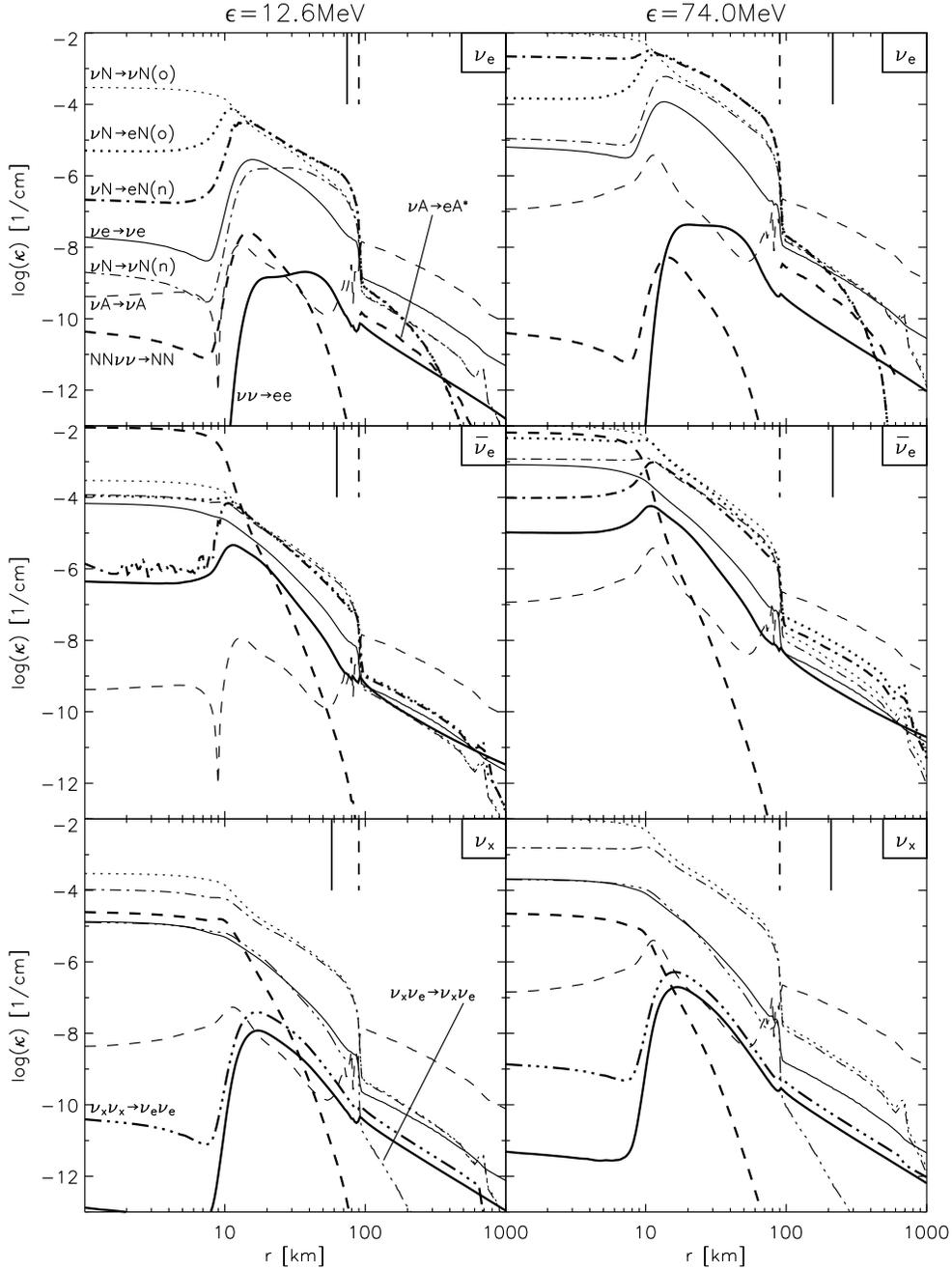} 

  \caption[]{
  Opacities for the neutrino interactions taken into account in Model
  s15Gio\_1d.b at $t=11.6$ms post-bounce for two representative
  neutrino energies. Neutrino (``$\nu$'') reactions with electrons
  (``e'', solid), nucleons (``N'', dash-dotted), nuclei (``A'',
  dashed), and electron-type neutrinos (``$\nue$'',
  dash-triple-dotted, for $\nu_\mathrm{x}$ only) are shown. For
  comparison the ``standard'' opacities for the neutrino-nucleon
  interactions are also displayed (dotted). Thin lines represent
  scattering processes, thick lines correspond to absorption
  processes. Note that different from this convention nucleon-nucleon
  bremsstrahlung (important only below the shock) is also represented
  by thick dashed lines. (See also caption of Fig.~\ref{fig:kappa_1}).
  }\label{fig:kappa_0}
\end{figure*}

\begin{figure*}[!]
  \centering

  \includegraphics[width=13cm]{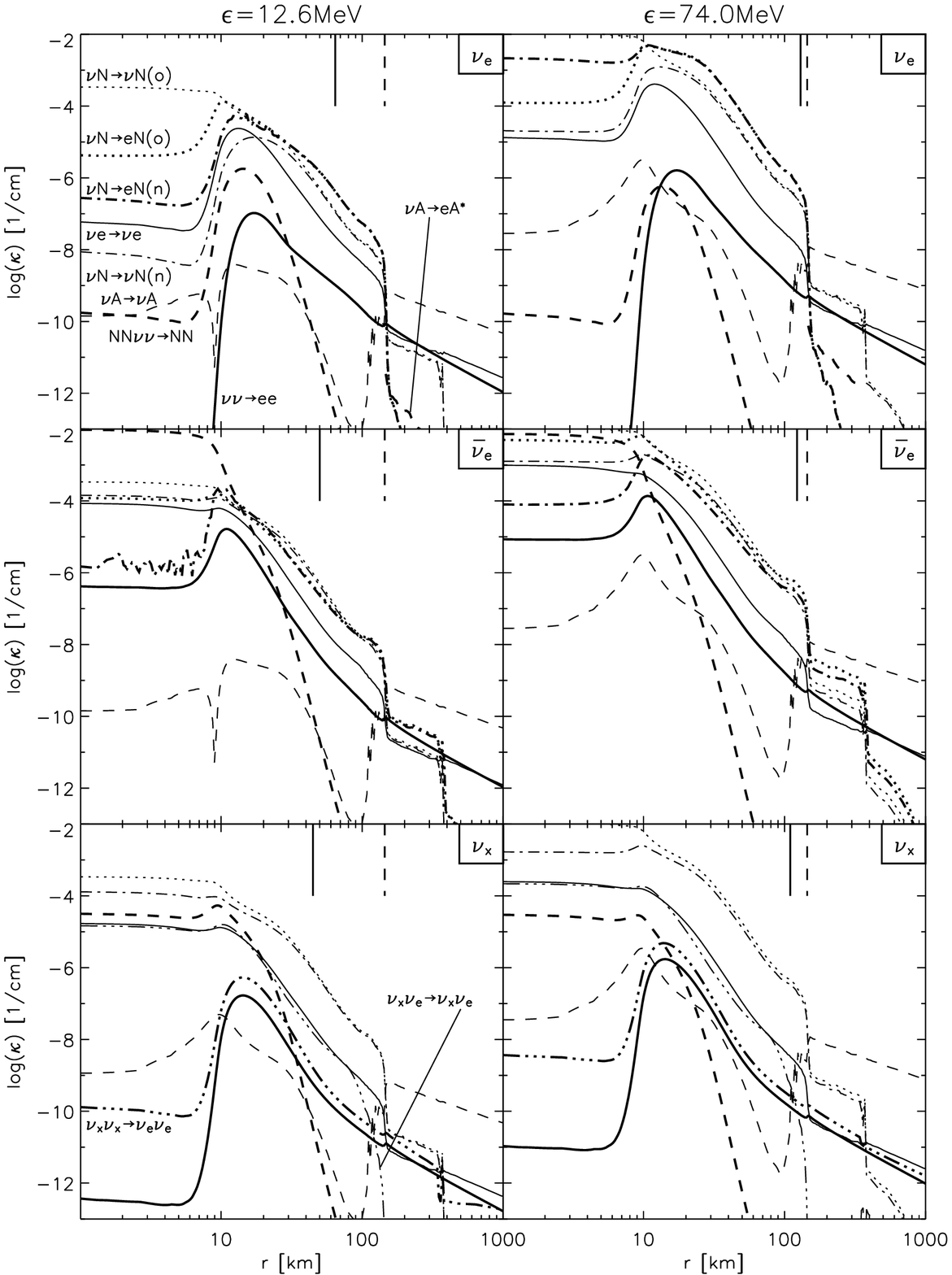} 

  \caption[]{
  Same as Fig.~\ref{fig:kappa_0}, but for $t=73.4$ms after bounce. The
  opacities include processes with all possible interaction partners:
  ``N'' stands for nucleons and can represent protons and/or neutrons,
  ``e'' can be electrons and/or positrons, ``A'' represents all
  nuclei, ``$\nu$'' can be any type of neutrino or antineutrino,
  ``$\nue$'' can be electron neutrinos and/or antineutrinos, and
  ``$\nu_\mathrm{x}$'' can be any type of heavy lepton neutrino or
  antineutrino. The opacities include all phase-space blocking
  factors. The short vertical lines on the top of the plots indicate
  the radii of the neutrinosphere for the given neutrino type and
  energy (solid) and of the shock (dashed).
  }\label{fig:kappa_1}
\end{figure*}

In order to compare the properties of the different interaction rates
we have plotted in Figs.~\ref{fig:kappa_0} and \ref{fig:kappa_1} the
opacities (inverse mean free paths) of all reactions included in Model
s15Gio\_1d.b for $\nue$, $\nuae$, and heavy-lepton neutrinos ($\nux$)
for two different energies and for the conditions present in the model
at two different post-bounce times. For completeness, we also show the
rates for neutrino scattering and absorption on nucleons as used in
Case so, i.e.~the rates with the approximations of ignoring nucleon
recoil and thermal motions, weak magnetism, effective nucleon masses,
and nucleon-nucleon correlations. During the discussion, remember that
absorption and emission are linked by detailed balance. Phase space
blocking of neutrinos (and other leptons) in the final channels of the
reactions is included in all reactions.


Charged-current absorption processes with nucleons are the dominant
mode of destroying $\nue$ and $\nuae$ after bounce, when the
post-shock material is essentially fully dissociated. However, we can
see that in the PNS core, the improved rates are reduced significantly
compared to the ``standard'' rates of Case so, for $\nu_\mathrm{e}$ by a
factor of up to 30, for $\bar\nu_\mathrm{e}$ even by up to 100. In the
region around the neutrinosphere, including the gain layer, recoil and
weak magnetism reduce the $\bar\nu_\mathrm{e}$ absorption rate by
10--20\%, while the effects nearly compensate each other in case of
$\nu_\mathrm{e}$ (compare also Appendix \ref{sec:neuopa} and
\citealp{horli99}). Absorption on nuclei is dwarfed by nucleon
absorption even in the pre-shock material. However, our rates for
$\nue$ absorption by nuclei are still rather approximative (not taking
into account recent advances of electron capture rate calculations by
\citealp{lanmar01}) and $\nuae$ capture by nuclei is
neglected. Improving this treatment is desirable.

Neutrino-antineutrino pairs can be produced and destroyed in different
charged- and neutral-current reactions. One of these is
nucleon-nucleon bremsstrahlung, which contributes significantly to the
emitted flux of heavy-lepton neutrinos $\nux$
\cite[][]{thobur00,keiraf03,thobur03} because this process dominates
the production of $\nu_\mathrm{x}$ at high densities. Interestingly,
bremsstrahlung is also the dominant absorption reaction for
$\bar\nu_\mathrm{e}$ in the PNS core, where the electron degeneracy is
very high and therefore $\nue$ are orders of magnitudes more abundant
than $\nuae$. It is also the dominant production rate since positrons
are very rare in the degenerate region, leading to a strongly
suppressed rate of $\mathrm{e}^++\mathrm{n}\rightarrow \nuae +
\mathrm{p}$. The second pair process is the annihilation of
$\nu\bar\nu$ pairs to $\mathrm{e}^+\mathrm{e}^-$ pairs. The inverse
reaction dominates bremsstrahlung in creating $\nuae$ at low densities
and is also a very important process for the generation of
$\nu_\mathrm{x}$. This rate, however, is surpassed by the rate of
$\nu\bar\nu$ pair conversion between different flavors, which in most
regimes is approximately a factor of two more important for producing
$\nux\bar\nu_\mathrm{x}$ pairs than $\mathrm{e}^+\mathrm{e}^-$
annihilation. Both rates dominate the $\nu_\mathrm{x}$ production
around the neutrinosphere, thus being crucial in the spectrum
formation process of these neutrinos
(cf.~\citealp{keiraf03,burjan03:nunu}). For higher energies, the
leptonic pair production rates become increasingly important relative
to bremsstrahlung.

Finally, scattering processes contribute to the neutrino
opacity. Scattering of neutrinos on nuclei dominates the scattering
rates above the shock. Scattering on nucleons are the dominant rates
in the NS core and postshock layer. Dense medium effects (see Appendix
\ref{sec:neuopa}), however, reduce the rate by a factor of 2--3 at
high densities below the neutrinosphere and phase space blocking
effects lead to a dramatic reduction for scattering rates for $\nue$
in case of nucleon recoil being included. Again, weak magnetism and
recoil produce a 10\% difference for $\bar\nu_\mathrm{e}$-nucleon
scattering around the neutrinosphere, and a much smaller effect for
$\nue$ (see Appendix \ref{sec:neuopa}). Scattering off electrons and
positrons and off $\nue$, $\nuae$ are of similar importance for $\nux$
(see also \citealp{burjan03:nunu})\footnote{Note that the strong
decrease of the $\nu_\mathrm{x}\nu \rightarrow \nu_\mathrm{x}\nu$ rate at
large radii ($\ga 80$km) where the process is essentially
unimportant is due to an artificial suppression of this rate with a
factor of $\l(1+\rho_{10}^{-3}\r)^{-1}$ where $\rho_{10} = \rho \l/
\l(10^{10}\gcm\r)\r.$. This procedure was necessary because our rate
implementation works correctly only when electron neutrinos are in
chemical equilibrium.}.

At higher neutrino energies the different opacities show similar
behaviour (Fig.~\ref{fig:kappa_0}). The suppression of neutrino
absorption on nucleons at high densities is weaker (because phase
space blocking is less strong), and bremsstrahlung processes are less
important compared to other processes. At later times the temperature
in the postshock layer increases and $\mathrm{e}^-$ degeneracy is
reduced so that the $\mathrm{e}^+\mathrm{e}^-$ pair process and $\nu
\mathrm{e}^\pm$ scattering become more important (compare
Fig.~\ref{fig:kappa_1} with Fig.~\ref{fig:kappa_0}).

The influence of the neutrino interactions can be seen in dynamical
simulations with different input physics
(Figs.~\ref{fig:spos_all},~\ref{fig:dtE_all},~\ref{fig:rnu_all},~\ref{fig:eav_all},~\ref{fig:lum_all},~\ref{fig:lum2_all}). In
addition to the above presented Model s15Gio\_1d.b with our set of
``improved'' opacities, we have performed two dynamical simulations
s15Gso\_1d.b and s15Gso\_1d.b*, which both were run with the
traditional approximations for the opacities. In the latter case
bremsstrahlung was also neglected in order to use the rate input of
previous simulations by other groups. The influence of nucleon-nucleon
bremsstrahlung was already discussed by \cite{keiraf03} and
\cite{thobur03}, the importance of interactions between neutrinos of
different flavors was pointed out by \cite{burjan03:nunu}.

The collapse phase is hardly influenced by the use of our improved
description of neutrino-nucleon interactions, because free nucleons
are not very abundant during core collapse
($Y_\mathrm{n}+Y_\mathrm{p}\simeq 10^{-3}$)\footnote{During core
collapse until bounce, our simulations were run without
$\bar\nu_\mathrm{e}$ and $\nu_\mathrm{x}$, both being irrelevant
during this phase.}. The improvement of the electron capture rate on
protons slightly increases the $\nu_\mathrm{e}$ production below the
trapping density (see Fig.~\ref{fig:opac_prof}). With the coherent
scattering rate, which dominates the opacity during infall, not being
changed, more $\nu_\mathrm{e}$ are created and escape before the
matter becomes optically thick. As a consequence, the simulation with
improved opacities yields slightly lower entropies ($|\Delta s| \simeq
0.06$) and a slightly decreased lepton fraction ($|\Delta
Y_\mathrm{lep}| \la 0.01$) in the core after trapping compared to the
simulation with standard opacities
(Fig.~\ref{fig:corecoll}). Therefore the collapse time also decreases
($|\Delta t_\mathrm{coll}|\simeq2.3\mathrm{ms}$) and the homologous
core becomes slightly smaller, which reduces the mass enclosed by the
shock formation radius ($|\Delta M_\mathrm{sc}|\simeq 0.02$).

After bounce only the first 50ms are suited for comparing the models,
in this early phase the structure of the PNS (and the above lying
layer where the neutrinospheres are located) is still very similar in
all general relativistic models.

The lowest panel in Fig.~\ref{fig:eav_all} shows that the models
cluster into two categories concerning the mean $\nux$ energies. The
decrease by $\sim 2$MeV can be explained mainly by the effects of the
neutrino-nucleon interactions (without bremsstrahlung) when energy
transfer by nucleon recoil in $\nu$--n, p scattering is taken into
account (see also \citealp{raf01}). Bremsstrahlung (compare models
s15Gso\_1d.b$^\ast$ and s15Gso\_1d.b) and the flavor-coupling neutrino
reactions (compare models\footnote{Note that the omission of
$\beta$-dependent terms has no noticeable influence on the layers of
neutrino emission at early post-bounce times.} s15Gio\_1d.a and
s15Gio\_1d.b) have a small effect on $\l<\ene_{\nux}\r>$ ($\sim
0.5$MeV). The mean electron type neutrino energies are hardly changed
by the improvements of the neutrino opacities, only the mean energy of
$\bar\nu_\mathrm{e}$ increases slightly by $\sim0.7$MeV.

\begin{figure}[!]
  \includegraphics[width=0.95\hsize]{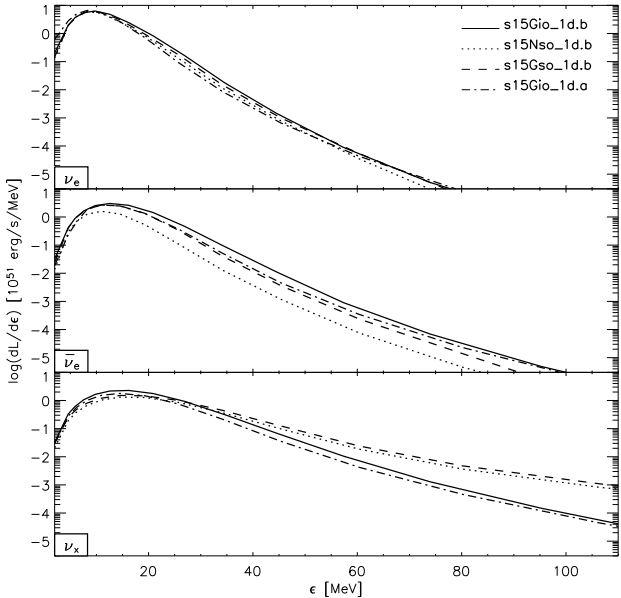} 
  \caption[]{
  Neutrino flux spectra for an observer at rest, evaluated at a radius
  of 400km at a time of 30ms after bounce. The different panels
  display results for $\nue$, $\nuae$, and $\nux$, respectively. The
  different lines correspond to different 1D models as labeled in the
  plot.
  }\label{fig:spectra_four}
\end{figure}

Similarly, the $\nue$ spectrum is not altered by the improvements of
the opacities, see Fig.~\ref{fig:spectra_four}. For $\nuae$ the
high-energy flux increases by a factor of three, being consistent with
the increase of $\l<\ene_{\nuae}\r>$. The most significant change can
again be seen for $\nux$. We observe a pinching of the spectrum
(higher peak, steeper drop at high energies).

%

Concerning the luminosities, see Fig.~\ref{fig:lum_all}, we again find
a clustering of the models in case of the $\nux$ luminosity. Here,
however, the increase by 25--30\% comes from the interactions between
the neutrinos of different flavors, whereas the improved
neutrino-nucleon interactions have a minor ($<10\%$) influence
(compare s15Gso\_1d.b$^\ast$ with s15Gio\_1d.a). Again, the effects on
$\nue$ and $\nuae$ are much smaller. For both $\nue$ and $\nuae$ the
luminosities increase by less than $10\%$.

As a direct consequence of the increased luminosities, the cooling
rate of the PNS, $\partial_tE_\mathrm{cool}$ in Fig.~\ref{fig:dtE_all},
is 25\% higher in case the interactions between neutrinos of different
flavors are turned on. The effect on the evolution of the shock,
however, is very weak and becomes visible only later than $100$ms
after bounce when the enhanced contraction of the PNS in models
including the flavor-coupling neutrino interactions has altered the
structure of the star between neutrinosphere and shock.

\paragraph{Newtonian vs.~general relativistic simulations}

Finally we discuss the differences arising between simulations with
our approximative treatment of general relativity (GR) and Newtonian
calculations by comparing Models s15Gso\_1d.b with s15Nso\_1d.b. In
the Newtonian simulation, the gravitational potential is less deep and
the onset of core collapse is not supported by relativistic
destabilization. Hence the collapse time (until shock formation) of
196.1ms is longer compared to 177.7ms for the relativistic
potential. Nevertheless the central lepton fraction after trapping is
nearly the same (it is smaller by only $\Delta Y_\mathrm{lep}\simeq
0.002$ in the Newtonian case, see Fig.~\ref{fig:corecoll}), suggesting
a very similar enclosed mass at shock formation. However, due to the
lower gravitational forces and infall velocities the shock forms at an
enclosed mass of $M_\mathrm{sc}=0.62\msol$ and a radius of
$r_\mathrm{sc}=12.5\mathrm{km}$ in the Newtonian case, whereas it is
launched at $M_\mathrm{sc}=0.51\msol$ and
$r_\mathrm{sc}=10.8\mathrm{km}$ in the GR simulation.

\begin{figure*}[!]
\centering
  \includegraphics[width=17cm]{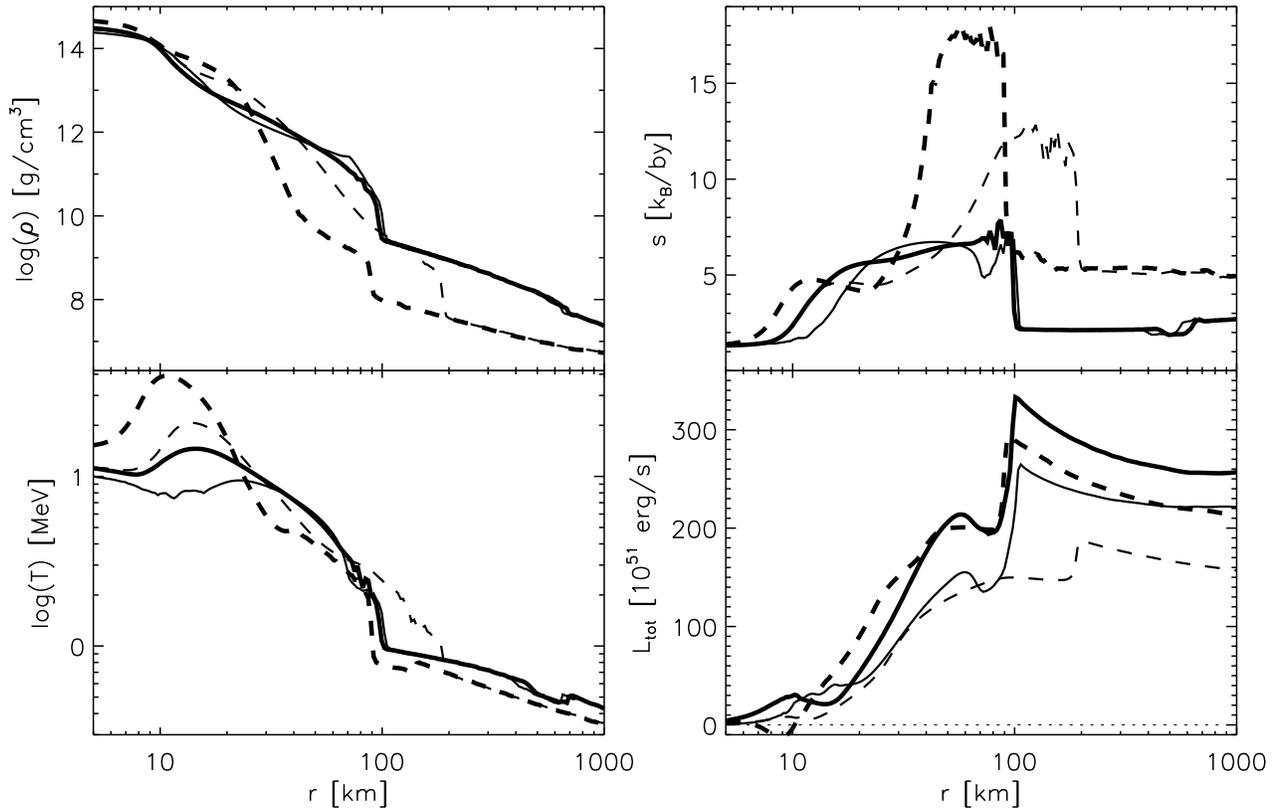} 
  \caption[]{
  Comparison of radial profiles for different variables at post-bounce
  times of 12.3ms (solid) and 177.7ms (dashed) for Models s15Nso\_1d.b
  (thin) and s15Gso\_1d.b (thick). $L_\mathrm{tot}$ is the total neutrino
  luminosity, i.e.~the sum of contributions from neutrinos and
  antineutrinos of all flavors.
  }\label{fig:newt}
\end{figure*}

The less compact structure of the PNS in the run with the Newtonian
gravitational potential strongly influences the post-bounce evolution:
The neutrinospheres and the accretion shock stand at larger radii,
the increase of temperature and entropy in the gain layer is slower,
and the neutrino luminosities are lower, see Figs.~\ref{fig:spos_all},
\ref{fig:rnu_all}, \ref{fig:lum_all}, and \ref{fig:newt}. These
results agree with the more elaborate discussion of differences
between Newtonian and GR simulations by \cite{liemez01}.

\subsubsection{Velocity-dependent terms in the first order moments equation}
\label{sec:veloc}

Terms which contain $\beta$ in the first order moments equation of the
transport equation (except for the $\beta\partial/\partial r$-terms) are
usually considered to be small and therefore negligible for many
problems when $v/c \ll 1$ (\citealt{mihmih84}; see also the discussion
in \citealt{ramjan02}). However, in the presented supernova
simulations, we found that the velocities are sufficiently large in
the neutrino decoupling and heating layers, which are crucial for the
explosion, that these ``$\beta$-terms'' can make a big effect.  
With ``$\beta$-terms'' we mean here all terms depending on $\beta_r$,
except the advection terms, which are proportional
to $\beta_r\partial/\partial r$ and which are included in all
calculations. 
The omission of these $\beta$-terms in the neutrino momentum equation
turned out to change the dynamical evolution of the supernova and to
lead to an artificial explosion in 1D and 2D
(Sects.~\ref{sec:osci} and \ref{sec:expl}). 
Having in mind that 
in the ${\cal O}(v/c)$ approach all velocity-dependent first-order 
terms in the radiation energy equation are known to be relevant
(see \citealt{mihmih84}) and the importance of the 
$\beta_\vartheta$-terms was recognized in Sect.~\ref{sec:latadvec}, we 
therefore conclude that all terms of first order in the fluid velocity
must be taken into account in supernova simulations\footnote{ 
We point out that some
authors think that going even beyond the ${\cal O}(v/c)$ approach of
our work is desirable (cf.~\citealp{carlen05}).}.
Doing so, we found that our transport code produces
results in very good agreement with spherically symmetric simulations
with the Oak Ridge/Basel AGILE-BOLTZTRAN code for Newtonian as well as
relativistic gravity (\citealp{lieram05,mardim05}).

\begin{figure*}[!]
\centering
  \begin{tabular}{lr}
    \put(0.9,0.3){{\Large\bf a}}
    \includegraphics[width=8.5cm]{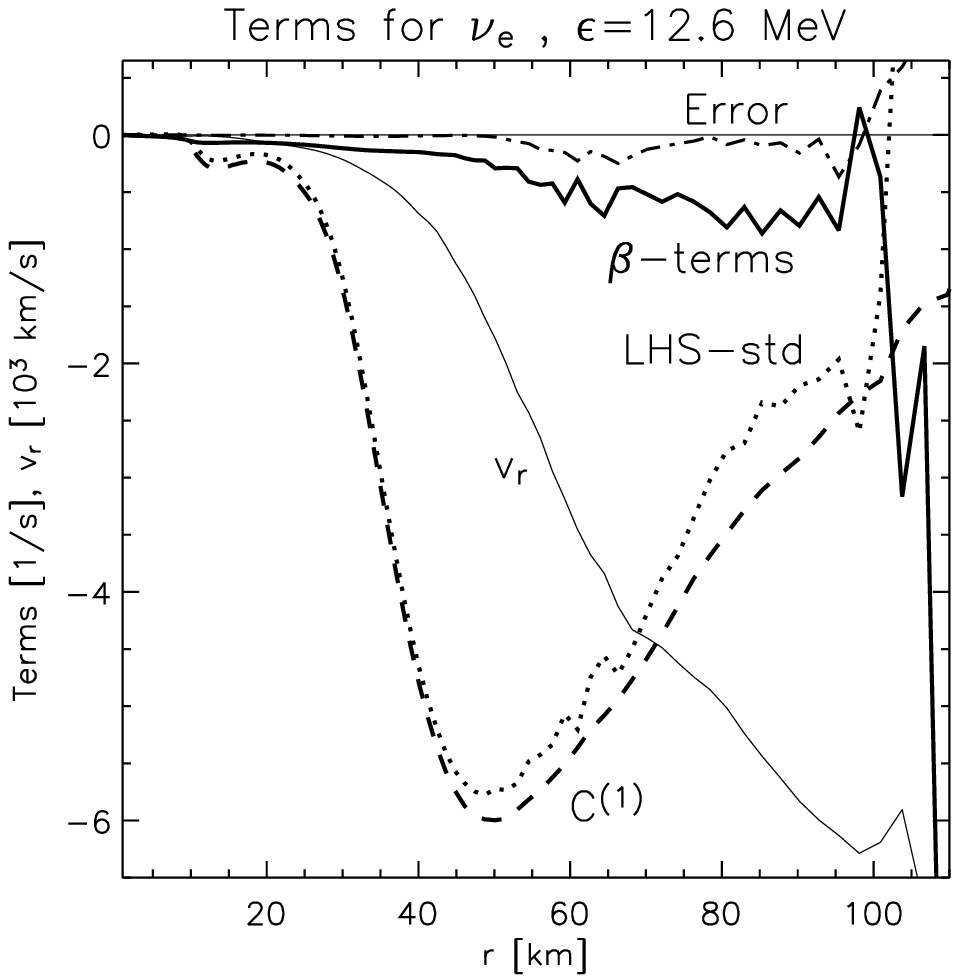} &  
    \put(0.9,0.3){{\Large\bf b}}
    \includegraphics[width=8.5cm]{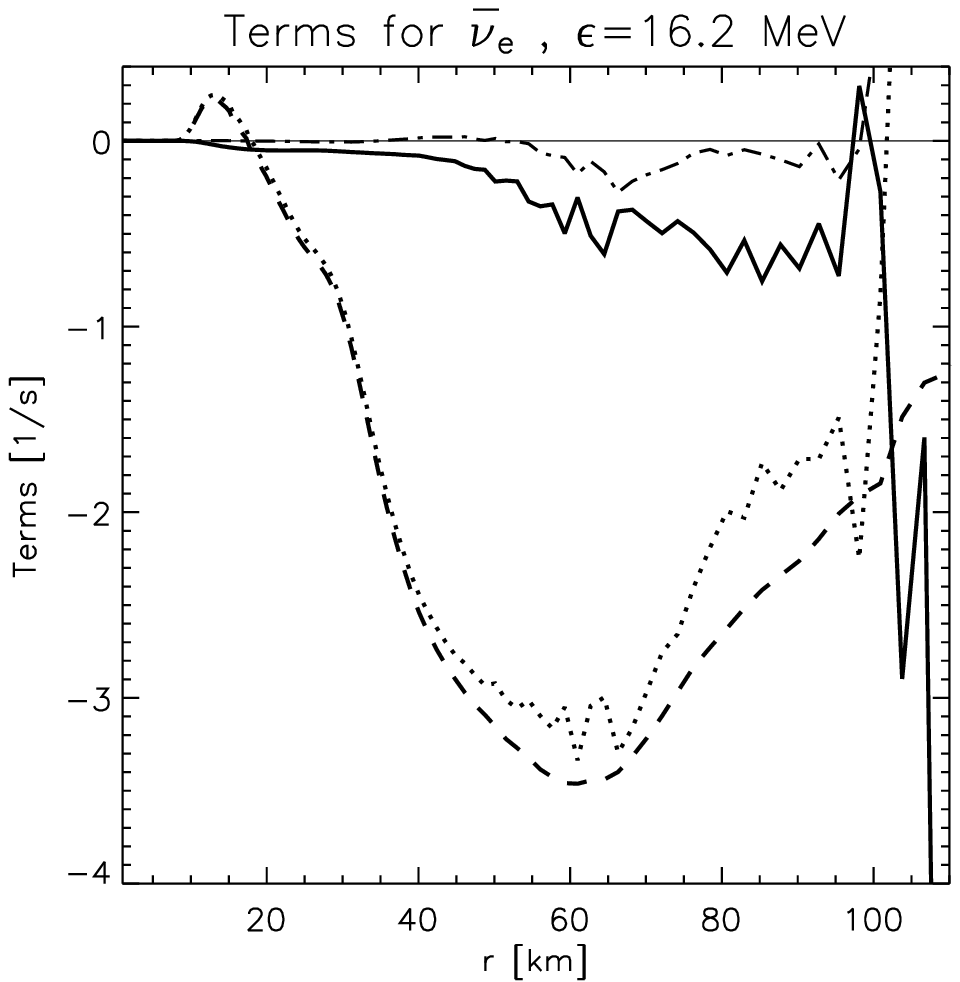} \\ 
    \put(0.9,0.3){{\Large\bf c}}
    \includegraphics[width=8.5cm]{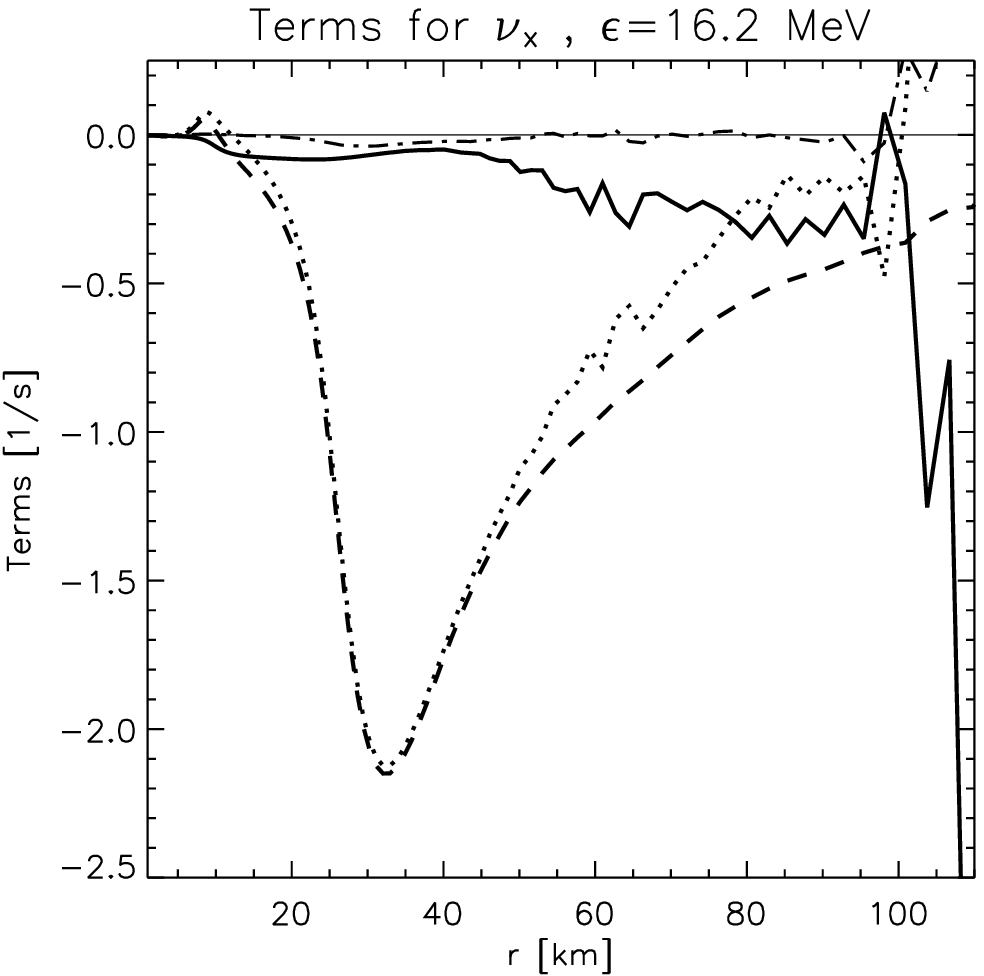} &  
    \put(0.9,0.3){{\Large\bf d}}
    \includegraphics[width=8.5cm]{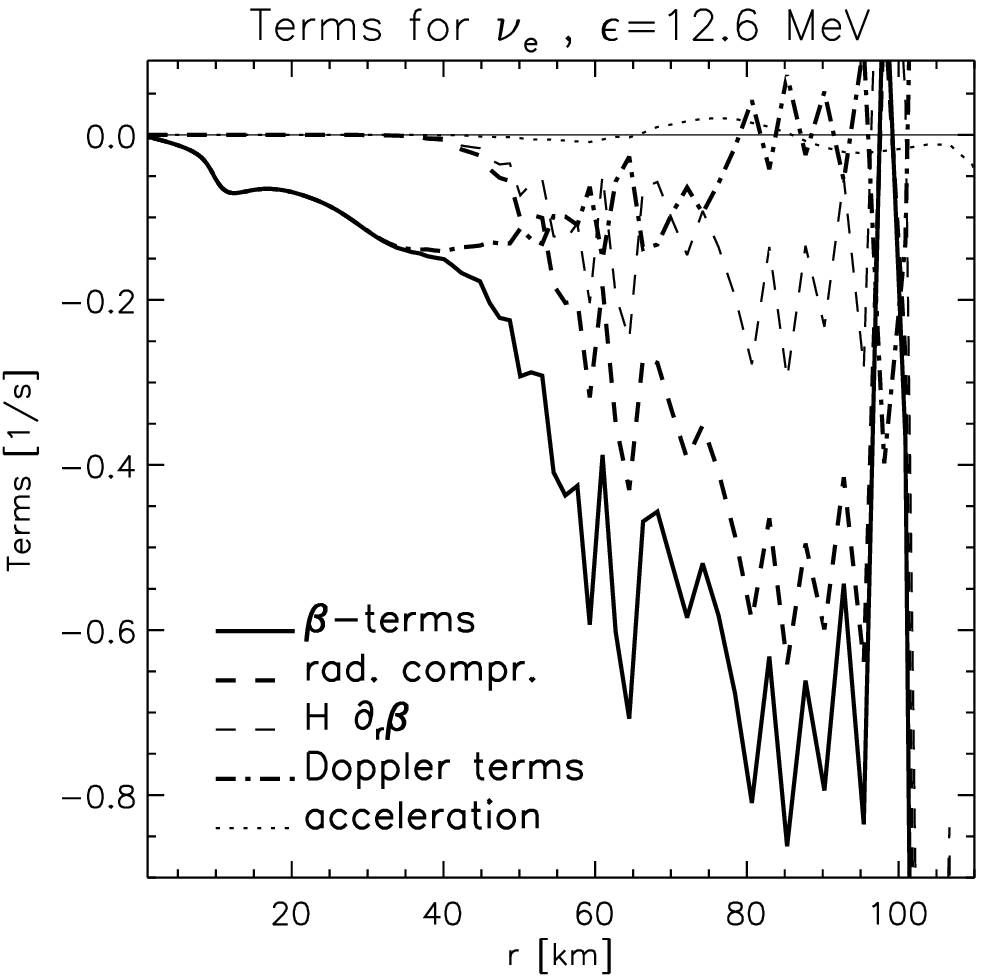}    
  \end{tabular}
  \caption[]{
  Comparison of the different terms in the neutrino momentum equation
  (Eq.~\ref{eq:momeqe2}, multiplied by $m_\mathrm{by}/\rho$) for Model
  s15Gio\_1d.b at $t_\mathrm{pb}= 114\mathrm{ms}$ for $\nue$ (top
  left), $\nuae$ (top right), and heavy-lepton neutrinos, $\nux$
  (bottom left), at the indicated energies. The different lines
  represent the $\beta$-terms (thick solid), the remaining terms on
  the LHS of Eq.~(\ref{eq:momeqe2}) (dotted), and the source term on
  the RHS (dashed). We have also plotted the sum of all three
  (dash-dotted), which reveals minor deviations from zero as a
  consequence of the post-processing evaluation of the discretized
  gradients. Finally, the thin solid line in the upper left plots
  shows the velocity profile.
  Panel d displays for $\nue$ the different $\beta_r$-terms in the neutrino
  momentum equation (Eq.~\ref{eq:momeqe2}, multiplied by
  $m_\mathrm{by}/\rho$) individually. These correspond to the physical effects
  of radiation compression, $H r^{-2} \partial_r(r^2 \beta_r)$ (thick
  dashed) and the term $H \partial_r \beta_r$ (thin dashed), Doppler
  effects (thick dash-dotted), and fluid acceleration (thin dotted).
  }\label{fig:st_Hterms}
\end{figure*}

We want to investigate the role of the $\beta$-terms in the neutrino
momentum equation, Eq.~(\ref{eq:momeqe2}), more closely 
with a snapshot of Model
s15Gio\_1d.b at $t_\mathrm{pb}= 114\mathrm{ms}$, for which we have
evaluated these terms at energies near the spectral maxima of
neutrinos and antineutrinos, see Fig.~\ref{fig:st_Hterms}. For $\nue$
and $\nuae$ the $\beta$-terms account for up to 20--30\% of the LHS of
Eq.~(\ref{eq:momeqe2}) in the neutrino-heated region between 70 and
110km; for muon neutrinos, which do not contribute to the heating at a
significant level, the $\beta$-terms can even dominate the LHS of
Eq.~(\ref{eq:momeqe2}). The individual size of these
terms at 114 ms after core bounce is visible in
Fig.~\ref{fig:st_Hterms}d. One can see that the different terms have  
different relative importance in different regions of the star. We 
emphasize that it can therefore not be concluded that any of these terms
can be safely ignored for all situations. 

The role of the $\beta$-terms can be made more
transparent if we write them on the RHS and compare them with the
source term for momentum exchange. In regions with negative velocity
and negative velocity gradient, the $\beta$-terms have the opposite
sign of the source terms in Eq.~(\ref{eq:momeqe2}) and therefore
effectively \emph{reduce} the coupling between matter and neutrino
flux. Thus, neglecting the $\beta$-terms has the opposite effect,
leading to a less rapid streaming of neutrinos and an increased energy
density of neutrinos above and below the gain radius. Note that for
higher energies, the velocity terms become less important.

\begin{figure*}[!]
\centering
  \begin{tabular}{c}
    \put(0.9,0.3){{\Large\bf a}}
    \includegraphics[width=14.0cm]{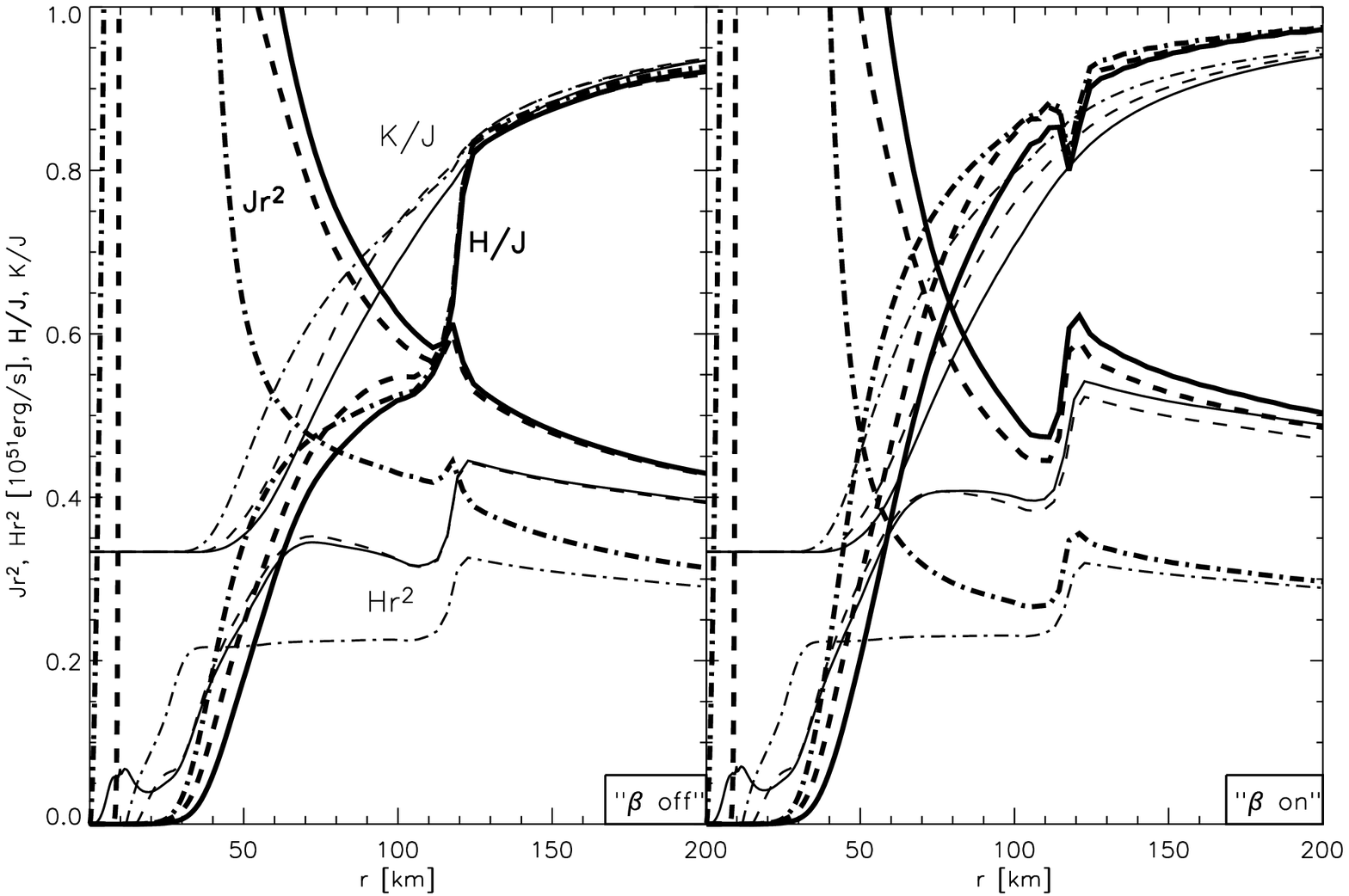} \\ 
    \put(0.9,0.3){{\Large\bf b}}
    \includegraphics[width=14.0cm]{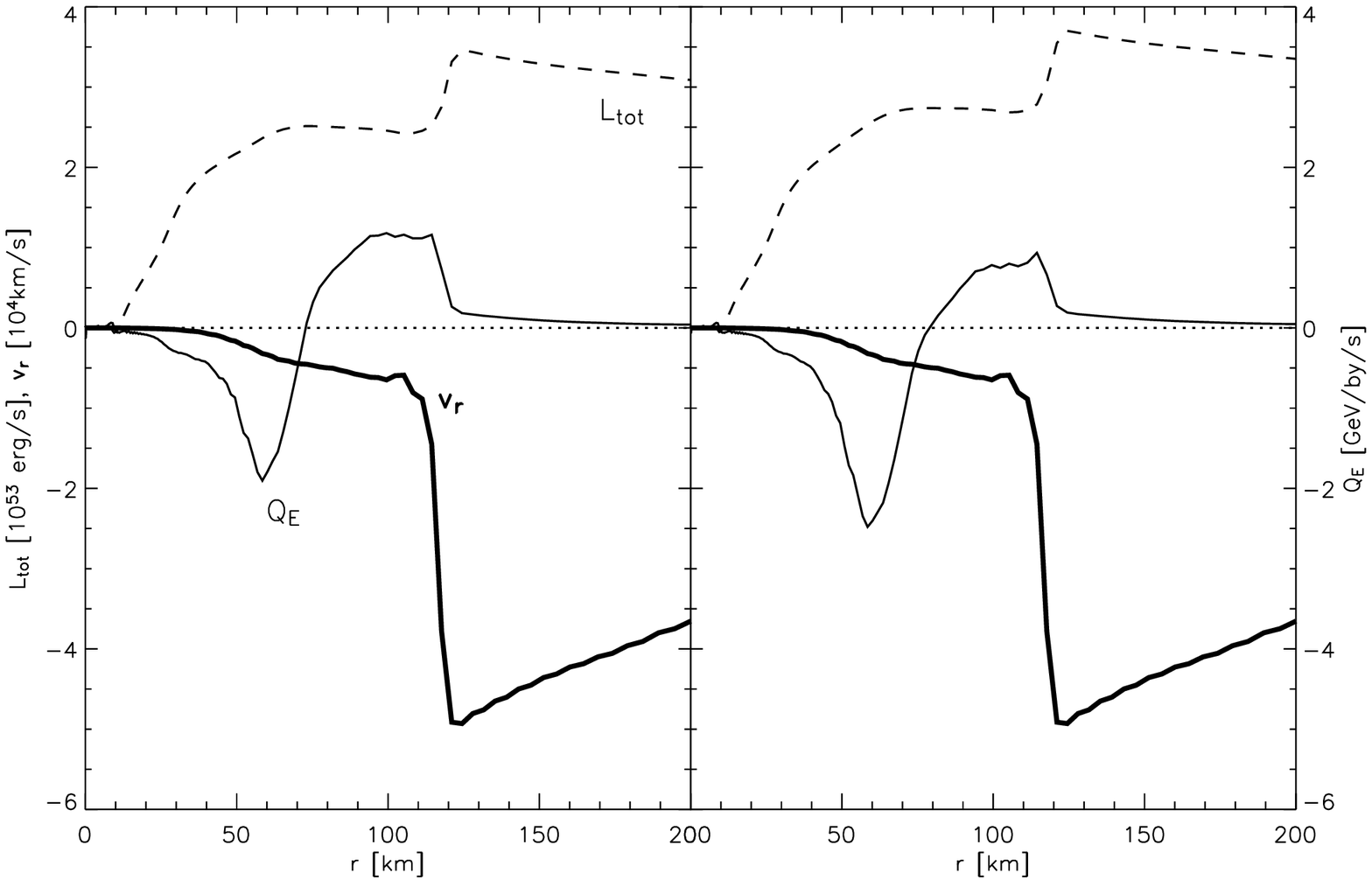} 
  \end{tabular}

  \caption[]{
  {\bf a} Comparison of steady-state transport solutions without
  (left) and with the $\beta$-terms in the neutrino momentum equation,
  computed for a background profile from Model s15Gio\_1d.b at
  $t_\mathrm{pb}= 114\mathrm{ms}$. The neutrino energy density $J$
  (times $r^2$), the flux factor $f_{H}=H/J$ (both with thick lines),
  the neutrino energy flux $H$ (times $r^2$), and the variable
  Eddington factor $f_\mathrm{ed}=K/J$ (both thin) are given for
  electron neutrinos (solid), electron antineutrinos (dashed) and muon
  neutrinos (dash-dotted) for an observer comoving with the stellar
  fluid. The kink of $f_H$ within the shock is a numerical artifact
  produced by the interpolation of $J$ and $H$ between the grid zones
  of a staggered mesh.
  {\bf b} Corresponding velocity profile (thick), total neutrino
  luminosities (as the sum of the luminosities of neutrinos of all
  flavors; dashed) and the energy source terms (neutrino heating minus
  cooling; thin solid line).
  }\label{fig:st_HJ}
\end{figure*}

These statements are confirmed by Fig.~\ref{fig:st_HJ}, where we show
stationary transport solutions calculated for the considered stellar
profile both with and without the $\beta$-terms. The neutrino flux $H$
and the flux factor $H/J$ are larger in the solution with the
$\beta$-terms, and the neutrino density is smaller, which directly
decreases the neutrino heating (Fig.~\ref{fig:st_HJ}b). From the
plots, it can also be seen that the solution with the $\beta$-terms is
consistent with physical requirements: Since the neutrino quantities
are given in a comoving frame, both luminosity and energy density
reveal a larger blue-shifting ahead of the shock where the infall
velocities are higher. The corresponding ``step'' at the shock,
however, is absent in case of $J$ in the solution without
$\beta$-terms (Fig.~\ref{fig:st_HJ}a, left panel). The flux factor is
affected by these differences. Without $\beta$-terms it clearly
disagrees with results of transport calculations with an $S_N$ solver
for the Boltzmann equation \cite[see ][]{lieram05}. In contrast, the
variable Eddington factor $f_\mathrm{ed}=K/J$ turns out not to reveal the
same degree of sensitivity to the $\beta$-terms.
\begin{figure*}[!]
\centering
  \begin{tabular}{c}
    \put(0.9,0.3){{\Large\bf a}}
    \includegraphics[width=14cm]{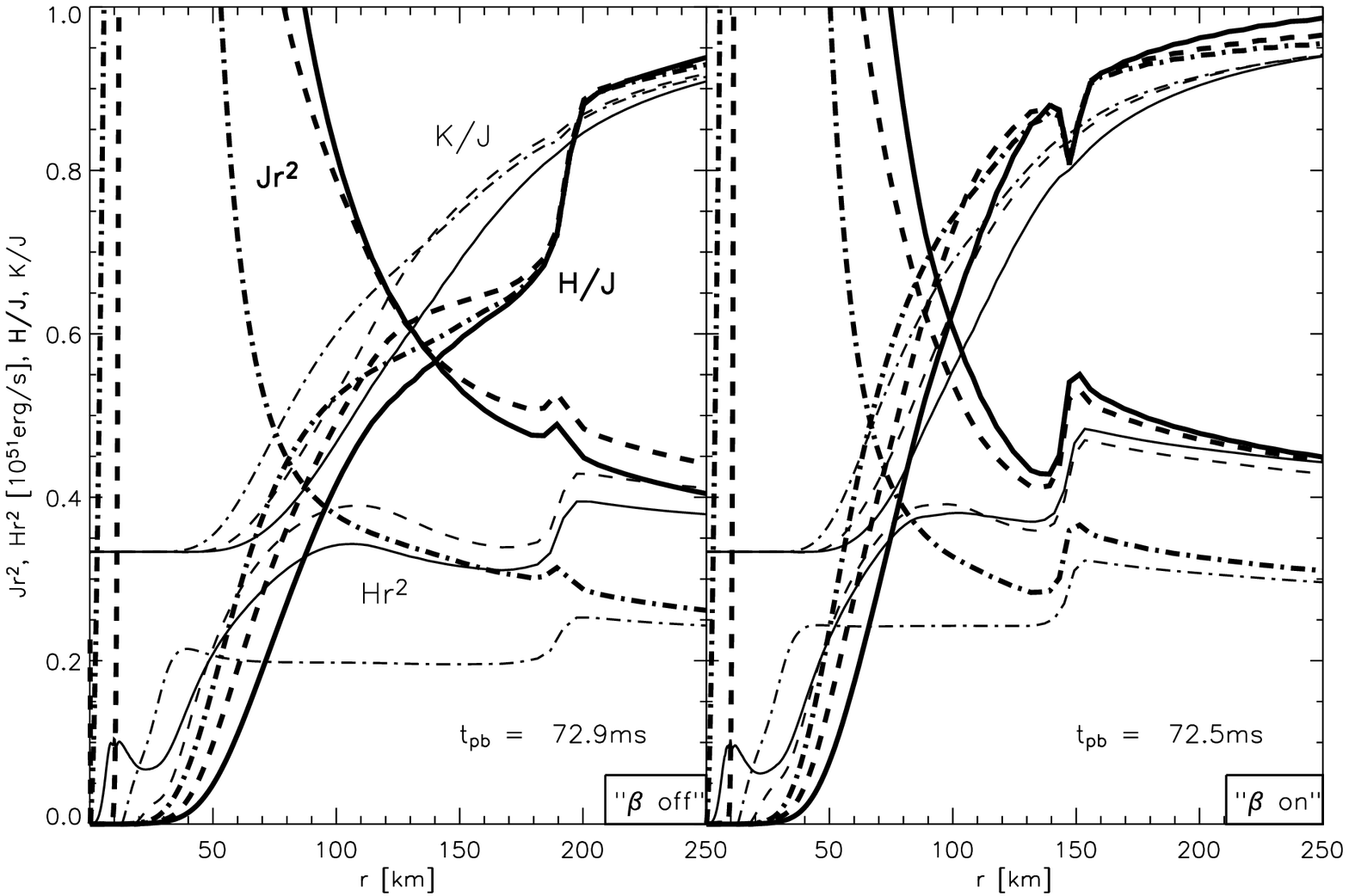} \\ 
    \put(0.9,0.3){{\Large\bf b}}
    \includegraphics[width=14cm]{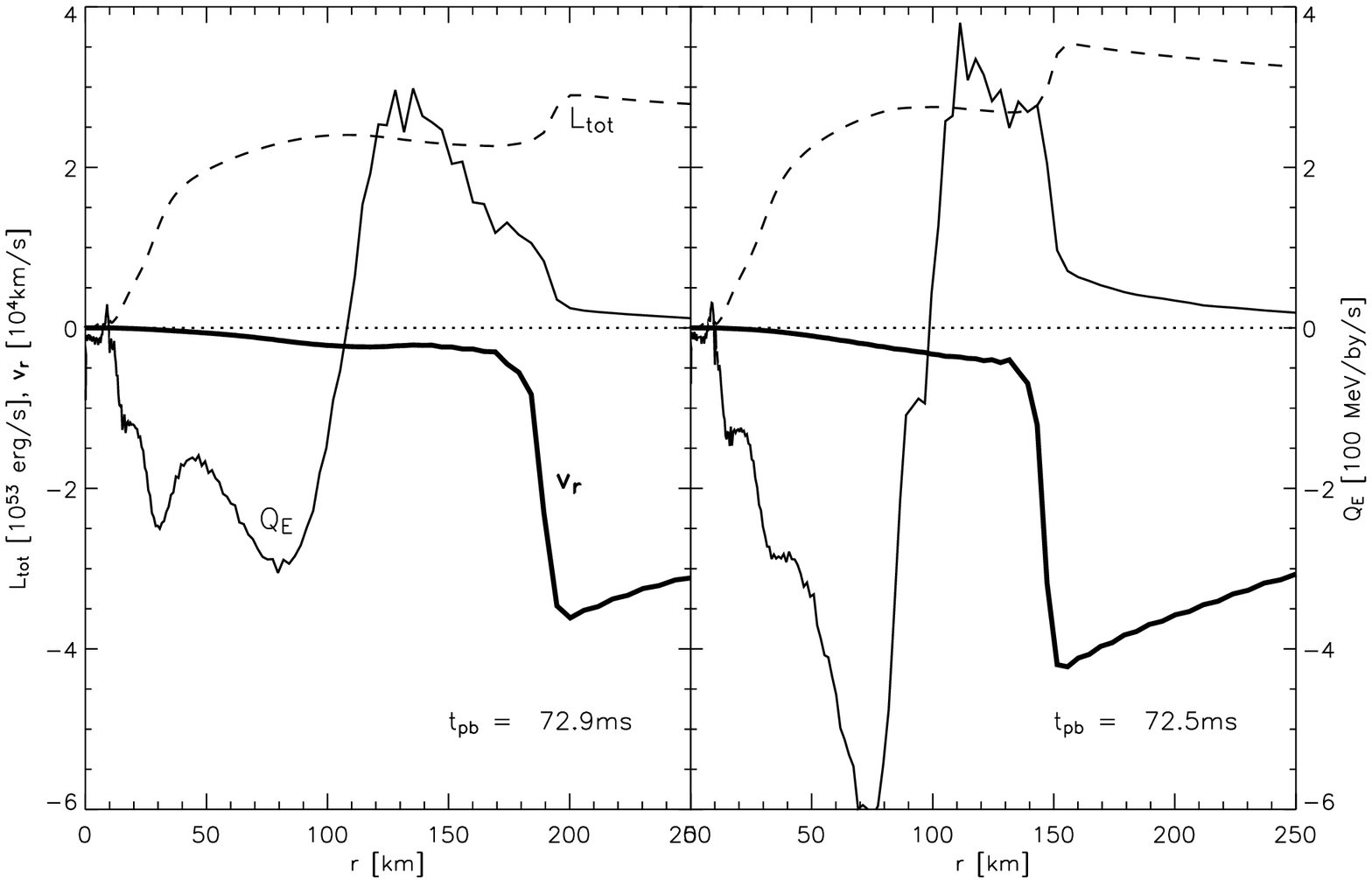} \\ 
  \end{tabular}

  \caption[]{
  {\bf a} Transport results at $t_\mathrm{pb}= 73\mathrm{ms}$ for
  Model s15Gio\_1d.a where the $\beta$-terms were dropped, and for
  Model s15Gio\_1d.b, which was computed with all terms in the
  neutrino momentum equation. The neutrino energy density $J$ (times
  $r^2$) and the flux factor $f_{H}=H/J$ are plotted with thick lines,
  the neutrino energy flux $H$ (times $r^2$) and the variable
  Eddington factor $f_\mathrm{ed}=K/J$ with thin lines for electron
  neutrinos (solid), electron antineutrinos (dashed) and muon
  neutrinos (dash-dotted).
  {\bf b} The corresponding profiles of velocity (thick), total
  luminosity of neutrinos and antineutrinos of all flavors (dashed)
  and the source term for energy exchange between neutrinos and
  stellar medium (thin solid).
  }\label{fig:dy_HJ}
\end{figure*}
\begin{figure*}[!]
\sidecaption
  \includegraphics[width=12cm]{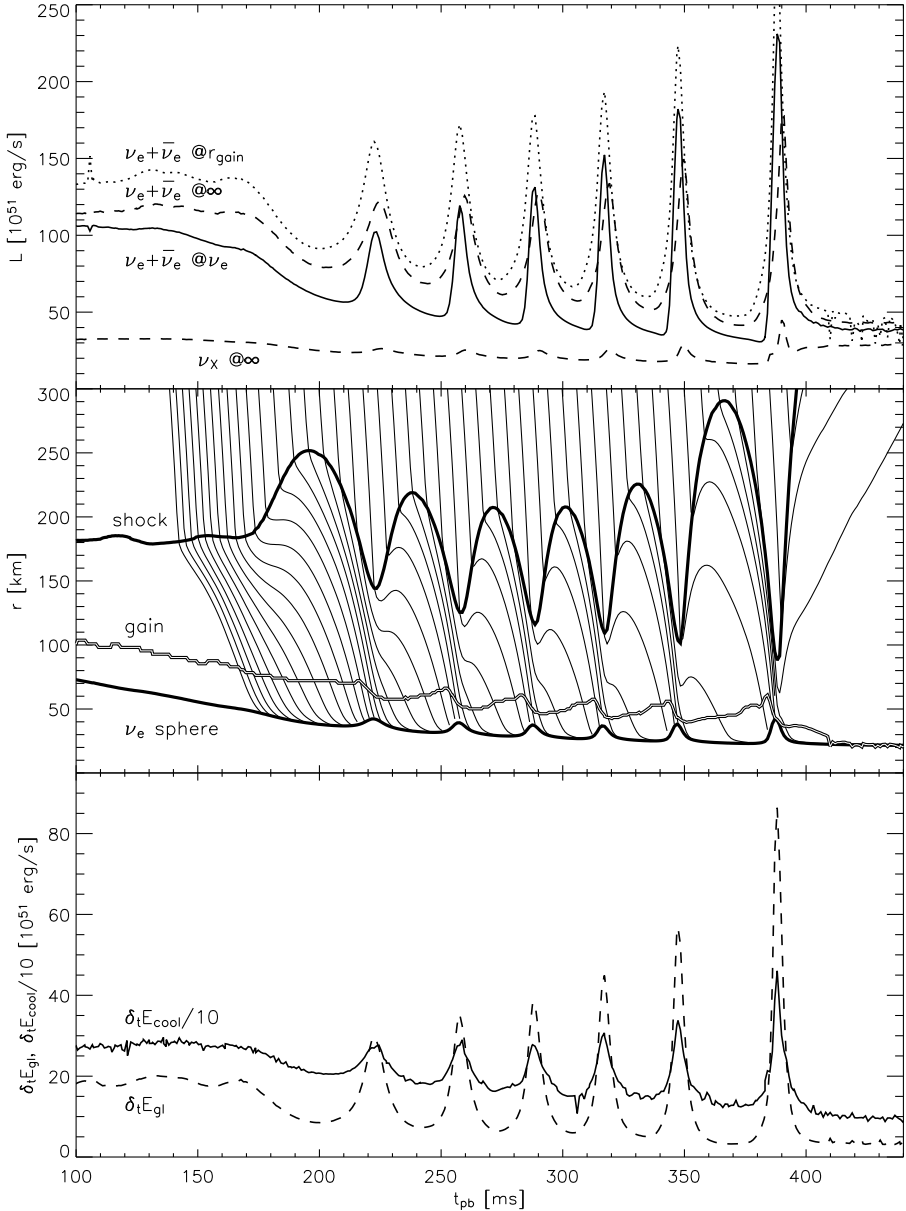} 
  \caption[]{
  Post-bounce evolution of Model s15Gio\_1d.a. The upper panel shows
  the summed comoving-frame luminosity of electron neutrinos and
  antineutrinos at the neutrinosphere of $\nu_\mathrm{e}$ (solid) and
  at the gain radius (dotted), and this luminosity at 400km for an
  observer at rest (dashed). The lower dashed line corresponds to the
  (individual) luminosity for neutrinos of $\mu$ or $\tau$ type, also
  for an observer at rest at 400km. The middle panel displays the
  evolution of the shock (upper thick line), gain radius, and
  $\nu_\mathrm{e}$-sphere (lower thick line). The thin lines represent
  mass shells, starting at $M=1.41\msol$ and increasing in mass in
  steps of $0.002\msol$. The lines were truncated at the
  $\nu_\mathrm{e}$-sphere. The lower panel depicts the total energy
  loss rate in neutrinos from the cooling layer (solid, divided by 10)
  and the net energy deposition rate by neutrinos in the gain layer
  (dashed). Note that here, in contrast to Fig.~\ref{fig:dtE_all}, the
  curves were \emph{not} smoothed over time intervals.
  }\label{fig:osci_mecha}
\end{figure*}
\begin{figure*}[!]
\centering
  \includegraphics[width=17cm]{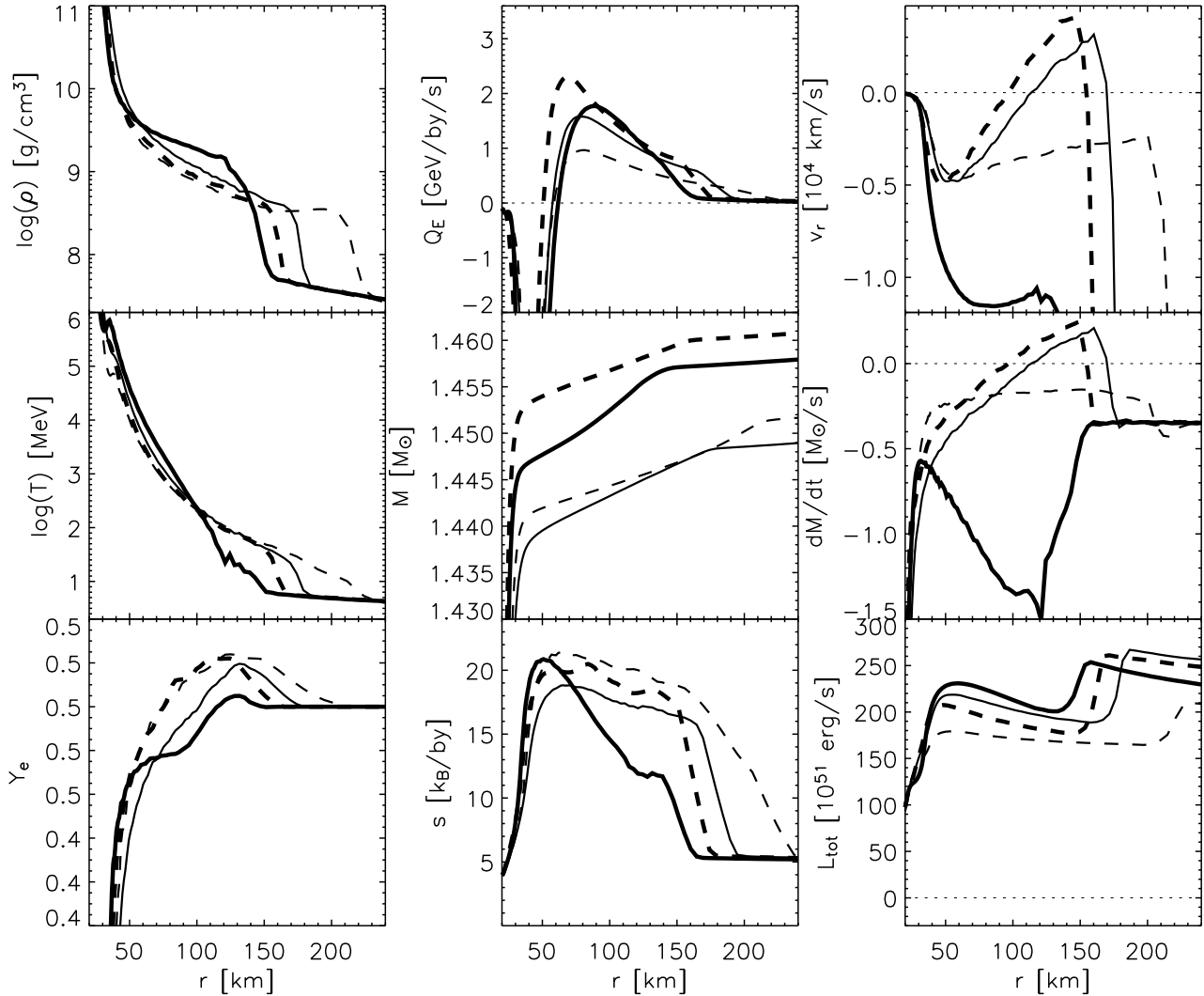} 
  \caption[]{
  Radial structure of Model s15Gio\_1d.a at post-bounce times of
  228.6ms (solid), 236.4ms (thin dashed), 254.4ms (thick solid), and
  262.6ms (thick dashed). At the first and last time, the post-shock
  velocities reach their maximum positive values, at the second time
  the shock reaches its local maximum, and at the third time the
  post-shock velocities reach their minimum negative values. At 254ms
  the maximum cooling rate reaches $Q_\mathrm{E} = -30.1\mathrm{GeV/s}$ per
  baryon.
  }\label{fig:1d_snaps_osci}
\end{figure*}

The $\beta$-terms were found to have a dramatic dynamical effect. In
Fig.~\ref{fig:spos_all} we see the evolution of Models s15Gio\_1d.a
and s15Gio\_1d.b, which mostly differ in the fact that only Model
``b'' includes the $\beta$-terms\footnote{The disregard of
flavor-coupling neutrino processes in Model s15Gio\_1d.a is not very
relevant in the present context.}. We see that the shock positions
evolve increasingly differently. Initially, in Model ``a'' the
post-shock heating by neutrinos is higher, see
Fig.~\ref{fig:dtE_all}. At later times, the increased heating has
driven the shock farther out, thus increasing the size and mass of the
gain layer. The resulting positive feedback between increasing gain
layer and thus increasing heating on the one hand and expanding shock
radius on the other leads to a much larger shock radius than in Model
``b'', where the contracting shock causes the gain layer to become
more and more narrow. During the phase of rapid accretion until about
180ms post-bounce, the total heating rate in the gain layer is larger
in Model s15Gio\_1d.b by a factor of about two, and even larger at
later times.

This analysis is confirmed when Models s15Gio\_1d.b (with
$\beta$-terms) and s15Gio\_1d.a are compared at $t=73$ms post-bounce
(Fig.~\ref{fig:dy_HJ}): The flux factor $f_H\equ H/J$ in the former
model increases much faster in the regions of effective neutrino
cooling and heating than in Model s15Gio\_1d.a. The lower neutrino
energy density corresponds to a significantly reduced integral rate of
neutrino heating and a much reduced rate of cooling below the gain
radius.

\subsubsection{A model with marginal behaviour}
\label{sec:osci}

Although Model s15Gio\_1d.a was computed with reduced accuracy in the
neutrino transport (the $\beta$-terms in the first order moments
equation were neglected), its evolution deserves a discussion. It
reveals an oscillating shock behaviour with growing amplitude (similar
to what was previously seen in models of the Livermore group, see
e.g.~\citealp{wilmay86}), which finally leads to an explosion.

As discussed in Sect.~\ref{sec:veloc}, the omission of the
$\beta$-terms causes increased neutrino heating behind the stalled SN
shock. The shock is therefore driven farther out before it stagnates
at about 180km. When the interface between silicon layer and
oxygen-rich silicon shell arrives at the shock, a sudden decrease of
the ram pressure occurs and instigates a sequence of periodic phases
of shock expansion and contraction (Figs.~\ref{fig:osci_mecha},
\ref{fig:1d_snaps_osci}). Whenever the shock expands and the mass
infall is decelerated, the neutrino luminosity drops with a slight
delay as a consequence of the decreased mass accretion rate towards
the cooling region. In response to the decreased luminosity the
heating almost simultaneously drops, partly also because the accreted
material expands behind the shock and stays away from the gain radius
where heating is most efficient. With the energy input being quenched
the matter behind the shock quickly falls inward. Being drained of its
pressure support the shock expansion comes to a halt again and the
shock retreats. The sudden increase of the mass flow into the cooling
region raises the accretion luminosity extremely, and the heating rate
reaches up to 20\% of the cooling rate, also enhanced by more mass
moving into the region just above the gain radius. The shock bounces
and is driven out in a new phase of expansion. This cycle is repeated
several times. Obviously, the feedback of the cycle is positive, since
the amplitude of the shock expansion increases with time and the cycle
finally leads to an explosion.

It is beyond the scope of this work to determine exactly the criterion
when such a feedback cycle is obtained and why Model s15Gio\_1d.a but
no other of our models shows such a vertiginous fate. The fact that the
$\beta$-terms are neglected certainly increases the heating, and a
sufficiently large driving force seems crucial for entering the
oscillatory mode.

We see two possibilities for explaining the phenomenon, however make
no effort to perform a detailed analysis of any of these. On the one
hand the physical conditions which characterize the expansion and
contraction phases resemble the conditions that allow for
non-adiabatic vibrational instability ($\kappa$-mechanism) in stellar
atmospheres where the outward-going heat flow is modulated by the
rhythm of the pulsation and instability occurs if during adiabatic
compression the absorption coefficient increases \cite[see,
e.g.~][~Chapter 39.1]{kipwei90}.

On the other hand, the model might reveal the action of the so-called
acoustic-advective cycle proposed by \cite{fogtag00} for adiabatic
accretion flows to black holes. In this scenario, acoustic waves
created at the PNS surface propagate to the shock and cause entropy
fluctuations there. When these fluctuations are advected to the PNS
surface by the accretion flow, they create new acoustic waves and so
on. If the feedback is positive, an $l=0$ mode can be built up, as
seen in our model. A preliminary analysis by T.~Foglizzo (2003,
personal communication) revealed that our oscillating model has
favorable conditions for developing an $l=0$ instability by the
advective acoustic cycle. However, in our model neutrino heating and
cooling dominate and the accretion flow is not at all adiabatic. A
more reliable analysis has to account for this fact and remains to be
done.

\subsection{Two-dimensional models}
\label{sec:tdm}

\begin{figure}[!]
  \begin{tabular}{c}
    \put(0.9,0.3){{\Large\bf a}}
    \resizebox{0.9\hsize}{!}{\includegraphics{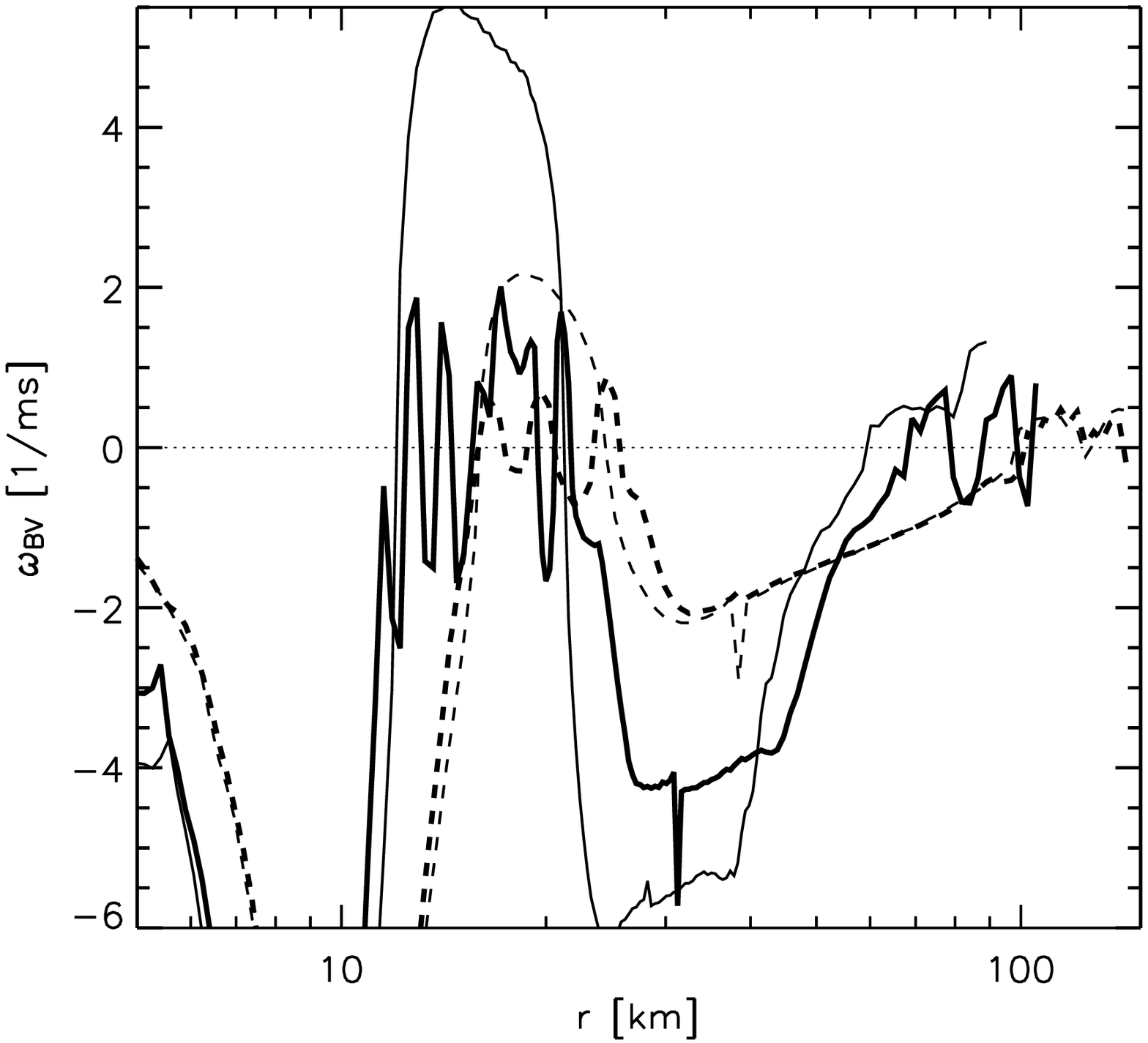}} \\ 
    \put(0.9,0.3){{\Large\bf b}}
    \resizebox{0.9\hsize}{!}{\includegraphics{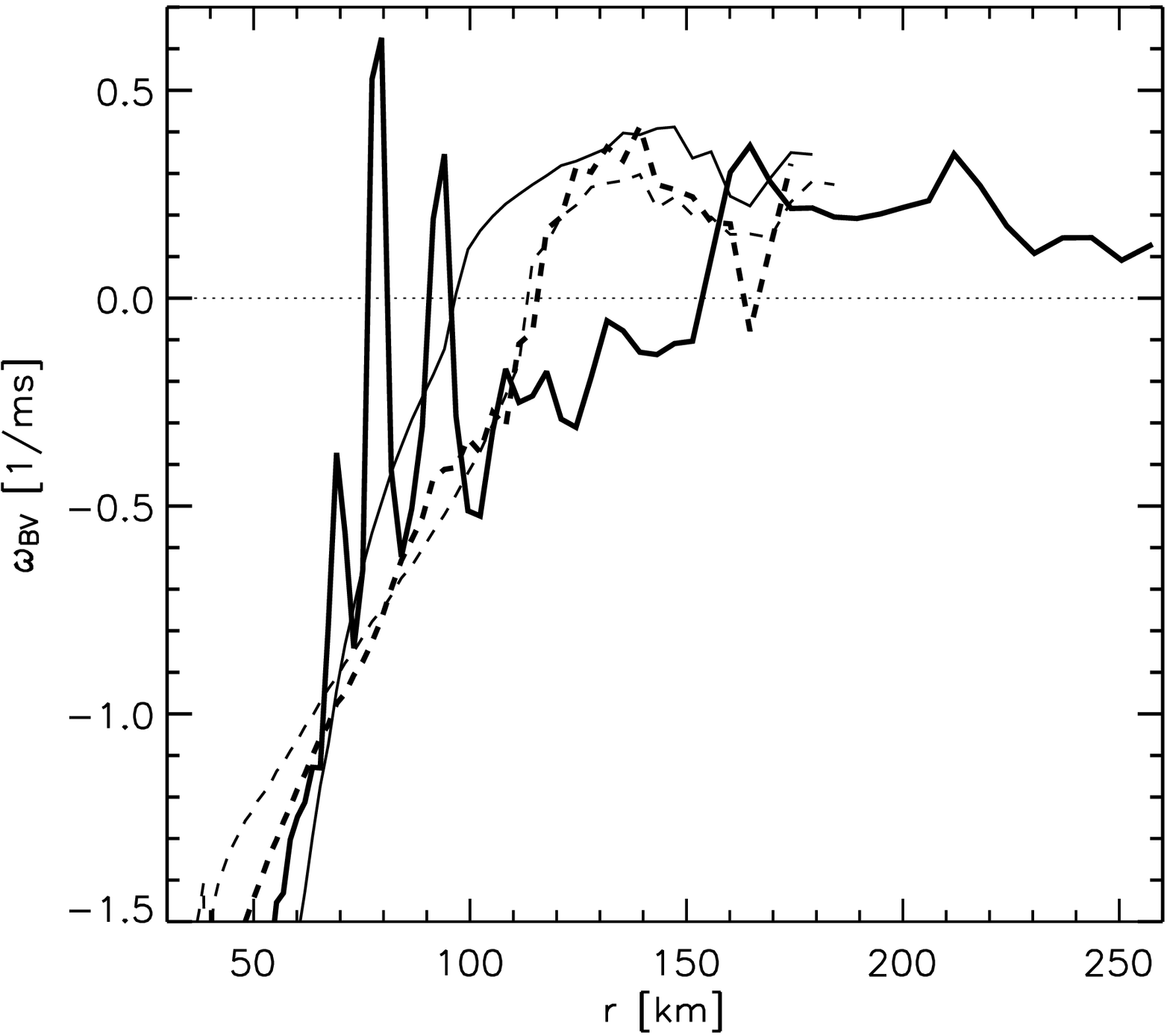}} \\ 
  \end{tabular}

  \caption[]{
  {\bf a} Brunt-V\"ais\"al\"a frequency as defined in
  Sect.~\ref{sec:latadvec} at the post-bounce times of 72.3ms (dashed)
  and 181.4ms (solid) for Model s15Gio\_1d.b (thin) and s15Gio\_32.b
  (thick). The lines are plotted only downstream of the
  shock. Positive values indicate instability. For the 2D model the
  evaluation was performed with laterally averaged quantities.
  {\bf b} Same as {\bf a}, but for models s15Gio\_1d.a (thin) and
  s15Gio\_32.a (thick) at post-bounce times of 72.2ms
  (dashed) and 150.1ms (solid).
  }\label{fig:ledoux_HB_dud}\label{fig:led_HB_expl}
\end{figure}

An inspection of the stability criterion (see Sect.~\ref{sec:latadvec}
for our specific definition of the stability criterion) for the 1D
models (Fig.~\ref{fig:ledoux_HB_dud}) tells us that the gain layer as
well as an extended region in the PNS are convectively unstable, so
that multi-dimensional simulations become mandatory. Here, we present
the first two-dimensional simulations of convection in supernova cores
using a multi-frequency treatment of neutrino transport. The two
Models s15Gio\_32.a and s15Gio\_32.b correspond to the one-dimensional
Models s15Gio\_1d.a and s15Gio\_1d.b, respectively, presented in
Sects.~\ref{sec:std} and \ref{sec:osci}. Both 2D simulations include
our most elaborate description of neutrino-matter interactions (Case
``io'') as described in Appendix \ref{sec:neuopa} of this paper and an
approximative, spherically symmetric treatment of general relativity
(Sect.~\ref{sec:hydro}). Remember that Model ``b'' is the one with the
full treatment of neutrino transport, while Model ``a'' incorrectly
neglects the $\beta$-terms in radial direction (see
Sect.~\ref{sec:veloc}). In the angular direction both models employ 32
zones with a resolution of $2.7\degr$, covering a wedge around the
equator from $-43.2\degr$ to $+43.2\degr$. Both models were
followed through core collapse in one dimension and mapped to the 2D
grid around 7ms after the bounce. Simultaneously, perturbations were
seeded by randomly changing the radial velocity by up to $\pm$1\% in
the entire star. The treatment of the lateral neutrino transport was
carried out as described in Sect.~\ref{sec:2dneutra}. Note that in
Model s15Gio\_32.b the inner 2km were treated spherically
symmetrically to maintain a reasonably large CFL time step. Model
s15Gio\_32.a did not include the lateral momentum transfer to the
fluid, Eq.~(\ref{eq:collint_theta}), for which reason the core inside
a radius of 25km had to be calculated in spherical symmetry. Thus
artificial hydrodynamic instability could be avoided, but also the
physical PNS convection would not be followed in a layer below the
neutrinospheres. Note that the entropy wiggles in the postshock layer
discussed in detail in Sect.~\ref{sec:swiggles} were present in both
2D simulations presented here and introduce artificial perturbations
of the stellar models in radial directions. More recent calculations
with our improved scheme of treating rest-mass effects in the EoS
(Sect.~\ref{sec:swiggles} and Appendix \ref{app:noff}), however,
showed that these numerical perturbations on the entropy profile did
not have an appreciable influence on the development and growth rates
of convection in the hot bubble region.

\subsubsection{s15Gio\_32.b: A model with full transport treatment}
\label{sec:acme}

\begin{figure*}[!]
\sidecaption
  \includegraphics[width=11cm]{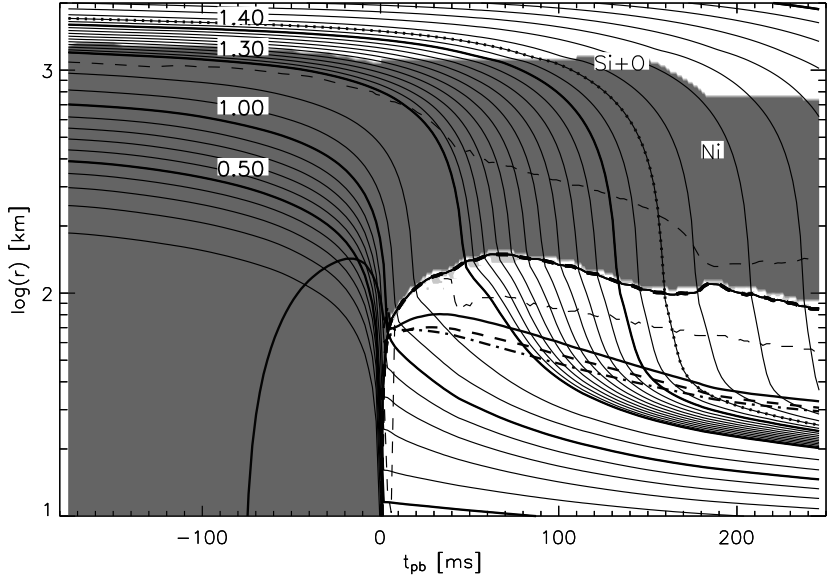} 
  \caption[]{
  ``Mass shell'' trajectories for Model s15Gio\_32.b. The shells
  plotted with bold solid lines are labelled by the corresponding
  enclosed masses. The plot also shows the transport neutrinospheres
  for $\nu_\mathrm{e}$ (thick solid), $\bar \nu_\mathrm{e}$ (thick dashed),
  and $\nu_\mathrm{x}$ (thick dash-dotted), the mass shell at which the
  silicon shell becomes oxygen-rich (knotted solid line, at
  1.42$\msol$), and the angle-averaged shock radius (solid line with
  dashes). Further, we have marked the regions which contain a mass
  fraction of more than 60\% in Fe-group elements (dark shaded) and of
  more than 30\% in $^4\mathrm{He}$ (light shaded, visible in the
  vicinity of the shock). Finally, the inner thin dashed line marks
  the gain radius, while the outer one marks the interface between the
  regimes of the low-density and high-density EoSs
  ($\rho=6\times10^7\gcm$).
  }\label{fig:mas_acme}
\end{figure*}

Model s15Gio\_32.b represents a 2D supernova computation with the
presently most complete and detailed implementation of spectral
neutrino transport in such models. It fails to explode. In
Fig.~\ref{fig:mas_acme} we show the evolution of shock trajectory and
``mass shells'' of this model. Note that a ``mass shell'' trajectory
of a 2D simulation does not represent the evolution of Lagrangian mass
elements as in spherical symmetry but describes the time evolution of
the radius of the spherical shell enclosing a certain value of the
mass. After the prompt expansion, the shock is pushed slowly outward
by accumulating mass until it reaches a maximum radius of 150km around
60--70ms, and then starts to slowly retreat again. This is quite
similar to the 1D case (c.f.~Fig.~\ref{fig:spos_all}). Convection in
the neutrino-heated layer and in the PNS shifts the shock position
farther out by just several 10km.

The initial hope in performing 2D calculations was that the ``hot
bubble'' (HB) convection in the gain layer would strengthen the shock
sufficiently to lead to an explosion, similar to what was found
previously by \cite{herben94} and \cite{burhay95} with neutrino
transport being described by grey, flux-limited diffusion. Hot bubble
convection was found to be helpful for shock revival by means of
transporting energy to the shock, thus enhancing the postshock
pressure and increasing the efficiency of neutrino energy deposition:
Strongly heated material close to the neutrinosphere is carried to the
shock in bubbles rising due to buoyancy forces. At the same time
narrow downflows passing besides the bubbles feed the region of strong
neutrino heating with cool material. 2D calculations also get rid of
another big disadvantage of 1D calculations: When the shock starts
expanding in 1D calculations, the accretion of fresh material into the
cooling region is quenched, the sudden drop in luminosity and thus
heating reduces further support for the expanding shock and thus can
be fatal for an explosion.

\begin{figure}[!]
  \begin{tabular}{c}
    \put(0.9,0.3){{\Large\bf a}}
    \resizebox{0.9\hsize}{!}{\includegraphics{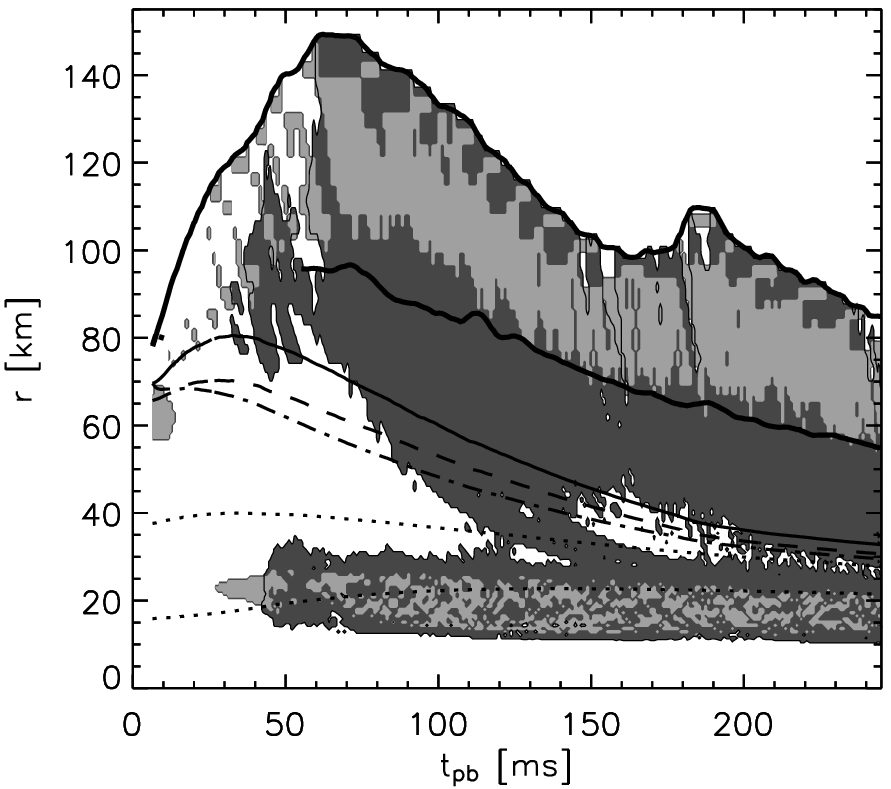}} \\ 
    \put(0.9,0.3){{\Large\bf b}}
    \resizebox{0.9\hsize}{!}{\includegraphics{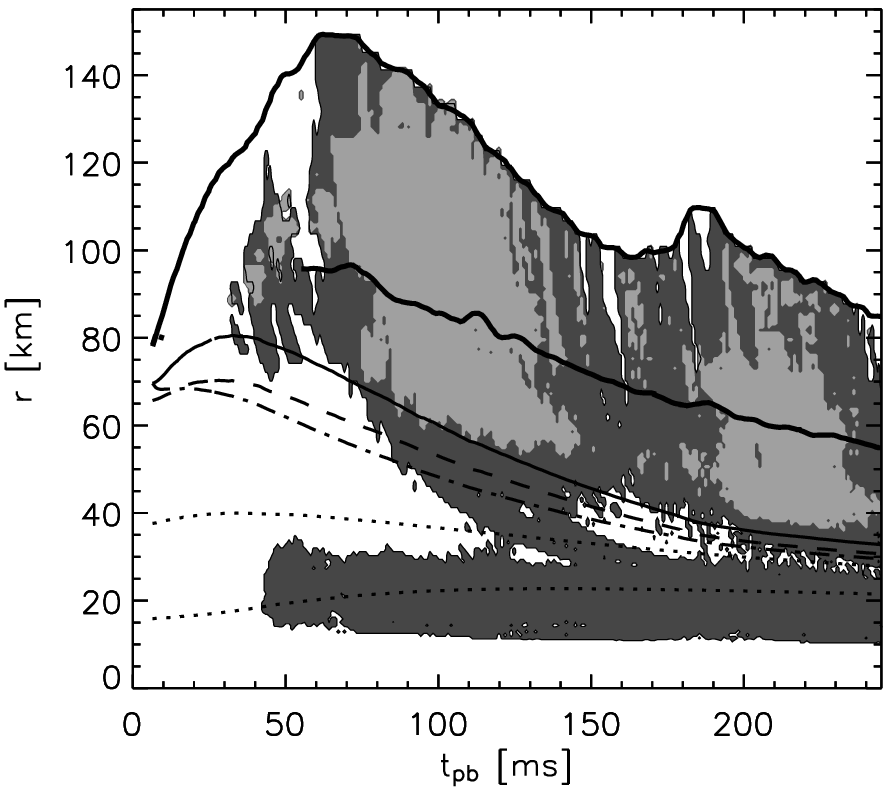}} 
  \end{tabular}

  \caption[]{
  {\bf a} Convective regions for Model s15Gio\_32.b. Dark shading
  indicates regions where the lateral velocities reach more than
  700km/s, light shading regions where the stability criterion
  described in Sect.~\ref{sec:latadvec} predicts instability. The
  contour lines correspond to the boundaries of the dark-shaded
  regions. The evaluation was performed only downstream of the shock
  (upper thick solid line) with polar averages of the variables. The
  lower thick solid line depicts the gain radius. The thin lines
  depict the neutrinospheres for $\nue$ (solid), $\nuae$ (dashed), and
  $\nux$ (dash-dotted). Finally, the dotted lines show where the
  densities are $10^{12}\gcm$ (upper line) and $10^{13}\gcm$ (lower
  line).
  {\bf b} Same as {\bf a} but with the light shading marking regions
  where the standard deviation of entropy variations (defined in
  analogy to Eq.~\ref{eq:sigma}) is $\sigma_{s}>0.02$.
  }\label{fig:conv_HB_dud}
\end{figure}

\begin{figure}[!]
  \begin{tabular}{c}
    \put(0.9,0.3){{\Large\bf a}}
    \resizebox{0.95\hsize}{!}{\includegraphics{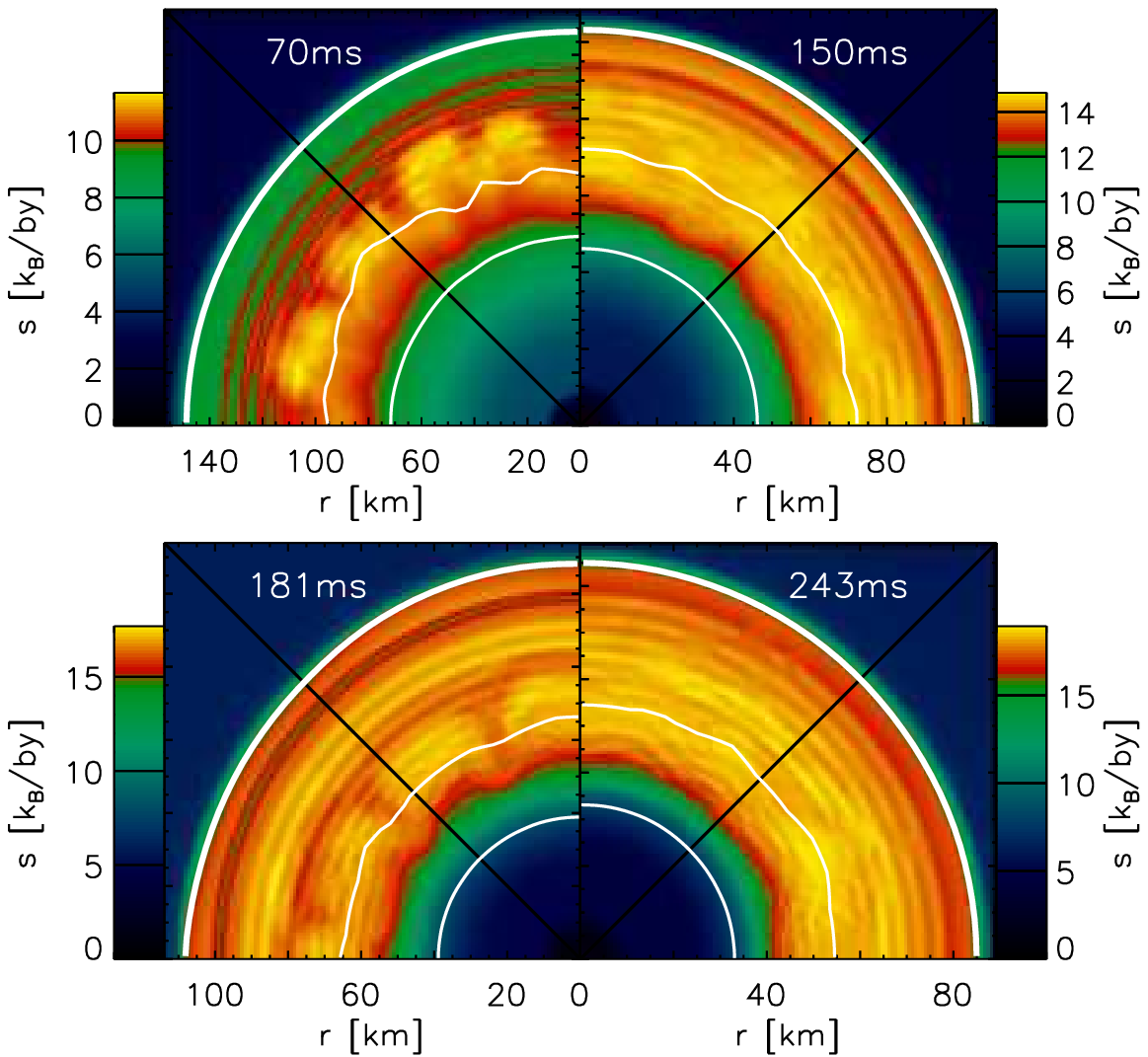}} \\ 
    \put(0.9,0.3){{\Large\bf b}}
    \resizebox{0.95\hsize}{!}{\includegraphics{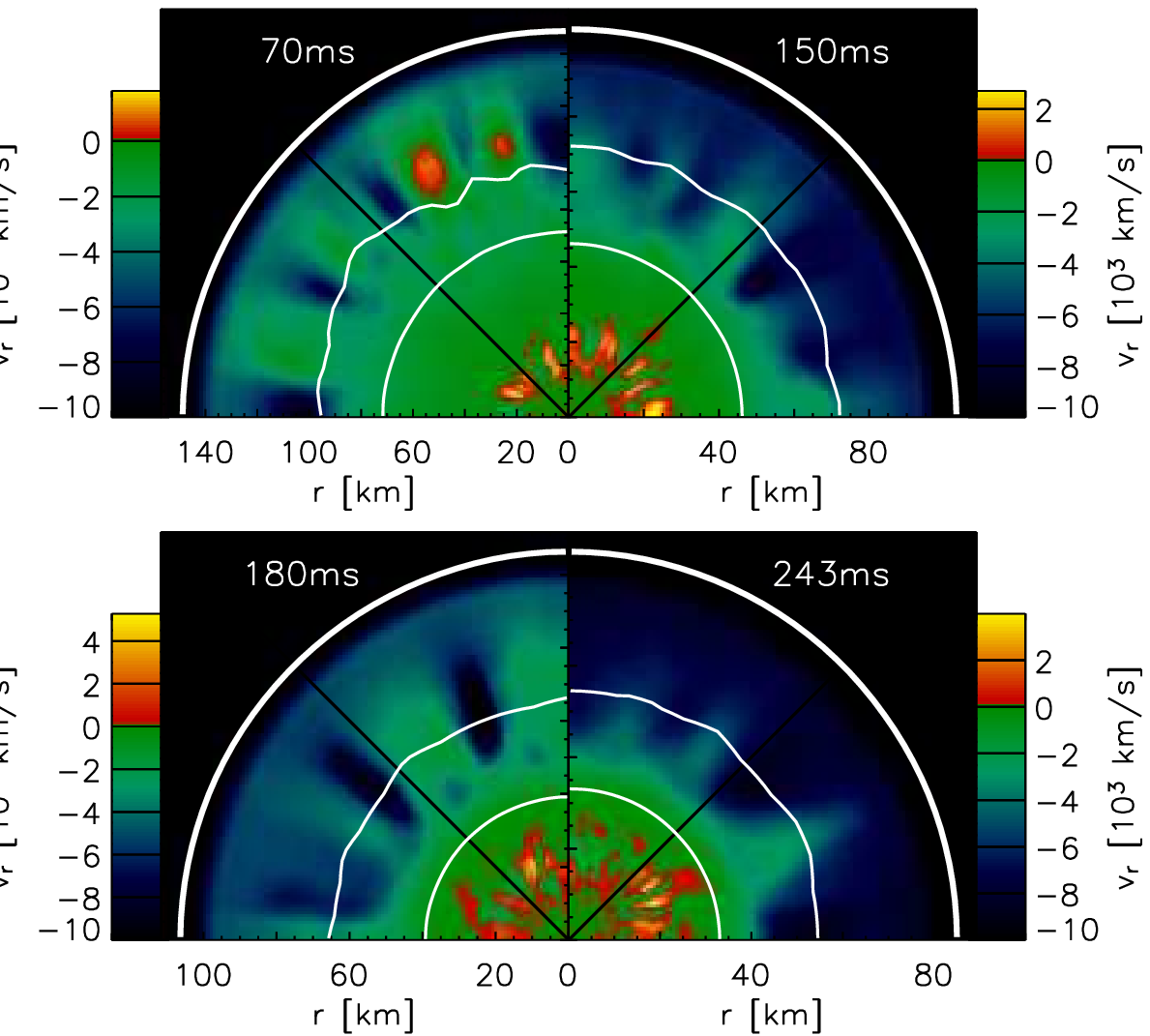}} 
  \end{tabular}

  \caption[]{
  Snapshots of Model s15Gio\_32.b with convection in the ``hot
  bubble'' region. The white lines mark the shock position (thick),
  the gain radius (outer thin), and the $\nu_\mathrm{e}$-sphere (inner
  thin). The solid black line denotes the equatorial plane of the
  polar grid used in the simulation. {\bf a} These panels show colour
  coded the entropy $s$. {\bf b} All panels show colour coded the
  radial velocity $v_r$.
  }\label{fig:snaps_HB_dud}
\end{figure}

In the current model, although the whole gain layer is convectively
unstable from $t_\mathrm{pb}=40$ms on (see Fig.~\ref{fig:conv_HB_dud}a),
the convection has difficulties developing: Becoming distinct at 60ms
after bounce, the HB convection (Fig.~\ref{fig:snaps_HB_dud}) has not
yet grown strong and formed large-scale structures when the shock
begins retreating again. Therefore HB convection is unable to push the
shock significantly farther out relative to the shock in the 1D
simulation. Also during the later evolution the HB convection cannot
develop strength because within only a few milliseconds the matter
accreted by the shock is advected into the cooling layer and buoyancy
forces are not strong enough to enforce significant rise against the
rapid infall of the postshock flow. The shock itself shows almost no
response to the convection and remains spherically symmetric
throughout the run (Fig.~\ref{fig:snaps_HB_dud}). The HB convection in
this model leads to only small quantitative differences compared to
the one-dimensional results.

Eventually, the transient shock expansion correlated with the sudden
decrease of the ram pressure when the oxygen-rich silicon shell meets
the shock allows the hot bubble convection to strengthen because the
accretion velocities behind the shock are reduced. At this time
(180ms), however, the shock has already retreated to below 100km, and
the infall velocity is too large to be inverted by the outward
acceleration of buoyant, neutrino-heated matter.

Fig.~\ref{fig:conv_HB_dud}b is suitable for showing the effects
arising from convection. While entropy fluctuations develop due to the
convective activity after convection has become distinct
about 60ms after bounce, convection is too weak to noticeably
perturb the entropy profile about 80ms later. The perturbations grow
again when convection is transiently revived after the Si-SiO
interface has passed the shock. Interestingly, the entropy
perturbations persist for some time when they are advected downstream
into the convectively stable cooling region (compare with
Fig.~\ref{fig:conv_HB_dud}a).

A feature visible in Fig.~\ref{fig:conv_HB_dud} betwen 30 and 50km
needs to be explained: Large lateral velocities associated with matter
advected downward from the HB to a convective zone in the PNS show up
between 100 and 150ms post-bounce. This phenomenon is caused by the
periodic lateral boundary conditions applied in our simulation, which
allow rings of uniform, lateral velocity to develop and to be
stable. 
In case of reflecting boundary conditions such an effect would be
suppressed. This reminds us that in two-dimensional simulations the
chosen lateral boundary conditions determine the results to some
extent. In the present case the kinetic energy stored in the rings of
lateral motion is negligible compared to the energy needed to trigger
an explosion and we are therefore confident that this numerical
phenomenon has no macroscopic dynamical consequences.

Fig.~\ref{fig:conv_HB_dud} shows that a second region of convective
activity is present inside the PNS and persistently encompasses a
layer between 10--15km and about 30km. The layer of large lateral
velocities is somewhat more extended in the radial direction than the
layer of convective instability due to over- and undershooting of
buoyant matter. The absence of visible entropy fluctuations in
Fig.~\ref{fig:conv_HB_dud}b is explained by the very high efficiency
with which the PNS convection transports and redistributes
entropy. Therefore the entropy gradient becomes extremely flat in this
region. Despite of leading to changes of the PNS structure and
affecting the neutrino emission, this convection inside the PNS has no
major consequences for the NS evolution during the simulated periods
of time. We will discuss these effects in greater detail in our
forthcoming Paper II.

\begin{figure}[!]
  \begin{tabular}{c}
    \put(0.9,0.3){{\Large\bf a}}
    \resizebox{0.9\hsize}{!}{\includegraphics{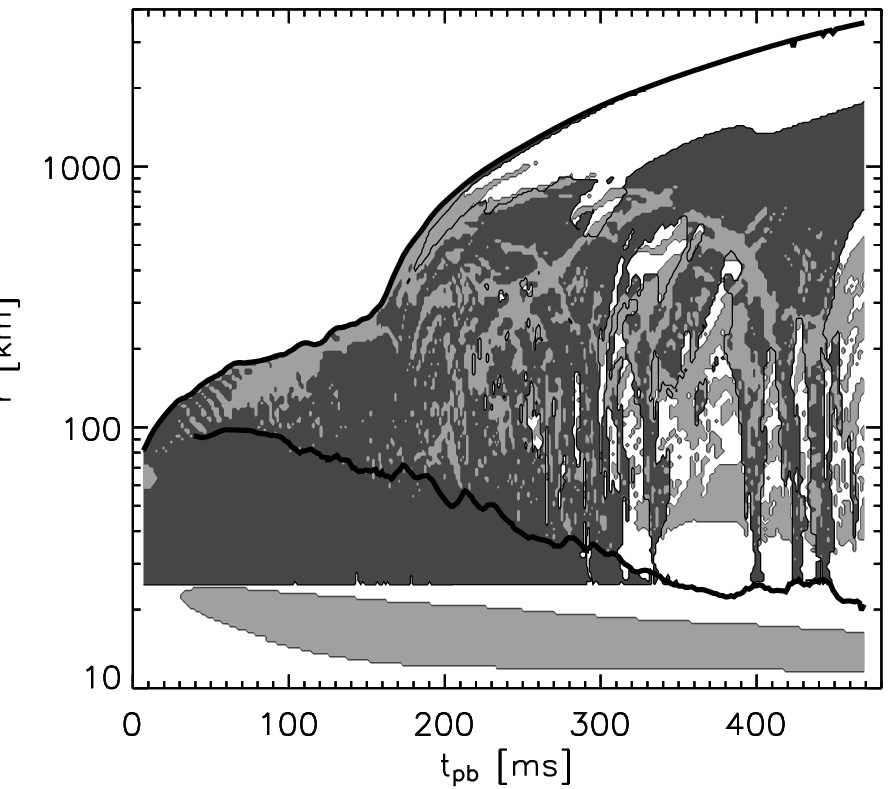}} \\ 
    \put(0.9,0.3){{\Large\bf b}}
    \resizebox{0.9\hsize}{!}{\includegraphics{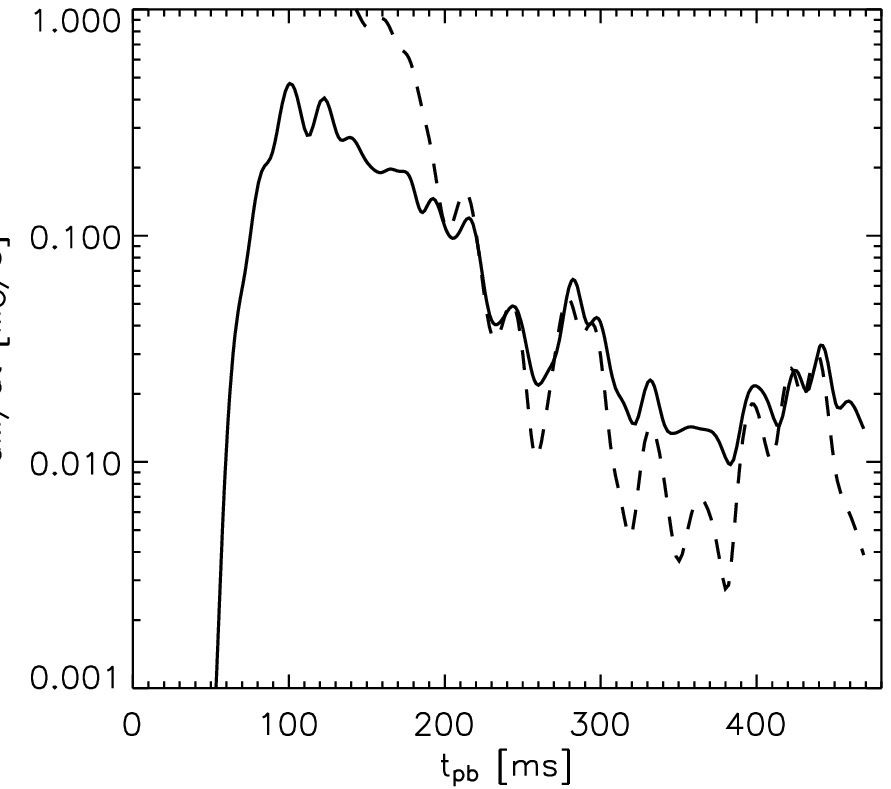}} 
  \end{tabular}

  \caption[]{
  {\bf a} Convective regions for Model s15Gio\_32.a. Dark shading
  indicates where the mass inflow rate (evaluated for zones where $v_r
  < 0$) exceeds $10^{-2}\msol/\mathrm{s}$. Light shading denotes regions
  where $\omega_\mathrm{BV} > 30/\mathrm{s}$ (see Sect.~\ref{sec:latadvec}
  for the definition of the stability criterion). The contour lines
  mark the boundaries of the dark-shaded regions. The evaluation was
  performed only downstream of the shock (upper thick solid line) with
  polar averages of the variables. The lower thick solid line
  corresponds to the gain radius. Note that the core of this model
  inside of 25km was computed in spherical symmetry.
  {\bf b} The solid line represents the mass outflow rate (evaluated
  for zones where $v_r > 0$) at $r=100$km versus time. The dashed line
  shows the corresponding mass inflow rate. The lines were smoothed
  over time intervals of 5ms.
  }\label{fig:conv_HB_expl}
\end{figure}
\begin{figure}[!]
  \begin{tabular}{c}
    \put(0.9,0.3){{\Large\bf a}}
    \resizebox{0.95\hsize}{!}{\includegraphics{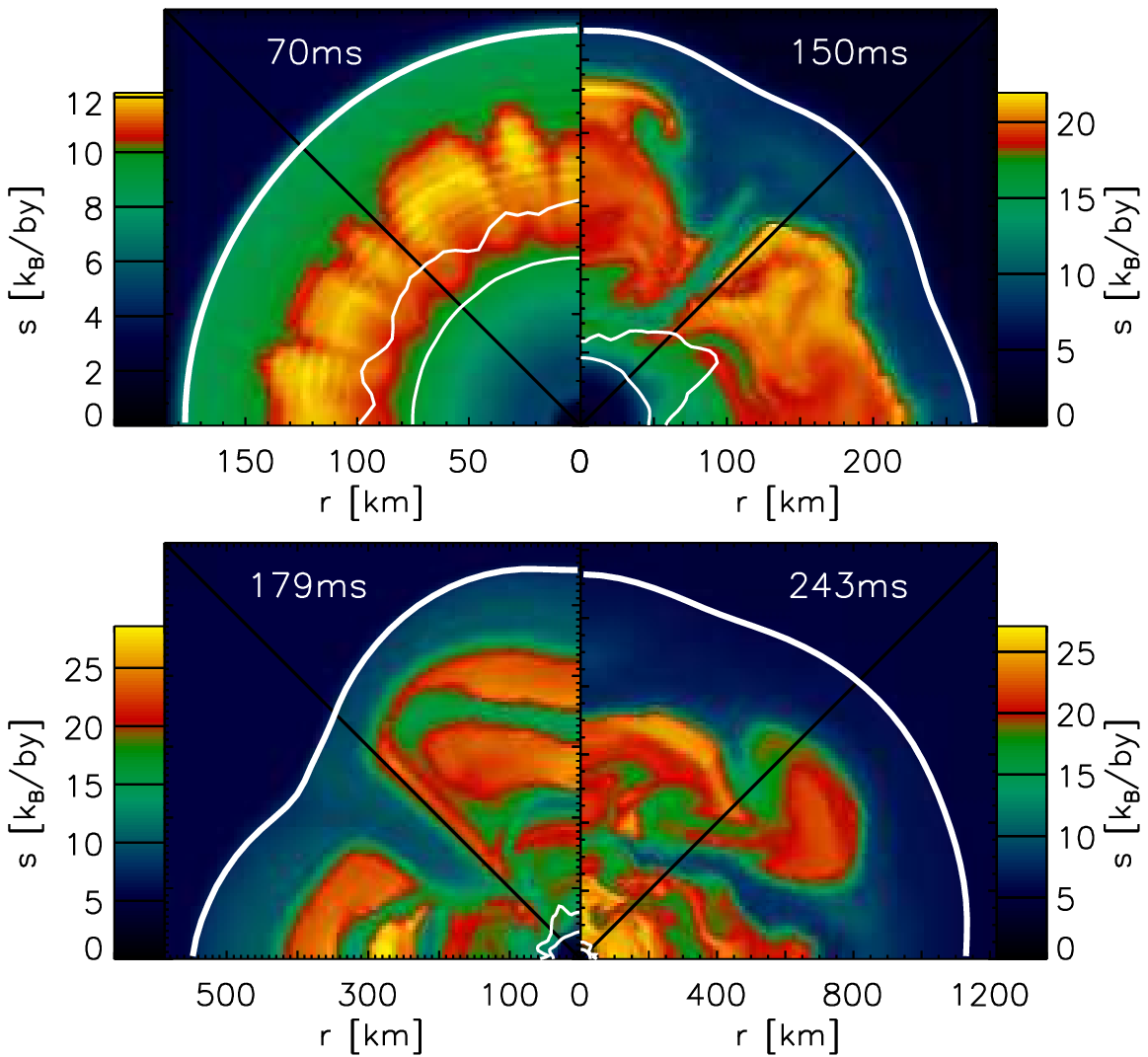}} \\ 
    \put(0.9,0.3){{\Large\bf b}}
    \resizebox{0.95\hsize}{!}{\includegraphics{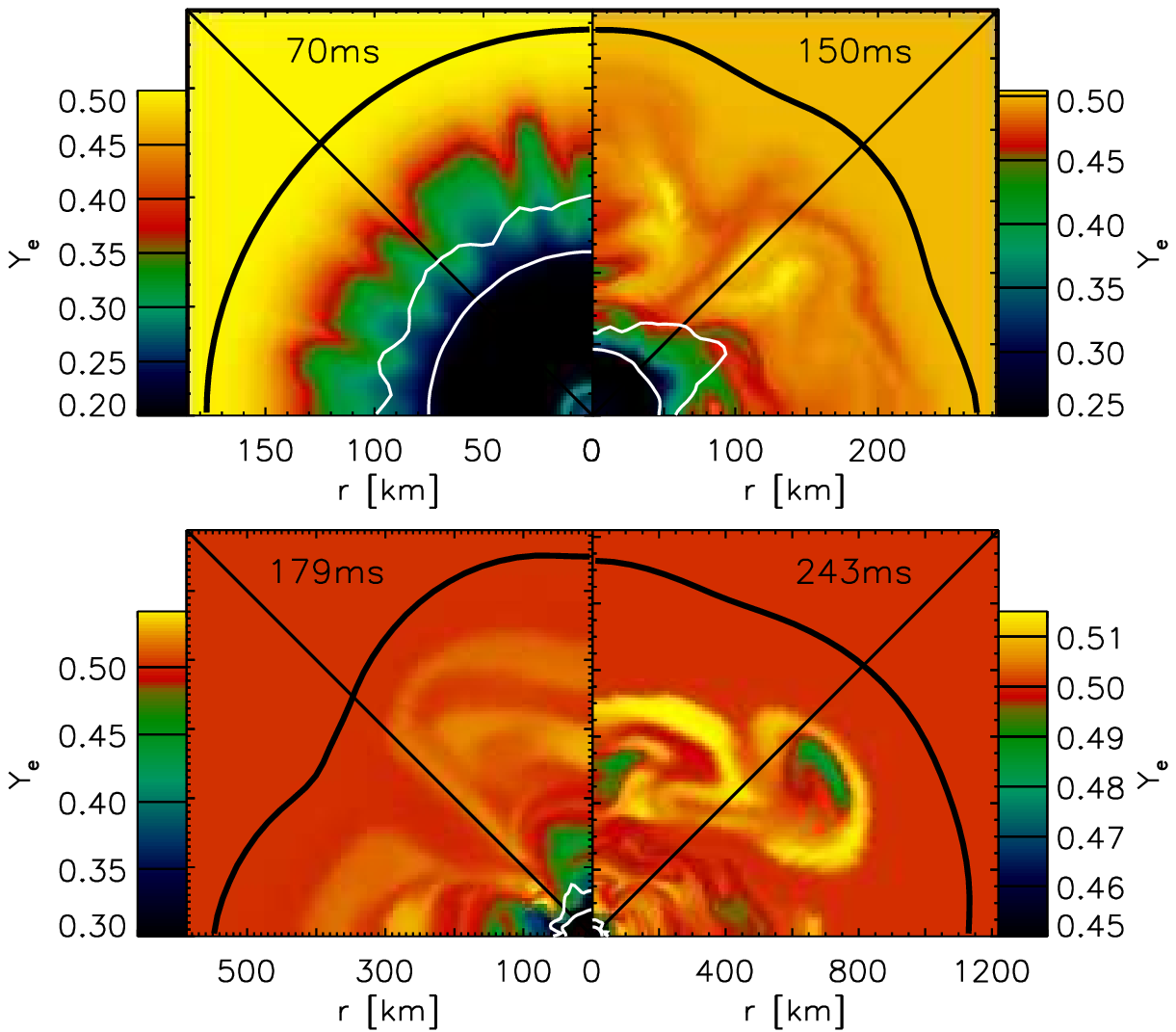}} 
  \end{tabular}

  \caption[]{
  2D snapshots of Model s15Gio\_32.a for four different post-bounce
  times. {\bf a} shows the entropy $s$, {\bf b} shows the electron
  fraction $Y_\mathrm{e}$. The diagonal black lines mark the equator
  of the polar grid, the horizontal axis corresponds to
  $\cos(\vartheta)=-0.67$, the perpendicular axis to
  $\cos(\vartheta)=+0.67$. The shock radius is indicated by a thick
  black or white line, and the gain radius and $\nu_\mathrm{e}$-sphere
  by thin white lines.
  }\label{fig:snapsh_expl}
\end{figure}

\subsubsection{s15Gio\_32.a: A model with artificial explosion}
\label{sec:expl}

A comparison of the simulations performed for this work shows that
none of our models with the full implementation of relevant physics
(state-of-the-art description of neutrino-matter interactions, general
relativistic gravity, variable Eddington factor closure of coupled set
of neutrino moments equations and model Boltzmann equation) yields
explosions, neither in spherical symmetry nor in 2D with convective
processes taken into account. But as we already have seen in
connection with Model s15Gio\_1d.a, the omission of the
$\beta_r$-dependent terms (but not the advection term 
$\beta_r\partial/\partial r$, which is always included) 
in the neutrino momentum equation
(Eq.~\ref{eq:momeqe2}) makes a big difference and favors
explosions. Model s15Gio\_32.a is the counterpart of this 1D model in
two spatial dimensions and accounts for the fact that the
neutrino-heated ``hot bubble'' region behind the shock is convectively
unstable according to the Ledoux criterion
(Fig.~\ref{fig:led_HB_expl}b). However the reader should keep in mind
that the omission of the $\beta$-terms in Eq.~(\ref{eq:momeqe2}) is an
approximation which alters the neutrino transport in an unacceptable
way. Thus, the explosion obtained in Model s15Gio\_32.a has to be
considered as artificial.

Note that Model s15Gio\_32.a was calculated with a 2D polar grid for
$r\ge 25$km, but spherical symmetry was assumed for radii smaller than
25km, thus ignoring the existence of a convectively unstable layer
below the neutrinosphere. However, convection inside the PNS does not
have a significant influence on the explosion which develops in the
discussed model; it is important for the cooling of the nascent PNS
only on a timescale much longer than 100ms.

For continuing the simulation to late stages with acceptable
requirements of CPU time we have mapped the 2D calculation on a 1D
grid after 468ms of post-bounce evolution. At this time,
multi-dimensional processes (transient downflows of low-entropy
matter) affect the onset of the neutrino-driven wind below 200km,
while the immediate postshock layer is convectively only weakly
unstable (Fig.~\ref{fig:conv_HB_expl}a) and convective structures in
the bulk of expanding matter between this layer and the developing
wind are going to freeze out and expand essentially
self-similarly. The mapping required to set the lateral velocity
components as well as the corresponding momentum and kinetic energy
terms to zero. Along with this change we turned on the $\beta$-terms
in the neutrino momentum equation to follow the evolution of the
neutrino-driven wind from the proto neutron star more realistically,
and switched to the more sophisticated treatment of nuclear
statistical equilibrium at low densities by a 4-species Saha solver as
described in Sect.~\ref{sec:eos}. The latter replacement was necessary
because the previous approximative description of NSE in the
low-density EoS of \citet[Appendix B]{ramjan02} did not solve the
Saha equations but described nuclear dissociation and recombination by
temperature-dependent composition changes which do not guarantee
entropy conservation and thus cause numerical artifacts during the
later phases of wind evolution.

\begin{figure*}[!]
\sidecaption
  \includegraphics[width=11cm]{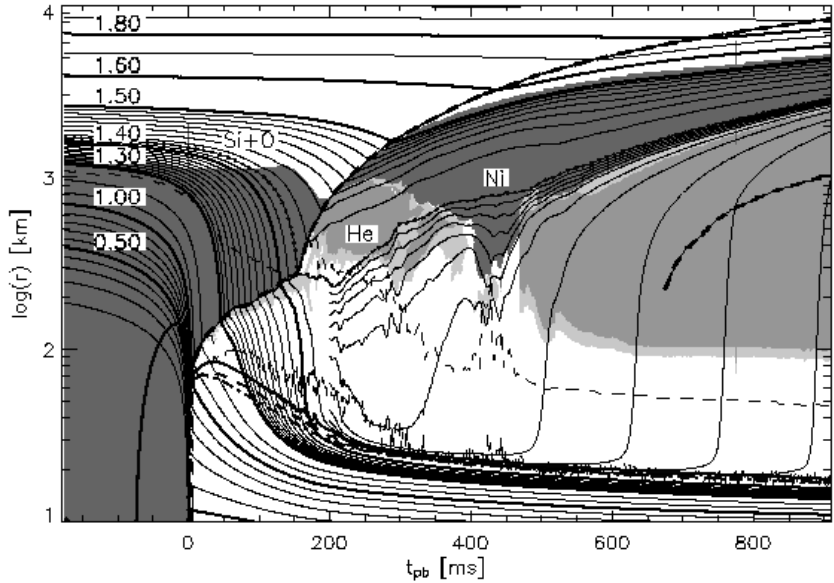} 
  \caption[]{
  ``Mass shell'' trajectories for the exploding 2D Model
  s15Gio\_32.a. ``Shells'' plotted by bold lines are labelled with
  their corresponding enclosed masses. Also shown are the transport
  neutrinospheres for $\nu_\mathrm{e}$ (thick solid), $\bar
  \nu_\mathrm{e}$ (thick dashed), and $\nu_\mathrm{x}$ (thick
  dash-dotted), the interface between the silicon shell and the
  oxygen-rich silicon shell (knotted solid line, at 1.42$\msol$), and
  the shock and reverse shock or wind termination shock (thick solid
  lines with dashes). Moreover, the regions are marked which contain a
  mass fraction of more than 60\% of Fe-group elements (dark shaded)
  or more than 60\% of $^4\mathrm{He}$ (shaded). Light shading
  indicates other regions which have a mass fraction of more than 30\%
  of $^4\mathrm{He}$. The lower dashed line denotes the gain radius
  and the upper one marks the interface between the low-density and
  high-density EoSs (i.e.~this line corresponds to
  $\rho=6\times10^7\gcm$).
  }\label{fig:mas_expl}
\end{figure*}

Note that the steepening density gradient at the PNS surface with
ongoing time made it necessary to refine our Eulerian grid in this
region. Rezoning was done at 206ms, 468ms, and 759ms. We should also
mention that Model s15Gio\_32.a did only take into account Si burning,
but not C, O, Mg, and Ne burning as described in \cite{ramjan02}. A
post-processing estimate, however, showed that oxygen burning would
have contributed less than $0.011\times 10^{51}$erg to the explosion
energy, while the ignition conditions for the other burning processes
were not reached in our model.

\begin{figure}[!]
\sidecaption
  \includegraphics[width=\hsize]{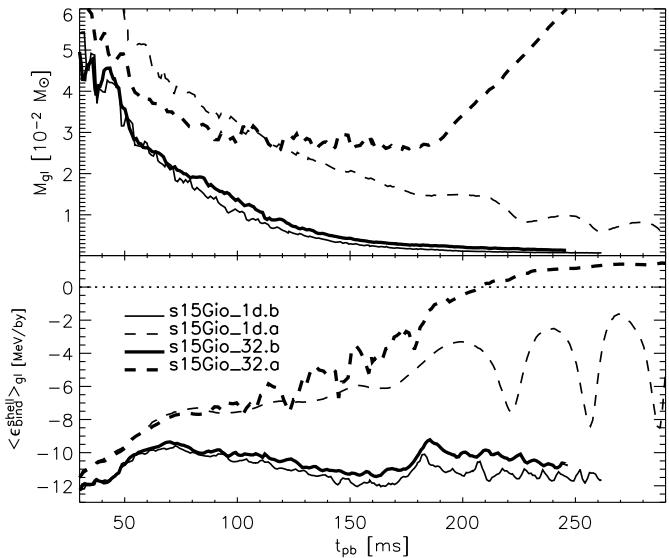} 
  \caption[]{
  Time evolution of the total mass in the gain layer and the average
  local specific binding energy in the gain layer for four models,
  including the exploding Model s15Gio\_32.a.
  }\label{fig:mgeg}
\end{figure}
\begin{figure}[!]
  \resizebox{\hsize}{!}{\includegraphics{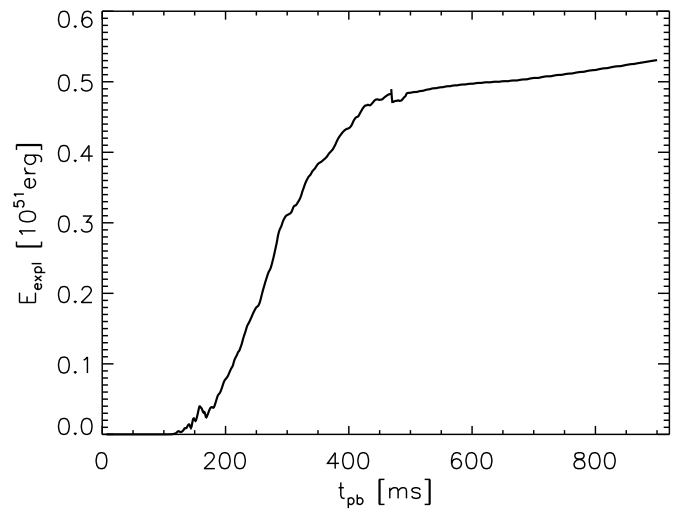}} \\ 
  \caption[]{
  Explosion energy for Model s15Gio\_32.a as defined in
  Eq.~(\ref{eq:eexpl_expl}). The small jump at
  $t_\mathrm{pb}=468~\mathrm{ms}$ is a consequence of the mapping of
  the 2D model to a 1D grid, and the slow increase afterwards is
  associated with the power of the neutrino-driven baryonic wind.
  }\label{fig:eexpl_expl}
\end{figure}

\paragraph{Evolution}

Model s15Gio\_32.a develops powerful convection. Like in Model
s15Gio\_32.b, convective activity in the gain layer starts at about
60ms after bounce (see Fig.~\ref{fig:snapsh_expl}). However, in this
model neutrino heating is stronger (the heating integral is higher
roughly by a factor of 2, cf. $\delta_t E_\mathrm{gl}$ in
Fig.~\ref{fig:dtE_all}) due to the omission of the $\beta$-terms,
leading to a steeper entropy gradient, which in turn means a stronger
destabilization of the gain layer. Furthermore, the stronger heating
is sufficient to prevent the shock contraction seen in Model
s15Gio\_32.b after 70ms post-bounce (see Fig.~\ref{fig:dtE_all}). As
a consequence the convective region between gain radius and shock
remains extended and the convective cells can grow and merge to larger
structures, leading to the largest scale possible on the grid for the
boundary conditions imposed in our simulation (i.e.~a wedge of
$\pm43.2\degr$ around the equator with periodic boundary
conditions): After about 150ms postbounce one big convective cell with
a rising bubble and one downflow has formed and transports energy
efficiently to the shock (Fig.~\ref{fig:snapsh_expl}). Convection in
this model becomes strong enough to push the shock farther out; the
gain layer, and therefore the convective region, continuously expand
after convection has set in (Figs.~\ref{fig:spos_all},
\ref{fig:mas_expl}). Finally, convection supports an explosion,
although the corresponding 1D model, s15Gio\_1d.a, did not
explode at such an early time.

\begin{figure*}[!]
\centering
  \includegraphics[width=17cm]{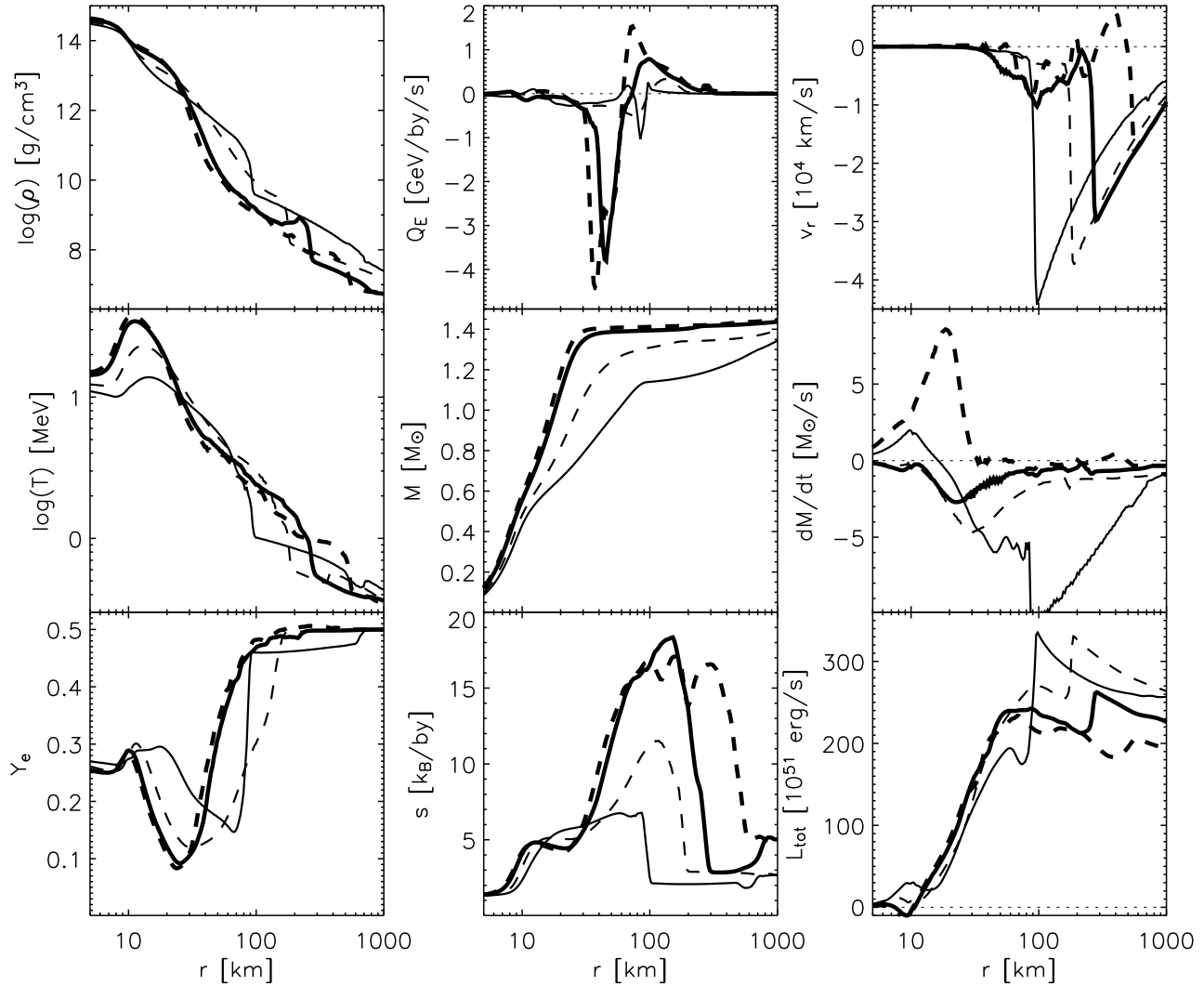} 
  \caption[]{
  Radial profiles (laterally averaged) for different quantities at
  post-bounce times of 11.0ms (solid), 72.2ms (dashed), 150.1ms
  (thick), and 178.5ms (thick dashed) for the exploding Model
  s15Gio\_32.a.
  }\label{fig:1d_snaps_expl}
\end{figure*}

The onset of the explosion occurs when the infalling oxygen-rich
silicon shell arrives at the shock at 260km and 150ms after
bounce. The interface between the silicon layer and the O-rich silicon
shell is characterized by a steep entropy jump and the preshock
density drops to a much lower value. This leads to a sudden decrease
of the mass accretion rate and thus of the ram pressure. The shock
reacts to that by expansion and rapid acceleration. Although the
explosion starts at this moment, the existence of the composition
interface may not be crucial for the explosion of Model
s15Gio\_32.a. Favorable conditions for an explosion can be seen even
before the composition interface has reached the shock: For one the
shock is already expanding (Fig.~\ref{fig:mas_expl}). Second, the mass
contained in the gain layer remains constant between 70 and 170ms
after bounce in contrast to all other simulations, including Model
s15Gio\_1d.a, where $M_\mathrm{gl}$ is always decreasing, see
Fig.~\ref{fig:mgeg}. (Note that the difference in $M_\mathrm{gl}$
between Models s15Gio\_32.a and s15Gio\_1d.a during the first 70ms
after bounce arises from the omission of the $\nu\nu\leftrightarrow
\nu\nu$ rates in the latter model). Third, the matter in the gain
layer is accumulating energy: The whole gain layer is continuously
getting closer to become gravitationally unbound (Fig.~\ref{fig:mgeg})
and the energy contained in grid zones with positive local specific
binding energy (``explosion energy''; Eq.~\ref{eq:eexpl_expl}) already
starts growing after 120ms post-bounce
(Fig.~\ref{fig:eexpl_expl}). Also, looking at the profiles at
$t_\mathrm{pb}=150$ms in Fig.~\ref{fig:1d_snaps_expl}, one can see
that the jump in entropy and density has not yet arrived at the shock,
but that the average radial velocities behind the shock have almost
become positive. All these facts suggest that the convective overturn
in the gain layer has been successful in transporting energy from the
gain radius to the shock and thus has increased the heating
efficiency.

\begin{figure*}[!]
\centering
  \includegraphics[width=17cm]{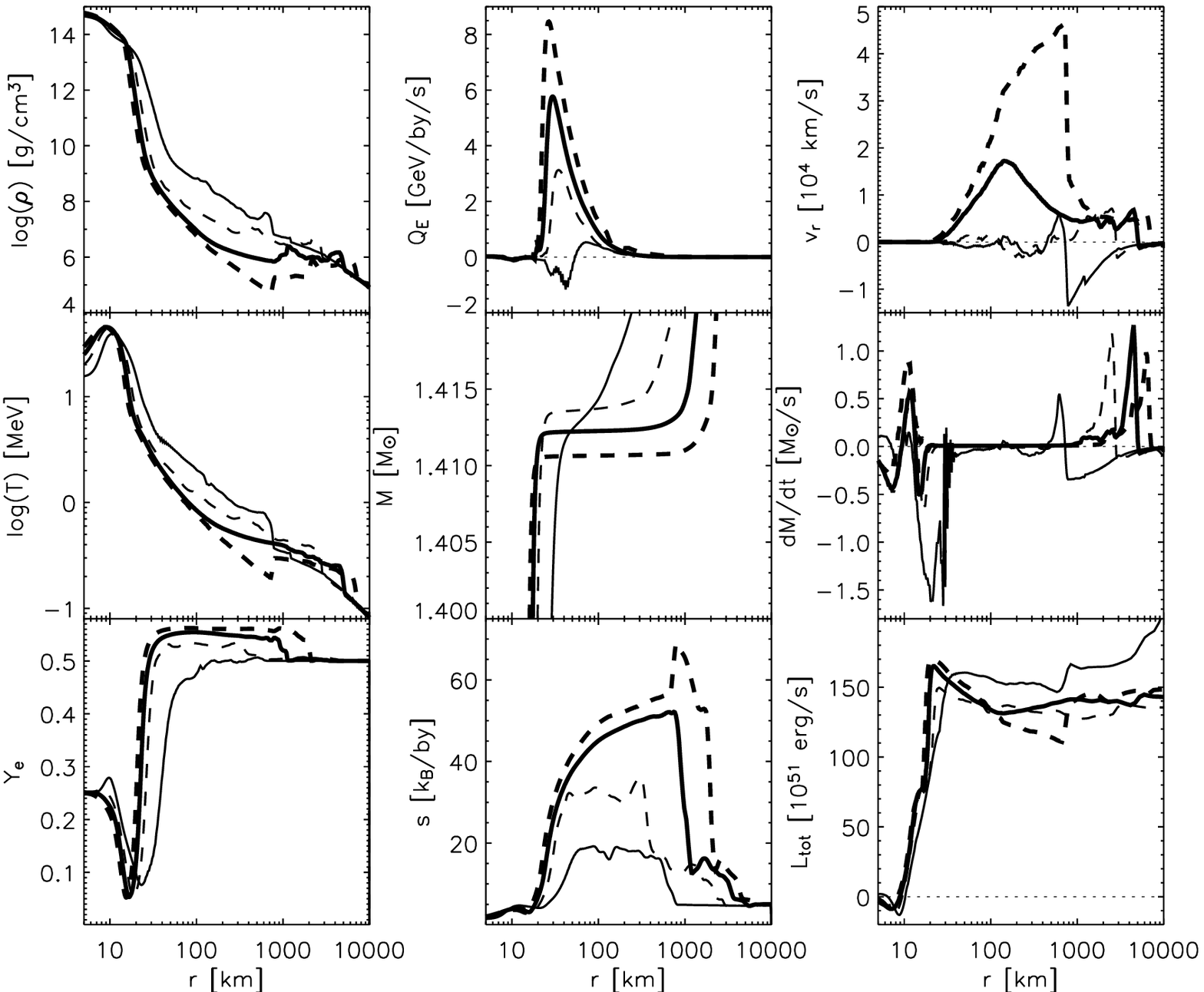} 
  \caption[]{
  Radial profiles for different quantities at post-bounce times of
  200.2ms (solid), 400.6ms (dashed), 600.5ms (thick solid), and
  800.9ms (thick dashed) for the exploding Model s15Gio\_32.a. For
  200.2ms and 400.6ms laterally averaged information is shown, the two
  later times correspond to the phase of the evolution after $t=468$ms
  which was followed in spherical symmetry.
  }\label{fig:1d_snaps_expl2}
\end{figure*}

Shortly after the launch of the outgoing SN shock the density in the
region between dense post-shock shell and PNS drops
(Fig.~\ref{fig:1d_snaps_expl2}) and the gain radius retreats to a
position just outside the neutrinosphere, which it reaches about 300ms
after bounce (Fig.~\ref{fig:mas_expl}). The onset of the
neutrino-driven wind after the start of the explosion is visible in
Fig.~\ref{fig:1d_snaps_expl2} as a rapid (in our simulation
spherically symmetric), baryonic outflow with velocities up to
$\sim5\times 10^9 \mathrm{cm/s}$. It develops after $\sim 500$ms
post-bounce (see also Fig.~\ref{fig:mas_expl}) and catches up with the
slower and denser shell of early SN ejecta. In the deceleration region
a reverse shock (wind termination shock) develops
\cite[][]{janmue95,tomkac04} which is a very prominent feature in the
density, velocity, temperature, and entropy profiles of
Fig.~\ref{fig:1d_snaps_expl2} after $\sim 600$ms post-bounce (see also
Fig.~\ref{fig:mas_expl}).

Figure \ref{fig:conv_HB_expl}a shows the convective activity during
the 2D phase of the simulation. While convection in the PNS is
inhibited to develop despite of convective instability because we
calculate with a spherically symmetric grid below 25km, large parts of
the gain layer are unstable and show convective motions
(Fig.~\ref{fig:snapsh_expl}). On the other hand we see that in the
matter swept up by the expanding shock convective instability is so
weak (the condition $\omega_\mathrm{BV} < 30/\mathrm{s}$ applies) that
convective activity between about 1000 and 3000km does not set
in. Around 320ms after bounce the downflows of mass to the PNS cease
(see also Fig.~\ref{fig:conv_HB_expl}b) and the neutrino-driven
baryonic outflow begins to develop (Fig.~\ref{fig:mas_expl}). However,
a massive fallback in the form of several downflows occurs between
390ms and 440ms after bounce and destroys this neutrino-driven
``wind''. The wind begins to recover again afterwards
(Fig.~\ref{fig:mas_expl}). In Fig.~\ref{fig:conv_HB_expl}b we see that
the mass inflow (at 100km) becomes almost equal to the mass outflow
after the onset of explosion, meaning that the net mass flow is small
compared to the overturn. Consistent with
Fig.~\ref{fig:conv_HB_expl}a, the outflow dominates between 300ms and
390ms post-bounce and after 440ms post-bounce, while the transient
downflows between 390ms and 440ms lead to strong overturn activity
also in the vicinity of the nascent NS.

\paragraph{Explosion energy}

To calculate the explosion energy at a given time, we add up the local
specific binding (LSB) energy $\varepsilon_\mathrm{bind}^\mathrm{shell}$,
defined in Eq.~(\ref{eq:eexpl_bind_def}), for all grid zones between
neutrinosphere and shock where this energy and the radial velocity are
both positive:
\be
E_\mathrm{expl} = \sum_{r,\vartheta,\mathrm{cond}}
   \varepsilon_\mathrm{bind}^\mathrm{shell}(r,\vartheta)
  \rho(r,\vartheta)
   \Delta V(r,\vartheta)\,.
\label{eq:eexpl_expl}
\ee
Note that the gravitational potential
$\Phi_\mathrm{1D}^\mathrm{Newt,~enclosed}(r)$ used to calculate the
LSB energy was calculated assuming a spherically symmetric mass
distribution (2D effects are only caused by convective mass motions
and the associated density fluctuations are modest and therefore play
a minor role for the gravitational potential).

The explosion energy as a function of time is shown in
Fig.~\ref{fig:eexpl_expl}. Its increase with time has several reasons:
The deposition of energy by neutrinos is the dominant contribution
during the early stages of the explosion after it has taken the LSB
energy in the gain layer to a value close to zero even before rapid
expansion has begun and the explosion has set in
(Fig.~\ref{fig:mgeg}). When the shock reaches a radius of about 400km
and expands farther, the post-shock matter cools sufficiently that
neutrons and protons start recombining to $^4\mathrm{He}$ (or matter
accreted through the shock is not sufficiently heated to dissociate to
free nucleons), which releases/saves about 7.5MeV per baryon and
causes the energy in the gain layer to become positive. Subsequently
helium recombines further to Fe-group elements, releasing another
$\sim1.3$MeV per baryon. When the shock reaches about 800km
(corresponding to an enclosed mass of 1.45--1.46$\msol$), the material
accreted by the shock is no longer dissociated to later recombine, but
instead, silicon burns in the shock to $^{56}\mathrm{Ni}$. This process,
however, ends when the shock reaches about 1700km ($1.527\msol$) and
does not heat the stellar gas to Si-burning temperatures any more. At
late times, the explosion energy is increased by the energy input from
the baryonic wind which is driven by neutrino energy deposition just
outside of the neutrinosphere (starting at $t\ga300$ms after
bounce). The wind material initially also fully recombines, releasing
additional nuclear binding energy, but with decreasing densities and
increasing entropy the recombination becomes incomplete and a large
number of $\alpha$-particles are left ($\alpha$-rich freeze-out). The
time evolution of the composition in the infalling and exploding
layers is indicated by the grey shadings in Fig.~\ref{fig:mas_expl}.

On the other hand, the shock also loses energy by lifting accreted
material of the outer layers in the gravitational potential of the
star.

The explosion energy in Fig.~\ref{fig:eexpl_expl} is already positive
when the interface between Si-layer and O-rich Si-shell passes the
shock at 150ms. We then notice a steep rise after the onset of the
explosion due to neutrino heating, nucleon recombination, and
burning. At 300ms post-bounce burning ceases and the energy increase
becomes less steep. Finally, around 400ms, the $\nu$-driven wind sets
in, which contributes energy at a much smaller rate: The energy
increases only very slowly. The transient little drop in the explosion
energy between 468ms and 500ms results from the mapping of the model
from a 2D to a 1D grid.

At the end of our simulation, the shock has reached an enclosed mass
of 1.9$\msol$. From Fig.~\ref{fig:prog_prof} we can thus say that the
shock will still lose an energy of 0.16FOE (FOE = ten to the fifty-one
erg) for unbinding the outer layers of the star. The final value of
the explosion energy will also be changed by further input from the
$\nu$-driven wind, and possibly by fallback of material onto the
PNS. Assuming that wind power and gravitational binding energy of the
outer layers roughly compensate each other, a value of 0.5FOE seems to
be a good estimate for the final explosion energy of our model. This
is comparable to the energy that has been estimated for SN1999br
\cite[][]{ham03} and must certainly be considered as ``weak''
explosion.

The neutron star left behind by the explosion has an initial baryonic
mass of about $1.4\msol$. Figure \ref{fig:1d_snaps_expl2} shows that
it is $1.41\msol$ at 800ms after bounce, which can decrease by some
$0.01\msol$ due to losses in the neutrino-driven wind, and later
increase due to possible fallback.

\paragraph{Mass and composition of ejecta}

The discussed explosion model reveals interesting properties also with
respect to the neutron-to-proton ratio in the ejecta. Since we did not
carry along marker particles in our Eulerian 2D simulation, the
$Y_\mathrm{e}$ distribution of the ejecta was evaluated in a
post-processing step: To determine the total amount of matter that is
ejected as function of $\ye$, we integrate over time the flux of mass
(for given value of $\ye$) through a surface of fixed radius $r_0$,
resulting in the expression
\ba
&&
M(Y_i<Y_\mathrm{e}\leq Y_{i+1}) =
\frac{2}{\cos(\vartheta_\mathrm{min})
-\cos(\vartheta_\mathrm{max})} \nn\\ &&\sum_n \sum_{\vartheta, \rm cond}
\Delta A(r_0,\vartheta) \cdot v_r(r_0,\vartheta,t_n) \cdot
\rho(r_0,\vartheta,t_n) \cdot (t_{n+1} - t_{n}).
\label{eq:expl:massye}
\ea
Here, $\Delta A=2\pi r_0^2 (\cos\vartheta_{j+1} -\cos\vartheta_j)$ is
the cell surface through which the matter flows and the factor in
front of the sum renormalizes the result from the wedge used in our
simulation to the whole star. The first sum accomplishes the
integration over time, while the second sum integrates over those
angular bins where the following condition is fulfilled:
$v_r(r_0,\vartheta,t_n)>0 ~\wedge~
Y_i<Y_\mathrm{e}(r_0,\vartheta,t_n)\leq Y_{i+1}$ for a given bin
$[Y_i,Y_{i+1}]$.

Choosing a suitable value for $r_0$ for this evaluation is not a
trivial task: On the one hand, it should be as large as possible so
that the matter advected through the surface at $r_0$ is no longer
affected by neutrino processes and thus $\ye$ has reached its terminal
value. On the other hand, the advection of material on our Eulerian
grid and the associated mixing of matter tends to destroy small
patches with extreme values of $Y_\mathrm{e}$. Therefore we would like to
choose $r_0$ as small as possible to keep this numerical diffusivity
as low as possible.

\begin{figure}[!]
  \resizebox{0.95\hsize}{!}{\includegraphics{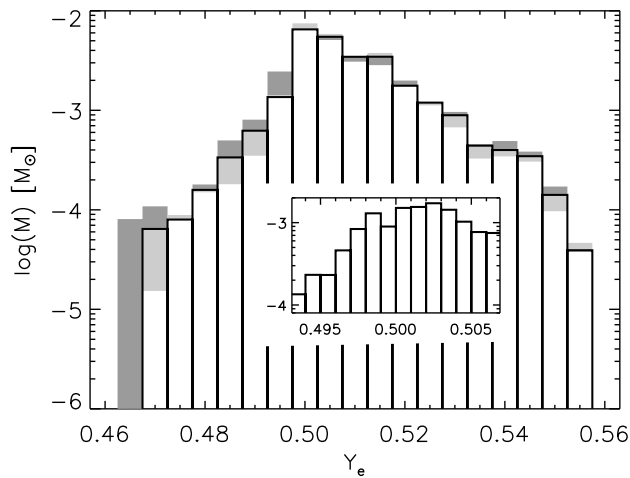}} \\ 
  \caption[]{
  Mass of neutrino-processed ejecta as a function of the terminal
  value of the electron fraction $Y_\mathrm{e}$ for Model
  s15Gio\_32.a. The masses are calculated with
  Eq.~(\ref{eq:expl:massye}). The solid bars correspond to an
  evaluation of the outward directed mass flow in the 2D model through
  a sphere at a radius of $r_0=350\mathrm{km}$, while the dark and
  light shading indicate the variation obtained when the analysis is
  performed at $r_0=300\mathrm{km}$ and $r_0=400\mathrm{km}$,
  respectively. The insert shows the mass distribution for values of
  $Y_\mathrm{e}$ around 0.5 with increased resolution (figure from
  \citealp{pruwoo05}).
  }\label{fig:massye_expl}
\end{figure}

The results shown in Fig.~\ref{fig:massye_expl} were obtained with
$r_0=350\mathrm{km}$.  The time resolution of the processed data lies
between one and two ms, the post-processing for
Fig.~\ref{fig:massye_expl} has ended at $t_\mathrm{pb}=468$ms. To
estimate the reliability of our analysis, we tested the differences
caused by moving $r_0$ to $300\mathrm{km}$ (shift indicated by dark
shading in Fig.~\ref{fig:massye_expl}) or $400\mathrm{km}$ (light
shading). In the latter case, the advection between grid zones has
partly ``washed away'' small regions of, in particular, low
$Y_\mathrm{e}$, while in the former case we miss some neutrino
processing of matter at radii between 300km and 350km, which has the
tendency of increasing $Y_\mathrm{e}$. The differences between the
three values of $r_0$ can be interpreted as error bars. The integrated
masses in the three cases with $r_0=350$, $300$, and $400$km are
$M_\mathrm{tot} = 0.0267$, $0.0273$, and $0.0274 \msol$, respectively.

Another problem is given by material that falls back through the
surface at $r_0$, and then reemerges with a different value of
$Y_\mathrm{e}$; this material is counted twice in our analysis. For
example, in Fig.~\ref{fig:mas_expl} we can see that between 400ms and
425ms, some material, about $10^{-3}\msol$, temporarily falls back
below the radius of 350km. The amount of such double processed
material, however, is small compared to the total integrated
mass. Also note that the distribution of neutrino processed ejecta of
Fig.~\ref{fig:massye_expl} is not the final one because it is based on
an analysis of the explosion only until 468ms, excluding the
subsequent neutrino-driven wind. For $\ye<0.5$, however, the plot
presents upper limits of the ejected mass at early times of the
explosion, because the following ejecta have $Y_\mathrm{e}>0.5$
(although at very late times the neutrino-driven wind might become
neutron-rich again).

The production of proton-rich ejecta ($\ye>0.5$) during the early
phase of the supernova explosion seems to be a generic result of
models with an elaborate spectral treatment of the neutrino transport
either by solving the Boltzmann equation
\cite[][]{liemez03,thiarg03,frohau04} or by the variable Eddington
factor closure of the moments equations used in this work \cite[see
also][]{pruwoo05}. According to \cite{hofwoo96} the amount of
supernova ejecta with $\ye\la0.47$ must be less than about
$10^{-4}\msol$ per event in order to avoid overproduction of closed
neutron shell ($N=50$) nuclei, for example $^{88}$Sr, $^{89}$Y, and
$^{90}$Zr, compared to observed galactic abundances of these
elements. Our explosion model fulfills this constraint
(Fig.~\ref{fig:massye_expl}; possible late-time ejection of $\ye<0.5$
material in the neutrino-driven wind is unlikely to be sufficiently
massive to be in conflict with the observations) in contrast to
previous models with partly crude or simple approximations of the
neutrino transport \cite[][]{herben94,burhay95,janmue96}.

\begin{figure*}[!]
\sidecaption
  \includegraphics[width=12cm]{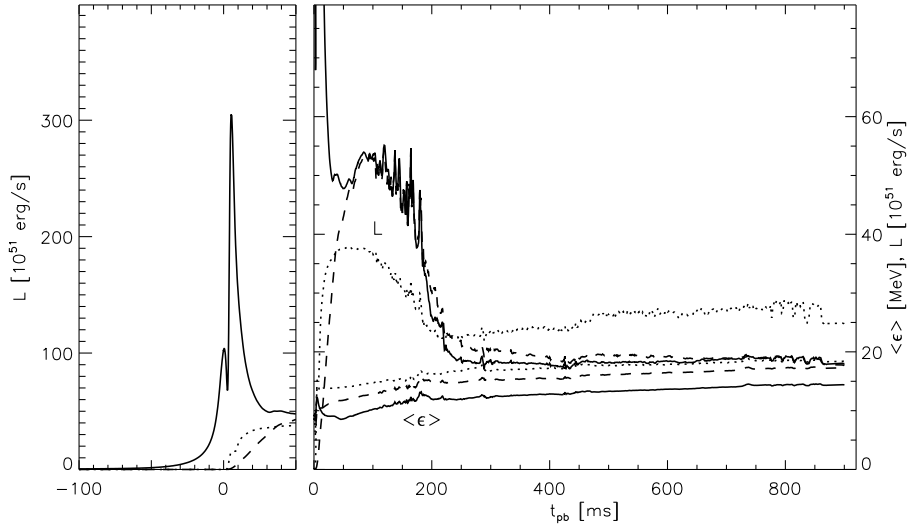} 
  \caption[]{
  Evolution of the luminosities for $\nu_\mathrm{e}$ (solid),
  $\bar\nu_\mathrm{e}$ (dashed), and $\nu_\mathrm{x}$ (dotted) in Model
  s15Gio\_32.a around core bounce (left) and during the later
  post-bounce phases (right), evaluated at a radius of 400km for an
  observer at rest. For the post-bounce phase, we also plot the
  average energies of the emitted neutrinos.
  }\label{fig:lum_expl}
\end{figure*}
\begin{figure*}[!]
\sidecaption
  \includegraphics[width=12cm]{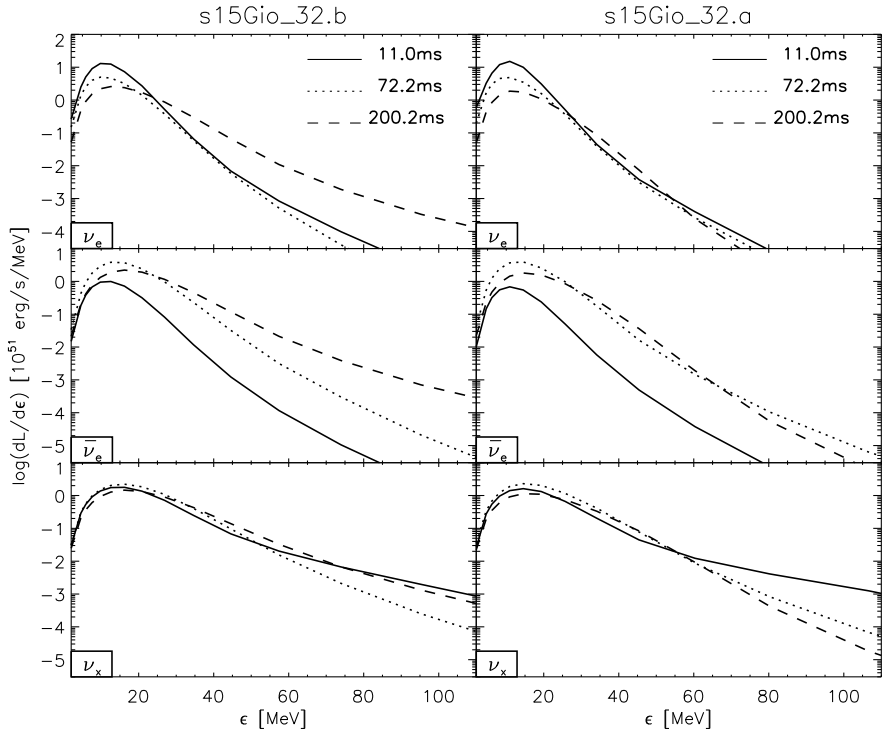} 
  \caption[]{
  Luminosity spectra for $\nue$, $\nuae$, and $\nux$ for the 2D Models
  s15Gio\_32.a (right) and s15Gio\_32.b (left) for an observer at rest
  at 400km. The spectra (given as averages over all directions) are
  shown for three different post-bounce times: Shortly after the peak
  of the $\nue$ shock breakout burst, when the shock reaches its   
  maximum radius in Model s15Gio\_32.b, and shortly after Model
  s15Gio\_32.a has exploded.
  }\label{fig:spectra_both}
\end{figure*}

The neutrino-processed outflow becomes proton-rich because the
neutron-to-proton mass difference favors positron captures,
$\mathrm{e}^+ + \mathrm{n} \rightarrow \mathrm{p} + \nuae$, against
electron captures, $\mathrm{e}^- + \mathrm{p} \rightarrow \mathrm{n} +
\nue$, when the electron degeneracy is low, i.e.~the electron chemical
potential $\mu_\mathrm{e}\la kT$ \cite[][]{bel03,liemez03}. At large
distance from the nascent neutron star, before weak processes freeze
out, neutrinos are more energetic and more abundant than electrons and
positrons and therefore $\nue$ and $\nuae$ captures on neutrons and
protons, respectively, may be the dominant weak reactions (in
particular during the neutrino-driven wind phase). If the $\nue$ and
$\nuae$ spectra are not too dissimilar, the neutron-to-proton mass
difference as well as weak magnetism and recoil corrections of the
neutrino absorption cross sections \cite[][]{horli99} favor a value of
$\ye\ga0.5$ (by reducing the $\nuae$ absorption opacity relative to
the one of $\nue$, see Sect.~\ref{sec:vars},
Figs.~\ref{fig:kappa_0},~\ref{fig:kappa_1} and Appendix
\ref{sec:neuopa}, Fig.~\ref{fig:opac_cc}) before the weak processes
finally freeze out when the expansion timescale becomes shorter than
the reaction timescales. The corresponding nucleosynthesis in these
neutrino-processed ejecta has been recently investigated in more
detail for our Model s15Gio\_32.a by \cite{pruwoo05} and independently
for 1D explosion models by \cite{frohau04}.

An interesting number is the amount of $^{56}$Ni produced by the
explosion. Since the simulation was run with a strongly simplified
treatment of NSE and nuclear burning, and not with a nuclear reaction
network, we can only come up with a crude estimate here. Integrating
the mass of all iron-group nuclei at the end of the simulation yields
$M_\mathrm{heavy}=0.12\msol$. This value contains the $^{56}$Ni produced
by $^{28}$Si burning in the shock-heated layers as well as the
material ejected from the neutrino-processed region which recombines
to iron-group nuclei. Adding the $0.009\msol$ of $^{56}$Ni produced by
oxygen burning in the O-rich Si-layer we obtain a total estimate of
$\sim0.13\msol$ of $^{56}$Ni. Since we assume here that Si burns to Ni
completely and that Ni is the favored nucleus produced by nucleon
recombination in the ejecta (ignoring isotopes and other elements),
our number is certainly an upper limit for the true $^{56}$Ni
yield. Some of this matter is also likely to fall back to the PNS
instead of being ejected in our fairly weak explosion, reducing the
final Ni yield even more below our estimate.

\paragraph{Neutrino emission}

The evolution of the neutrino luminosities and mean energies is
plotted in Fig.~\ref{fig:lum_expl}. We can distinguish four phases:
The collapse phase, the prompt burst of electron neutrinos, the
accretion phase and, after the onset of the explosion, the beginning
of the Kelvin-Helmholtz neutrino cooling phase of the nascent neutron
star. The latter is marked by the sudden and pronounced drop of the
luminosities which occurs when the accretion of matter onto the PNS is
terminated by the start of the explosion. During this early phase of
PNS cooling the neutron star deleptonizes, contracts, and heats
up. For this reason the mean energies of the emitted neutrinos are
still rising and the neutrino luminosities level off to a nearly
constant value with no sign of long-time decay. We, however, point out
here that our results during this phase depend on the employed
approximation for relativistic gravity. The effective potential
introduced by \cite{ramjan02} turned out to be a bit too strong and to
overestimate the PNS contraction and heating and thus the energies and
luminosities of the emitted neutrinos \cite[see][]{lieram05}. It also
can not ensure the correct Newtonian limit at large radii and thus
leads to an overestimation of the gravitational force and fluid
acceleration in the infall region ahead of the shock. This may be the
reason why the mass accretion rate at late post-bounce times (in
nonexploding models) reveals a slight tendency of increase
(Fig.~\ref{fig:mdot}). The improvement of the treatment of gravity
suggested by \cite{mardim05} should perform better in this respect and
might modify the results during the long-time PNS cooling to some
degree.

Also note the transient drop of the neutrino luminosity shortly before
the $\nue$ burst sets in (around the bounce time), see
Fig.~\ref{fig:lum2_all}, upper panel, and Fig.~\ref{fig:lum_expl}, left
panel. A detailed analysis of the numerical simulations reveals that
this feature correlates with the moment of the most extreme
compression of the homologous core and the onset of shock
formation. The luminosity minimum is caused by a drop of the density
and neutrino emission in the semitransparent regime exterior to the
shock formation radius and thus in the non-homologously collapsing
layers near the mass shell where the infall velocities reach the
maximum value.

The evolution of the neutrino spectra also changes due to the onset of
the explosion. Before the explosion sets in, the spectra become
gradually harder with time (Fig.~\ref{fig:spectra_both}). This is a
consequence of the contraction of the PNS and the persisting inflow
and compression-heating of fresh material accreted through the shock.
These structural changes lead to an increase of the temperature at the
radii at which the neutrinos of given energy decouple from the
medium. The increase of this frequency-dependent neutrinospheric
temperature is stronger for high-energy neutrinos, leading to an
enhanced flux at high energies and thus to a harder spectrum.

While in Model s15Gio\_32.b this tendency continues until late times,
the spectral evolution changes in Model s15Gio\_32.a when the
explosion sets in: The mass accretion onto the PNS is abruptly
stopped. Now that the inflow of compression-heated matter has ceased,
the layer near the PNS surface where the high-energy neutrinos
decouple quickly cools and the ``neutrinospheric'' temperature for
high-energy neutrinos does not increase as much as with ongoing
accretion (it does not drop either because the energy and lepton
number loss lead to contraction).  The neutrino spectra no longer
harden. This is not in contradiction with the gradual rise of the
average neutrino energies which are determined by the spectra around
their maxima ($\ene \la 20$MeV). The details of this differential
behaviour in different spectral bands are sensitive to the stellar
structure in the neutrino decoupling layer and to the interaction
rates and therefore depend also on the neutrino flavor. A deeper
analysis goes beyond the scope of this work.

\begin{figure}[!]
  \begin{tabular}{c}
    \put(0.9,0.3){{\Large\bf a}}
    \resizebox{0.95\hsize}{!}{\includegraphics{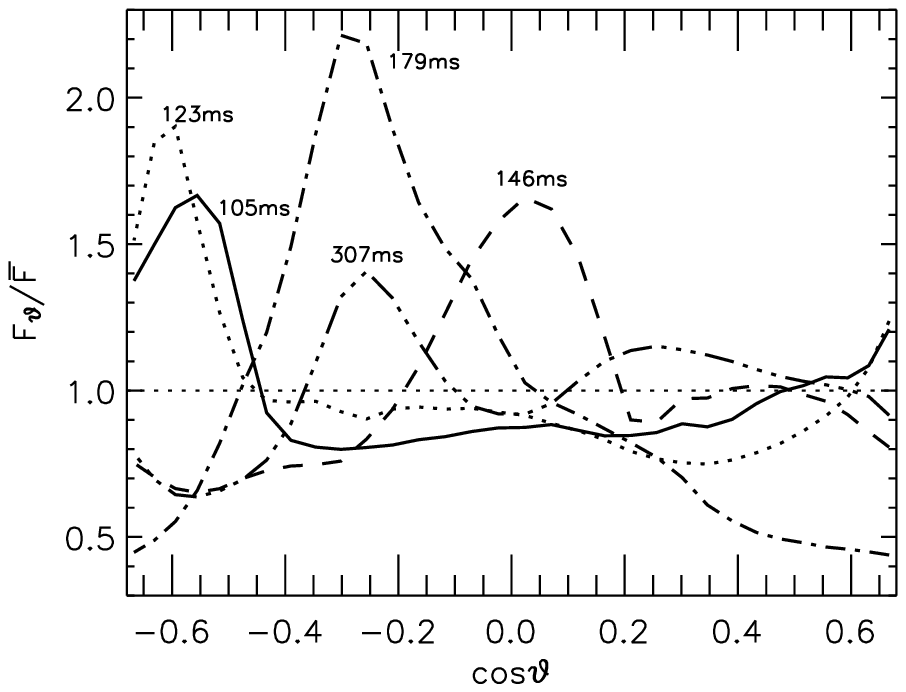}} \\ 
    \put(0.9,0.3){{\Large\bf b}}
    \resizebox{0.95\hsize}{!}{\includegraphics{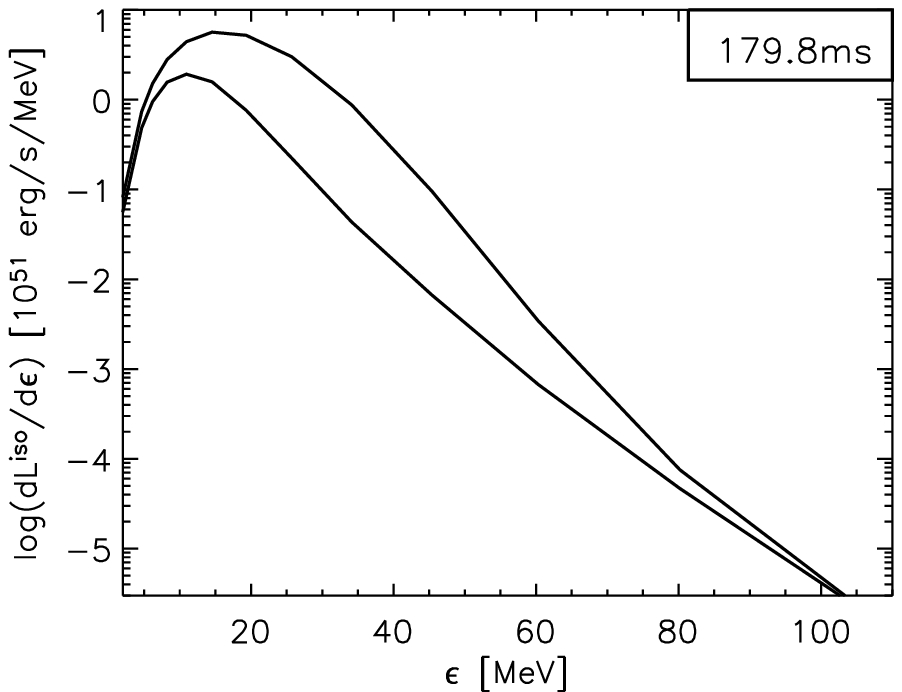}} 
  \end{tabular}
  \caption[]{
  {\bf a} Energy flux $F_\vartheta$ of $\nue$ normalized to the
  average energy flux $\overline{F}$ as function of polar angle
  $\vartheta$ for different post-bounce times in Model
  s15Gio\_32.a. The results are evaluated for an observer at rest at a
  radius of 400km, and the times are picked for showing maximal
  angular variations.
  {\bf b} For $t=179$ms after bounce the $\nue$ flux spectra are shown
  for the lateral directions with maximal and minimal fluxes, again
  given for an observer at rest at a radius of 400km. Here,
  $\mathrm{d}L^\mathrm{iso}/\mathrm{d}\ene$ is the spectrum corresponding
  to the ``isotropic equivalent luminosity'' $L^\mathrm{iso}$, calculated
  by assuming that the flux in one polar direction were representative
  for all other directions.
  }\label{fig:devlum_expl}
\end{figure}

The violent hot bubble convection behind the shock also affects the
neutrino emission. When strong downflows hit the cooling region,
transient lateral variation of the electron neutrino and antineutrino
flux of up to a factor two can occur (Fig.~\ref{fig:devlum_expl}a),
corresponding to a total luminosity increase in $\nue$ or $\nuae$ by
up to about 20\%. Note that our ``ray-by-ray'' transport scheme tends
to overestimate the anisotropy and lateral variability of the neutrino
emission (flux and flux spectra, see Fig.~\ref{fig:devlum_expl})
produced by the narrowly collimated downflows. A fully two-dimensional
treatment of the transport would certainly reveal more isotropic
neutrino emission even when the neutrinos are produced mostly in well
localized regions (``hot spots'') near the points where the downflows
dive into the cooling layer \cite[see e.g.][]{walbur05}.

Interestingly, sizeable lateral flux variations show up even long
after the onset of the explosion, e.g.~at 307ms, caused by downflows
which still manage to penetrate inward towards the PNS against the
expanding supernova ejecta.

\section{Summary and conclusions}
\label{sec:summary}

In this paper (Paper I of a series) we have described the employed
equations and numerical implementation of our ``ray-by-ray plus''
method for multi-group neutrino transport coupled to hydrodynamics in
2D models of stellar core collapse and supernova explosions. The
existing code, named \textsc{MuDBaTH}, can be extended to a 3D version
in a straightforward way. Starting from the basic assumption that the
neutrino phase space distribution is azimuthally symmetric around the
radial direction, which implies that nonradial flux components are
zero, we could directly build up the 2D spectral transport scheme from
the 1D code version \textsc{Vertex} of \cite{ramjan02}.  In course of
this, the ${\cal O}(v/c)$ moments equations for neutrino number,
energy, and momentum were extended by the remaining terms containing
lateral derivatives, which mediate the coupling between the different
angular directions of the polar grid and depend on the lateral
component of the fluid velocity. These terms are integrated in an
operator splitting step.  Closure of the set of moments equations is
achieved by using variable Eddington factors, which are computed from
the solution of a model Boltzmann equation as in spherical symmetry
\cite[cf.][]{ramjan02}. The current code offers the options to do this
separately for each angular direction of the grid or -- CPU-time
saving -- once on a laterally averaged stellar background.

In addition to the velocity-dependent terms with lateral derivatives
in the transport part of the code, we also take into account the
lateral components of neutrino pressure gradients in the hydrodynamics
equations at optically thick conditions.

We provided a compilation of tests for our numerical scheme and input
physics, by which we attempted to assess potential limits of its
applicability or accuracy in the context of core collapse
calculations. In particular the following points were addressed:

\begin{itemize}

\item[(1)] Our approximation of general relativity (i.e., replacing
Newtonian gravity by an effective relativistic gravitational potential
and including redshift and time dilation in the neutrino transport;
cf.~\citealp{ramjan02}) was tested against fully relativistic
calculations in spherical symmetry \cite[][]{lieram05,mardim05}. Very
good agreement in all relevant quantities was found for cases with
moderately relativistic fluid velocities (up to $\sim\,$10--20\% of
the speed of light) at least until several 100$\,$ms after
bounce. Moreover, by 2D simulations with different polar wedges and
lateral boundary conditions we demonstrated empirically that linear
momentum conservation, which cannot be strictly guaranteed with the
use of our effective relativistic potential in multi-dimensional
simulations \cite[see][]{mardim05}, is satisfactorily well fulfilled
in practical applications.

\item[(2)] We explored the role of the lateral derivatives in the
moments equations and of the lateral components of neutrino pressure
gradients in the hydrodynamics equations.  Both turned out to be
necessary extensions when non-equilibirium transport is applied to the
multidimensional case and artificial fluid acceleration and buoyancy
shall be avoided.  The latter may occur in situations where neutrinos
contribute significantly to the gas pressure and are strongly coupled
to stellar medium so that their advection with moving fluid elements
cannot be ignored. Our treatment of neutrino-hydrodynamics in two
dimensions produces results in agreement with the expectations from a
stability analysis of the stellar fluid, which we performed with a
``Quasi-Ledoux'' criterion that tries to take into account (in a
simple way) the lepton number exchange by neutrino diffusion between
moving fluid elements and their surroundings.  This at least shows
that our scheme does not artificially instigate convective activity,
although, of course, we cannot exclude that disregarding lateral
components of the neutrino number and energy fluxes in our code
underestimates neutrino diffusion and thus hampers or suppresses the
development of doubly diffusive instabilities below the neutrinosphere
(in particular since their occurrence seems to be very sensitive to
the efficiency of diffusion; cf.~\citealp{brural04}).

\item[(3)] Our transport treatment, favoring radial streaming,
potentially overestimates the magnitude of asymmetry and angular
contrast of the radiation field of neutrinos that stream off from the
neutrinosphere or cooling layer. When downflows or convective eddies
create hot spots with enhanced neutrino emission, material at the
involved latitudes exterior to these emission regions receives more
radiation, whereas the exposure of neighboring lateral directions is
underestimated. Our hydrodynamical simulations revealed that the
corresponding angular anisotropy is nonstationary and varies quickly
in space and time on timescales of several milliseconds.  Short,
transient, local flux enhancements reach typically some 10\% up to
occasionally a factor of two compared to the average value. Performing
a postprocessing test by making the extreme assumption that the
neutrino radiation field is spherically symmetric (with the same value
of the total luminosity) we found that the total heating rate in the
gain layer is not significantly changed at any time. The average
heating rate is changed only by a few percent and the heating rate per
baryon even less. Distinguishing in the postprocessing evaluation
between downflows and rising hot bubbles, we determined a decrease or
increase, respectively, of roughly 10\% in the time-averaged net
heating. We therefore conclude that underestimating the spherical
symmetry of the neutrino emission is not very likely to produce
dynamical consequences for the supernova evolution in our simulations.

\item[(4)] We performed a number of 1D tests to assess the
consequences of an error in the EoS of Lattimer \& Swesty, which is
usually used in our core collapse simulations at NSE conditions and at
densities above $6\times 10^7\gcm$. This error leads to
an underestimation of the mass fraction of $\alpha$ particles and
corresponding effects in the other thermodynamical quantities like
pressure, entropy, and temperature for given density, internal energy
density, and electron fraction.  Replacing the Lattimer \& Swesty EoS
below a density of $10^{11}\gcm$ by different versions of
a low-density EoS with 4 or 17 nuclear species (neutrons, protons,
$\alpha$'s and one or 14 heavy nuclei, respectively) in NSE reveals
the expected differences of the dependent thermodynamical variables
behind the shock, but negligibly small effects in the post-bounce
evolution of the models, e.g., nearly identical shock trajectories.
We determined two reasons for this insensitivity: Firstly, the
postshock entropies in our simulations with relativistic gravity are
so high that $\alpha$ particles play a relatively unimportant role and
therefore the EoS differences are rather small (of order
10\%). Secondly, the pressure profile in the essentially hydrostatic
layer behind the standing accretion shock is tightly constrained and
thus determined by the gravitational field of the nascent neutron star
on the one hand, and by the mass accretion rate and jump conditions at
the shock on the other hand. Differences of density and temperature in
the gain layer, of course, affect the neutrino heating, which,
however, is dynamically not very relevant in our models which are
``quite far'' from an explosion.

\item[(5)] Finally, we investigated the implications of wiggles in the
entropy profile (of about 10\% amplitude), which we discovered in the
deceleration layer behind a standing accretion shock, when a physical
EoS is employed in which the energy density includes contributions
from the particle rest masses. This numerical phenomenon seems to be
linked to the use of a Riemann solver in our hydrodynamics code and
turned out to be sensitive to the ratio of pressure to energy density
fed into the Riemann solver. The artificial entropy wiggles
disappeared after we applied an algorithmic procedure which allows us
to extract the rest-mass contributions from the energy density of the
EoS so that the true internal energy is evolved in the hydrodynamics
part of the code. Comparative calculations revealed that the ugly
entropy fluctuations did not lead to any significant dynamical effects
in 1D as well as 2D supernova simulations.

\end{itemize}

In the present paper our new neutrino-hydrodynamics code was applied
to 1D and 2D simulations of stellar core collapse and post-bounce
evolution of a 15$\,\msol$ progenitor star. Our aim was to
investigate the effects of different input physics.  We also explored
modifications in the neutrino transport, which change the
neutrino-matter coupling in the heating and cooling layers outside of
the neutrinosphere.  Our findings can be summarized as follows:

\begin{itemize}

\item[(a)] None of our simulations with the most complete
implementation of neutrino transport yields an explosion, neither in
spherical symmetry nor in two dimensions. The 2D models were computed
with a lateral wedge of about 90 degrees around the equatorial plane
and periodic boundary conditions, a setup which is similar to the one
used in the first generation of models of convectively supported,
neutrino-driven supernova explosions
\cite[][]{herben94,burhay95,fry99}. Our models with the accurate
spectral description of neutrino transport and neutrino-matter
interactions are therefore unable to reproduce the success of these
previous simulations, in which a simplified treatment of neutrinos by
grey, flux-limited diffusion was applied. Since the transport is the
major difference between the older models and our current ones (of
course, there are other differences, too, numerical as well as in the
input physics), we interpret our negative results as a confirmation of
concerns raised by \cite{mezcal98:ndconv}, who suspected that the
transport simplifications might have favored the explosions found by
\cite{herben94} and \cite{burhay95}.

\item[(b)] Using a 90$\degr$ wedge with periodic boundary
conditions, however, imposes constraints on the size of the nonradial
structures (i.e., the wavelengths of the modes) which are allowed to
exist. It also prevents one from studying the growth of low (dipole,
$l = 1$, and quadrupole, $l=2$) modes in the postshock flow as seen in
neutrinoless (adiabatic) simulations of \cite{blomez03} and predicted
as a consequence of the vortical-acoustic cycle \cite[][]{fog02}.  In
Paper~II we will demonstrate by 2D simulations with a full 180 degree
grid that a thus less constrained flow can indeed develop a dominant
low-mode pattern behind a highly deformed shock front, which performs
bipolar oscillations.  This can lead not only to quantitative but even
qualitative differences in the post-bounce evolution of the collapsing
stellar core, deciding about explosion or non-explosion.

\item[(c)] The sensitivity to the treatment of the neutrino transport
became evident once more from two simulations which we performed
without the terms in the neutrino momentum equation, which depend on
the radial velocity (the terms responsible for neutrino advection
were, however, included).  In contrast to the corresponding
simulations with the most complete transport implementation these two
models produced explosions, caused by roughly a factor of two less
energy loss and more neutrino heating due to 20--30\% higher neutrino
energy densities in the cooling and heating layers,
respectively. These layers between neutrinosphere and stalled shock in
the 1D model underwent a number of nonlinear, radial oscillations with
growing amplitude and final runaway, similar to the $\kappa$ mechanism
of pulsational instability of stellar atmospheres. The 2D model
developed very strong postshock convection and started to explode when
the interface between silicon shell and oxygen-rich silicon shell at
1.42$\,\msol$ reached the shock about 150$\,$ms after core bounce. The
net explosion energy was rather low, roughly $0.5\times
10^{51}\,$erg. The explosion left behind a nascent neutron with an
initial baryonic mass of about 1.41$\,\msol$ and produced less than
$\sim\,$0.13$\,\msol$ of iron-group elements. The latter two numbers
may change by the effects of the subsequent neutrino-driven baryonic
wind and later fallback. Most interesting of this ``artificial''
explosion, however, is the electron fraction $Y_\mathrm{e}$ in the
neutrino-heated ejecta. Most of this material has values above 0.5 and
less than about $10^{-4}\,\msol$ have $Y_\mathrm{e}\la 0.47$, in
agreement with constraints derived from observed galactic chemical
abundances \cite[][]{hofwoo96}.  The model therefore does not show the
enormous overproduction of $N=50$ closed neutron shell nuclei found in
previous simulations \cite[see also][]{pruwoo05}. Values of the
electron fraction above 0.5 are a consequence of our accurate,
spectral description of electron neutrino and antineutrino transport,
enhanced by the inclusion of weak magnetism corrections in the
charged-current reactions with nucleons.  These corrections reduce the
absorption of $\bar\nu_\mathrm{e}$ relative to that of
$\nu_\mathrm{e}$ in the neutrino-heated ejecta
\cite[cf.][]{horli99}. The proton richness should not depend on the
manipulation of the neutrino transport which caused the explosion,
because electron neutrinos and antineutrinos were affected in the same
way. Moreover, values of $Y_\mathrm{e}$ above 0.5 were also obtained
in the early phase of the neutrino-driven wind, which set in (at
$\sim\,$300$\,$ms p.b.)  after the supernova explosion had been
lauched and which we followed from $\sim\,$470$\,$ms to more than a
second after bounce in 1D with the velocity-dependent terms in the
neutrino momentum equation turned on again. Finally, our result is
supported by \cite{frohau04}, who used spherically symmetric general
relativistic simulations in which the numerical treatment of the
transport differed from ours and in which explosions were achieved by
other modifications of the transport than in our work.

\item[(d)] Our 1D models reveal the existence of a
(Quasi-)Ledoux-unstable layer below the neutrinosphere. In the
corresponding 2D simulations convective activity inside the nascent
neutron star sets in about 30$\,$ms after bounce and persists until
the end of the simulated evolution. Since the convective layer nearly
reaches with the muon and tau neutrinosphere, the luminosity of these
neutrinos is enhanced.  A detailed analysis of the consequences of
this convective activity for neutrino emission and supernova evolution
will be given in Paper~II.

\item[(e)] Improving neutrino-nucleon interactions by the effects of
nucleon thermal motions and recoil and weak-magnetism corrections, and
including pair annihilation and scattering between neutrinos of
different flavors and nucleon-nucleon bremsstrahlung, we found
important differences in many aspects of the transport (in spectra as
well as luminosities) for neutrinos of all flavors. This leads to
quantitative differences of the post-bounce dynamics of the collapsing
stellar core and to differences in the shock evolution, without,
however, causing a qualitative change of the outcome of the
simulations.

\end{itemize}

In Paper~II we shall present simulations for different
progenitors between 11$\,\msol$ and 25$\,\msol$, among which we
will also consider a case with rotation. We will explore the role of
pre-collapse random perturbations in the core, will study the effects
of convection below the neutrinosphere, and will investigate the
implications of hydrodynamic instabilities -- convective and
vortical-acoustic -- in the layer between neutrinosphere and stalled
supernova shock by varying the angular resolution and the size of the
lateral wedge of the polar grid. Some results of these models were
already published in \cite{burram03} and
\cite{janbur04:nuccos,janbur04:ringb,janbur05}.

The spectral treatment of neutrino transport and neutrino-matter
interactions applied in these models means a new level of refinement
and accuracy in two-dimensional hydrodynamical supernova simulations.
But our transport scheme has advantages as well as disadvantages. On
the positive side is the fact that it was developed on grounds of a
well tested code for spherically symmetric problems and thus
guarantees to produce results of well-known good quality in an
important limit. Moreover, our neutrino-hydrodynamics code possesses
good coarse level parallelism, which allows the efficient use of a
large number of processors. Due to the fully implicit
time-differencing and the corresponding matrix inversions in the
transport part, however, systems with a low-latency, high-bandwidth
memory are preferable. As we argued in this paper and presented tests
for, we believe that our transport treatment is a reasonably good
approach for problems in which local macroscopic anisotropies occur
but which still have an overall spherical geometry. The applicability
of our method, however, becomes questionable in situations with large,
global deformation, because in this case our basic assumption of (on
average) azimuthal symmetry of the neutrino intensity is certainly
inadequate.  On the negative side there are also substantial CPU-time
requirements. The disregard of neutrino flux components in the angular
directions tends to cause an overestimation of the angular asymmetry
of neutrinos streaming out from the nascent neutron star. This may be
a handicap when accurate information about the corresponding
directional variation is required, e.g., for estimating pulsar kicks
or gravitational wave signals associated with the anisotropic neutrino
emission. Future tests will also have to clarify the question whether
the disregard of nonradial neutrino flux components leads to a
suppression of doubly diffusive instabilities inside the neutron
star. A comparison with other transport methods, which might employ
alternative, complementary approximations to deal with the high
dimensionality of transport in multi-dimensional problems, is
therefore highly desirable and in fact indispensable for verification
and validation of our approach. We are therefore looking forward to
compare our results with those obtained by other groups.

\begin{acknowledgements}
We are grateful to Matthias Keil, Georg Raffelt, and Leonhard Scheck
for helpful discussions, to Kohji Takahashi for adapting the neutrino
rates of \cite{bursaw98,bursaw99} for use in our code, and to Charles
Horowitz for providing us with the weak magnetism corrections in
neutral and charged-current neutrino-nucleon interactions. We also
thank Christian Cardall for pointing out missing terms in the
hydrodynamics equations written in
Eqs.~(\ref{eq:hydro.v_r})--(\ref{eq:hydro.v_p}). RB would especially
like to thank Francisco Kitaura for helping to solve the problem of
the entropy wiggles. We thank the Institute for Nuclear Theory of the
University of Seattle for its hospitality during an early stage of
this work. Support by the Sonderforschungsbereich 375 on
``Astro-Particle Physics'' of the Deutsche Forschungsgemeinschaft is
acknowledged. The computations were performed on the NEC SX-5/3C and
the IBM p690 ``Regatta'' system of the Rechenzentrum Garching, and on
the Cray T90 of the John von Neumann Institute for Computing (NIC) in
J\"ulich.
\end{acknowledgements}

\bibliographystyle{aa}
\bibliography{lit}

\Online
\onecolumn
\appendix

\section{Neutrino opacities}
\label{sec:neuopa}

\begin{table}[!t]
\centerline{
\begin{tabular}{rclll}
\hline \hline
Reaction  & & & Implementation described in & References \\
\hline
$\nu  \mathrm{e}^{\pm}$ &$ \rightleftharpoons $ & $\nu \mathrm{e}^{\pm}$   &
\cite{ramjan02}               & 
\cite{mezbru93:nes,cer94}  \\
%
$\nu  A$       &$ \rightleftharpoons $ & $\nu A $        & 
\cite{ramjan02}        &
 \cite{hor97,brumez97}               \\
%
$\nu  \mathrm{N}$     &$ \rightleftharpoons $ & $\nu \mathrm{N} $      & 
Sect.~\ref{sect:neutrino-nucleon}       & 
\cite{bursaw98}                              \\
%
$\nue \mathrm{n}$       &$ \rightleftharpoons $ & $\mathrm{e}^- \mathrm{p} $        & 
Sect.~\ref{sect:neutrino-nucleon}      & 
\cite{bursaw99}          \\
%
$\nuae \mathrm{p}$       &$ \rightleftharpoons $ & $\mathrm{e}^+ \mathrm{n} $        & 
Sect.~\ref{sect:neutrino-nucleon}      &
 \cite{bursaw99}         \\
%
$\nue A'$       &$ \rightleftharpoons $ & $\mathrm{e}^- A $      & 
\cite{ramjan02}         & 
\cite{bru85,mezbru93:coll}                              \\
%
$\nu\bar\nu$   &$ \rightleftharpoons $ & $\mathrm{e}^- \mathrm{e}^+ $      & 
\cite{ramjan02}              &
 \cite{bru85,ponmir98}                              \\
%
$\nu\bar\nu \,\mathrm{N}\mathrm{N}$  &$ \rightleftharpoons $ & $\mathrm{N}\mathrm{N} $      & 
\cite{ramjan02}  &
\cite{hanraf98}                              \\
%
$\nu_{\mu,\tau}\bar\nu_{\mu,\tau}$   &$ \rightleftharpoons $ & $\nue\nuae $      & 
\cite{burjan03:nunu}              &
 see \cite{burjan03:nunu}                              \\
%
%
$\overset{\scriptscriptstyle (-)}{\nu}_{\mu,\tau}\overset{\scriptscriptstyle (-)}{\nu}_\mathrm{e}$   &$ \rightleftharpoons $ &$\overset{\scriptscriptstyle (-)}{\nu}_{\mu,\tau}\overset{\scriptscriptstyle (-)}{\nu}_\mathrm{e}$  & 
\cite{burjan03:nunu}              &
 see \cite{burjan03:nunu}                             \\
\hline
\end{tabular}
}
\caption[]{
  Overview of all neutrino-matter interactions currently implemented
  in our Supernova code as our set of ``improved opacities'' (in
  contrast to the set of opacities listed in \citealp{ramjan02}, which
  we call our ``standard case'' containing a number of approximations
  removed in the current improved treatment). For each type of
  interaction we refer to the relevant section in this work or point
  to references where the fundamental aspects of the calculation of
  the corresponding rate are summarized and details of its numerical
  implementation are given.  The third column lists references where
  comprehensive information can be found about the physics and the
  approximations employed in the rate calculations.  In the first
  column the symbol $\nu$ represents any of the neutrinos
  $\nue,\nuae,\nu_\mu,\bar\nu_\mu,\nu_\tau,\bar\nu_\tau$, the symbols
  $\mathrm{e}^-$, $\mathrm{e}^+$, $\mathrm{n}$, $\mathrm{p}$, and $A$
  denote electrons, positrons, free neutrons and protons, and heavy
  nuclei, respectively. The symbol $\mathrm{N}$ means $\mathrm{n}$ or
  $\mathrm{p}$.
} \label{tab:reactions}
\end{table}

The description of the neutrino opacities which we call our
``standard'' set closely follows \cite{bru85} and
\cite{mezbru93:code,mezbru93:nes} with the only exception that we, in
addition, take into account neutrino pair processes due to
nucleon-nucleon bremsstrahlung. A complete list of the considered
reactions, corresponding pointers to the literature and details of the
numerical implementation into our transport code can be found in the
appendix of \cite{ramjan02}.

The ``improved description'' of neutrino opacities which is used in
our latest simulations comprises the reactions summarized in
Table~\ref{tab:reactions}.  Specifically, the so-called iso-energetic
or elastic approximation \cite[cf.][]{bru85,redpra98}, which is the
standard simplification for calculating rates of neutral-current
neutrino scatterings off free nucleons and charged-current absorption
reactions on nucleons, has been abandoned in order to take into
account energy exchange due to nucleon recoil and thermal motions
(e.g.~\citealp{sin90}) as well as nucleon-nucleon correlations in the
dense medium \cite[]{bursaw98,bursaw99,redpra98,redpra99}.
Modifications of the neutrino opacities due to the weak magnetism
corrections associated with the large anomalous magnetic moments of
the proton and the neutron are also accounted for \cite[cf.][]{hor02}.
Moreover, we employ in the reaction cross sections a density dependent
effective mass of the nucleon, which at nuclear densities is different
from its vacuum value, and also take into account the possible
quenching of the axial-vector coupling in nuclear matter
\cite[]{carpra02}. In the following we shall describe the numerical
handling of these neutrino-nucleon interactions in some detail.

Note that the complete list of considered interactions also includes
neutrino pair production by nucleon-nucleon bremsstrahlung and in
particular also the flavor-coupling neutrino interactions (last two
lines of Table~\ref{tab:reactions}), which until recently have
received only little attention from core-collapse supernova
modelers. The significance of these reactions as well as
implementation details were discussed elsewhere
\cite[]{burjan03:nunu}. These processes are therefore not included in
the following discussion.

\subsection{Neutrino-nucleon interactions}\label{sect:neutrino-nucleon}

\subsubsection{Neutrino-nucleon scattering ($\nu  \mathrm{N}
  \rightleftharpoons \nu \mathrm{N}$)}\label{sect:NNS}

Differential rates for inelastic scattering of neutrinos off free
nucleons are calculated according to Eq.~(38) of \cite{bursaw98}. For
incorporating this type of neutrino-matter interactions into our
transport code the formalism developed for inelastic scattering of
neutrinos off electrons (NES) can be exploited \cite[see][
Appendix~A]{ramjan02}.  Different from NES, however, the angular
integration (let $\omega$ denote the cosine of the scattering angle)
of the scattering kernels $R(\ene,\ene',\omega)$ which yields the
coefficients $\phi_l(\ene,\ene')$ of the corresponding Legendre
expansion \cite[cf.][ Appendix~C]{bru85} cannot be performed
analytically in the case of neutrino-nucleon scattering.  Moreover,
the characteristic width of the kernels as a function of in- and
outgoing neutrino energies $\ene$, $\ene'$ can be small (but finite)
compared to the numerical resolution of the neutrino spectrum in the
transport scheme, in which currently only of the order of 20 energy
bins with a resolution of $\Delta\ene/\ene\simeq 0.3$ can be afforded.
Hence, in order to adequately sample the scattering kernels on such a
coarse energy grid it is not sufficient to simply evaluate the
functions $\phi_l(\ene,\ene')$ at each combination of energies
$(\ene_{j+\hlf},\ene_{j'+\hlf})$.
Instead, we introduce for each energy bin $[\ene_j,\ene_{j+1}]$ a
numerical sub-grid of $N_{\ene_j}$ neutrino energies $\ene_{j}\le \ene
\le \ene_{j+1}$ (and equivalently for the final-state energies
$\ene'$), and $N_\omega$ angle cosines $-1\le\omega\le 1$ to compute
$R(\ene,\ene',\omega)$ as given by \citet[][ Eq.~38]{bursaw98} for all
such combinations of $\ene$, $\ene'$, and $\omega$. For fixed values
of $\ene$ and $\ene'$ we then numerically integrate
$R(\ene,\ene',\omega)$, appropriately weighted with the Legendre
polynomials $P_l(\omega)$ over angles to obtain $\phi_l(\ene,\ene')$.
Averaging $\phi_l(\ene,\ene')$ over all subgrid energies $\ene$ in
energy bin $[\ene_{j},\ene_{j+1}]$ and summing over all energies
$\ene'$ in bin $[\ene_{j'},\ene_{j'+1}]$ the final, binned Legendre
coefficients $\phi_l(\ene_{j+\hlf},\ene_{j'+\hlf})$ are
computed. These are then employed in our neutrino transport scheme.
In practice a six-point Gauss-quadrature scheme is used here for the
angular and energy integrations.
In order to correctly reproduce the low-temperature (in the SN case
therefore low-density) and low-neutrino energy limit, where only tiny
energy transfers $\ene-\ene'$ between neutrinos and nucleons occur and
the coefficients $\phi_l$ become increasingly narrowly peaked
functions of $\ene-\ene'$, we calculate $\phi_l$ according to the
(analytically tractable) iso-energetic approximation
\cite[]{bru85,redpra98} if the density drops below $10^8\gcm$.  Note
that in this case the cross section is corrected a posteriori for
nucleon recoil effects (see end of Sect.~\ref{sect:WM}) but
non-vanishing energy exchange between high-energy neutrinos and
nucleons are ignored. At densities below $10^8\gcm$, however, the
typical neutrino energies are moderate ($\l<\ene_\nu\r> < 30$MeV) and
the scattering rate with nucleons becomes very low.

Given $\phi_l^\mathrm{out}(\ene_j,\ene_{j'})$ for
$\ene_j\le\ene_{j'}$, detailed balance arguments 
\cite[see][]{cer94} are exploited in order to compute 
$\phi_l^\mathrm{out}(\ene_j,\ene_{j'})$ for $\ene_j>\ene_{j'}$ and the coefficients
$\phi_l^\mathrm{in}(\ene_j,\ene_{j'})=\phi_l^\mathrm{out}(\ene_{j'},\ene_j)$,
corresponding to the inverse reaction \cite[for details, see][Appendix~A]{ramjan02}.

Once the Legendre coefficients $\phi_0$ and $\phi_1$ are known, the
contribution of neutrino-nucleon scattering to the collision integral
of the Boltzmann equation and its angular moments is calculated in
exactly the same way as described for NES in \citet[][
Appendix~A]{ramjan02}.

Note that evaluating Eq.~(38) of \cite{bursaw98} requires the
knowledge of nucleon-nucleon interaction potentials. Within the
framework pursued by \cite{bursaw98} the latter can be expressed in
terms of the Fermi-liquid parameters, which, in turn, are directly
related to macroscopic observables such as the incompressibility
modulus $K_\mathrm{s}$ and the symmetry energy $S_\mathrm{v}$ of bulk
nuclear matter \cite[see][]{redpra99}. Accordingly, we adopt values
for the Fermi-liquid parameters which are consistent with the
parameters chosen for the nuclear EoS we use for our simulations
\cite[][ with $K_\mathrm{s}=180~\mev$,
$S_\mathrm{v}=29.3~\mev$]{latswe91}.

\subsubsection{Absorption of electron-flavor neutrinos by free
  nucleons ($\nue \mathrm{n} \rightleftharpoons 
  \mathrm{e}^- \mathrm{p} $, 
  $\nuae \mathrm{p} \rightleftharpoons 
  \mathrm{e}^+ \mathrm{n} $ )}\label{sect:NNA}

The calculation of the inverse mean free path $1/\lambda(\ene)$ 
for absorption of electron (anti)neutrinos by free neutrons (protons) 
is based on Eq.~(2) of \cite{bursaw99}. 
For the numerical implementation we employ the same
techniques which we have already described for the neutral-current
reactions in Sect.~\ref{sect:NNS}.  
Different from neutrino-nucleon scattering, however, the outgoing lepton is 
a charged lepton, which, in our context is assumed to be in local
thermodynamic equilibrium within the stellar
medium. Hence, the dependence of the interaction kernels on the
energy $\ene'$ of the outgoing lepton does not need to be retained and the 
individual sums over energy bins $[\ene_{j'},\ene_{j'+1}]$
are consequently replaced by the integral over the entire spectrum of
energies $\ene'$ of the outgoing charged lepton.  Given
$1/\lambda(\ene)$, the absorption opacity corrected for stimulated
absorption $\kappa^*(\ene)$ which enters our neutrino transport scheme
can be calculated in a straightforward way \cite[cf.][
Appendix~A]{ramjan02}.

\subsubsection{Effective mass of the nucleons}

Following \cite{bursaw99} we substitute an effective mass $m^*$ for
the nucleon mass $m$, wherever the latter appears explicitly in the
formalism. Notably, $m^*$ is also used for inverting the relation
between particle density and chemical potential which is used to
compute the chemical potentials of protons and nucleons (displaced by
the nuclear interaction potential) for given number densities
\cite[cf.][~Eq.~11]{bursaw98}.

For the effective mass we adopt the parametric density dependence \cite[]{redpra99}
\begin{equation}\label{eq:effmass}
m^*(\rho)=\frac{m}{1+\alpha\cdot\rho/\rho_\text{nuc}}
\,,
\end{equation}
where $m$ is the vacuum mass of the nucleon and the constant $\alpha$
is defined by the value of the effective mass at nuclear saturation
density ($\rho_\text{nuc}/m=0.16\,\text{fm}^{-3}$) being
$m^*(\rho_\text{nuc})=0.8\,m$.

\subsubsection{Weak magnetism}\label{sect:WM}

Correction factors for the iso-energetic cross-sections accounting
both for the effects of weak magnetism and nucleon recoil effects are
provided by \cite{hor02}. Both effects appear at the level of
$\mathcal{O}(\ene/m)$ with $\ene$ being the neutrino energy and $m$
the nucleon mass.
For our purposes, we have to disentangle the contributions of both
effects to the final rates in order to obtain a pure weak magnetism
correction for the formalism described above, since the latter already
includes recoil and also nucleon-nucleon correlations, but disregards
weak magnetism.  This is achieved by analytically averaging the
differential rates which include weak magnetism as well as recoil and
those for recoil alone over the scattering angle and using the
resulting angle-independent ratio of both as a weak magnetism
correction factor to the rates of \cite{bursaw98,bursaw99}.
For the charged-current interactions the angular reduction is
performed by simply integrating the differential rates over all
directions of motion of the charged lepton in the final state.
In case of neutral-current scattering a $(1-\omega)$-weighting for the
differential rate is used which is motivated by the definition of the
transport opacity \cite[for details, see][]{hor02}.

In detail, the Legendre coefficients for neutral-current scatterings
as described in Sect.~\ref{sect:NNS} are modified according to
$\phi_l^\mathrm{out}(\ene,\ene')\to
\xi^\text{nc}(\ene)\cdot\phi_l^\mathrm{out}(\ene,\ene')$, where
\begin{equation}\label{eq:corr_nc}
\xi^\text{nc}(\ene)\equ 
\frac{
 n_\text{n}\cdot \mathcal{X}^\text{nc:n}_\text{WM,Rec}(\ene)
+n_\text{p}\cdot \mathcal{X}^\text{nc:p}_\text{WM,Rec}(\ene)
}
{
 n_\text{n}\cdot \mathcal{X}^\text{nc:n}_\text{Rec}(\ene)
+n_\text{p}\cdot \mathcal{X}^\text{nc:p}_\text{Rec}(\ene)
}
\end{equation}
is the ratio of the full correction factor \cite[ Eq.~32]{hor02}
\begin{eqnarray}\label{eq:r_wmrec_nc}
\mathcal{X}^\text{nc:n/p}_\text{WM,Rec}(\ene) =\left\{ C_\mathrm{V}^2 \left[ {\frac{e-1}{2e^3}} \ln \zeta + 
{\frac{3+ 12e + 9e^2 - 10e^3}{3e^2 \zeta^3}}
\right] \right. 
+C_\mathrm{A}^2 \left[ {\frac{1+e}{2e^3}} \ln \zeta - 
{\frac{10e^3 +27e^2 +18e +3}{3e^2 \zeta^3}}\right] \nonumber \\
\pm  \left(C_\mathrm{V} +F_2\right) C_\mathrm{A} \left[ {\frac{1}{e^2}} \ln \zeta - 
\frac{2+10e+ {\frac{28}{3}}e^2}{e \zeta^3} \right] 
+ C_\mathrm{V} F_2 \left[\frac{1}{e^2}\ln \zeta -\frac{2}{3} 
\left( \frac{3+15e +22e^2}{e \zeta^3}\right)\right]\nonumber \\
\left.\left. + F^2_2 \left[\frac{1}{4e^2} \ln \zeta + 
\frac{8e^3 -22e^2 - 15e-3}{6e \zeta^3} \right] \right\}\right/ 
\left\{ \frac{2}{3} \left(C_\mathrm{V}^2 +5C_\mathrm{A}^2 \right) \right\} \,, 
\end{eqnarray}
to the correction factor for recoil alone \cite[ Eq.~29]{hor02}
\begin{equation}\label{eq:r_rec_nc}
\mathcal{X}^\text{nc:n/p}_\text{Rec}= 
\left.\left\{
 C_\mathrm{V}^2 \left[ \frac{e+1}{e^3}\ln \zeta
                    -\frac{2}{e^2}
              \right]
       +C_\mathrm{A}^2 \left[ \frac{2 e^2-1-e}{e^3 \zeta} \ln \zeta
                    +\frac{2}{e^2 \zeta}
              \right] 
\right\} \right/ 
\left\{\frac{2}{3} (C_\mathrm{V}^2+5 C_\mathrm{A}^2)
\right\}\,.
\end{equation}
In Eqs.~(\ref{eq:r_wmrec_nc}, \ref{eq:r_rec_nc}) we have used the
abbreviations
\begin{equation}
e(\ene)\equ\frac{\ene}{m\,c^2}\hspace{1cm}
\text{and} \hspace{1cm}\zeta(e)\equ 1+2e \,.
\end{equation}
Numerical values of the coupling constants $C_\mathrm{V}$,
$C_\mathrm{A}$ and $F_2$ can be found in \citet[][
Table~I]{hor02}. Since the values of $C_\mathrm{V}$, $C_\mathrm{A}$
and $F_2$ are different for the reactions $\nu
\mathrm{p}\rightleftharpoons\nu \mathrm{p}$ and $\nu
\mathrm{n}\rightleftharpoons\nu \mathrm{n}$, the individual correction
factors for protons $\mathcal{X}^\text{nc:p}$ and neutrons
$\mathcal{X}^\text{nc:n}$ are weighted with the corresponding number
densities $n_\text{n}$ and $n_\text{p}$, respectively, to yield the
final correction factor $\xi^\text{nc}$ for neutral-current
scatterings as defined in Eq.~(\ref{eq:corr_nc}). This factor is then
applied to the combined rate of neutrino scatterings off neutrons and
protons in our code.
\medskip

Analogously, we modify the inverse mean free path for
charged-current reactions (see Sect.~\ref{sect:NNA}) according to
$1/\lambda(\ene)\to\xi^\text{cc}(\ene)\cdot 1/\lambda(\ene)$, where
\begin{equation}\label{eq:corr_cc}
\xi^\text{cc}(\ene)\equ \frac{\mathcal{X}^\text{cc}_\text{WM,Rec}(\ene)}{\mathcal{X}^\text{cc}_\text{Rec}(\ene)}\,,
\end{equation}
with 
\begin{eqnarray}\label{eq:r_wmrec_cc}
\mathcal{X}^\text{cc}_\text{WM,Rec}(\ene) = 
\left\{ C_\mathrm{V}^2 \left( 1 + 4e + \frac{16}{3} e^2 \right)
+ 3C_\mathrm{A}^2 \left( 1 + \frac{4}{3}e \right)^2 \pm 
4\left( C_\mathrm{V} + F_2 \right) C_\mathrm{A} e \left(1+ \frac{4}{3}e \right)
\right. \nonumber \\
\left.\left. + \frac{8}{3} C_\mathrm{V} F_2 e^2 + \frac{5}{3} e^2 \left( 1+ \frac{2}{5} e
\right) F^2_2  \right\} \right/ \left\{ \left(C_\mathrm{V}^2 + 3 C_\mathrm{A}^2 \right)
 \zeta^3 \right\} \,,
\end{eqnarray}
as given by \citet[ Eq.~22]{hor02}, and (ibid, Eq.~19)
\begin{equation}
\begin{split}
\mathcal{X}^\text{cc}_\text{Rec}(\ene)&= 
\left.\left\{
  C_\mathrm{V}^2 \left[ \frac{1}{e}-\frac{1}{2e^2}\ln \zeta \right]
 +C_\mathrm{A}^2 \left[ \frac{4 e^2 - 2 e + \zeta\ln \zeta}{2 e^2 \zeta}
              \right] 
\right\} \right/ 
\left\{C_\mathrm{V}^2+3 C_\mathrm{A}^2\right\} \,.
\end{split}
\end{equation}
The $\pm$-symbol in Eqs.~(\ref{eq:r_wmrec_nc}, \ref{eq:r_wmrec_cc})
translates to a positive sign for neutrinos and a negative sign for
antineutrinos, respectively.
In our transport code, we presently do not discriminate between
$\nu_{\mu,\tau}$ and the corresponding antiparticles. For these
flavors we therefore adopt the arithmetic mean of the correction
factors of Eq.~(\ref{eq:r_wmrec_nc}) for neutrino and antineutrino
scattering which in effect means that the terms with the $\pm$-symbol
cancel for the heavy-lepton neutrinos.

\medskip

Recall that we switch from the description of \cite{bursaw98,bursaw99}
to the iso-energetic approximation \cite[]{bru85} if $\rho <
10^8\gcm$ (see Sect.~\ref{sect:NNS}).  Consequently, in order to take
into account nucleon recoil and weak magnetism also at these
conditions the correction factors defined in Eqs.~(\ref{eq:corr_nc},
\ref{eq:corr_cc}) are replaced by
$\xi^\text{nc}=
( n_\text{n}\cdot \mathcal{X}^\text{nc:n}_\text{WM,Rec}
 +n_\text{p}\cdot \mathcal{X}^\text{nc:p}_\text{WM,Rec})/
( n_\text{n} + n_\text{p})$, and
$\xi^\text{cc} = \mathcal{X}^\text{cc}_\text{WM,Rec}$, respectively,
whenever the density falls below $10^8\gcm$.

\subsection{Quenching of the axial-vector coupling}
\label{sect:quench}

In the calculations of all neutrino-matter interactions we replace the
axial-vector coupling $g_\text{A}=1.254$ by an effective, "quenched"
value $g^*_\text{A}$, which depends on the baryon density according to
\begin{equation}\label{eq:gaquench}
g^*_\text{A}(\rho)=
g_\text{A}\,\left(1-\frac{\rho}{4.15\,(\rho_\text{nuc}+\rho)}\right)
\,,
\end{equation}
as suggested by \citet[][ Eq.~13]{carpra02}.
In effect, opacities are reduced roughly by a factor
$(g^*_\text{A}/g_\text{A})^2$ at densities $\rho\gtrsim\rho_\text{nuc}$.

\subsection{Discussion}\label{app:opac_dis}

\begin{figure}[!htp]

  \epsfig{width=18cm,file=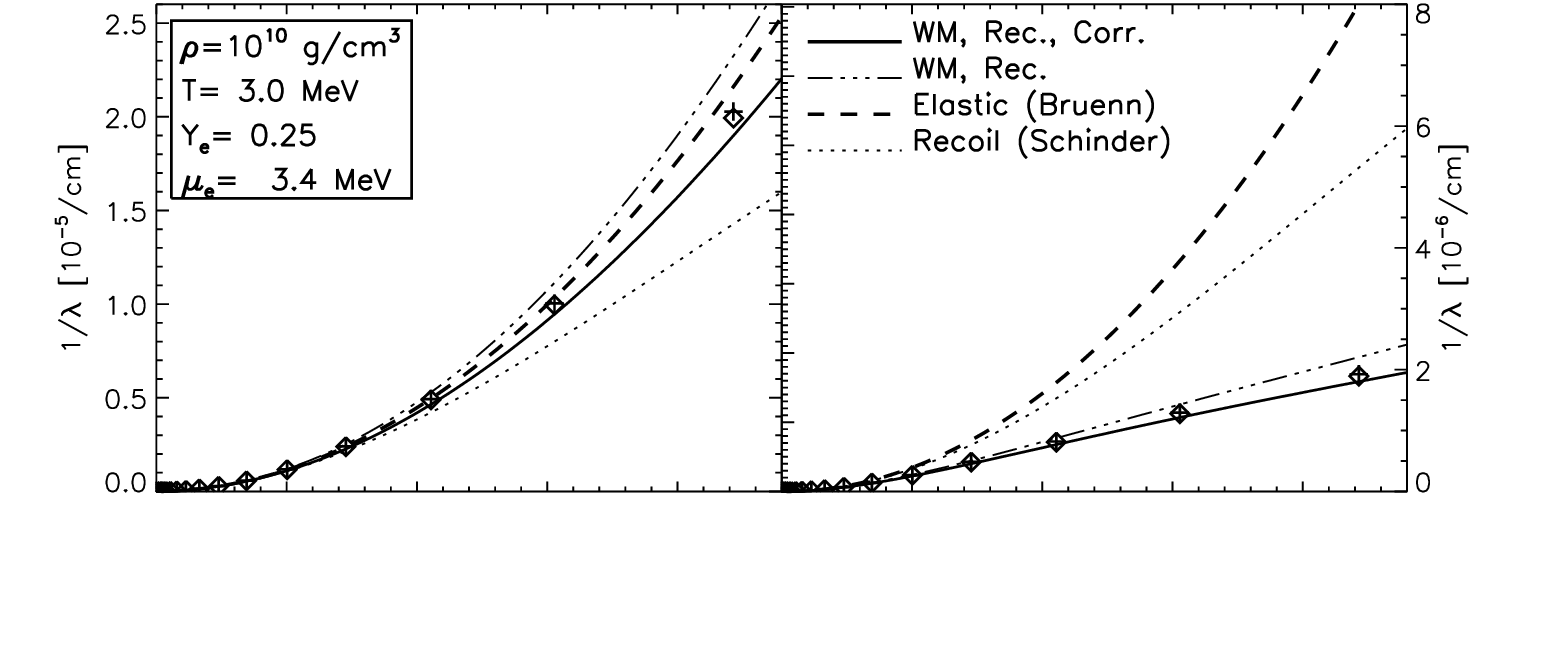}\vspace{-1.9cm}\\ 
  \epsfig{width=18cm,file=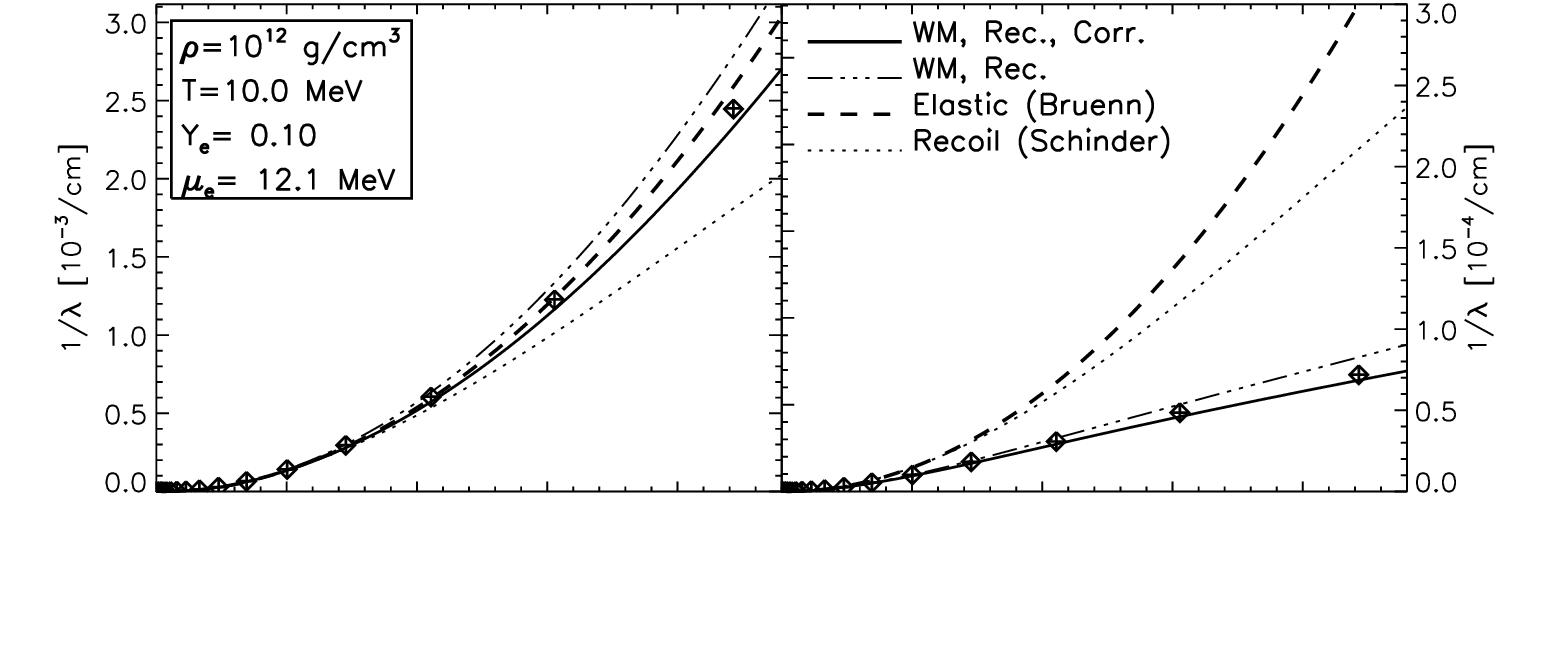}\vspace{-1.9cm}\\  
  \epsfig{width=18cm,file=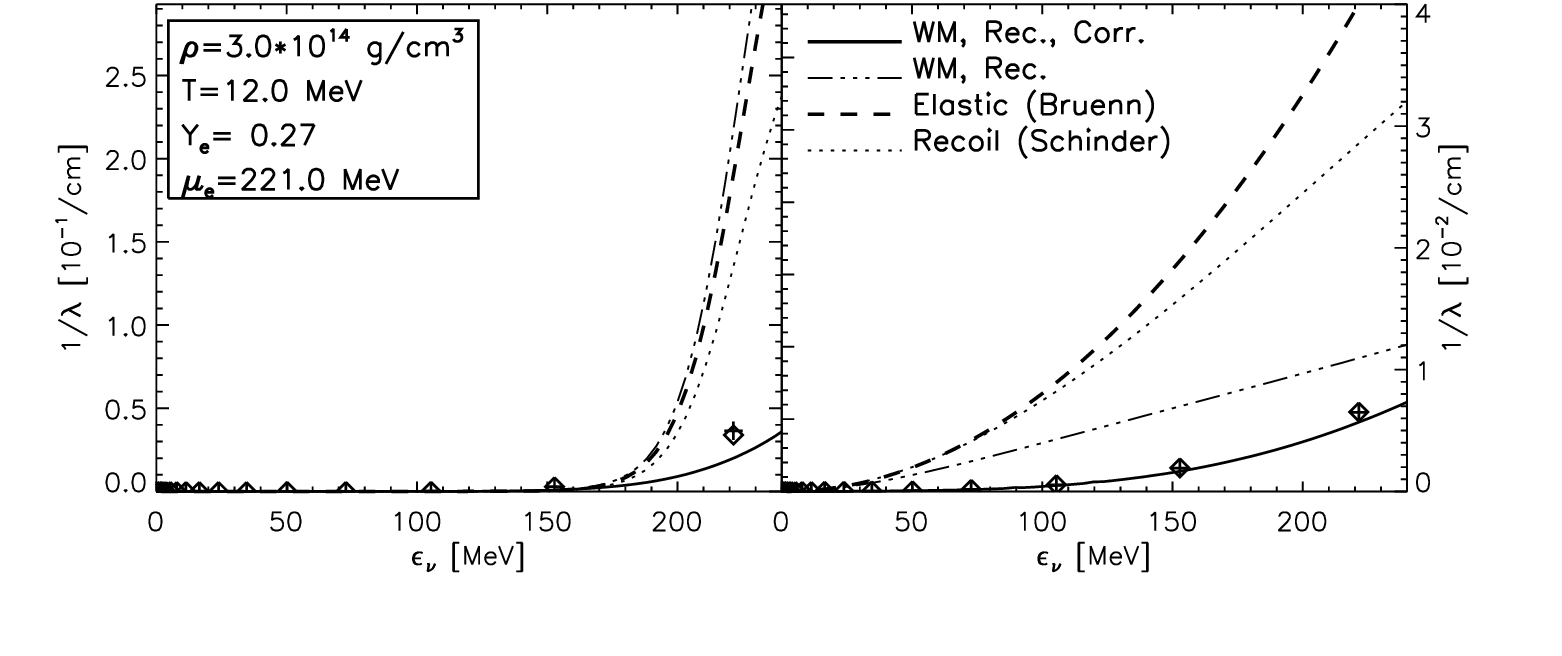}\vspace{-1cm} 

\caption[]{
  Inverse mean free paths as
  functions of neutrino energy for absorption of $\nue$
  (left column) and $\nuae$ (right column) by free nucleons. Note the
  different scales of the ordinates.  
  Solid lines and symbols are drawn for the charged-current reaction
  rates computed according to Sect.~\ref{sect:NNA}. 
  Thermodynamic
  conditions are given in the top left corners of the plots. For given
  density effective
  values for the nucleon mass $m^*$ and the axial-vector coupling
  $g^*_\text{A}$ were calculated using Eq.~(\ref{eq:effmass}) and
  Eq.~(\ref{eq:gaquench}), respectively. 
  Thin dashed-dotted lines show the weak magnetism plus recoil
  correction of \cite{hor02} applied to the iso-energetic cross section 
  of \cite{bru85}, which itself is drawn as a dashed line. 
  The dotted lines correspond to the recoil and
  thermal motion approximation of \cite{sin90}.
  In all cases the neutrino phase space was assumed to be empty.
  Lines interpolate values computed on an energy grid of 100 points which
  are equidistantly spaced between $0$ and $380~\mev$.
  For comparison, crosses show the corresponding rates on a geometrical
  energy grid with 17 bins and six-point 
  Gauss-quadratures for the angular and spectral integrations
  (relevant for the description given in Sect.~\ref{sect:NNA}).  
  This is the typical spectral resolution of our dynamical supernova
  simulations. 
  Results obtained by using 30-point Gauss-quadratures instead are
  displayed by open diamonds. See the end of Sect.~\ref{app:opac_dis} for a discussion.
}\label{fig:opac_cc} 
\end{figure}

\begin{figure}[!htp]
\epsfig{width=18cm,file=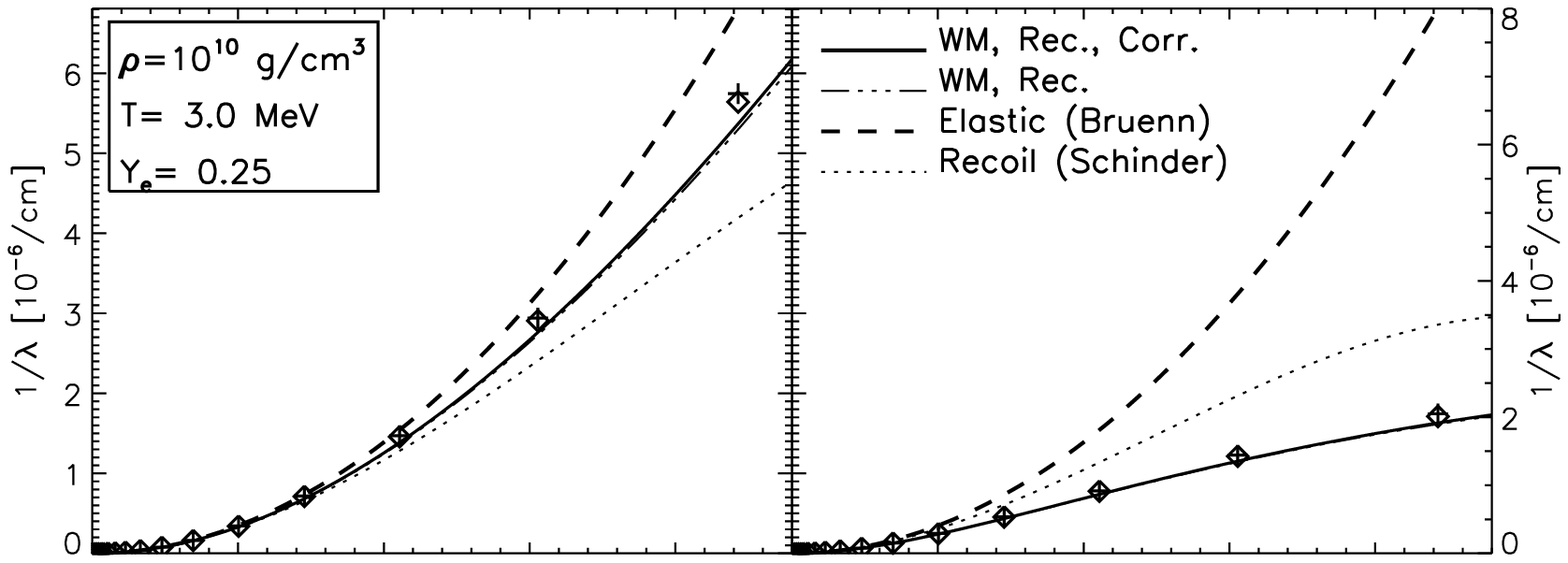}\vspace{-1.9cm}\\ 
\epsfig{width=18cm,file=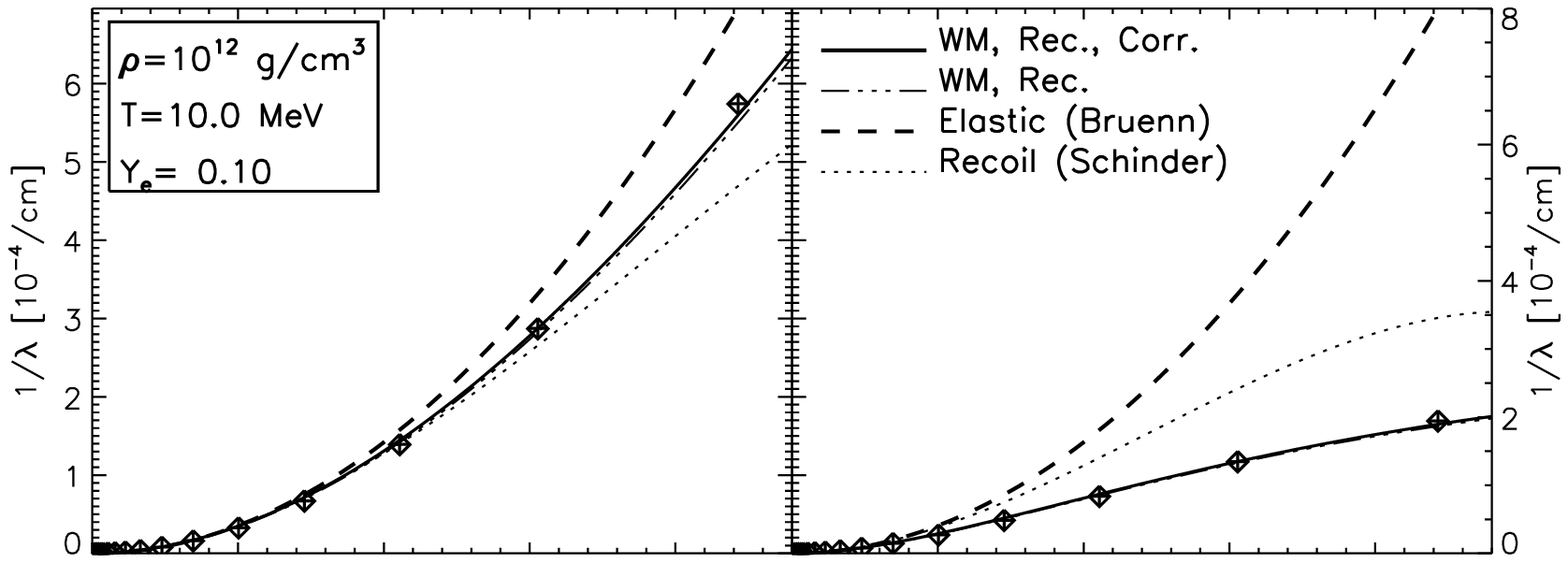}\vspace{-1.9cm}\\ 
\epsfig{width=18cm,file=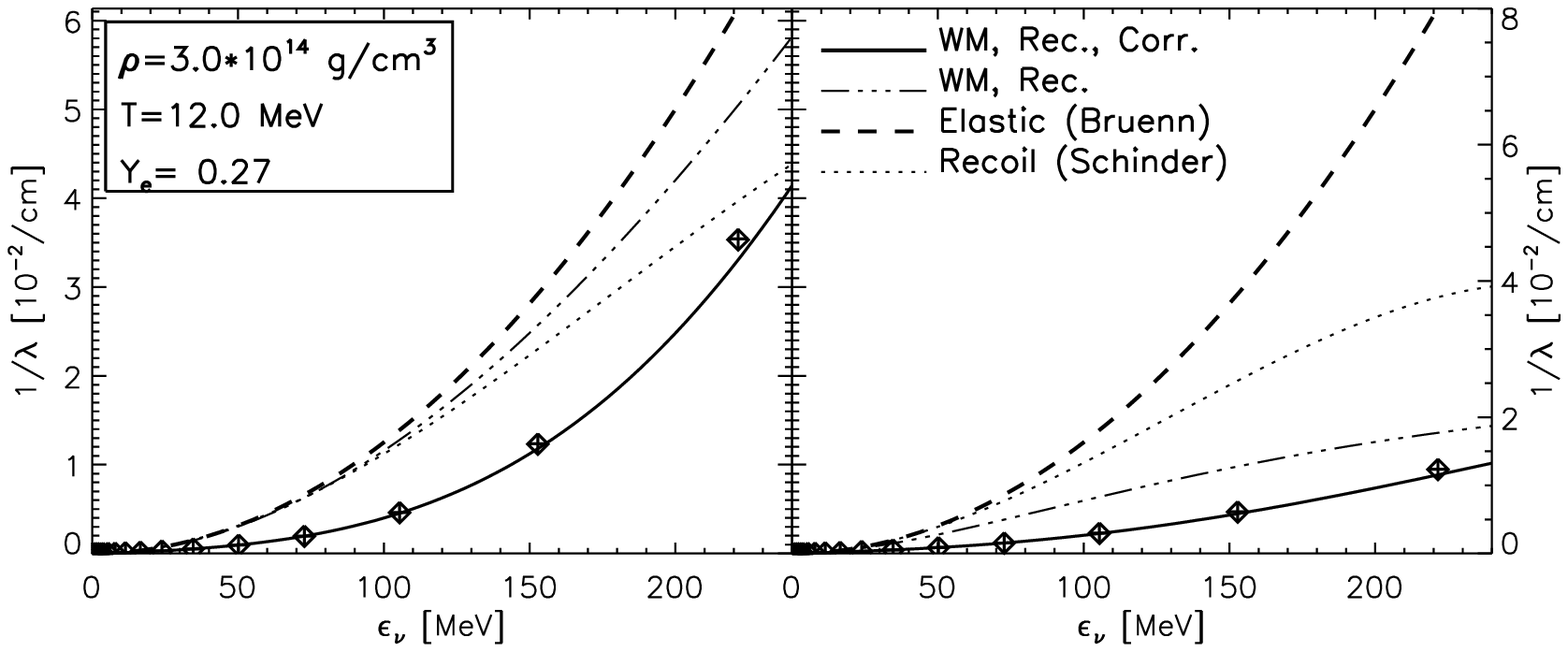}\vspace{-1cm} 
\caption[]{
  Same as Figure~\ref{fig:opac_cc}, but showing the inverse
  mean free paths for neutral-current scattering of neutrinos
  (left column) and antineutrinos (right column) as computed according 
  to Sect.~\ref{sect:NNS} (solid lines, symbols) in comparison with
  the conventional (``standard'') description (dashed lines) and
  different approximations to the complete physics (dotted and
  dash-dotted).
  In order to obtain the quantity $1/\lambda(\ene)$ the differential
  scattering rate $R(\ene,\ene',\omega)$ was integrated over the
  spectrum of final neutrino energies $\ene'$ and all values of the
  cosine $\omega$ of the scattering angle.
}\label{fig:opac_nc}
\end{figure}

\begin{figure}[!htp]
\begin{tabular}{lr}
\epsfig{width=11cm,file=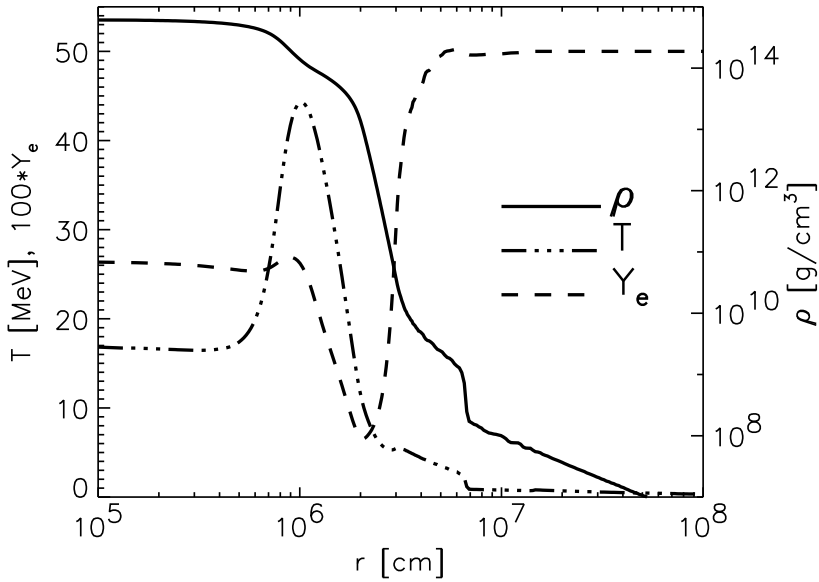} & 
\raisebox{9cm}{\parbox[t]{6cm}{
\caption[]{
  Stellar profile of a spherically symmetric postbounce model (Model
  s15Gio\_1d.b at $t_\text{pb}=0.257$~s).
  Density $\rho$ (solid line), temperature $T$ (dash-dotted), and
  electron fraction $\ye$ (dashed) as functions of stellar radius $r$
  were used to calculate the cross sections as functions of $\rho(r)$
  shown in Fig.~\ref{fig:opac_prof}.
}\label{fig:opac_star} }} \\
\end{tabular}
\end{figure}

\paragraph{Main effects:}
In Figures~\ref{fig:opac_cc}, \ref{fig:opac_nc} opacities for
neutrino-nucleon interactions
computed with the procedures described above are  
compared with the iso-energetic approximation adopted by \cite{bru85}
and also with the description of \cite{sin90}. 
The latter work approximately takes into account the reaction
kinematics (recoil, thermal motions and final-state blocking of the
nucleons) but disregards weak magnetism and nucleon-nucleon correlations.
According to our core-collapse simulations for the $15~\msol$ star
the chosen combinations of values for the density
$\rho$, temperature $T$, and electron fraction $\ye$ of the stellar
medium are characteristic for the conditions in the gain
layer, where neutrino heating is strongest
(top row of Figs.~\ref{fig:opac_cc}, \ref{fig:opac_nc}), 
the region where most of the
neutrino luminosity is produced (middle row), and the interior of the
forming proto neutron star (bottom row). 

\begin{figure}[!htp]

\epsfig{width=17.77cm,file=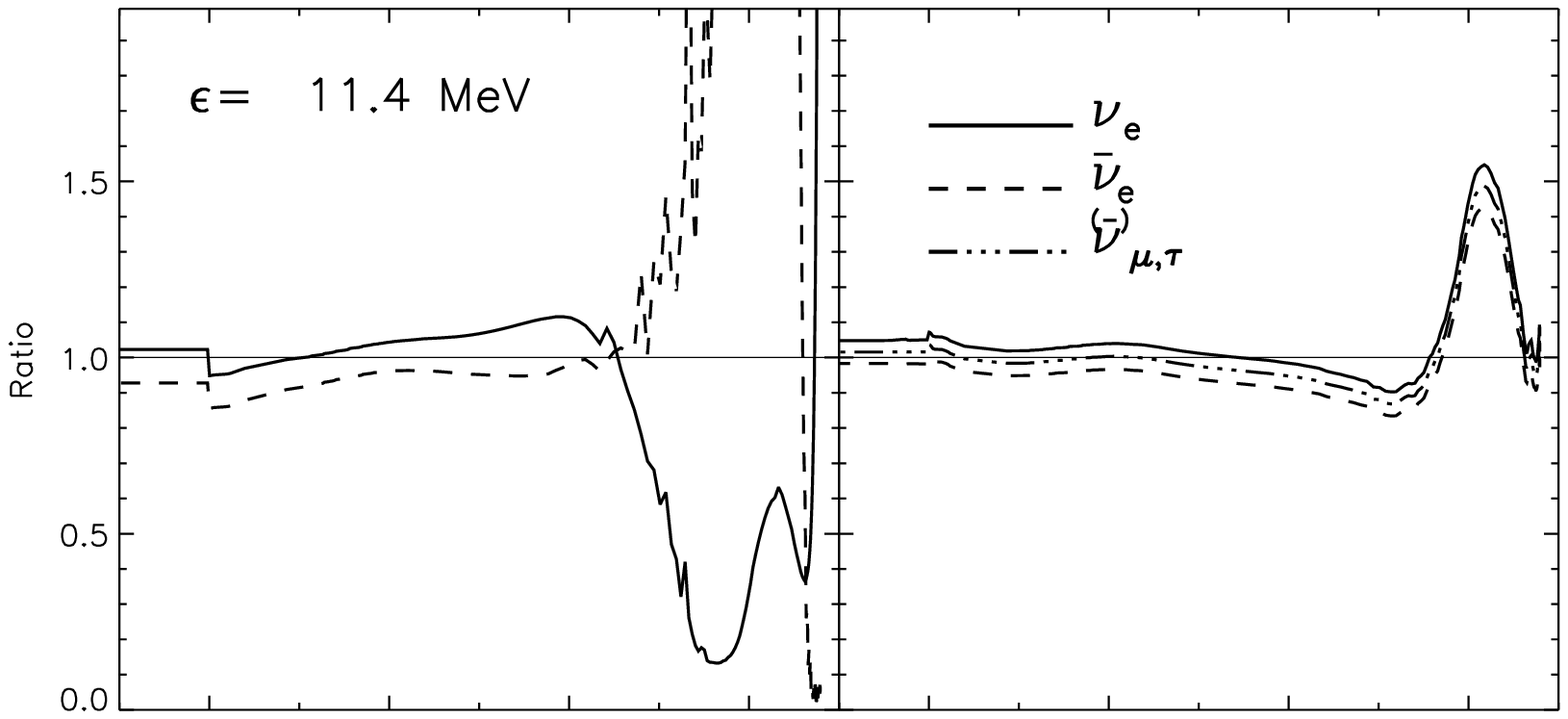}\vspace{-2.4cm}\\ 
\epsfig{width=17.77cm,file=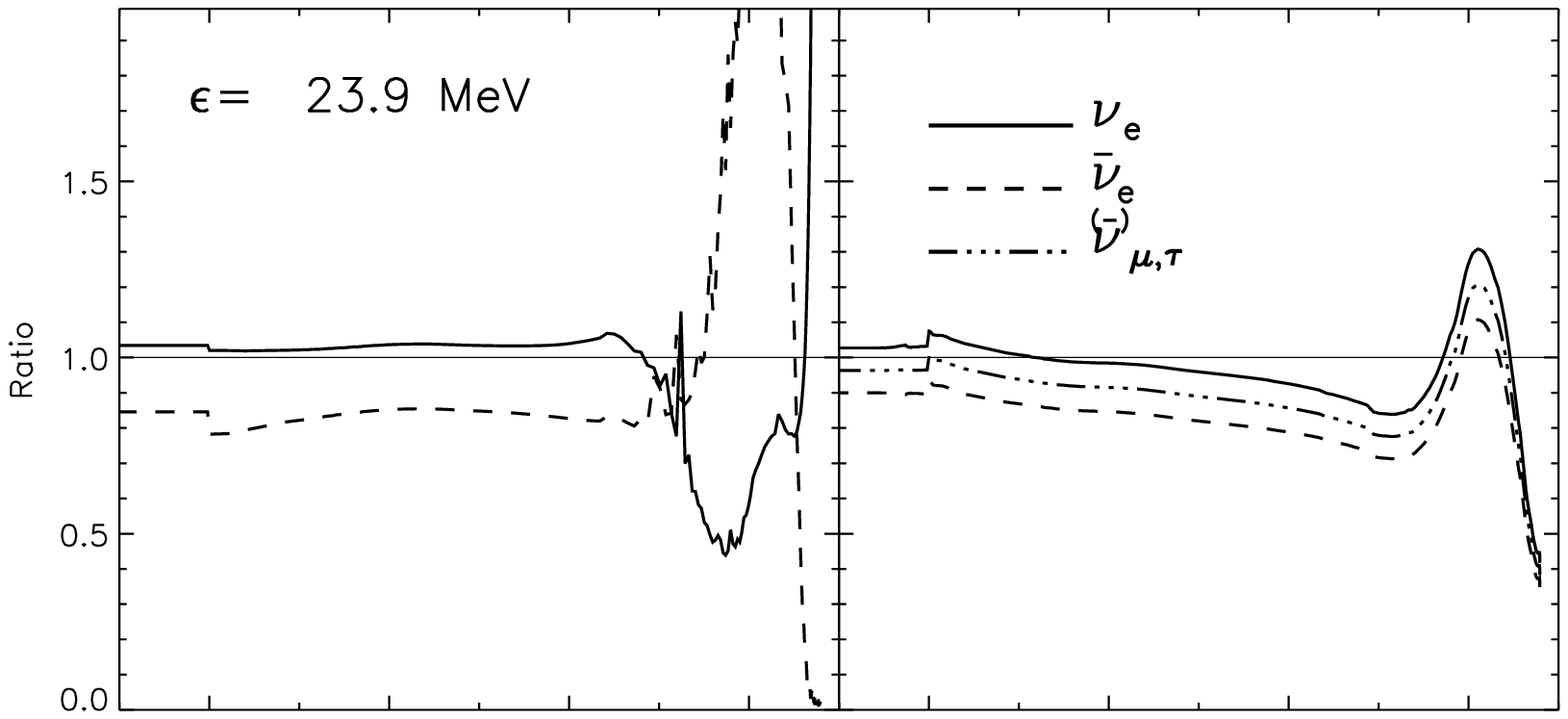}\vspace{-2.4cm}\\ 
\epsfig{width=17.77cm,file=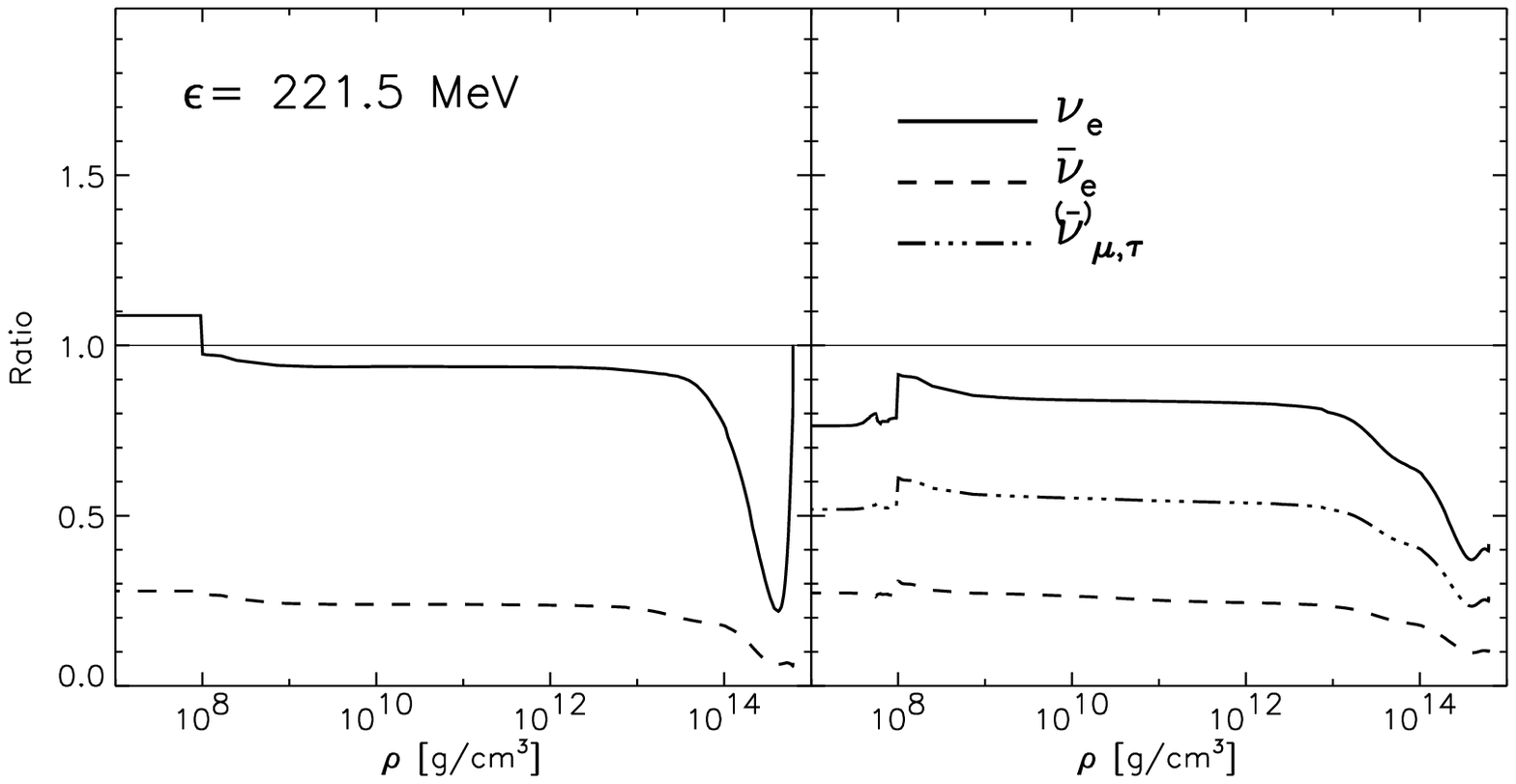}\vspace{-1.5cm} 

\caption[]{
  Ratios of cross sections as computed according to
  Sects.~\ref{sect:neutrino-nucleon}, \ref{sect:quench} to the
  conventional (``standard'') approximation for given neutrino energies as
  functions of density $\rho$ (the input quantities
  $T$($\rho$) and $\ye$($\rho$) were
  obtained from the stellar profile shown in
  Fig.~\ref{fig:opac_star}).
  In the left column we show the ratio for charged-current  
  absorption of $\nue$ (solid lines) and $\nuae$ (dashed lines). 
  The right column shows the ratio for neutral-current
  scattering of $\nue$, $\nuae$, and heavy-lepton neutrinos 
  (dash-dotted lines) off free nucleons. According to our
  equal treatment of heavy-lepton neutrinos and antineutrinos
  the latter quantity is computed using arithmetic averages of the
  particle and antiparticle cross sections (see
  Sect.~\ref{sect:WM}). The neutrino phase space was assumed to be empty.
}\label{fig:opac_prof}  
\end{figure}

As discussed in detail by \cite{hor02} the effects of weak magnetism
and nucleon recoil
counterbalance each other for $\nue$ while in the case of $\nuae$ both
effects add up leading to an appreciable net reduction of the standard opacities
\cite[computed according to][]{bru85}. This can be seen in
Figs.~\ref{fig:opac_cc}, \ref{fig:opac_nc} by comparing the bold,
dashed lines (``elastic approximation'') with the thin, dotted (``recoil
only'') and dash-dotted (``recoil plus weak magnetism'') lines. 
While for given neutrino energy, the relative effects of the weak magnetism
and recoil corrections are independent of the nucleon density, a
sizeable additional reduction of both $\nue$ and $\nuae$ opacities due to
nucleon-nucleon correlations shows up at nuclear and supranuclear 
densities \cite[see][ for a thorough discussion of the underlying
  physics]{bursaw98,bursaw99,redpra98,redpra99,hor02}. 

This is also illustrated by Fig.~\ref{fig:opac_prof} which displays
the ratio of the rates computed with the procedures described above to
the iso-energetic approximations adopted from \cite{bru85}, for a
typical stellar (post-bounce) profile. Using the density as a
coordinate (and taking $T$($\rho$) and $\ye$($\rho$) from the stellar
profile shown in Fig.~\ref{fig:opac_star}) we plot this ratio for the
charged-current (left column) and neutral-current interactions (right
column) as function of density for three different values of the
neutrino energy.
Up to densities of $10^{13}\gcm$ one observes a nearly constant ratio
of the interaction rates, which is weakly dependent on the
thermodynamic conditions of the stellar medium but increases for
higher neutrino energies.
Extant variations can be attributed to small but finite effects of the
reaction kinematics (the latter are disregarded for $\rho <
10^{8}\gcm$ where the ratios deviate from unity because of weak
magnetism and recoil corrections of the interaction cross sections in
the numerators; cf.~Sects.~\ref{sect:NNS}, \ref{sect:WM}).
While sizeable cross section reduction due to weak magnetism and
recoil of the nucleons is visible for $\nuae$ (dashed lines) the
aforementioned counterbalancing effect leads to only small deviations
from the iso-energetic approximation in case of $\nue$ (solid lines).
For densities between $10^{13}\gcm$ and $10^{14}\gcm$ the prominent
peak in the adopted temperature profile of the stellar medium
(see Fig.~\ref{fig:opac_star}) is reflected in the density variation of
the interaction rates plotted in Fig.~\ref{fig:opac_prof}
due to the temperature dependence of the
reaction kinematics and nucleon-nucleon correlations. Interestingly,
for low neutrino energies this leads to an \emph{increase} of the
cross sections compared to the iso-energetic approximation (top and
middle row of Fig.~\ref{fig:opac_prof}). 
In the considered density and temperature range, however, neutrinos are in
thermodynamic equilibrium with the stellar
medium and therefore precise values of the cross sections are not very
relevant for neutrinos of low energies.
In the high-energy part of the spectrum, however, which carries the
significant part of the neutrino flux, our description of the
neutrino-matter interactions leads to an overall \emph{reduction} of
the opacities for neutrinos and antineutrinos of all flavors (bottom
row of Fig.~\ref{fig:opac_prof}) and hence facilitates transport of
neutrinos out of the proto neutron star.

\paragraph{Numerical considerations:}

For plotting the solid lines in Figs.~\ref{fig:opac_cc},
\ref{fig:opac_nc} the corresponding cross sections were evaluated 
on an equidistant energy grid consisting of 100 points. 
Such spectral resolution, however, can currently not be afforded 
in our dynamical supernova simulations. 
For comparison we therefore supplement the plots with values (shown by
crosses) computed on the comparably coarse energy grid of the
simulations (geometrical grid of 17 bins).
\begin{figure}[!htp]
\begin{tabular}{lr}
\epsfig{width=11cm,file=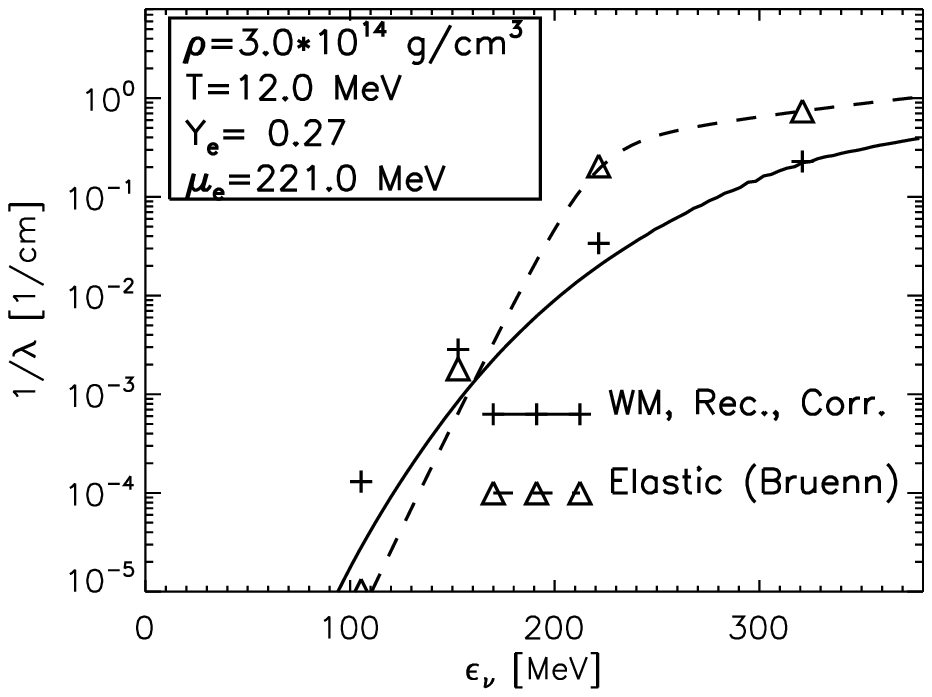} & 
\raisebox{9cm}{\parbox[t]{6cm}{
\caption[]{
Inverse mean free path for the absorption of $\nue$ 
by free neutrons for the electron-degenerate conditions given in the last row of 
Fig.~\ref{fig:opac_cc}. The reference curve (solid line) 
was obtained with high overall spectral resolution whereas 
crosses show averages over the bins of a comparably 
coarse energy grid which is typically used in our dynamical 
simulations. For comparison the same 
averaging procedure was also applied to the iso-energetic 
approximation of \cite{bru85}. The results are shown as a dashed line with
open triangles marking the corresponding coarse-grid average values.  
}\label{fig:opac_fermi} 
}} \\
\end{tabular}
\end{figure}
For a wide range of conditions we find good overall agreement with 
the corresponding "reference" solutions.
Notable deviations show up in regions with a steep variation of the
opacity with the neutrino energy, which can be caused, e.g., by the
final state blocking of the 
degenerate electron gas in the reaction $\nue + \mathrm{n} \to
\mathrm{e}^- + \mathrm{p}$ (see Fig.~\ref{fig:opac_fermi}). 
Obviously, this phenomenon is not specific to the
adopted physical description (see, e.g.~the open triangles in
Fig.~\ref{fig:opac_fermi}) but is simply due to the fact that 
the rates are calculated as averages over the finite 
width of an energy bin (cf.~Sects.~\ref{sect:NNS}, \ref{sect:NNA}).
Increasing the number of quadrature weights for the corresponding numerical
integrations within the spectral (and angular) bins by a factor of
five each, i.e.~employing 30-point Gauss-quadratures (open diamonds in
Figs.~\ref{fig:opac_cc}, \ref{fig:opac_nc}), we find very good
agreement with the standard six-point results (crosses). 
Hence, the relevant discretization error of the dynamical 
simulations is dominated by the employed overall spectral 
resolution (relative width of energy bins) and not by the 
specific sub-gridding used for evaluating the neutrino opacities.

\section{Moments equations in three dimensions}
\label{app:momeq3d}

To order $\mathcal{O}(v/c)$ of the fluid velocities (the so-called
Newtonian approximation) the full three-dimensional moments equations
are given by \cite[][ correcting a number of misprints]{kanmor84}:
\be
\frac{\rho}{c}\frac{D}{Dt}\l(\frac{J}{\rho}\r) 
 + {\bf \nabla \cdot H}
 + \frac{1}{c}\pd{}{t} \l(\boldsymbol{\beta} \cdot {\bf H}\r)
 - \frac{\ene}{c} \pd{\boldsymbol{\beta}}{t} \cdot \pd{\bf H}{\ene}
 - \ene {{\bf \nabla} \boldsymbol{\beta}} : \pd{\bf P}{\ene}
 = \Cj
\ee
\be
 \frac{\rho}{c}\frac{D}{Dt}\l(\frac{\bf H}{\rho}\r) 
   + {\bf \nabla \cdot P}
   + {\bf H} \cdot {\bf \nabla} \boldsymbol{\beta}
   + \frac{1}{c}\pd{}{t} \l({\boldsymbol{\beta} \cdot {\bf P}}\r)
   + \frac{J}{c} \pd{\boldsymbol{\beta}}{t}
   - \frac{1}{c} \pd{\boldsymbol{\beta}}{t} \cdot \pd{(\ene {\bf P})}{\ene}
   - {{\bf \nabla} \boldsymbol{\beta}} : \pd{(\ene {\bf N})}{\ene}
 = {\bf C}^{(1)}\, ,
\ee
where $\boldsymbol{\beta}=\boldsymbol{v}/c$. In general spherical
coordinates, the moments are defined by
\be J = \ofp\int\int I \dlin{\Omega},~ 
    {\bf H} = \ofp\int\int I {\bf n} \dlin{\Omega},~ 
    {\bf P} = \ofp\int\int I {\bf n}{\bf n} \dlin{\Omega},~
    {\bf N} = \ofp\int\int I {\bf n}{\bf n}{\bf n} \dlin{\Omega}.
\label{appeq:moments}\ee
Here, ${\bf
n}=(n_r,n_\vartheta,n_\varphi)=(\cos\Theta,\sin\Theta\cos\omega,
\sin\Theta\sin\omega)$, $\mu=\cos\Theta$, and $\Theta$ and $\omega$
are the two angles defining the direction of propagation. The
functional dependences
$\boldsymbol{\beta}=\boldsymbol{\beta}(t,r,\vartheta,\varphi)$,
$J=J(t,r,\vartheta,\varphi,\ene)$, \dots, are suppressed in the
notation. The moments equations written in coordinate form are the
following:
\ba
&& \frac{\rho}{c}\frac{D}{Dt}\l(\frac{J}{\rho}\r)
   + \frac{1}{r^2} \pd{(r^2 \Hr)}{r}
   + \irst \pd{(\st \Ht)}{\vartheta}
   + \irst \pd{\Hp}{\varphi}
   + \oc \pd{}{t}
     \l(\br \Hr + \bt \Ht + \bp \Hp \r)
\nn\\&&{}
   - \frac{\ene}{c} \l(
      \pd{\br}{t} \pd{\Hr}{\ene} +
      \pd{\bt}{t} \pd{\Ht}{\ene} + 
      \pd{\bp}{t} \pd{\Hp}{\ene} \r)
\nn\\&&{}
   - \ene
     \l\{ \pd{\Prr}{\ene} \l( \pd{\br}{r} \r)
        + \pd{\Ptr}{\ene} \l( \ir\pd{\br}{\vartheta} - \frac{\bt}{r} \r)
        + \pd{\Ppr}{\ene} \l( \irst\pd{\br}{\varphi} - \frac{\bp}{r} \r)
        + \pd{\Prt}{\ene} \l( \pd{\bt}{r} \r)
        + \pd{\Ptt}{\ene} \l( \ir\pd{\bt}{\vartheta} + \frac{\br}{r} \r)
\r.\nn\\&&{}
        + \l.\pd{\Ppt}{\ene} \l( \irst\pd{\bt}{\varphi} 
                                            - \frac{\bp\ct}{r\st} \r)
        + \pd{\Prp}{\ene} \l( \pd{\bp}{r} \r)
        + \pd{\Ptp}{\ene} \l( \ir\pd{\bp}{\vartheta} \r)
        + \pd{\Ppp}{\ene} \l( \irst\pd{\bp}{\varphi} + \frac{\br}{r}
                                            + \frac{\bt\ct}{r\st} \r)
\r\} 
   = \Cj \, , 
\label{J_spher}\ea
\ba
&& \frac{\rho}{c}\frac{D}{Dt}\l(\frac{\Hr}{\rho}\r) 
     - \frac{\Ht\bt}{r} - \frac{\Hp\bp}{r}
     + \frac{2}{r}\Prr + \pd{\Prr}{r} +
     \frac{\ct}{\rst}\Ptr + \ir\pd{\Ptr}{\vartheta} +
     \irst\pd{\Ppr}{\varphi} - \frac{\Ptt+\Ppp}{r} \nn\\
&&{} + \Hr \l( \pd{\br}{r} \r) +
     \Ht \l( \ir\pd{\br}{\vartheta} - \frac{\bt}{r} \r) +
     \Hp \l( \irst\pd{\br}{\varphi} - \frac{\bp}{r} \r) \nn\\
&&{} + \oc\pd{}{t}\l(\br\Prr+\bt\Ptr+\bp\Ppr\r)
     + \frac{J}{c}\pd{\br}{t}
     - \oc\pd{\br}{t}\pd{(\ene\Prr)}{\ene} -
     \oc\pd{\bt}{t}\pd{(\ene\Ptr)}{\ene} -
     \oc\pd{\bp}{t}\pd{(\ene\Ppr)}{\ene} \nn\\
&&{} -
     \l\{ \pd{(\ene\Nrrr)}{\ene} \l( \pd{\br}{r} \r)
        + \pd{(\ene\Nrtr)}{\ene} \l( \ir\pd{\br}{\vartheta} - \frac{\bt}{r} \r)
        + \pd{(\ene\Nrpr)}{\ene} \l( \irst\pd{\br}{\varphi} - \frac{\bp}{r} \r)
\r.\nn\\
&&{}    + \pd{(\ene\Ntrr)}{\ene} \l( \pd{\bt}{r} \r)
        + \pd{(\ene\Nttr)}{\ene} \l( \ir\pd{\bt}{\vartheta} + \frac{\br}{r} \r)
        + \pd{(\ene\Ntpr)}{\ene} \l( \irst\pd{\bt}{\varphi} 
                                                 - \frac{\bp\ct}{r\st} \r)
\nn\\
&&{} \l.  + \pd{(\ene\Nprr)}{\ene} \l( \pd{\bp}{r} \r)
        + \pd{(\ene\Nptr)}{\ene} \l( \ir\pd{\bp}{\vartheta} \r)
        + \pd{(\ene\Nppr)}{\ene} \l( \irst\pd{\bp}{\varphi} + \frac{\br}{r}
                                                 + \frac{\bt\ct}{r\st} \r)
\r\} 
= \Cr \, ,
\label{Hr_spher}\ea
\ba
&& \frac{\rho}{c}\frac{D}{Dt}\l(\frac{\Ht}{\rho}\r)
     - \frac{\Hr\bt}{r} + \frac{\Hp\bp\ct}{\rst}
     + \frac{3}{r}\Prt + \pd{\Prt}{r} +
     \frac{\ct}{\rst}\Ptt + \ir\pd{\Ptt}{\vartheta} +
     \irst\pd{\Ppt}{\varphi} - \frac{\Ppp\ct}{\rst} \nn\\
&&{} + \Hr \l( \pd{\bt}{r} \r) +
     \Ht \l( \ir\pd{\bt}{\vartheta} + \frac{\br}{r} \r) +
     \Hp \l( \irst\pd{\bt}{\varphi} - \frac{\bp\ct}{r\st} \r) \nn\\
&&{} + \oc\pd{}{t}\l(\br\Prt+\bt\Ptt+\bp\Ppt\r)
     + \frac{J}{c}\pd{\bt}{t}
     - \oc\pd{\br}{t}\pd{(\ene\Prt)}{\ene} -
     \oc\pd{\bt}{t}\pd{(\ene\Ptt)}{\ene} -
     \oc\pd{\bp}{t}\pd{(\ene\Ppt)}{\ene} \nn\\
&&{} -
     \l\{ \pd{(\ene\Nrrt)}{\ene} \l( \pd{\br}{r} \r)
        + \pd{(\ene\Nrtt)}{\ene} \l( \ir\pd{\br}{\vartheta} - \frac{\bt}{r} \r)
        + \pd{(\ene\Nrpt)}{\ene} \l( \irst\pd{\br}{\varphi} - \frac{\bp}{r} \r)
\r.\nn\\
&&{}      + \pd{(\ene\Ntrt)}{\ene} \l( \pd{\bt}{r} \r)
        + \pd{(\ene\Nttt)}{\ene} \l( \ir\pd{\bt}{\vartheta} + \frac{\br}{r} \r)
        + \pd{(\ene\Ntpt)}{\ene} \l( \irst\pd{\bt}{\varphi} 
                                                 - \frac{\bp\ct}{r\st} \r)
\nn\\
&&{} \l.  + \pd{(\ene\Nprt)}{\ene} \l( \pd{\bp}{r} \r)
        + \pd{(\ene\Nptt)}{\ene} \l( \ir\pd{\bp}{\vartheta} \r)
        + \pd{(\ene\Nppt)}{\ene} \l( \irst\pd{\bp}{\varphi} + \frac{\br}{r}
                                                 + \frac{\bt\ct}{r\st} \r)
\r\} 
 = \Ct \, ,
\label{Ht_spher}\ea
\ba
&& \frac{\rho}{c}\frac{D}{Dt}\l(\frac{\Hp}{\rho}\r)
     - \frac{\Hr\bp}{r} - \frac{\Ht\bp\ct}{\rst}
     + \frac{3}{r}\Prp + \pd{\Prp}{r} +
     \frac{2\ct}{\rst}\Ptp + \ir\pd{\Ptp}{\vartheta} +
     \irst\pd{\Ppp}{\varphi} \nn\\
&&{} + \Hr \l( \pd{\bp}{r} \r) +
     \Ht \l( \ir\pd{\bp}{\vartheta} \r) +
     \Hp \l( \irst\pd{\bp}{\varphi} + \frac{\br}{r}
                                        + \frac{\bt\ct}{r\st} \r) \nn\\
&&{} + \oc\pd{}{t}\l(\br\Prp+\bt\Ptp+\bp\Ppp\r)
     + \frac{J}{c}\pd{\bp}{t}
     - \oc\pd{\br}{t}\pd{(\ene\Prp)}{\ene} -
     \oc\pd{\bt}{t}\pd{(\ene\Ptp)}{\ene} -
     \oc\pd{\bp}{t}\pd{(\ene\Ppp)}{\ene} \nn\\
&&{} -
     \l\{ \pd{(\ene\Nrrp)}{\ene} \l( \pd{\br}{r} \r)
        + \pd{(\ene\Nrtp)}{\ene} \l( \ir\pd{\br}{\vartheta} - \frac{\bt}{r} \r)
        + \pd{(\ene\Nrpp)}{\ene} \l( \irst\pd{\br}{\varphi} - \frac{\bp}{r} \r)
\r.\nn\\
&&{}      + \pd{(\ene\Ntrp)}{\ene} \l( \pd{\bt}{r} \r)
        + \pd{(\ene\Nttp)}{\ene} \l( \ir\pd{\bt}{\vartheta} + \frac{\br}{r} \r)
        + \pd{(\ene\Ntpp)}{\ene} \l( \irst\pd{\bt}{\varphi} 
                                                 - \frac{\bp\ct}{r\st} \r)
\nn\\
&&{} \l.  + \pd{(\ene\Nprp)}{\ene} \l( \pd{\bp}{r} \r)
        + \pd{(\ene\Nptp)}{\ene} \l( \ir\pd{\bp}{\vartheta} \r)
        + \pd{(\ene\Nppp)}{\ene} \l( \irst\pd{\bp}{\varphi} + \frac{\br}{r}
                                                 + \frac{\bt\ct}{r\st} \r)
\r\} 
 = \Cp \, .
\label{Hp_spher}\ea
Here $\br (t,r,\vartheta,\varphi) = v_r/c$, $\bt
(t,r,\vartheta,\varphi) = v_\vartheta/c$, and $\bp
(t,r,\vartheta,\varphi) = v_\varphi/c$. Note that using the continuity
equation yields
\ba 
\frac{\rho}{c}\frac{D}{Dt}\l(\frac{X}{\rho}\r) &=&
 \l(\oc\pd{}{t} + {\boldsymbol{\beta} \cdot {\bf\nabla}}\r)X + X{{\bf \nabla} \cdot
 \boldsymbol{\beta}} \nn\\
 &=& \l(\oc\pd{}{t} + \br\pd{}{r} + \frac{\bt}{r}\pd{}{\vartheta} +
    \frac{\bp}{r}\pd{}{\varphi}\r) X 
    + X \l(\frac{1}{r^2}\pd{(r^2\br)}{r}+\irst\pd{(\st\bt)}{\vartheta}
     +\irst\pd{\bp}{\varphi}\r)\, .
\label{continuityeq}\ea

The different moments defined in Eqs.~(\ref{appeq:moments}) are linked
by the following relations, which are generally valid:
\ba \Prr +\Ptt +\Ppp  &=& J \, ,\nn\\
    \Nrrr+\Nrtt+\Nrpp &=& \Hr \, ,\nn\\
    \Nrrt+\Nttt+\Ntpp &=& \Ht \, ,\nn\\
    \Nrrp+\Nttp+\Nppp &=& \Hp \, .\label{appeq:tensorrelations} \ea
Also note that ${\bf P}$ and ${\bf N}$ are completely symmetric
tensors, i.e.~the indices can be permutated arbitrarily. The number of
independent variables therefore is $\#(J,H,P,N)=(1,3,5,7)$ after
applying Eqs.~\ref{appeq:tensorrelations}.

In two dimensions the symmetry relation $I(-\omega,\mu)=I(\omega,\mu)$
holds which leads to
\be \Hp = \Prp = \Ptp = \Nrrp = \Nrtp = \Nttp = \Nppp \equ 0
\label{appeq:2dequivs}\ee
and therefore $\#(J,H,P,N)=(1,2,3,4)$; note that the last equation in
(\ref{appeq:tensorrelations}) becomes redundant. In one dimension one has
$I(\omega,\mu)=I(\mu)$ so that in addition
\be \Ht = \Prt = \Nrrt = \Nttt = \Ntpp \equ 0. \label{appeq:1dequivs} \ee
This time, the third equation in (\ref{appeq:tensorrelations}) becomes
redundant. Furthermore, one obtains two new equations from comparing
the relations in Eqs.~(\ref{appeq:moments}):
\be \Ptt=\Ppp, ~~~~ \Nrtt=\Nrpp, \label{appeq:1drels}\ee
which results in $\#(J,H,P,N)=(1,1,1,1)$. Now one can introduce the
definitions $H \equ \Hr, K \equ \Prr, L \equ\Nrrr$.

In our approximation of two-dimensional transport, i.e.~assuming
$I(\omega,\mu)=I(\mu)$,
Eqs.~(\ref{appeq:tensorrelations}--\ref{appeq:1drels}) apply, but in
contrast to the one-dimensional case not all terms with lateral
gradients or components vanish. The moments equations for the
remaining independent variables, $J$ and $H\equiv \Hr$, simplify from
Eqs.~(\ref{J_spher}) and (\ref{Hr_spher}) to
\ba
&& \l(\oc\pd{}{t} + \br\pd{}{r}
    \boldsymbol{ + \frac{\bt}{r}\pd{}{\vartheta}}\r) J
    + J \l(\frac{1}{r^2}\pd{(r^2\br)}{r}
    \boldsymbol{ +\irst\pd{(\st\bt)}{\vartheta}}
     \r)
     + \frac{1}{r^2} \pd{(r^2 H)}{r}
     + \frac{\br}{c}\pd{H}{t} \nn\\
&&{} - \pd{}{\ene}\l\{\ene\l[
	\oc\pd{\br}{t}H
	+ K \l( \pd{\br}{r} - \frac{\br}{r} 
            \boldsymbol{ -\frac{1}{2r\sin\vartheta}\pd{(\st\bt)}{\vartheta}}\r)
	+ J \l( \frac{\br}{r}
            \boldsymbol{ +\frac{1}{2r\sin\vartheta}\pd{(\st\bt)}{\vartheta}}\r)
\r]\r\}
\nn\\ &&{}
	+ K \l( \pd{\br}{r} - \frac{\br}{r} 
            \boldsymbol{ -\frac{1}{2r\sin\vartheta}\pd{(\st\bt)}{\vartheta}}\r)
	+ J \l( \frac{\br}{r}
            \boldsymbol{ +\frac{1}{2r\sin\vartheta}\pd{(\st\bt)}{\vartheta}}\r)
	+ \frac{2}{c} \pd{\br}{t}H
      = \Cj \label{eq:momeqe1}\,, 
\\ \nn \\
&& \l(\oc\pd{}{t} + \br\pd{}{r}
    \boldsymbol{ + \frac{\bt}{r}\pd{}{\vartheta}}\r) H
    + H \l(\frac{1}{r^2}\pd{(r^2\br)}{r}
    \boldsymbol{ +\irst\pd{(\st\bt)}{\vartheta}}
     \r)
     + \pd{K}{r} +
       \frac{3K-J}{r}
     + H \l( \pd{\br}{r} \r)
     + \frac{\br}{c}\pd{K}{t}
\nn\\ &&{}
     - \pd{}{\ene}\l\{\ene\l[
	\oc\pd{\br}{t}K
	+ L \l(\pd{\br}{r} - \frac{\br}{r}
             \boldsymbol{ -\frac{1}{2r\sin\vartheta}\pd{(\st\bt)}{\vartheta}}\r)
	+ H \l( \frac{\br}{r}
             \boldsymbol{ +\frac{1}{2r\sin\vartheta}\pd{(\st\bt)}{\vartheta}}\r)
         \r]\r\}
     + \oc\pd{\br}{t}(J+K)
 = \Cr
\, .
\label{eq:momeqe2}\ea
With ${\cal J}=J/\ene$, ${\cal H}=H/\ene$, ${\cal K}=K/\ene$, and
${\cal L}=L/\ene$, the moments equations describing the evolution of
neutrino number read
\ba
&& \l(\oc\pd{}{t} + \br\pd{}{r}
    \boldsymbol{ + \frac{\bt}{r}\pd{}{\vartheta}}\r) {\cal J}
    + {\cal J} \l(\frac{1}{r^2}\pd{(r^2\br)}{r}
    \boldsymbol{ +\irst\pd{(\st\bt)}{\vartheta}}
     \r)
     + \frac{1}{r^2} \pd{(r^2 {\cal H})}{r}
     + \frac{\br}{c}\pd{{\cal H}}{t}
\nn\\ &&{}
     - \pd{}{\ene}\l\{\ene\l[
	\oc\pd{\br}{t}{\cal H}
	+ {\cal K} \l( \pd{\br}{r} - \frac{\br}{r} 
            \boldsymbol{ -\frac{1}{2r\sin\vartheta}\pd{(\st\bt)}{\vartheta}}\r)
	+ {\cal J} \l( \frac{\br}{r}
            \boldsymbol{ +\frac{1}{2r\sin\vartheta}\pd{(\st\bt)}{\vartheta}}\r)
\r]\r\}
	+ \frac{1}{c} \pd{\br}{t}{\cal H}
      = {\cal C}^{(0)}\,,
\label{eq:momeqn1} \\ \nn \\
&& \l(\oc\pd{}{t} + \br\pd{}{r}
    \boldsymbol{ + \frac{\bt}{r}\pd{}{\vartheta}}\r) {\cal H}
    + {\cal H} \l(\frac{1}{r^2}\pd{(r^2\br)}{r}
    \boldsymbol{ +\irst\pd{(\st\bt)}{\vartheta}}
     \r)
     + \pd{{\cal K}}{r} +
       \frac{3{\cal K}-{\cal J}}{r}
     + {\cal H} \l( \pd{\br}{r} \r)
     + \frac{\br}{c}\pd{{\cal K}}{t}
\nn\\ &&{}
     - \pd{}{\ene}\l\{\ene\l[
	\oc\pd{\br}{t}{\cal K}
	+ {\cal L} \l(\pd{\br}{r} - \frac{\br}{r}
             \boldsymbol{ -\frac{1}{2r\sin\vartheta}\pd{(\st\bt)}{\vartheta}}\r)
	+ {\cal H} \l( \frac{\br}{r}
             \boldsymbol{ +\frac{1}{2r\sin\vartheta}\pd{(\st\bt)}{\vartheta}}\r)
         \r]\r\}
\nn\\ &&{}
	- {\cal L} \l( \pd{\br}{r} - \frac{\br}{r} 
            \boldsymbol{ -\frac{1}{2r\sin\vartheta}\pd{(\st\bt)}{\vartheta}}\r)
	- {\cal H} \l( \frac{\br}{r}
            \boldsymbol{ +\frac{1}{2r\sin\vartheta}\pd{(\st\bt)}{\vartheta}}\r)
     + \oc\pd{\br}{t}{\cal J}
= {\cal C}^{(1)}_r\,.
\label{eq:momeqn2}\ea
This system of moments equations (\ref{eq:momeqe1}--\ref{eq:momeqn2})
is very similar to the Newtonian, $\mathcal{O}(v/c)$ moments equations
in spherical symmetry \cite[see][Eqs.~7,8,30,31]{ramjan02}. We have
set the additional terms arising from our approximative generalization
to two dimensions in boldface.

\section{Specific energy without offset}
\label{app:noff}

Here we present our new procedure how the Eulerian equations of
hydrodynamics, Eqs.~(\ref{eq:hydro.rho}--\ref{eq:hydro.x_k}), together
with a general EoS that handles NSE as well as 
non-NSE conditions, can be numerically treated without producing 
spurious local fluctuations in dependent thermodynamic variables like
the entropy and pressure, a kind of numerical noise which was
described in Sect.~\ref{sec:swiggles}.

As discussed in that section, the wiggles observed in radial profiles
are linked to situations where the value of 
$\widetilde{\Gamma}_e = p/(e\rho) + 1$ drops below 4/3. The Riemann 
solver in our PROMETHEUS code, however, expects 
$\widetilde{\Gamma}_e$ to be thermodynamically
consistent with the adiabatic index $\Gamma_{\mathrm{ad}} = 
(\mathrm{d}\ln p/\mathrm{d} \ln \rho)_s$, in particular 
$\widetilde{\Gamma}_e$ should fulfill $\widetilde{\Gamma}_e \ge 4/3$, 
which implies that $p$ and $e$ should 
go to zero at the same time. This, however, cannot be guaranteed
with an arbitrary normalization of the energy $e$. On 
the other hand, there is freedom of choosing the normalization
of the specific energy in the EoS, and also the solution of the 
time-evolution equations of a system depends only on differences
of the specific energy and not on the adopted
zero point of this energy. In order to employ an EoS which makes
use of this natural freedom of the energy normalization 
in our hydrodynamics code, we therefore introduce a shift of the
reference level of the energy when the latter is transferred between
hydrodynamics code and EoS.
 
We find that the wiggles observed in the radial profiles of
dependent thermodynamic variables  
disappear when the total specific energy used in the hydrodynamics
solver PROMETHEUS is redefined compared to Sect.~\ref{sec:swiggles} as
\be
\varepsilon = e_\mathrm{int} + e_\mathrm{kin} \, ,
\label{eq:app:ehyd}\ee
where $e_\mathrm{int}$ is the internal specific energy and
$e_\mathrm{kin}=\frac{1}{2} (v_r^2+v_\vartheta^2+v_\varphi^2)$ is the
kinetic energy per unit of mass. Then the EoS index $\Gamma_e
\equiv p/[(\varepsilon - e_\mathrm{kin})\rho] + 1 =
p/[e_\mathrm{int}\rho] + 1$ adopts values in the physically meaningful
range ($\Gamma_e\geq 4/3$), which seems necessary for the
Riemann solver used in PROMETHEUS to produce smooth hydrodynamic
solutions. However, our high-density EoS, provided by \cite{latswe91},
employs an energy definition which includes the nucleon rest masses
and the rest masses of the electrons, the so-called ``relativistic''
specific energy, plus a constant energy offset:
\be
e_\mathrm{rel,0} = e_\mathrm{int} + e_\mathrm{rm} + e_0 \, ,
\label{eq:app:erel} \ee
where the total specific rest mass energy is defined as
\be
e_\mathrm{rm} = \sum_{k=1}^{n_Y} Y_k\cdot \frac{m_k}{m_\mathrm{by}} \cdot c^2 
  + Y_\mathrm{e} \cdot \frac{m_\mathrm{e}}{m_\mathrm{by}} \cdot c^2 \, .
\label{eq:app:def_e_rm}
\ee
The specific energy offset $e_0$ is an arbitrary constant (set equal
to $\l.-930.7731\mathrm{MeV} \times 1.602 \times 10^{-6}
\frac{\mathrm{erg}}{\mathrm{MeV}} \r/ m_\mathrm{by}$, see
\citealp{latswe91} for details). This energy definition was used in
the equation of energy solved in our hydrodynamics code hitherto,
i.e.~$\varepsilon = e_{\mathrm{rel},0} + e_\mathrm{kin}$. For reasons
explained below, we still wish to retain the energy definition,
Eq.~(\ref{eq:app:erel}), when evaluating the EoS and at the same time
use the new definition Eq.~(\ref{eq:app:ehyd}) when solving the
hydrodynamics. This is achieved by the following scheme.

The Euler equation for the specific energy as defined in
Eq.~(\ref{eq:app:ehyd}) can be derived from the Euler equation for the
specific energy as defined in our code hitherto,
Eq.~(\ref{eq:hydro.e}), by subtracting the normalized sum over the
Eqs.~(\ref{eq:hydro.x_k}) and further subtracting
Eq.~(\ref{eq:hydro.ye}), i.e.~``$(\mathrm{Eq.}~\ref{eq:hydro.e}) - \sum_k
(\mathrm{Eq.}~\ref{eq:hydro.x_k})_k m_k c^2 / m_\mathrm{by} -
(\mathrm{Eq.}~\ref{eq:hydro.ye}) m_e c^2 / m_\mathrm{by}$'':
\be
\frac{\partial}{\partial t} \left(\rho \varepsilon\right) +
\frac{1}{r^2}\frac{\partial}{\partial r}\left(r^2\left(\rho \varepsilon + p\right)\,v_r\right) +
\frac{1}{r\st}\frac{\partial}{\partial \vartheta}\Big(\left(\rho \varepsilon + p\right) \st \, v_\vartheta\Big)
= 
-\rho\left(v_r\frac{\partial\Phi}{\partial r}
      + \frac{v_\vartheta}{r} \frac{\partial\Phi}{\partial \vartheta}
    \right)
+Q_\mathrm{E} + Q_\mathrm{nuc}
\,,
\label{eq:app.hydro.e2}
\ee
where the nuclear source term is
\be
Q_\mathrm{nuc} = - \sum_{k=1}^{n_Y} \frac{m_k}{m_\mathrm{by}} c^2 R_k
              - \frac{m_\mathrm{e}}{m_\mathrm{by}} c^2 Q_\mathrm{N} \, .
\label{eq:app.nucsour}\ee
The abundance of each nuclear species $k$ evolves according to
Eq.~(\ref{eq:hydro.x_k}):
\begin{equation}\label{eq:app.hydro.x_k}
\frac{\partial}{\partial t} \left(\rho Y_k\right) +
\frac{1}{r^2}\frac{\partial}{\partial r}\left(r^2\rho Y_k\,v_r\right) +
\frac{1}{r\st}\frac{\partial}{\partial \vartheta}\left(\st \,
\rho Y_k\,v_\vartheta\right) = R_k
\, 
\end{equation}
and the electron fraction according to Eq.~(\ref{eq:hydro.ye}):
\begin{equation}\label{eq:app.hydro.ye}
\frac{\partial}{\partial t} \left(\rho \ye\right) +
\frac{1}{r^2}\frac{\partial}{\partial r}\left(r^2\rho \ye\,v_r\right) +
\frac{1}{r\st}\frac{\partial}{\partial \vartheta}\left(\st \,
\rho \ye\,v_\vartheta\right) =
Q_\mathrm{N}
\,.
\end{equation}

Starting at time step $n$, we calculate a hydrodynamics time step with
PROMETHEUS using the initial total specific energy at this time,
$\varepsilon^n$, as defined in Eq.~(\ref{eq:app:ehyd}). Remember that
PROMETHEUS only solves the LHS of the hydrodynamics equations (in this
implementation
Eqs.~\ref{eq:hydro.rho}--\ref{eq:hydro.v_p},\ref{eq:app.hydro.e2},\ref{eq:app.hydro.x_k},\ref{eq:app.hydro.ye}),
i.e.~the source terms and gravitational effects on the RHS of the
equations are ignored for the moment. Also note that the EoS is
\emph{not} solved simultaneously with the hydrodynamics so that for
instance composition changes, even in regions with NSE, originate
\emph{solely} from advection, Eqs.~(\ref{eq:app.hydro.x_k})!

The solution of the evolution equations, after additionally applying
the gravitational effects and all source terms except for
$Q_\mathrm{nuc}$, $Q_\mathrm{N}$, and $R_k$ yields $\varepsilon^\ast$
as well as the advected nuclear abundances $Y_k^\ast$ and the electron
fraction $Y_\mathrm{e}^\ast$. The asterisk denotes that the nuclear
reaction effects and the neutrino source term $Q_\mathrm{N}$ of time
step $n+1$ have not yet been taken into account.

Next, we transform the energy to
\be\label{eq:app:trafo1}
e_\mathrm{rel,0}^\ast = \varepsilon^\ast - e_\mathrm{kin}^{n+1}
   + \sum_{k=1}^{n_Y} Y_k^\ast\cdot \frac{m_k}{m_\mathrm{by}} \cdot c^2
   + Y_\mathrm{e}^\ast \cdot \frac{m_\mathrm{e}}{m_\mathrm{by}} \cdot c^2
   + e_0 \, ,
\ee
where $e_\mathrm{kin}^{n+1}$ is the specific kinetic energy after the
hydrodynamics step.

This energy transformation brings us back to the energy definition
originally used in Eq.~(\ref{eq:hydro.e}) so that the remaining source
terms of that equation and in the lepton number and nuclear abundance
equations can now be applied to the hydrodynamics variables. Then all
thermodynamic quantities and the composition can be updated by
applying the EoS and nuclear burning routines
\emph{exactly} in the same way
as we did in our former version of the code.
This results in the new
composition $Y_k^{n+1}$ and electron fraction $Y_\mathrm{e}^{n+1}$.

We now transform the energy back to
\be\label{eq:app:trafo2}
\varepsilon^{n+1} = e_\mathrm{EoS}^{n+1} + e_\mathrm{kin}^{n+1}
   - \sum_{k=1}^{n_Y} Y_k^{n+1}\cdot \frac{m_k}{m_\mathrm{by}} \cdot c^2
   - Y_\mathrm{e}^{n+1} \cdot \frac{m_\mathrm{e}}{m_\mathrm{by}} \cdot c^2
   - e_0 \, ,
\ee
where we use $e_\mathrm{rel,0}^{n+1} = e_\mathrm{rel,0}^\ast$. By this
simple treatment, we have implicitly taken into account the nuclear
source term $Q_\mathrm{nuc}$ since
\be
\varepsilon^{n+1} - \varepsilon^\ast =
   - \sum_{k=1}^{n_Y} (Y_k^{n+1} - Y_k^\ast)
                           \cdot \frac{m_k}{m_\mathrm{by}} \cdot c^2 
   - (Y_\mathrm{e}^{n+1} - Y_\mathrm{e}^\ast)
                           \cdot \frac{m_e}{m_\mathrm{by}} \cdot c^2
 = \frac{1}{\rho} Q_\mathrm{nuc} \Delta t^n \, ,
\ee
where $\Delta t^n$ is the size of the time step $n$.

In other words, because we take into account effects which transform
mass into internal energy and vice versa, i.e.~composition changes due
to nuclear transmutation and transformation of electrons to massless
neutrinos, at a stage where the energy is defined including the rest
masses of the particles which change abundance,
Eq.~(\ref{eq:app:erel}), this energy is not affected by the
transmutations and does not change. The conversion of energy from
$e_\mathrm{int}$ to $e_\mathrm{rm}$ or vice versa is not visible in
their sum, $e_\mathrm{rel} = e_\mathrm{int} + e_\mathrm{rm}$. Although
in NSE it is actually \emph{not} necessary to advect the nuclear
composition because it is unambiguously given for local
thermodynamical conditions $(\rho,T,\ye)$ by the EoS, we follow the
evolution equations (Eq.~\ref{eq:app.hydro.x_k}) for a set of discrete
nuclei which reproduce the baryon number and charge number of the
representative heavy nucleus present at NSE in the employed
high-density EoS (see Appendix B.1 in \citealp{ramjan02}, which scheme
was here extended by adding one more representative nucleus), in order
to be able to subtract the rest-mass contribution to
$e_\mathrm{rel,0}$ (c.f. Eq.~\ref{eq:app:trafo1}).

This procedure features several advantages: First, we do not need to
calculate any nuclear source terms. This includes the NSE regime,
where the nuclear composition and temperature are well defined
functions of $\rho$, $\ye$, and $e_\mathrm{rel,0}$ so that an
iteration between $T$, $e_\mathrm{int}$, and the composition becomes
unnecessary. Also in a regime where nuclear burning applies, the
conservation of $e_\mathrm{rel,0}$ ensures strict energy conservation
in accordance with the philosophy of PROMETHEUS. Second, problems
which may arise from the fact that the nucleon masses are not equal to
their masses in vacuum at high densities because of nucleon-nucleon
interactions are avoided by our implementation. For the energy
transformations, Eqs.~(\ref{eq:app:trafo1},\ref{eq:app:trafo2}), we
can safely use the vacuum rest masses, because we need only ensure
that $e_\mathrm{rel,0}$ is correct. Its correctness is necessary to
derive correct temperatures, composition, and the relativistic
gravitational potential (via the gravitational mass, which includes
terms depending on $e_\mathrm{rel} = e_\mathrm{int} + e_\mathrm{rm}$
in general relativity).  At the high densities in the PNS, however,
the use of $\varepsilon$ (Eq.~\ref{eq:app:ehyd}) without detailed
knowledge of the effective nucleon masses once more leads to the
problem that $\Gamma_e = {{p}\over{(\varepsilon -
e_\mathrm{kin})\rho}} + 1$ may adopt values which are not perfectly
consistent with the EoS index which relates gas pressure and internal
energy density. However, this discrepancy now occurs at conditions
where $\Gamma_e \gg 4/3$ and the relative error is much
smaller than at low densities. Our test runs with the procedure
described here confirm that the hydrodynamics results do not suffer
from spurious oscillations anywhere in the SN core.

\end{document}